\documentclass{report}
\usepackage{citesort}
\usepackage{fullpage,dalthesis,latexsym,amsfonts}
\usepackage{makeidx}
\usepackage[dvips]{graphicx}
\input epsf
\makeindex

\newsymbol \lesssim 132E
\newsymbol \gtrsim 1326
\newcommand{\bb}[1]{\textbf{#1}}

\newcommand{\di}{\partial}
\newcommand{\mR}{\ensuremath{\mathbb{R}}}

\newcommand{\mM}{\ensuremath{\mathbb{M}}}

\newcommand{\mZ}{\ensuremath{\mathbb{Z}}}
\newcommand{\benonumber}{\begin{displaymath}}
\newcommand{\eenonumber}{\end{displaymath}}
\newcommand{\be}{\begin{equation}}
\newcommand{\ee}{\end{equation}}
\newcommand{\eref}[1]{(\ref{#1})}
\newcommand{\grad}{\nabla}

\newcommand{\half}{\frac{1}{2}}
\newcommand{\third}{\frac{1}{3}}
\newcommand{\quart}{\frac{1}{4}}
\newcommand{\sixth}{\frac{1}{6}}

\newcommand{\bvect}[2][\xi]{\mbox{\boldmath${#1}$}_{(#2)}}
\newcommand{\sBox}[1]{\ensuremath{\makebox[0pt][l]{\tiny{\ \,${^{#1}}$}}\Box}}
\newcommand{\bv}[1]{\mbox{\boldmath${#1}$}}
\newcommand{\up}[1]{\!\,^{^{\mbox{\rm\tiny(#1)}}}\!}
\newcommand{\q}{\quad}
\newcommand{\gtwo}{\ensuremath{G_2}}

\newcommand{\calm}{{\cal M}}
\newcommand{\calv}{{\cal V}}
 
\newcounter{saveeqn}%

\newcommand{\beqn}{\setcounter{saveeqn}{\value{equation}}%
	\stepcounter{saveeqn}\setcounter{equation}{0}%
	\renewcommand{\theequation}
	{\mbox{\arabic{chapter}.\arabic{saveeqn}\alph{equation}}}%
	\begin{eqnarray} }%

\newcommand{\eeqn}{\end{eqnarray}\setcounter{equation}{\value{saveeqn}}%
	\renewcommand{\theequation}{\arabic{chapter}.\arabic{equation}}}%

\newcommand{\mainlabel}[1]{\renewcommand{\theequation}%
	{\arabic{chapter}.\arabic{saveeqn}}\label{#1}%
	\renewcommand{\theequation}{\mbox{\arabic{chapter}.\arabic{saveeqn}%
	\alph{equation}}}}%

\newcommand{\sappendix}%
    {%
     \appendix %
	\renewcommand{\beqn}{\setcounter{saveeqn}{\value{equation}}%
		\stepcounter{saveeqn}\setcounter{equation}{0}%
		\renewcommand{\theequation}
		{\mbox{\Alph{chapter}.\arabic{saveeqn}\alph{equation}}}%
		\begin{eqnarray} }%
	\renewcommand{\eeqn}{\end{eqnarray}\setcounter{equation}%
		{\value{saveeqn}}%
		\renewcommand{\theequation}{\Alph{chapter}.\arabic{equation}}}%
	\renewcommand{\mainlabel}[1] {	\renewcommand{\theequation}%
          	{\Alph{chapter}.\arabic{saveeqn} } \label{##1}%
		\renewcommand{\theequation} {%
	   	\mbox{\Alph{chapter}.\arabic{saveeqn}\alpha{equation}} }  }%
    }
\renewcommand{\vec}[1]{\stackrel{\rightharpoonup} {#1}}

\begin{document}
\phd
\dept{Physics}

\title{The Asymptotic Behaviour of Cos\-mo\-logi\-cal Mo\-dels Con\-taining Mat\-ter and Sca\-lar Fields}

\convocation{October}{1999}
\author{Andrew Philip Billyard}
\supervisor{Alan A. Coley}
\firstreader{H. J\"{u}rgen Kreuzer}
\secondreader{D.J. Wallace Geldart}
\examiner{Kayll Lake}

\dedicate{To Jane, \\for providing the bridges \\over the many
turbulent waters.}

\beforepreface
 
\prefacesection{Abstract} The asymptotic behaviour of two classes of
scalar field cosmological models are studied using the theory of dynamical
systems: general relativistic Bianchi models containing matter and a
scalar field with an exponential potential and a class of spatially
homogeneous string cosmological models.  The purpose of this thesis is to
examine some of the outstanding problems which currently exist in
cosmology, particularly regarding isotropization and inflation.  It is
shown that the matter scaling solutions are unstable to curvature
perturbations. It is then shown that the Bianchi class B exponential
potential models can alleviate the isotropy problem; an open set of models
within this class do isotropize to the future.  It is also shown that the
presence of an interaction term in the subclass of isotropic models can
lead to inflationary models with late-time oscillatory behaviour in which
the matter is not driven to zero.  Next, within the class of the string
cosmologies studied, it is shown that there is a subclass 
which do not inflate at late times in the post-big
bang regime.  Furthermore, all string models studied
typically do not have a late--time flatness problem. Indeed, it is shown
that curvature typically plays an important r\^{o}le only at intermediate
times in most models.  It is also shown that the presence of a
positive cosmological constant in the models studied can lead to
interesting physical behaviour, such as multi-bouncing universes.  A
mathematical equivalence between general relativistic scalar field
theories and scalar-tensor theories and string theories has been
extensively exploited and thus the results obtained from the string
analysis compliment the results obtained from the Bianchi class B
exponential potential analysis.

\prefacesection{Acknowledgements}

Looking back upon the last four eternal years, it is incredible to
realize how many people have influenced my life, both academically and
personally.  First, I would like to thank
my supervisor for his keen mathematical expertise and insight, for the
many enlightening conversations and for dragging me with him to his
old stomping grounds in England.  I would also extend my appreciation
to Jesus Ib\'a\~nez, Robert van den Hoogen, James Lidsey and Itsaso
Olasagasti for the interesting collaborative projects.  Further thanks
to Jesus for his invitation to Spain and his gracious hospitality
during my visit there.  Similar thanks to Bernard Carr for welcoming
me into his home during my stay in England.

Of course, my appreciation extends to Tracy Kerr, Carolyn Smyth,
Hossein Abolghasem, Patricia Benoit, James Flynn, Trevor Droesbeck and
Andrew Whinnett for the mutual sharing of both the simple pleasures
and the angst that come with being a graduate student.

My eternal gratitude naturally extends to Jane Groves for her patience
and understanding, especially near the end of my degree when all she
saw of me was the back of my head as I was glued to the computer;
thanks to her, I'm pleasantly reminded that physics isn't the only
important thing in life.  

I would like to thank my sister Eileen Wilson for the periodic
checkups on my well being and listening to all my rants.  I also thank
my old friends Daniel Oleskevich, Jordan Smith, Maninder Kalsi, Bill
Sajko, Tracy Kerr, Sandra De Souza, Greg Dick and Bruce Lloyd for
their steadfast friendship which has never faded over time or
distance.  Furthermore, my appreciation extends to Gordon Oakey and
Maggie Bell who gave me a place to stay when I first arrived and, more
importantly, gave me a great life outside of the Dunn building.  This
appreciation also extends to Kevin Wood and Alice Giddy for all their
ceaseless kindness and great hospitality extended to me througout my time here.

Finally, I would like to thank the staff in the Physics department,
namely Bridget Trim, Judy Hollett, Ruth Allen, Barbara Gauvin, and Jim
Chisholm,  whose value to the department can never be overstated.

\prefacesection{List of Symbols}

$a=e^\alpha$ \dotfill scale factor\\
$\{\tilde A\equiv\frac{\tilde a}{9H^2}, ~N_+\equiv\frac{n_+}{3H}\}$ \dotfill normalized curvature terms\\
$c$ \dotfill speed of light\\
$g_{\alpha \beta}$ \dotfill metric \\
$G$ \dotfill gravitational (Newton's) constant\\
$G_{\alpha \beta}$ \dotfill Einstein tensor\\
$H\equiv \dot\alpha=\third\theta$ \dotfill Hubble parameter\\
$k$ \dotfill exponential potential paramter $V=V_0 e^{k\phi}$\\
$\{\tilde K, K\}$ \dotfill curvature terms\\
$l=h^{-1}$ \dotfill Bianchi type VI$_h$ and VII$_h$ parameter\\
$p$ \dotfill pressure\\
$q$ \dotfill deceleration parameter\\
$Q$ \dotfill Ramond--Ramond parameter\\
$R$ \dotfill Ricci scalar\\
$R_{\alpha \beta}$ \dotfill Ricci tensor\\
$\{s, t, T, \tau, \eta\}$ \dotfill time\\
$T_{\alpha \beta}$ \dotfill energy-momentum tensor\\
$u_\alpha$ \dotfill fluid velocity\\
$U$ \dotfill scalar field potential in Jordan frame\\
$V$ \dotfill scalar field potential in Einstein frame\\
$\{u, U_i, v, V_i, x, X_i, y, Y_i, z, Z_i\}$ \dotfill normalized phase space variables\\

\noindent$\beta_m$ \dotfill modulus field\\
$\beta_s$ \dotfill shear\\
$\beta$ \dotfill $\dot\beta_s^2+6\dot\beta_m^2$\\
$\delta$ \dotfill interaction term \\
$\Delta$ \dotfill normalized shear and curvature term\\
$\gamma$ \dotfill equation of state parameter\\
$\Gamma^\alpha_{\beta\delta}$ \dotfill affine connection\\
$\{\zeta_i, \kappa, \lambda_i, \mu_i, \nu_i, \xi_i, \chi_i\}$ \dotfill phase space variables\\
$\Lambda$ \dotfill central charge deficit\\
$\Lambda_{\rm M}$ \dotfill cosmological constant\\
$\mu\equiv 3H\Omega$ \dotfill energy density\\
$\sigma$ \dotfill axion field\\
$\rho$ \dotfill axion field energy density \\
$\{\Sigma_+\equiv \frac{\sigma_+}{3H}, ~\tilde\Sigma\frac{\tilde \sigma}{9H^2}\}$ \dotfill normalized shear terms\\
$\phi$ \dotfill scalar field in Einstein frame \\
$\varphi\equiv \hat\Phi-3\alpha$ \dotfill shifted dilaton field\\
$\Psi\equiv \frac{\dot\phi}{\sqrt6 H}$ \dotfill normalized scalar field variable \\
$\Phi \equiv e^{-\hat\Phi}$ \dotfill scalar field in Jordan frame\\
$\Upsilon\equiv \frac{\sqrt V}{\sqrt3 H}$ \dotfill normalized potential variable

\afterpreface
 
\chapter{Introduction}

\section{Current Issues in Cosmology}
With the birth of the theory of relativity, ideas in cosmology
went through a great reformation.  The advent of
special relativity removed the notion of an absolute ``inertial''
spatial reference frame
in which the sequence of events were measured by a concept of time
identical for all observers.  Furthermore, when general relativity was
formulated the physical arena was now considered to be a
four-dimensional spacetime in which measurements of an event not only
depended on the relative motions of the observers, but also on the
gravitational field present, and gravity was now considered in
geometrical terms; objects moving under the influence of gravity were
considered to be following specific four-dimensional paths on a
manifold whose curvature is induced by the presence of matter.
Because special relativity brought to light the equivalence between
mass and energy, this matter source could include a bath of radiation
(in which the constituent components are massless).  The universe,
previously static with neither beginning nor end, was now modeled by a
dynamic theory formulated within the mathematical framework of general
relativity, namely pseudo-Rie\-man\-nian 
geometry\footnote{Strictly speaking, Rie\-man\-nian
geometry involves manifolds endowed with a positive definite metric
whereas the metric of a pseudo-Riemannian $N$-dimensional manifold has
a signature of $N-2$}.  However, the cosmological reformation that had
transpired at the birth of general relativity did not cease
thereafter, and the issues of cosmology lead to an active field of
research to this day, and hereby the issues of great importance are
discussed.

It is known that on the largest observable scales, the universe
appears to be spatially isotropic and spatially homogeneous.
Observational support for spatial isotropy comes from the Cosmic
Background Explorer (COBE) data, where measured anisotropies of the
microwave background radiation are only 1 part in $10^5$
\cite{Bennet1989a,Smoot1992a,Smoot1992b,Wright1992a,Bennet1992a,Hancock1994a}.
Support for spatial homogeneity come from galaxy distributions
\cite[p. 39]{Peebles1993a} and partially from nuclear abundances 
\cite{Fowler1960a,Seeger1970a,Sigl1995a,Copi1997a}.  However,
homogeneity is mostly a theoretical assumption based on the Copernican
principle \cite[p. 407]{Weinberg1972a} that we're not in a preferred
location in the universe; if the universe appears homogeneous from our
point of view, then it is assumed to be homogeneous from any point of
view.  If a space is isotropic about all points in a space-like
hypersurface then it {\em must} be spatially homogeneous
\cite[p. 109]{Rindler1977a}.  Partial support for homogeneity can also come from the COBE data.  Cosmological models with such properties are
described by the Friedmann--Robertson--Walker (FRW) models.  Once
these models are assumed to describe the universe, it is possible to measure 
the values of the Hubble parameter, $H_0$, (a measure of the universe's
expansion rate; here the subscript ``0'' refers to such values at the
current epoch) and the \index{deceleration parameter ($q$)} deceleration parameter, $q_0$ (a unitless
quantity which measures the expansion's acceleration; $q<0$ implies an
accelerated expansion whilst $q>0$ describes a decelerated expansion).
Such measurements lead to information about the universe, such as its
curvature or the value of the cosmological constant (if one is
present).  The FRW models have a lot of predictive power.  Not only
do they lead to the Hubble law at low redshift (which states that the ratio of
redshift to distance is equal to $H$, see equation (\ref{HubblesLaw}) on page
\pageref{HubblesLaw}), but they predict a hot Big-Bang which is
compatible with the observed microwave background radiation and observed
nucleosynthesis yields, and which predicts a possible neutrino
background \cite[Ch. 15.6]{Weinberg1972a}.  However, it would seem
that there are presently several potential problems that arise when
the FRW models are used.

\index{spatial isotropy!isotropy issue}
The primary issue associated with this thesis is that the FRW models
do not {\em explain} the observed isotropy of the universe, but rather
this property is assumed.  The isotropy condition may be relaxed and
one of the spatially homogeneous (but spatially anisotropic)
Bianchi models or Kantowski-Sacks model can be used.  In doing so, it
has been shown that these models do not always isotropize.  In
particular, models of Bianchi type IX recollapse \cite{Coley1997a} and
do not necessarily isotropize \cite{vandenHoogen1999a}.  Furthermore,
Collins and Hawking \cite{Collins1973a} have shown that there is a
zero measure of spatially homogeneous models containing ordinary
matter which can isotropize as $t\rightarrow\infty$; this implies that
the conditions of the universe currently observed are not generic
within this class of models.  This is referred to as the {\em isotropy
issue}, and will be one of the main foci of this thesis.  Recently it
has been shown \cite{Wainwright1998a} that for a particular Bianchi
VII$_0$ model the fluid's shear terms became negligible at late times,
but anisotropies in the Weyl curvature tensor became dominant and so
the model does not isotropize.  However, such models of this type are not 
considered within this thesis, and isotropy effectively occurs when the fluid's
shear vanishes.

\index{spatial homogeneity!cosmological issue}
The issue of spatial homogeneity also implies a few potential
problems.  First of all, small perturbations from spatial homogeneity
would be needed to seed the formation of galaxies and stars.  However,
FRW models without a cosmological constant predict that the largest
possible age of the universe is $H_0^{-1}$; using FRW models, data
suggests that the universe began less than $10^{10}/h_0$ years ago
(since data suggests values of $H_0\approx 50-100 km/s/Mpc$ the Hubble
parameter is often written $H_0=h_0\times 100 km/s/Mpc$ with $h_0 \in
[\half,1]$), yet some of the oldest globular clusters are in the age
range $10\times10^9$ to $13\times10^9$ years
\cite{Chaboyer1998a,Chaboyer1997a} (in \cite{Chaboyer1998a} it was
stated with 95\% confidence that the minimum age of the universe is
$9.5\times10^9$ years based on the ages of globular clusters) and so
there may be a discrepancy between observation and theory.
Recently, a value of $h_0=0.7\pm0.07$ (corresponding to an age of
$13\times10^9$ to $15\times10^9$ years in the FRW models) has been
suggested \cite{Richtler1999a,Astronomy1999a}, although this value is
under much debate \cite{Astronomy1999a}.  This potential discrepancy can be
easily resolved, even within FRW models, by including a scalar field
with a potential (or even just a cosmological constant); with such a
field present the models may go through a period of accelerated
expansion and hence $H_0^{-1}$ cannot be used to determine the upper
age of the model.  This type of behaviour is typical of the models
considered within this thesis.

\index{horizon problem}
Another problem with standard FRW models is
associated with particle horizons, which is the distance travelled by
light since the time of the Big Bang.  Since the ages of standard FRW
model universes are finite, there are regions of space which are
outside of each other's particle horizons and hence have never been in
causal contact, and consequently there is no reason why they should
have evolved in the same manner.  However, it is very easy
to observe two regions of space which are causally disconnected from each other
and yet appear the same; both have formed galaxies and clusters
of galaxies and super-clusters etc., the distributions of which appear
similar, and the measured anisotropies in the microwave background
radiation is the same.  Hence, it would seem natural to assume that
sometime in the past these regions had been in causal contact.  This
is known as the {\em horizon problem}.

\index{flatness problem}
Another issue is the {\em flatness problem} which involves the
\index{critical density parameter ($\Omega$)}critical density parameter, $\Omega$, which is proportional to the
ratio of energy density of the universe to it's expansion rate; for
$\Omega>1$ the energy content of the universe eventually causes a
recollapse, for $\Omega<1$ there is insufficient matter to halt the
expansion and for critical value $\Omega=1$ the universe stops
expanding as $t\rightarrow\infty$.  In FRW models, $\Omega$ is
linearly proportional to the universe's curvature $k$ (see section
\ref{formal}), and in particular $k=0$ corresponds to $\Omega=1$.  The
flatness problem is so called because the FRW models do not explain
why current data suggests $0.1<\Omega<10$; that is, near
$\Omega\approx 1$ ($k=0$).  Unless $k=0$ exactly, it would be expected for
the current epoch that $\Omega\approx 0$ for open ($k<0$) models or
$\Omega \gg 1$ for closed ($k>0$) models.

\index{inflation paradigm|(}
Note that these problems (isotropy, particle horizon and flatness) are
related to one another \cite{Ellis1971a}.  Many of these problems may
be alleviated within the inflationary scenario, a paradigm in which
the universe undergoes an accelerated expansion at an early epoch
($q<0$).  In such a scenario, a universe could very well be
inhomogeneous and anisotropic, but the current {\em observable}
portion of that universe was originally a small causally connected
region of the entire universe, which has inflated
\cite{Guth1981a,Linde1987a}.  Hence, inhomogeneities could have been
pushed outside our observable universe whilst allowing present,
causally disconnected parts of the universe to have been connected in
the past, solving the horizon problem.  This mechanism could also
explain the flatness problem, since it is possible for inflation to
drive the universe towards $\Omega=1$, although not all inflationary
models will do this (indeed, there are inflation models where this is
not the case \cite{Lidsey1991a}).  Furthermore, the age discrepancy
between theory and observation could be alleviated; since $H_0^{-1}$
will no longer necessarily be an upper limit to the model's age.
However, there is presently no direct observational evidence for
inflation (although this should not rule inflation out as a
possibility) and the exact mechanism for inflation is not known.

The first inflationary theory was proposed by Guth \cite{Guth1981a},
in which the early universe was in a ``false vacuum'' meta-stable
state.  The stress energy of this false vacuum was extremely large
(compared to regular matter's energy density), obeying the equation of
state $\mu =-p$.  This is equivalent to assuming that the components
of are negligible compared to a cosmological constant, $\Lambda$; 
therefore the cosmological constant (in this particular instance) is
the source for the false vacuum with $\mu_\Lambda $ = - $p_\Lambda =
\Lambda$.  Using these conditions for $k=0$ (FRW models always
asymptote towards $k=0$ models at early times), the scale factor of 
the universe is exponential;
\be 
a(t) = a_0 \exp\left(\sqrt{\third \Lambda}~ t \right).
\ee 
This is known as the de Sitter solution
and often arises when considering inflation.  In particular, Wald
\cite{Wald1983a} proved that all initially expanding, spatially
homogeneous models (except a subclass of Bianchi IX models
which recollapse \cite{Coley1997a}) with a cosmological constant
asymptotically approach this isotropic de Sitter solution.  This is
known as the cosmic no-hair theorem.

Even though Guth's model could solve the horizon and flatness problem,
it had shortcomings such as producing inhomogeneities which are too
large \cite{Ciufolini1995a} and having no graceful exit (the universe
 exponentially expands forever), and since then there have been
different models for inflation, such as extended inflation (inflation
in the Brans--Dicke theory), hyperextended inflation (inflation in
scalar--tensor theory, an extension to Brans--Dicke theory where the
theory's parameter $\omega$ is not constant - see chapter
\ref{SFtoST}) and chaotic inflation (whereby the universe contains a
scalar field with either a quadratic or quartic potential)
\cite{Linde1983a}, to name a few.

The mechanism for driving inflation is usually a self-interacting
scalar field, the potential of which acts as an effective cosmological
constant \cite{Lidsey1992a}.  A standard scenario is as follows
\cite{Linde1987a,Peebles1993a,Oliveira1998a}.  It is conjectured that
in the early universe shortly after the Planck era, the scalar field's
kinetic energy is much less than the potential energy.  Consequently,
the deceleration parameter, which is proportional to $\half\dot\phi^2
-V$, is negative and therefore the universe undergoes an accelerated
rate of expansion.  In order for inflation to last long enough to
alleviate the problems mentioned above, the scalar field is required
to slowly roll down its potential \cite{Oliveira1998a}.  Eventually, as
the scalar field reaches the minimum of its potential it will
oscillate and its energy is released in the form of light particles.
The final process, known as reheating, occurs when these light
particles thermalize and reheat the universe.  Since Guth's work,
there have been many types of inflationary scenarios, from models
which give too large initial density perturbations to ones in which
the perturbation size is correct, but do not solve the horizon problem
\cite{Linde1987a,Lukash1991a,Linde1982a,Albrecht1982a}.

The inflation scenario is also a major consideration in this thesis;
inflation can be a generic feature in models involving scalar fields
with a potential, and this thesis will also be concerned with the graceful
exit problem when inflation arises. \index{inflation paradigm}

\section{Formalism to General Relativity\label{formal}}

Although the theory of general relativity may be introduced from
several different viewpoints, it is convenient to begin with Einstein's field
equations, given by\index{Einstein's field equations (EFE)|bb}
\be
G_{\alpha \beta} = T_{\alpha \beta} \label{EFE},
\ee
where 
\be 
G_{\alpha\beta} \equiv R_{\alpha\beta}-\half g_{\alpha\beta}R \label{LHS},
\ee
(field equations
\eref{EFE} are the general relativistic analogue of Newton's equation
$\grad^2\varphi=4\pi G\rho$).  Geometerized units are used here, in
which $8\pi G =\hbar= c=1$, although appendix \ref{restore} lists all
quantities with the fundamental quantities restored.  The index
notation is to use lower-case Greek letters to denote indices which
range $0-3$, whilst lower-case Latin letters to denote indices which
range $1-3$.  In higher-dimensional theories (where mentioned)
uppercase Latin letters are used for the ``internal'' dimensions.  The
metric's signature in this thesis is $(-+++)$.

The left-hand side of equation
\eref{EFE} is completely geometric; for a manifold endowed with a
metric, $g_{\alpha\beta}$, the induced affine
connection is defined as
\be
\Gamma^\alpha_{\beta\gamma} \equiv \half g^{\alpha \mu} \left( \di_\gamma
g_{\beta\mu} +\di_\beta g_{\gamma\mu} -\di_\mu g_{\beta\gamma} \right),
\label{connection}
\ee
where $g^{\alpha\beta}$ is defined by $g_{\mu\beta}g^{\mu\gamma} =
\delta^\gamma_\beta$.  The symmetric Ricci tensor, $R_{\alpha \beta}$,
in \eref{EFE} is then defined as 
\be
R_{\alpha\beta}\equiv \pm \{\di_\mu\Gamma^\mu_{\beta\alpha} -
\di_\beta\Gamma^\mu_{\mu \alpha} + \Gamma^\mu_{\beta\alpha}
\Gamma^\nu_{\nu\mu} - \Gamma^\nu_{\mu\alpha} \Gamma^\mu_{\nu\beta}\},
\label{RicciT}
\ee
and the Ricci scalar, $R$, is just the contraction of \eref{RicciT},
$R\equiv R^\mu_\mu\equiv g^{\alpha \mu} R_{\alpha \mu}$.  There is an
ambiguity in the definition of (\ref{RicciT}) as demonstrated.
{\em Although both definitions are listed here, this thesis explicitly uses the
``$+$'' definition}.  The left-hand side of \eref{EFE} is
divergence-less since $\grad_\alpha R^\alpha_\beta =\half \grad_\beta
R$ and $\grad_\delta g_{\alpha\beta}=0$.

``Physics'' enters this formalism via the stress-energy tensor,
defined on the right-hand side of equation \eref{EFE}.  For instance,
this thesis assumes that the universe can, in part, be modeled after a perfect
fluid (i.e. no dissipation or heat conduction terms), described by the
stress-energy tensor \cite{Ellis1971a,MacCallum1971a}
\be 
 T_{\alpha\beta} = (\mu+p)u_\alpha u_\beta + p g_{\alpha\beta}, \label{fluid}
 \ee
where $u^\alpha\equiv dx^\alpha/d\tau$ is the observers four-velocity
($\tau$ is proper time), $\mu$ is the total energy density, and $p$ is the
pressure.  Since, by its construction, the left-hand side of
\eref{EFE} is automatically conserved, $\grad_\mu G_\alpha^\mu=0$, then
$\grad_\mu T_\alpha^\mu=0$ must be imposed, which lead to the equation
\be
\frac{D\mu}{Dt} + \left(\mu +p \right) \theta = 0
\label{conservation} 
\ee
(and an equation for $Du^\alpha/Dt$) where $\theta\equiv\grad_\mu
u^\mu$ describes the fluid's expansion and $D/Dt\equiv u^\mu
\grad_\mu$ defines the relativistic total derivative, analogous to the
$\di/\di t + \vec{u} \cdot \vec{\grad}$ used in Newtonian fluid
dynamics.  Equation \eref{conservation} is the relativistic
conservation of mass-energy of the system.

Other forms of energy may be included to couple to gravity, such
as scalar fields (section \ref{sf}) which has its own form for
$T_{\alpha\beta}$, or a cosmological constant ($T_{\alpha\beta}=\Lambda g_{\alpha\beta}$).  The stress energy tensors of each energy source
simply sum together.  It is apparent through Einstein's field
equations how non-linear the models of gravity are;
matter, as specified on the right-hand side of \eref{EFE}, induces
curvature in the spacetime (left-hand side) which in turn controls how
matter is to move (right-hand side).  As can be inferred from
\eref{connection} and \eref{RicciT}, equations \eref{EFE}
represent a system of coupled second order, non-linear partial differential
equations which are very difficult to solve.  Typically,
assumptions on the form of matter, its velocity field and
what sort of symmetries (such as those which lead to spatially
homogeneous and isotropic metrics) are required in order to simplify
the problem of solving \eref{EFE}.

\index{spatial homogeneity|(}
Spatial homogeneity is considered in the following mathematical manner.
Suppose there exists a time-like vector, $v^\alpha$, which is derived as the
gradient of a function, $t=t(x^\alpha)$, i.e. $v_\alpha\propto
\di_\alpha t$, so that for $\{t=constant\}$ there exist three--dimensional 
space-like hypersurfaces orthogonal to $v^\alpha$.  Now,
let this three-dimensional space be defined through the line element
$d\sigma^2=g_{ij}dx^idx^j$, where $x^i$ are the coordinates of each
${t=constant}$ hypersurface.  Spatial homogeneity implies that, for
fixed $t$, translations of the spatial coordinates,
$\tilde{x}^i=x^i+\epsilon\xi^i$, leaves the metric form-invariant,
$\tilde{g}_{ij}(\tilde{x}^k)=g_{ij}(x^k)$, where $\bv{\xi}$ are
Killing vectors\footnote{Note that $\bv{\xi}$ is the coordinate-free
representation of this vector; for any given basis $\{\bv{e}_i\}$,
the vector may be written $\bv{\xi} = \xi^i \bv{e}_i$.} defined by
\be 
\grad_i\xi_j+\grad_j\xi_i=0. \label{Killing} 
\ee 
Alternatively, the set $\{\bvect{A}, A=1,2,3\}$ are the generators
for the Lie group\footnote{For these groups there corresponds a unique
Lie algebra: a finite-dimensional space with (1) a multiplication
operation $[\bvect{A},\bvect{B}]$ (2) obeying
$[\bvect{A},\bvect{B}]=-[\bvect{B},\bvect{A}]$ and (3) obeying the Jacobi
identities $[\bvect{A},[\bvect{B},\bvect{C}]] + [\bvect{C},[\bvect{A},\bvect{B}]] +
[\bvect{B},[\bvect{C},\bvect{A}]]=0$ } of motions $G_3$ which preserves the
three-dimensional metric.  Because this set of vectors span the vector
space, then
\be
[\bvect{A},\bvect{B}]=C^C_{AB} \bvect{C}, \label{spanned}
\ee
where $C^C_{AB}=-C^C_{BA}$ are the structure constants of the group
and depend on all four coordinates (however, equation
\eref{spanned} will determine the spatial dependence and leave the
structure constants as functions of time).  It may be shown
\cite{MacCallum1971a,Kramer1980a} that equations \eref{spanned} reduce
to
\beqn 
\nonumber
\left[ \bvect{1} , \bvect{2} \right] & = & A \bvect{2} + N_3\bvect{3},\\ 
\left[ \bvect{1} , \bvect{3} \right] & = & N_1 \bvect{1}, \\ \nonumber
\left[ \bvect{2} , \bvect{3} \right] & = & N_2\bvect{2} -A \bvect{3}, 
\eeqn
\noindent where $AN_1=0$.  The sets $\{A,N_1,N_2,N_3\}$ give rise to 
nine types \cite{Kramer1980a} of spatially homogeneous cosmological
models (listed in table 1), as first examined by Bianchi
\cite{Bianchi1897a} (note that some Bianchi models are subsets to another; for example, Bianchi type V models are Bianchi type VII$_h$ models when $h\rightarrow\infty$).  It should be stressed that this classification
does not {\em solve} \eref{EFE}, but is used to help simplify the
geometrical terms of \eref{EFE} by reducing them from partial
differential equations to ordinary differential equations (second
order and non-linear).
\index{spatial homogeneity!Bianchi models|bb}
\begin{table}[htp]
\caption{Classification of spatial homogeneous models \label{Bianchi}}
\begin{center}
\begin{tabular}{|c|c|c|c|c|c|} \hline
Bianchi Class & $A$ & $N_1$ & $N_2$ & $N_3$ & Comment \\ \hline \hline
I & 0 & 0 & 0 & 0 & \\ 
II & 0 & $+$ & 0 & 0 & \\ 
III & $A^2=-N_2N_3$ & 0 & $+$ & $-$ & same as type VI$_h$ for $h=-1$\\ 
IV & $+$ & 0 & 0 & $+$ & \\ 
V & $+$ & 0 & 0 & 0 & \\ 
VI & $A^2=hN_2 N_3$ & 0 & $+$ & $-$ & $h<0$ \\ 
VII & $A^2=hN_2 N_3$ & 0 & $+$ & + & $h>0$ \\ 
VIII & 0 & $+$ & $+$ & $-$ & \\ 
IX & 0 & $+$ & $+$ & $+$ &\\  \hline
\end{tabular}\end{center}\end{table}
\index{spatial homogeneity|)}

\index{spatial isotropy|bb}
Isotropy can be defined as follows.  At each point $p$ in the
$\{t=constant\}$ orthogonal-hypersurfaces, there is a group of motions
which maps $p$ to itself but which rotates any vector in the tangent
space at $p$ into another vector in the same tangent space (i.e.,
spherical symmetry about p).  These two assumptions (spatial
homogeneity and isotropy) reduce the three-dimensional metric to the
form
\be
d\sigma^2 =\frac{dx^2+dy^2+dz^2}{\left[1 + \frac{k}{4} \left(x^2+y^2+z^2 \right) \right]^2}, \label{maximal}
\ee where $k$ is a constant and can be normalized to $\pm 1$ if
non-zero. The line element can be written
\be
ds^2=-dt^2+a(t)^2 d\sigma^2, \label{FRW}
\ee 
which is the FRW \index{Friedmann--Robertson--Walker (FRW)|bb}line element.  Any model
which uses \eref{FRW} will be referred to as an FRW model.  With the
aforementioned assumptions, along with the further assumption that the
universe can be modeled as a perfect fluid \eref{fluid} (with the
possibility of a cosmological constant present) equations \eref{EFE}
lead to the two equations:
\beqn
\mu(t) & = & 3 \left(\frac{\dot{a}}{a}\right)^2+3\frac{k}{a^2} 
-\Lambda, \label{F1} \\
p(t) & = & -2\frac{\ddot{a}}{a}-\left(\frac{\dot{a}}{a} 
\right)^2 -\frac{k}{a^2} +\Lambda, \label{F2}
\eeqn
where $\dot{a} \equiv da/dt$ and $\ddot{a}\equiv d^2a/dt^2$.
Equation \eref{conservation} yields
\be
\dot{\mu}+3\left(\mu+p\right) \frac{\dot{a}}{{a}} = 0.
\ee
This is also obtained from combining \eref{F1} and \eref{F2} and so
there are only two independent equations for three unknowns: $\mu$,
$p$ and $a$.  An equation of state is therefore needed (e.g.,
for barotropic matter: $p=(\gamma-1)\mu$) in order close the system.

This section ends with introducing the measurable quantities in
cosmology, and showing how they are defined in FRW models when no
cosmological constant is present.  First, with the definition
$3\dot{a}^2/a^2\equiv \mu_{crit}$, \eref{F1} can be written as\index{critical density parameter ($\Omega$)|bb}
\benonumber
\frac{k}{\dot{a}^2} = \frac{\mu}{\mu_{crit}}-1 \equiv \Omega-1.
\eenonumber
If $k>0$
then $\mu>\mu_{crit}$ ($\Omega>1$) and the universe eventually
recollapses if $p\ge 0$ and $T_{00}\ge \|T_{ij}\|$ (for any
orthonormal basis)\cite{Ciufolini1995a}.  Conversely, $\Omega<1$
($\mu<\mu_{crit}$) for $k<0$ and the universe expands forever.  The
critical value occurs for $k=0$ when $\Omega=1$, or $\mu=\mu_{crit}$,
in which case the universe will still expand forever, but asymptotes
towards $\dot{a}= 0$.  Comparison of the universe's actual energy
density compared to the critical energy density, $\mu_{crit}$, would
glean insight into the late time fate of an FRW universe.  The critical
density parameter can be measured in two ways; by either measuring
$\mu_{crit}$ and $\mu$ or by inference from measuring the deceleration
parameter,\index{deceleration parameter ($q$)|bb}
\be
q\equiv -\frac{\ddot{a}a}{\dot{a}^2} = \left(\frac{3}{2}\gamma-1\right) \Omega.
\ee
To obtain a value for $\mu_{crit}$, the Hubble parameter, $H\equiv=\third
\theta= \dot{a}/{a}H$, needs to be measured.  This parameter is used
to determining the upper age of the universe in the absence of a
cosmological constant.  Any FRW model contains an initial singularity
for $\gamma > 2/3$ (or more generally with $\mu+3p>0$), and for $\ddot{a}<0$
and $a>0$ this singularity occurs in the past at time $t_0 <
H_0^{-1}$, where $H_0$ is the present value of the Hubble parameter
 (note that FRW models are not the only models which contain
singularities, and theorems \cite{Hawking1974a} indicate that
singularities are quite generic in general relativity although the
nature of which  can be quite different).  This result is
obtained by combining
\eref{F1} and \eref{F2} to obtain $3\ddot{a}=-2 a(\mu+3p)$ which is
always negative for $a(\mu+3p)>0$.  It may be shown that if a galaxy's
redshift is due solely to the universe's expansion, then there is a
correlation between its luminosity distance (distance obtained from
apparent/absolute luminosity of the galaxy), $d_L$, and its redshift,
$z$, namely (to second order)
\be 
z = H_0 d_L -\half
(1-q_0) \left(H_0d_L\right)^2 + ... \label{HubblesLaw}\ee 
The values of $H_0$ and $q_0$ can be
determined from graphs of $z$ vs. $d_L$ \cite{Ciufolini1995a}.  Table
\ref{Handq} 
presents various measured values of $q_0$ and 
$H_0$ found in the literature, listed in chronological order.  This is not 
intended to be a comprehensive list, but rather an indicator of past and 
present values of each of the parameters.

\begin{table}[htp]
\caption{Observed values of $H_0$ and $q_0$ quoted in the literature\label{Handq}}
\begin{center}
\begin{tabular}{|c|c|c|} \hline
Parameter & Value & Reference \\ \hline\hline
$q_0$ (unitless) & $3.7 \pm0.8$  & \cite{Humason1956a}\\
  & $1.0 \pm 0.5$ & \cite{Baum1957a} \\
  & $0.2 \pm 0.5$ & \cite{Sandage1961a} \\ 
  & $1.2 \pm 0.4$ & \cite{Sandage1966a}\\
  & $1.5 \pm 0.4$ & \cite{Peach1970a} \\
  & $0.7^{+0.5}_{-0.3}$ & \cite{Sandage1971a} \\
  & $0.5$ & \cite{Kaiser1991a,Kellermann1993a,Hamilton1993a} \\ \hline
$H_0$ (km/s/Mpc) & $526$ & \cite{Hubble1936a} \\
  & $260$ & \cite{Baade1952a} \\
  & $180$ & \cite{Humason1956a} \\
  & $98$ & \cite{Sandage1961a} \\
  & $75.3^{+19}_{-15}$ & \cite{Sandage1968a} \\
  & $80\pm 17$ & \cite{Freedman1994a,Pierce1994a} \\ 
  & $70\pm 7$ & \cite{Richtler1999a,Astronomy1999a} \\ \hline
\end{tabular}
\end{center}
\end{table}

Having discussed some of the issues associated with cosmology, it is
convenient to discuss theories that augment general relativity in one (or
more) ways.  Einstein's field equations \eref{EFE} may be derived from
the Einstein-Hilbert action\index{action|bb}
\be
S = \int d^4x \sqrt{-g}\left(R+\Lambda + {\cal L}_m \right),
\ee
(where ${\cal L}_m$ is the matter Lagrangian density related by 
\be
T_{\alpha\beta} = g_{\alpha\beta}{\cal L}_m -\frac{\di{\cal L}_m}{\di g^{\alpha\beta}}
\ee
to the stress-energy tensor), by varying the action with respect to
$g^{\alpha \beta}$.  Each section begins by
describing each of the following theories in terms of its governing
action, because correlations between the
theories becomes quite apparent through these actions.

\section{Scalar Field Gravitational Theories\label{sf}}

There are two prevalent theories involving scalar
fields: general relativity minimally coupled to a scalar field (herein
referred to as either ``GR scalar field theories'' or ``the Einstein frame'') and
scalar-tensor theory (herein referred to as either ``ST'' or ``the
Jordan frame'').  

\subsection{The Einstein Frame}
The action for GR scalar field theories can be expressed as \index{action!scalar field}
\be
\up{sf}S = \int d^4x \sqrt{-\up{sf}g}\left\{\half\up{sf}R -
\half \left(\grad\phi\right)^2 - V(\phi) 
+ {\cal \up{sf}L}_m \right\}, \label{sfaction}
\ee
which leads to the field equations
\beqn
\up{sf}G_{\alpha \beta} &=& \up{sf}T_{\alpha \beta} + \grad_\alpha\phi 
	\grad_\beta\phi -\up{sf}g_{\alpha \beta}\left[\half
	\left(\grad\phi\right)^2 +V\right],  \\
\grad^\alpha \up{sf}T_{\alpha \beta} &+ & \grad_\beta\phi\left(\Box 
	\phi - \frac{dV}{d\phi} \right)=0.
\eeqn
Historically, scalar field
cosmology became topical within the context of inflation.  For
example, in a chaotic inflation model, the scalar field is far from
its potential minimum in the early universe such that $V \gg
\dot{\phi}^2$.  In this way, the potential acts as a cosmological
constant and drives the universe into an inflationary state which ends
when the potential approaches its minimum, and then oscillates away
into radiation.

Heusler
\cite{Heusler1991a} showed that for any concave-up potential whose
minimum value is zero, the only Bianchi models that may isotropize at
late times are those whose Lie group admit FRW models.  Similar to
Wald's cosmic no hair theorem, Kitada and Maeda
\cite{Kitada1992a,Kitada1993a} have shown that for $V
\propto\exp(k\phi)$ most initially expanding, spatially homogeneous
models containing a scalar field will approach an isotropic, power-law
inflationary solution (power-law inflation: $a\propto t^b$ for
$b(b-1)>0$) for $k^2<2$ (this is only true in the Bianchi IX case
under certain conditions - see \cite{Coley1997a}).  For such
exponential potentials, Coley {\em et al.} \cite{Coley1997a}
generalized these results by showing that it is possible for Bianchi
I, V, $\mbox{VII}_h$ and IX models to isotropize for $k^2>2$, thereby
showing that there exists an open set in the set of all spatially
homogeneous spacetimes which {\em do} isotropize, unlike in general
relativity where isotropizing solutions are of measure zero.

These models are well motivated, especially from string theory (see
subsection \ref{StringtoEinstein} starting on page
\pageref{StringtoEinstein}), arise naturally from higher--dimensional
reduction (see appendix \ref{KKreduced}) and are clearly relevant in
cosmology as they can lead to isotropization.  Furthermore, it has
been demonstrated within GR scalar field models  that any potential present needs to be
exponential in order for the theory to be scale invariant (i.e., the
action remains invariant under the scaling $g_{\alpha\beta}\rightarrow
e^f g_{\alpha\beta}$)
\cite{Guendelman1999a,Guendelman1999b,Diaz1999a}.

When matter is included with such scalar fields there exist ``matter
scaling solutions'' in which the scalar field energy density tracks
that of a perfect fluid (at late times neither field is negligible)
\cite{Wetterich1988a,Wands1993a}.  In particular, in
\cite{Copeland1997a} a phase-plane analysis of the spatially flat
FRW) models showed that these scaling
solutions are the unique late-time attractors whenever they exist.
The cosmological consequences of these scaling models have been
further studied in \cite{Ferreira1997b,Ferreira1998a,Wetterich1995a}.
For example, in these models a significant fraction of the current
energy density of the universe may be contained in the scalar field
whose dynamical effects mimic cold dark matter \cite{Ferreira1998a};
the tightest constraint on these cosmological models comes from
primordial nucleosynthesis bounds on any such relic density
\cite{Wetterich1988a,Wands1993a,Copeland1997a,Ferreira1997b,Ferreira1998a,Wetterich1995a}.

\subsection{The Jordan Frame and String Theory}
The action for ST theory is written as \index{action!scalar-tensor}
\be
\qquad\up{st}S = \int d^4x \sqrt{-\up{st}g}\left\{ \half\left[\Phi \up{st}R-
\frac{\omega(\Phi)}{\Phi} \left(\grad\Phi\right)^2 
-2U(\Phi)\right] + {\cal L}_m \right\}, \label{jbdaction}
\ee
which leads to the field equations
\beqn
\up{st}G_{\alpha \beta} &=& \frac{\up{st}T_{\alpha \beta}}{\Phi} 
	 + \frac{\omega}{\Phi^2} \left[ 
	\grad_\alpha\Phi \grad_\beta\Phi- \half \up{st}g_{\alpha \beta}
	 \grad_\gamma\Phi \grad^\gamma\Phi \right] 
	+\frac{\grad_\alpha\grad_\beta\Phi}{\Phi} \nonumber \\ 
	&&-\up{st}g_{\alpha \beta}
	\left(\frac{V}{\Phi}+\frac{\Box\Phi}{\Phi} \right),\\
\frac{1}{\Phi} \grad^\alpha \up{st}T_{\alpha \beta} & + & 
	\half \frac{\grad_\beta\Phi}{\Phi} \left[ 2\omega\frac{\Box\Phi}{\Phi}
	+\frac{\grad_\gamma\Phi \grad^\gamma\Phi}{\Phi}\left(
	\frac{d\omega}{d\Phi} -\frac{\omega}{\Phi}\right) -2 \frac{dU}{d\Phi}
	+\up{st}R\right].
\eeqn
This theory is a variable-G theory in
which Newton's ``constant'' is proportional to $\Phi^{-1}$.

Scalar-tensor theories, \eref{jbdaction}, in the early guise of the so
called Jordan-Brans-Dicke theory in which $\omega=\omega_0=$constant,
were first studied by Jordan \cite{Jordan1949a,Jordan1959a}, Fierz
\cite{Fierz1956a}, and by Brans and Dicke \cite{Brans1961a}.  The
generalized forms of these theories were first studied by Bergmann
\cite{Bergmann1968a}, Nordtvedt \cite{Nordtvedt1970a} and Wagoner
\cite{Wagoner1970a}. The observational limits on these theories {\em
without} a potential arising from solar system tests
\cite{Buchmann1996a,Abramovici1992a,Hough1992a,Bradaschia1990a}, as
well as cosmological tests such as Big Bang nucleosynthesis
calculations \cite{Barrow1987a,Serna1996a}, all constrain present
values when $\omega$ is assumed constant by $\omega
\gg 500$.  Also, Damour
and Nordtvedt \cite{Damour1993a} showed that for any FRW scalar-tensor
model without a potential and with an $\omega(\Phi)$ satisfying
\be
\omega\rightarrow \infty \mbox{ and }
\omega^{-3}\frac{d\omega}{d\Phi} \rightarrow 0
\ee 
as $t\rightarrow\infty$, the theory will asymptote to ordinary general
relativity at late times.

Scalar-tensor theories, \eref{jbdaction}, have also been widely
studied in recent years \cite{Barrow1990a,Barrow1994a,Barrow1993a,Barrow1993b,Berkin1994a,Damour1993a,Damour1993a,Gerard1995a},
partially due to their relationship with the low energy limit of
various unified field theories such as superstring theory
\cite{Green1987a} (see below); in particular, the dimensional reduction of
higher-dimensional gravity results in an effective scalar-tensor
theory \cite{Freund1982a,Holman1991a} (see appendix \ref{KKreduced}).
Some studies on the possible isotropization of spatially homogeneous
scalar-tensor cosmological models have also been made.  For example,
Chauvet and Cervantes-Cota
\cite{Chauvet1995a} have studied the possible isotropization of Bianchi
models of types $I$, $V$ and $I\!X$ within the context of
Jordan-Brans-Dicke theory without a scalar potential but with
matter with a linear equation of state, $p=(\gamma -1) \mu$, by studying exact solutions at
late times.  Also, Mimoso and Wands \cite{Mimoso1995a} have studied
this theory in the presence of matter with a linear equation of state (but with no scalar
field potential) and, in particular, gave values of $\omega$ under
which Bianchi $I$ models isotropize.  There is a
formal equivalence between such a theory (with $\gamma\neq 2$) and a
scalar-tensor theory with a potential but without matter \cite{Billyard1999a}, via the
field redefinitions
\beqn
V & \equiv & (2-\gamma)\mu \\
\omega \grad_a\Phi\grad_b\Phi & \rightarrow &\omega
\grad_a\Phi\grad_b\Phi -\gamma\mu \Phi \delta^0_a\delta^0_b.
\eeqn

Of great relevance, ST theories are used in the context of
higher-dimensional string theory, wherein $\omega=-1$.  There are five
anomaly--free perturbative superstring theories: type I, type IIA,
type IIB, Heterotic with gauge group $E_8\times E_8$ and Heterotic
with gauge group {\em Spin}$(32)/\mZ_2$
\cite{Gibbons1998a}.  It is now widely believed that these theories
represent different perturbative expansions, in a weakly coupled
limit, of a more fundamental non--perturbative eleven--dimensional
theory, referred to as M--theory. The original formulation of
M--theory was given in terms of the strong coupling limit of the type
IIA superstring.

There has been considerable interest recently in the cosmological
implications of string theory.  The evolution of the very early
universe below the string (Planck) scale is determined by the
low--energy effective action.  All five perturbative string theories
contain a dilaton (i.e. $\Phi$), a graviton and a two--form gauge
potential in the Neveu--Schwarz/Neveu--Schwarz (NS--NS) sector of the
theory and a three--form gauge potential in the Ramond--Ramond (R--R)
sector of the theory (both the NS--NS and the R--R sectors are bosonic
sectors of the theory \cite{Gibbons1998a}). String theory is of
great physical interest since it is generally believed that the full,
non-perturbative theory (as yet developed) will incorporate energies
at the Planck scale and smooth out the initial singularity; these
models incorporate the concept of a pre-Big Bang where solutions for
$t<0$ are dual to the those for $t>0$ by letting $t\rightarrow-t$ and
$a\rightarrow a^{-1}$.

A typical scheme in string theory is to compactify from ten dimensions
onto an isotropic six--torus of radius $e^{\beta_m}$ to obtain the
effective four-dimensional action, given by
\begin{eqnarray}
\label{NSaction_i}
\up{st}S&=&\int d^4 x \sqrt{-\up{st}g} \left\{e^{-\hat\Phi} \left[ \up{st}R 
+\left(\grad\hat\Phi\right)^2  -6 \left(\grad\beta_m\right)^2
-\frac{1}{2\cdot 3!} H_{\mu\nu\lambda} H^{\mu\nu\lambda} -2\Lambda 
\right]\right. \nonumber \\ 
	&& \left. -\frac{1}{2\cdot 4!}e^{6\beta_m}F_{\mu\nu\lambda\kappa}
	F^{\mu\nu\lambda\kappa} -\Lambda_{\rm M}\right\}.
\end{eqnarray}
where the dilaton field, $\hat\Phi=-\ln\Phi$, parameterizes the string
coupling, $g_s^2 \equiv e^{\hat\Phi}=\Phi^{-1}$, $H_{\mu\nu\lambda}
\equiv \partial_{[\mu} B_{\nu\lambda ]}$ is the field strength of the
two--form potential $B_{\mu\nu}$, and
$F_{\mu\nu\lambda\kappa}\equiv\partial_{[\mu} A_{\nu\lambda\kappa ]}$
is the field strength of the three--form potential
$A_{\mu\nu\lambda}$. The constant $\Lambda$ is determined by the
central charge deficit of the NS--NS sector in type II and heterotic
superstring models
\cite{Callan1985a,Lovelace1986a,Fradkin1985b} and may be viewed as a
cosmological constant in the gravitational sector of the theory. The
$\Lambda_{\rm M}$ term may be interpreted as a perfect fluid matter
stress in the matter sector of the theory with an equation of state
$p=-\mu$. It could be generated by a slowly moving scalar field, with
a kinetic energy contribution dominated by a self--interaction
potential, $p \approx -V \approx -\mu$.  The Ramond--Ramond
three--form potential arises from the R--R sector of type IIA
supergravity \cite{Romans1986a,Bergshoeff1996a}.  Note that the
Ramond--Ramond term and the central charge deficit arise in different
string theories (although they've been included in one action above
for brief notation) and in the context of string theory will be
considered separately).  However, the central charge deficit will be
included in a portion of the analysis including the Ramond--Ramond term
as a perturbation parameter.

In four dimensions, a p--form is dual to a (4$-$p)--form and so the
ans\"{a}tze 
\beqn
\label{sigma_i}
H^{\mu\nu\lambda} &\equiv& e^{\hat\Phi} \epsilon^{\mu\nu\lambda\kappa} 
	\nabla_{\kappa} \sigma \label{H_ansatz}\\
F^{\mu\nu\lambda\kappa} & \equiv & Qe^{-6\beta_m}\epsilon^{\mu\nu\lambda\kappa}
	\label{F_ansatz}
\eeqn
(where $\epsilon^{\mu\nu\lambda\kappa}$ is the covariantly constant
four--form, and $Q$ is a constant) are used to solve the field
equations associated with the forms $H^{\mu\nu\lambda}$ and
$F^{\mu\nu\lambda\kappa}$, namely
\beqn
\grad_\mu\left(e^{-\hat\Phi} H^{\mu\nu\lambda}\right) = 0 
	&\Longleftrightarrow & \grad_\mu\left(e^{\hat\Phi}\grad^\mu
	\sigma\right)=0,\label{H_eqn}\\
\grad_\mu\left(e^{6\beta_m} F^{\mu\nu\lambda\kappa} \right)=0,\label{F_eqn}
\eeqn
respectively.  Here, (\ref{F_eqn}) is trivially satisfied using
(\ref{F_ansatz}) whereas equations (\ref{H_eqn}) are reduced to one
equation for $\sigma$ by (\ref{H_ansatz}).  With ans\"{a}tze
(\ref{H_ansatz}) and (\ref{F_ansatz}), the action (\ref{NSaction_i})
further reduces to the action \index{action!string}
\begin{eqnarray}
\label{sigmaaction_i}
\up{st}S&=&\int d^4 x \sqrt{-\up{st}g} \left\{e^{-\hat\Phi} \left[
	\up{st}R +\left( \nabla \hat\Phi  \right)^2  
	-6 \left( \nabla \beta_m \right)^2 
	-\frac{1}{2} e^{2 \hat\Phi} \left( \nabla \sigma \right)^2 -2\Lambda
	\right]\right. \nonumber \\
	&& \qquad -\left.\half Q^2e^{-6\beta_m} -\Lambda_{\rm M}\right\}
\end{eqnarray}
where $\sigma$ may be interpreted as a pseudo--scalar `axion' field.
Although, the $\Lambda_{\rm M}$ term is here considered in the context
of a cosmological constant in the matter sector, Billyard {\em et al}
\cite{Billyard1999d} have discussed how such a term and the $Q^2$ term
can be related via field redefinitions and so cases in which both
$Q\neq 0$ and $\Lambda_{\rm M}\neq 0$ will not be studied.

\section{{\em Quo Animo}}

The main goal of this thesis is to explore the genericity of isotropy
in cosmological models containing scalar fields, either in the
Einstein frame or the Jordan frame.  There are formal mathematical
equivalences between the two frames, although each has a different
physical representation.  For instance, in the Einstein frame the
scalar field is associated with the rest mass of particles, whereas in
the Jordan frame the scalar field is related to Newton's ``constant''
$G$.  Chapter \ref{SFtoST} discusses the mathematical equivalences
between the two frames, and discusses the asymptotic behaviour in ST
theory which leads to exponential potentials in GR scalar field theories.  Chapter
\ref{SFtoST} also includes discussions on other formal mathematical
relationships which allow comparisons to be made between various
theories in either frame; for example, a relationship between a ST
theory with matter but without a potential and a ST theory without
matter but with a potential, and a relationship between string theories
and GR scalar field theories with both exponential potentials {\em and}  matter with a non-linear equation of state ($p\not\propto\mu$).
In such transformations, there is the freedom of choosing a particular
theory (either out of physical interest or mathematical convenience)
and relate the asymptotic results to another theory.

Chapter \ref{scaling} provides the introduction to scaling solutions
in the context of GR scalar fields theories, by considering the
stability of these solutions
to curvature and shear perturbations.  The results contained therein
provide the motivation to Chapter \ref{BianchiB} which consider
scaling solutions in a much larger class of homogeneous models, namely
within models of Bianchi class B.  Considering such models is well motivated;
scaling solutions arise in the simplest models containing both a scalar
field and matter (separately conserved) and the scalar
field in such models may provide a plausible mechanism for dark
matter.  Furthermore, exponential potential models are of interest
since they can avoid the conditions of Wald's no-hair theorem (in which
exponential inflationary is the late-time behaviour), and can lead
to isotropization.

Chapter \ref{reheat} considers scalar field (with an exponential
potential) models containing an interaction term between the scalar
field and the matter content, in order to determine whether such terms
can greatly affect the dynamics of the system.  In particular, it will
be demonstrated (chapter \ref{BianchiB}) that for $k^2<2$ all models
asymptote to a power-law inflationary model in which the matter is
driven to zero.  It will be determined if its possible for this
solution to become a saddle solution rather than a source by the
introduction of an interaction term; hence models can inflate for an
indefinite period of time, but then evolve away towards other
attracting solutions where matter is not driven to zero.  Furthermore,
it will be determined if the same interaction terms can lead to
conditions necessary for reheating (namely, an oscillating scalar
field).  In general, there are no canonical forms (say, from particle
physics) for an interaction term between ordinary matter and a scalar
field, although there are a some {\em ad hoc} interaction terms in the
the literature which will be considered.

Chapters \ref{string} to \ref{Qsection} are devoted to string theory
in the context of a class of Bianchi models.  By comparing various
fields in the theory, four-dimensional phase spaces are obtained and
the asymptotic behaviours are determined.  Not only are string
cosmologies of great physical interest because it is believed that the
(yet to be formulated) full, non-perturbative theory will smooth out
the initial singularity, but the solutions in this thesis can be
transformed to a GR scalar field theory with an exponential potential {\em and}
matter with a non-linear equation of state, thereby complimenting the previous chapters.
In string theory, inflation typically occurs in the pre-Big Bang
scenario and not in the post-Big Bang, although this inflationary
behaviour has been questioned \cite{Turner1997b,Kaloper1999a} and in each case
there will be a comment on the r\^{o}le of
inflation.

It is important to stress that all the models considered are classical
and caution is imperative when entering quantum regimes (such as
when energies near the Planck energy).  Although the analysis is
performed for all times and all energies, it is assumed that the
models {\em will} break down when energies approach the appropriate
energy scales.  However, this does not hamper the arguments that will
be made by the dynamical systems analysis; if solutions asymptote into
the future towards one common solution, that solution may be invalid
after a certain time due to energy limitations, but it can still be
asserted that within the classical regime the solutions {\em do}
asymptote towards that solution.  The concept for reheating is also
semi-classical \cite{Abbott1984a}.  The reheating mechanism is quite
complicated and approximations are often needed in order to perform
the appropriate calculations.  However, the issues that this thesis
will address is whether the interaction terms included will allow the
model to evolve to a point where the governing equations are
equivalent to those found in reheating discussions.  

The main mathematical tool used in this thesis is theory of dynamical
systems, in which the field equations are rewritten as first order,
autonomous ordinary differential equations (ODE) and then the
early--time and late--time asymptotic behaviours are determined.
Although the early- and late-time behaviours are important,
intermediate behaviour can also be significant since trajectories in
the relevant phase spaces spend an indefinite period of time about
saddle points representing solutions that may be physically relevant
to the universe at the current epoch.  A brief survey of dynamical
systems theory can be found in appendix
\ref{OD}.  Although this thesis uses geometerized units, appendix 
\ref{restore} restores the fundamental constants to help clarify the units 
of each entity discussed within this thesis.  Finally appendix
\ref{KKreduced} discusses how exponential potentials arise from the
reduction of higher-dimensional theories.  This is a generalization of
the work found in Billyard and Coley \cite{Billyard1997a} in which it
was shown that there is a mathematical equivalence between
higher-dimensional vacuum Kaluza-Klein theory and vacuum solutions in
either ST theory or GR scalar field theory.  The purpose of
\cite{Billyard1997a} was to elucidate the fact that previously
solutions in one theory had been ``discovered'' after the
corresponding solutions in another theory had been previously known.
These mathematical relationships to higher--dimensional theories are
relevant within the context of string theory which are either ten or
eleven dimensional theories.

Much of the work in this thesis has been published with co-workers,
although that which appears in this thesis represents my contribution
of that work.  Specifically, the majority of chapter \ref{SFtoST}
represents research conducted with Alan Coley and Jesus Ib\'{a}\~{n}ez
(see ref. \cite{Billyard1999a}).  At the end of this chapter, the
discussion mentions how the transformation between the frames can be
extended to inhomogeneous $G_2$ models, which is based on research
with Alan Coley, Jesus Ib\'{a}\~{n}ez and Itsaso Olasagasti (see
ref. \cite{Billyard1999g}).  Chapter
\ref{scaling} is based on research with Alan Coley and Robert van den
Hoogen \cite{Billyard1998a} and chapter \ref{BianchiB} is based on
work with Alan Coley, Robert van den Hoogen, Jesus Ib\'{a}\~{n}ez and
Itsaso Olasagasti \cite{Billyard1999f}.  Chapters
\ref{string} - \ref{NSNSRR} are based on collaborative work with Alan
Coley and Jim Lidsey
\cite{Billyard1999b,Billyard1999c,Billyard1999d,Billyard1999e}, and
chapter \ref{Qsection} is recent work with Alan Coley, Jim Lidsey and
Ulf Nilsson.  There have been other papers written by Billyard (and
co-authors) during the time of this thesis
\cite{Billyard1996c,Billyard1997b,Billyard1997c,Agnese1999a},
but which are peripheral to the work contained herein.

\chapter{Scalar-Tensor Asymptopia from General Relativity with Scalar Fields and Exponential Potentials\label{SFtoST}}
In a recent paper \cite{Coley1997a} (see also \cite{Ibanez1995a} and
\cite{vandenHoogen1997a}), cosmological models containing a scalar
field with an exponential potential were studied.  In particular, the
asymptotic properties of the spatially homogeneous Bianchi models, and
especially their possible isotropization and inflation, were
investigated.  Part of the motivation for studying such models is that
they can arise naturally in alternative theories of gravity
\cite{Burd1988a}; for example, Halliwell \cite{Halliwell1987a} has shown that
the dimensional reduction of higher-dimensional cosmologies leads to
an effective four-dimensional theory coupled to a scalar field with an
exponential self-interacting potential (see Appendix \ref{KKreduced} for an explicit derivation).  

The action for
the general class of scalar-tensor theories (in the so-called
Jordan frame) is given by \cite{Bergmann1968a,Wagoner1970a},
\be 
~\up{st}S=
\int \sqrt{-\up{st}g} \left[ \Phi \up{st}R - \frac{\omega(\Phi)}{\Phi}\up{st}g^{\alpha\beta}\grad_\alpha\Phi \grad_\beta\Phi-2U(\Phi) + 2\up{st}{\cal{L}}_m \right] d^4x. \label{bdaction}
\ee
However, under the conformal transformation and field redefinition
\cite{Barrow1994a,Liddle1992a,Mimoso1995a}
\beqn
\mainlabel{SFtoSTtrans}
 \up{sf}g_{\alpha\beta} & = & \Phi \up{st}g_{\alpha\beta} \label{trans1} \\
\frac{d\phi}{d\Phi} & = & 
      \frac{\pm\sqrt{\omega(\Phi)+3/2}}{\Phi}, \label{trans2}
\eeqn
the action becomes (in the so-called Einstein frame)
\be
\up{sf}S=\int \sqrt{-\up{sf}g}\left[ \up{sf}R-\up{sf}g^{\alpha\beta}
\grad_\alpha\phi\grad_\beta\phi-2\frac{U(\Phi)}{\Phi^2} + 2\frac{
\up{st}{\cal{L}}_m}{ \Phi^2} \right] d^4x, \label{eaction}
\ee
which is the action for general relativity (GR) containing a scalar field
$\phi$ with the potential
\be
V(\phi)= \frac{U(\Phi(\phi))}{\Phi^2(\phi)}, \label{vstar}
\ee
and a matter Lagrangian
\be
\up{sf}{\cal{L}}_m=\frac{\up{st}{\cal{L}}_m}{ \Phi^2}.
\ee

The aim here is to exploit the results in previous work
\cite{Coley1997a} to study the asymptotic properties of scalar-tensor
theories of gravity with action \eref{bdaction} which under the
transformations \eref{SFtoSTtrans} transform to general relativity with a
scalar field with the exponential potential given by
\be
V = V_0 e^{k\phi} \label{exp},
\ee
where $V_0$ and $k$ are positive constants.  That is, since
the asymptotic behaviour of spatially homogeneous Bianchi models
with action \eref{eaction} with the exponential potential \eref{exp} is known,
the asymptotic properties of the corresponding
scalar-tensor theories under the transformations can be deduced
\eref{SFtoSTtrans}\footnote{The possible isotropization of spatially
homogeneous scalar-tensor theories which get transformed to a model
with an effective potential which passes through the origin and is
concave up may be deduced from the results of Heusler
\cite{Heusler1991a}} (so long as the transformations are not
singular!).  In particular, the possible
isotropization and inflation of such scalar-tensor theories will be considered.

The outline of this chapter is as follows.  In Section 2.1, 
the framework within which GR and a scalar field with a potential $V$
(Einstein frame) is formally equivalent to a scalar-tensor theory with
a potential $U$ (Jordan frame) is reviewed, concentrating on both the exact and
approximate forms for the parameters $U$ and $\omega$ in the Jordan
frame.  In particular, the explicit example discussed is of the
Brans-Dicke theory with a power-law potential and the
conditions which lead to appropriate late-time behaviour, as dictated
by solar system and cosmological tests, are discussed.  In Section 2.2,
the conformal transformations to Bianchi models studied in the
Einstein frame is applied to produce exact solutions which represent the
asymptotic behaviour of more general spatially-homogeneous models in
the Jordan frame (for $\omega=\omega_0$, a constant).  These
Brans-Dicke models are self-similar and the corresponding homothetic
vectors are also exhibited.  Because there is considerable interest in
string theory, Section 2.3 discusses how the string field equations in
the Jordan frame can be written in the Einstein frame as the field
equations of a scalar field with an exponential potential and a matter
field.  Section 2.4 concludes with a summary.

\section{ Analysis}

For scalar field Bianchi models the conformal factor in \eref{trans1}
is a function of $t$ only (i.e., $\Phi=\Phi(t)$), and hence under
(non-singular) transformations \eref{SFtoSTtrans} the
Bianchi type of the underlying models does not change (i.e., the
metrics $\up{st}g_{\alpha\beta}$ and $\up{sf}g_{\alpha\beta}$ admit three space-like Killing vectors
acting transitively with the same group structure).  In general, in
the class of scalar-tensor theories represented by \eref{bdaction}
there are two arbitrary (coupling) functions $\omega(\Phi)$ and
$U(\Phi)$.  The models which transform under \eref{SFtoSTtrans} 
to an exponential potential model, in which the two
arbitrary functions $\omega$ and $V$ are constrained by \eref{trans2}
and \eref{vstar}, viz., 
\be \frac{\Phi}{U}\frac{dU}{d\Phi}=2 \pm k
\sqrt{\frac{3}{2}+\omega(\Phi)} ,\label{Vdefine}
\ee
is a special subclass with essentially one arbitrary function.
Although only a subclass of models obey this constraint, this subclass
is no less general than massless scalar field models ($V=0$; see, for example
\cite{Mimoso1995a}) or Brans-Dicke models with a potential
($\omega=\omega_0$, constant), which are often studied in the
literature.  Indeed, the asymptotic analysis in this chapter is valid
not only for ``exact'' exponential models, but also for scalar-tensor
models which transform under \eref{SFtoSTtrans} to a
model in which the effective potential is a linear combination of
terms involving exponentials in which the dominant term asymptotically
is a leading exponential term; hence the analysis here is rather more
general (the next section will re-address this).  For the
remainder of this chapter, excluding section 2.3,  ordinary matter shall 
not be explicitly 
considered; i.e., the matter Lagrangians in \eref{bdaction} and
\eref{eaction} will be set to zero.  Matter can be included in a 
straightforward way \cite{Heusler1991a,Mimoso1995a,Kitada1992a}.

\subsection{ Exact Exponential Potential Models \label{exact}}

Scalar-tensor models which transform under \eref{SFtoSTtrans} to a model with an 
exact exponential potential satisfy
equations \eref{trans2} and \eref{vstar} with \eref{exp}, viz.,
\beqn
\mainlabel{potentialtransform}
\frac{d\phi}{d\Phi} & = & \pm \frac{\sqrt{\omega(\Phi)+3/2}}
{\Phi} \\
V_0 e^{k\phi} & = & \frac{U(\Phi)}{\Phi^2}.
\eeqn
So long as the transformations \eref{SFtoSTtrans} remain non-singular
determine the asymptotic properties of the underlying scalar-tensor
theories from the asymptotic properties of the exact exponential
potential model can be determined.  These properties were studied in \cite{Coley1997a}.
Note that the asymptotic behaviour depends crucially on the
parameter $k$ (in \eref{exp}) which will be related to the various
physical parameters in the scalar-tensor theory \eref{bdaction}.

In particular, in \cite{Coley1997a} it was shown that all scalar field
Bianchi models with an exponential potential \eref{exp} (except a
subclass of the Bianchi type $I\!X$ models which recollapse) isotropize to the future
if $k^2\leq 2$ and, furthermore, inflate if $k^2<2$; if $k=0$ these
models inflate towards the de Sitter solution and in all other cases
they experience power-law inflationary behaviour.  If $k^2 > 2$, then
the models cannot inflate, and can only isotropize to the future if
the Bianchi model is of type $I$, $V$, $V\!I\!I$, or $I\!X$.  Those
models that do not isotropize typically asymptote towards a
Feinstein-Ib\'{a}\~{n}ez anisotropic model \cite{Feinstein1993a}.
Bianchi $V\!I\!I_h$ models with $k^2>2$ can indeed isotropize
\cite{Coley1997a} but do not inflate, while generically the 
ever-expanding Bianchi $I\!X$ models do not isotropize \cite{vandenHoogen1998a}.

Therefore, at late times and for each specific choice of
$\omega(\Phi)$ both the asymptotic behaviour of the models and
the character of the conformal transformation \eref{SFtoSTtrans} may be
determined by the behaviour of the scalar field $\phi$ at the
equilibrium points of the system in the Einstein frame. Recently this
behaviour has been thoroughly investigated \cite{Coley1997a}. Only those aspects relevant to this study  shall be summarized.  The existence of
GR as an asymptotic limit at late times is also determined by the asymptotic
behaviour of the scalar field; this issue will be re-addressed in section
\ref{gr}.

For spatially homogeneous space-times the scalar field $\phi$ is
formally equivalent to a perfect fluid, and so expansion-normalized
variables can be used to study the asymptotic behaviour of Bianchi
models \cite{Coley1997a,Wainwright1997a}.  The scalar field variable,
$\Psi$, is defined by
\begin{equation}
\Psi\equiv \frac{\dot\phi}{\sqrt{6} \up{sf}H} \label{jesus1},
\end{equation}
where $\up{sf}H$ is the expansion of the timelike congruences
orthogonal to the surfaces of homoge\-neity\footnote{Note that
$\up{sf}H>0$ for all Bianchi models except those of type $I\!X$.}. At
the finite equilibrium points of the reduced system of autonomous
ordinary differential equations, where $\Psi$ is a finite constant, it
has been shown \cite{Wainwright1997a} that
$\up{sf}H=\up{sf}H_0/t^{*}$, where $t^{*}$ is the time defined in
the Einstein frame:
\begin{equation}
dt^{*}=\pm \sqrt{\Phi}\, dt. \label{jesus2}
\end{equation}
From equation  \eref{jesus1} it follows that $\dot\phi \propto 1/t^{*}$,
whence upon substitution into the Klein-Gordon equation
\begin{equation}
\ddot\phi+\up{sf}H\dot\phi+{\partial V\over \partial \phi}=0,
\label{jesus3}
\end{equation}
it can be shown that at the finite equilibrium points
\begin{equation}
\phi (t^{*})=\phi_0-{2\over k}\ln t^{*};\qquad k\neq 0,
\label{jesus4}
\end{equation}
where $\phi_0$ is a constant. Hence, from equation \eref{trans2} one
can obtain $\Phi$ as a function of $t^{*}$, provided a particular
$\omega(\Phi)$ is given.  From equation \eref{jesus2} the
relationship between $t^{*}$ and $t$ can be found, and consequently obtain $\Phi$
as a function of $t$, and hence determine the asymptotic behaviour of
$\Phi(t)$ for a given theory with specific $\omega(\Phi)$ (in the
Jordan frame).  Specifically, the
possible isotropization and inflation of a given scalar-tensor theory
in a very straightforward way can be determined.

As mentioned above, the behaviour determined from the key equation
\eref{jesus4} is not necessarily valid for all Bianchi models.  For
the Bianchi models in which the phase space is compact, the 
equilibrium points represent models that do have the behaviour described by
\eref{jesus4}, as do the finite equilibrium points in Bianchi models with
non-compact phase spaces.  It is possible that the infinite equilibrium
points in these non-compact phase spaces also share this behaviour,
although this has not been proven.  Finally, from equations \eref{SFtoSTtrans}
it can be shown that since the asymptotic behaviour is governed by
\eref{jesus4}, the corresponding transformations are non-singular and
this technique for studying the asymptotic properties of spatially
homogeneous scalar-tensor theories is valid.

\subsection{ An Example}

Consider a Brans-Dicke theory with a power-law potential, viz.,
\beqn
\omega(\Phi) & = & \omega_0 \label{bdw} \\
U & = & \beta \Phi^\alpha \label{power}
\eeqn
(where $\beta$ and $\alpha$ are positive constants), then \eref{trans2} integrates to yield \be
\Phi=\Phi_0 \ \exp\left(\frac{\phi-\phi_0}{\bar{\omega}}\right), \label{powerphi} 
\ee where \be
\bar{\omega}\equiv\pm\sqrt{\omega_0+3/2},\label{barw} \ee and hence
\eref{vstar} yields
\be
V = V_0 e^{\bar{k}\phi},
\ee
where the critical parameter $\bar{k}$ is given by 
\be \bar{k}=\frac{\alpha-2}{\bar{\omega}}. \ee
From \cite{Coley1997a} the asymptotic behaviour of
the models in the Einstein frame can now be determined, as discussed in
section \ref{exact}, for a given model with specific values of $\alpha$
and $\bar{\omega}$ (and hence a particular value for $\bar{k}$).

The possible isotropization of the given scalar-tensor theory can now
be obtained directly (essentially by reading off from the proceeding
results - see subsection 3.1).  For example, the inflationary
behaviour of the theory can be determined from equations
\eref{trans1}, \eref{jesus2} and \eref{jesus4}.  The asymptotic
behaviour of the corresponding scalar-tensor
theories (in the Jordan frame) can be further analyzed.
For instance, it can be shown from equations
\eref{jesus2}, \eref{jesus4} and \eref{powerphi}  that
asymptotically
\be
\Phi =\Phi_0 \left[\pm (t-t_0) \left(1+\frac{1}{k\bar{\omega}}\right)\right]^{-2/(1+k\bar{\omega})},
\ee
where the $\pm$ sign is determined from \eref{jesus2}.  Both this sign
and the signs of $\bar{\omega}$ and $1+k\bar{\omega}$ are crucial in
determining the relationship between $t^*$ and $t$; i.e., as $t^*
\rightarrow \infty$ either $t\rightarrow \pm \infty$ or $t\rightarrow
t_0$ and hence either $\Phi \rightarrow 0$ or $\Phi \rightarrow
\infty$, respectively, as $\phi \rightarrow -\infty$.

\subsubsection{ A Generalization}

Suppose again that $\omega=\omega_0$, so that \eref{powerphi} also
follows, but now $U$ is a sum of power-law terms of the form
\be
U = \sum^m_{n=0} \beta_n \Phi^{\alpha_n},
\ee
where $m>1$ is a positive integer.  Then \eref{vstar} becomes
\beqn
\nonumber
V & = & \sum^m_{n=0} \beta_n \Phi^{\alpha_n-2} \\
    & = & \sum^m_{n=0} \bar{\beta}_n \exp(\bar{k}_n \phi); \ \ \ 
\bar{k}_n = 
            \frac{\alpha_n-2}{\bar{\omega}}.
\eeqn
For example, if 
\be
U = U_0 + \half m \Phi^2 + \lambda \Phi^4,
\ee
then
\be
V = V_0 e^{-2\phi/\bar{\omega}} + \half \bar{m} 
    + \bar{\lambda}e^{2\phi/\bar{\omega}}
\ee
(with obvious definitions for the new constants), which is a linear
sum of exponential potentials.  Asymptotically one of these potentials
will dominate (e.g., as $\phi \rightarrow +\infty$, $V \rightarrow
\bar{\lambda}e^{2\phi/\bar{\omega}}$) and the asymptotic properties can
be deduced as in the previous section.

\subsubsection{ Approximate Forms}

In the last subsection there was a  comment upon the asymptotic properties of
a scalar-tensor theory with the forms for $\omega$ and $V$ given by
\eref{bdw} and \eref{power}.  Consider now a scalar-tensor
theory with forms for $\omega$ and $V$ which are approximately given
by \eref{bdw} and \eref{power} (asymptotically in some well-defined sense)
in order to discuss whether both theories will have the same
asymptotic properties.  In doing so, it is hoped to determine whether the
techniques discussed in this chapter have a broader applicability.

Assume that $\omega$ and $V$ are analytic at the asymptotic values
of the scalar field in the Jordan frame in an attempt to
determine whether their values correspond to the appropriate forms for
$\phi$ and $V$ in the Einstein frame, namely whether
$\phi\rightarrow -\infty$ and the leading term in $V$ is of the
form $e^{k\phi}$.

Consider an analytic expansion for $\Phi$ about $\Phi=0$:
\beqn
\omega & = & \sum_{n=0}^{\infty}\omega_n \Phi^n \\ U & = &
\sum_{n=0}^{\infty} U_n \Phi^n,
\eeqn
where all $\omega_n$ and $U_n$ are constants.  Using \eref{SFtoSTtrans} one
finds, up to leading order in $\Phi$, that for $\omega_0 \neq -3/2$
\be \phi-\phi_0 \approx \bar{\omega} \ln\Phi, \ee so
that $\phi \rightarrow \pm \infty$ (depending on the sign in
\eref{barw}) for $\Phi\rightarrow 0$.  The potential in the Einstein
frame is (to leading order) \be V \approx \exp\left\{
-\frac{2(\phi-\phi_0)}{\bar{\omega}} \right\}.  \ee Hence,
the parameter $k$ of \eref{exp} is defined here as $k\equiv
-2/\bar{\omega}$.  For $\omega_0=-3/2$
\beqn
(\phi-\phi_0)^2 & \approx & 4 \omega_1 \Phi \\ V & \approx &
\frac{16 \omega_1^2}{(\phi-\phi_0)^4},
\eeqn
so that $\phi \not{\!\!\rightarrow} -\infty$ as $\Phi\rightarrow
0$.

Next, consider an expansion in $1/\Phi$, valid for
$\Phi\rightarrow \infty$:
\beqn
\omega & = & \sum_{n=0}^{\infty}\frac{\omega_n}{ \Phi^n}, \\
U & = & \sum_{n=0}^{\infty} \frac{U_n }{\Phi^n}.
\eeqn
For $\omega_0\neq -3/2$, the results are similar to the $\Phi=0$ expansion:
\beqn
\phi-\phi_0 & \approx & - \bar{\omega} \ln\Phi \\
V & \approx & \exp\left\{ \frac{2(\phi-\phi_0)}{\bar{\omega}} \right\},
\eeqn
where now $\phi \rightarrow \mp \infty$ as $\Phi\rightarrow \infty$.
When $\omega_0=-3/2$, 
\beqn
(\phi-\phi_0)^2 & \approx & \frac{4 \omega_1}{ \Phi} \\
V & \approx & \frac{(\phi-\phi_0)^4}{16 \omega_1^2}.
\eeqn
It is apparent that the sign of $\bar{\omega}$ is important in
determining whether $\Phi\rightarrow \infty$ or $\Phi\rightarrow 0$ in order 
to obtain the appropriate form for $\phi$, as was
exemplified at the end of section 2.2.

Finally, in the event that $\omega$ and $V$ are analytic about some
finite value of $\Phi$, namely $\Phi_0$, it can be shown that $\phi
\rightarrow \phi_0$ as $\Phi \rightarrow \Phi_0$.  Hence, if one
insists that $\omega$ remain analytic as $\omega\rightarrow\omega_0$ in
the limit of $\phi\rightarrow -\infty$, then $\Phi$ must either
vanish or diverge, and the GR limit is not obtained.  This would
then suggest that if $\phi\rightarrow -\infty$ is imposed for
$\Phi \rightarrow \Phi_0$ then $\omega$ would not be analytic about
$\Phi=\Phi_0$.

\subsection{ Constraints on Possible Late-Time Behaviour\label{gr}}
In this chapter the goal is to obtain the possible asymptotic behaviour
of cosmological models in scalar-tensor theories of gravity.  However,
there are physical constraints on acceptable late-time behaviour (as $t^*
\rightarrow \infty$; see equation \eref{jesus2}).  For example, such
theories ought to have GR as an asymptotic limit at late times (e.g.,
$\omega\rightarrow \infty$ and $\Phi\rightarrow \Phi_0$) in order for
the theories to concur with observations such as solar system tests.
In addition, cosmological models must `isotropize' in order to be in
accord with cosmological observations.

Nordtvedt \cite{Nordtvedt1970a} has shown that for scalar-tensor
theories with no potential, $\omega(\Phi)\rightarrow\infty$ and
$\omega^{-3}d\omega/d\Phi \rightarrow 0$ as $t\rightarrow \infty$ in
order for GR to be obtained in the weak-field limit. Similar
requirements for general scalar-tensor theories with a non-zero
potential are not known, and as will be demonstrated from the
consideration of two particular examples found in the literature, not
all theories will have a GR limit.  

The first example is the
Brans-Dicke theory ($\omega=\omega_0=constant$) with a power-law
self-interacting potential given by \eref{power} studied earlier in
subsection 2.2. In this case, $\Phi$ is given by equations
\eref{powerphi} and \eref{barw} and the potential is given by
\eref{power}, viz.,
\be
U(\Phi)=\beta\Phi^\alpha; \qquad \alpha=2\mp k \sqrt{\omega_0+3/2}.
\ee
The $\alpha=1$ case for FRW metrics was studied by Kolitch
\cite{Kolitch1996a} and the $\alpha=2$ ($k=0$) case, corresponding to a
cosmological constant in the Einstein-frame, was considered for FRW
metrics by Santos and Gregory \cite{Santos1996a}.  Earlier it was 
considered whether anisotropic models in Brans-Dicke theory with a
potential given by \eref{power} will isotropize. Assuming a large
value for $\omega_0$, as suggested by solar system experiments, one
concludes that for a wide range of values for $\alpha$ the models
isotropize. However, in the low-energy limit of string theory where 
$\omega_0=-1$ the models are only guaranteed to isotropize for $1<\alpha<3$.

Substituting \eref{jesus4} in \eref{powerphi} yields 
\begin{equation}
\Phi\sim (t^{*})^{\pm 2\delta}, \qquad \delta={1\over k}\;\sqrt{{2\over 
3+2\omega_0}}.
\label{jesus8}
\end{equation}
Now, substituting the above expression into equation \eref{jesus1}, yields
$t^{*}$ as a function of $t$ and hence
\begin{equation}
\Phi\sim t^{{\pm 2\delta\over 1\mp\delta}}.
\label{jesus9}
\end{equation}
Depending on the sign, it is deduced from this expression that for large
$t$ the scalar field tends either to zero or to infinity and so
this theory, with the potential given by \eref{power}, does not have a
GR limit.

In the second example it was assumed that
\begin{equation}
\omega(\Phi)+{3\over 2}= { {A \Phi^2}\over \left(\Phi-\Phi_0\right)^2},
\label{jesus10}
\end{equation}
where $A$ is an arbitrary positive constant. This form for
$\omega(\Phi)$ was first considered by Mimoso and Wands
\cite{Mimoso1995a} (in a theory without a potential). Now,
\begin{equation}
\Phi=\Phi_0+B\, e^{\mp {\phi\over \sqrt{A}}},
\label{jesus11}
\end{equation}
where $B$ is a constant, and the potential, defined by equation
\eref{Vdefine}, is given by
\begin{equation}
U(\Phi)=U_0\,\Phi^2\left(\Phi-\Phi_0\right)^{\mp\sqrt{A}k}.
\label{jesus12}
\end{equation}
As before, at the equilibrium points $\Phi$ can be expressed as a
function of $t^{*}$, which allows $t$ to computed as a function of
$t^{*}$. At late times
\begin{equation}
\Phi\sim \Phi_0+t^\beta,
\label{jesus13}
\end{equation}
where $\beta$ is a constant whose sign depends on $k$, $\omega_0$ and
the choice of one of the signs in the theory. What is important
here is that in this case, at late times, the scalar
field tends to a constant value for $\beta<0$, thereby yielding a GR
limit.  In both of the examples considered above, the conformal
transformation for the equilibrium points is regular.

Of course, these are not the only possible forms for a variable
$\omega(\Phi)$.  For example, Barrow and Mimoso \cite{Barrow1994a}
studied models with $2\omega(\Phi)+3 \propto \Phi^\alpha$ $(\alpha>0)$
satisfying the GR limit asymptotically.  (The GR limit
is only obtained asymptotically as $\Phi\rightarrow \infty$, although
for a finite but large value of $\Phi$ the theory can have a limit
which is as close to GR as is required).  However, by studying the
evolution of the gravitational ``constant'' $G$ from the full Einstein
field equations (i.e., not just the weak-field approximation),
Nordtvedt \cite{Nordtvedt1970a,Nordtvedt1968b} has shown that
\be
\frac{\dot{G}}{G} = -\left( \frac{3+2\omega}{4+2\omega} \right)
\left( 1+\frac{2 \omega'}{(3+2\omega)^2} \right),
\ee
where $\omega' = d\omega/d\Phi$
(so that the correct GR limit is only obtained as $\omega\rightarrow
\infty$ and $\omega' \omega^{-3}\rightarrow 0$).  Torres
\cite{Torres1995a} showed that when $2\omega(\Phi) + 3 \propto \Phi^\alpha$,
$G(t)$ decreases logarithmically and hence $G\rightarrow 0$
asymptotically.  In the above work, no potential was included.  For a
theory with $2\omega(\Phi) +3 \propto \Phi^\alpha$ and with a non-zero
potential satisfying equation \eref{Vdefine}, then 
\be
\frac{\Phi}{U}\frac{dU}{d\Phi} = A + B \Phi^\alpha
\ee
($\alpha\neq 0$; $A$ and $B$ constants), so that 
\be
U(\Phi) = U_0 \Phi^A e^{B\Phi^\alpha/\alpha}.
\ee
A potential of this form was considered by Barrow \cite{Barrow1993c}.

Finally, Barrow and Parsons \cite{Barrow1997d} have studied three
parameterized classes of models for $\omega(\Phi)$ which permit
$\omega\rightarrow\infty$ as $\Phi\rightarrow\Phi_0$ (where the
constant $\Phi_0$ can be taken as $\Phi$ evaluated at the present
time) and hence have an appropriate GR limit;
\begin{eqnarray*}
(i) & & 2 \omega(\Phi) + 3 = 2 B_1^2 \left| 1-\Phi/\Phi_0 \right|^{-\alpha} 
\qquad (\alpha>\half), \\
(ii) & & 2\omega(\Phi) + 3 = B_2^2 \left| ln(\Phi/\Phi_0) \right|^{-2| \delta |}
\qquad (\delta>\half), \\
(iii) && 2\omega(\Phi) + 3 = B_3^2 \left| 1-(\Phi/\Phi_0)^{|\beta|} \right|^{-1}
\qquad (\forall~\beta).
\end{eqnarray*}
Other possible forms for $\omega(\Phi)$ were discussed in Barrow and
Carr \cite{Barrow1996a} and, in particular, they considered models
$(i)$ above but allowed $\alpha<0$ in order for a possible GR limit to
be obtained also as $\Phi\rightarrow \infty$.  Schwinger
\cite{Schwinger1970a} has suggested the form
$2\omega(\Phi)+3=B^2/\Phi$ based on physical considerations.

\section{ Applications}

Consider the formal equivalence of the class of scalar-tensor
theories \eref{bdaction} with $\omega(\Phi)$ and $U(\Phi)$ given by
\be
\omega(\Phi) = \omega_0, \qquad U(\Phi)=\beta\Phi^\alpha, \label{omega}
\ee 
with that of GR containing a scalar field
and an exponential potential \eref{exp}.  Indeed, since the conformal
transformation \eref{trans1} is well-defined in all cases of interest,
the Bianchi type is invariant under the transformation and 
the asymptotic properties of the scalar-tensor theories can be deduced from
the corresponding behaviour in the Einstein frame.  Also, it can be shown that
\be k\equiv\frac{\alpha-2}{\bar{\omega}}, 
\qquad \bar{\omega}^2\equiv \omega_0
+\frac{3}{2}. \label{kval} \ee

Recall that at the finite equilibrium points in the Einstein frame
 that
\beqn
\up{sf}H & =  & \up{sf}H_0t_*^{-1}, \label{Etheta} \\
\phi(t^*) & = & \phi_0 - \frac{2}{k}\ln(t^*), \label{Ephi}
\eeqn where \be \up{sf}H_0 = 1+\frac{k^2}{2} e^{k\phi_0}. 
\ee
Integrating equation \eref{trans2} yields
\be
\Phi(t^*) = d \exp{\left(\bar{\omega}^{-1}\phi(t^*)\right)} = \Phi_0 t_* 
^{-2/k\bar{\omega}}, \label{Jphi} \ee 
where the constant $\Phi_0 \equiv d \exp(\phi_0/\bar{\omega})$ and
recall that $t$ and $t^*$ are related by equation \eref{jesus2}, 
and equation \eref{trans1} can be written as
\be
\up{st}g_{\alpha\beta}=\Phi^{-1} \up{sf}g_{\alpha\beta}. \label{intrans1}
\ee

\subsection{ Examples}
{\bf 1) \ \ }
All initially expanding scalar field Bianchi models with an
exponential potential \eref{exp} with $0<k^2<2$ within general
relativity (except for a subclass of models of type IX) isotropize to the
future towards the power-law inflationary flat FRW model
\cite{Kitada1992a}, whose metric is given by
\be
\up{sf}ds^2 = -dt_*^2 + t_*^{4/k^2}\left( dx^2 + dy^2 + dz^2 \right). \label{powerlaw}
\ee 
In the scalar-tensor theory (in the Jordan frame), $\Phi$ is given by equation 
\eref{Jphi} and from \eref{intrans1} the line element can be written
\be
\up{st}ds^2 = \Phi_0^{-1} t_*^{2/k\bar{\omega}} \left\{ds^2\right\}.
\label{conform} \ee
Defining a new time coordinate by 
\be t=ct_*^{\frac{1+k\bar{\omega}}{k\bar{\omega}}}; \qquad c\equiv 
\frac{k\bar{\omega}}{1+k\bar{\omega}} \Phi_0^{-\half} \label{newT} \ee
(where $k\bar{\omega}+1 \neq 0$; i.e., $\alpha\neq 1$), after a constant 
rescaling of the spatial coordinates the line element can be written
\be
\up{st}ds^2 = -dt^2+t^{2K}\left(dX^2+dY^2+dZ^2\right), \label{newds}
\ee
where
\be
K\equiv \frac{k^2+2k\bar{\omega}}{k^2(1+k\bar{\omega})}.
\ee
Finally, the scalar field is given by 
\be
\Phi=\Phi_0 c^\frac{2}{1+k\bar{\omega}} T^\frac{-2}{1+k\bar{\omega}} 
\equiv \bar{\Phi}_0 t^\frac{2}{1-\alpha}. \label{newphi}
\ee

Therefore, all initially-expanding spatially-homogeneous models in 
scalar-tensor theories obeying
\eref{omega} with $0<(\alpha-2)^2 < 2\omega_0+3$ (except for a
subclass of Bianchi IX models which recollapse) will asymptote towards
the exact power-law flat FRW model given by equations \eref{newds}
and \eref{newphi}, which will always be inflationary since
$K=\frac{1+\alpha+2\omega_0}{(\alpha-1)(\alpha-2)} > 1$ [note that
whenever $2\omega_0 > (\alpha-2)^2-3 = \alpha^2-4\alpha + 1$, then
$1+\alpha+2\omega_0 > \alpha^2-3\alpha+2 =(\alpha-1)(\alpha-2)$].

When $k^2>2$, the models in the Einstein frame cannot inflate and may
or may not isotropize.  Consider two following examples.

{\bf 2) \ \ }  Scalar field models of Bianchi type VI$_h$ with an exponential potential \eref{exp}  with $k^2 > 2$ asymptote to the future towards the anisotropic Feinstein-Ib\'{a}\~{n}ez model \cite{Feinstein1993a} given by ($m\neq 1$)
\be
\up{sf}ds^2=-dt_*^2 + a_0^2 \left(t_*^{2p_1} dx^2 + t_*^{2p_2} e^{2mx} dy^2 + 
t_*^{2p_3} e^{2x} dz^2 \right), 
\ee
where the constants obey
\beqn\nonumber
p_1 & = & 1, \\ 
p_2 & = & \frac{2}{k^2} \left(1 + \frac{(k^2-2)(m^2+m)}{2(m^2+1)} \right), \\
\nonumber
p_3 & = & \frac{2}{k^2} \left(1 + \frac{(k^2-2)(m+1)}{2(m^2+1)} \right).
\eeqn

In the scalar-tensor theory (in the Jordan frame), $\Phi$ is given by equation
\eref{Jphi} and the metric is given by \eref{conform}.  After defining the new
time coordinate given by \eref{newT}, the line element can be written
\be
\up{st}ds^2 = -dt^2 +A_0^2 \left(t^{2q_1} dX^2 + t^{2q_2} e^{2mX}dY^2 + t^{2q_3}
e^{2X} dZ^2 
\right), \label{finds}
\ee
where
\be q_i \equiv \frac{1+k\bar{\omega}p_i}{1+k\bar{\omega}} ~~~~(i=1,2,3); 
\qquad A_0^2 =a_0^2 \Phi_0^{-1}c^{-2q_1}, \ee 
and $Y$ and $Z$ are obtained by a simple constant rescaling (and
 $X=x$).  Finally, the scalar field is given by equation
 \eref{newphi}.

The corresponding exact Bianchi VI$_h$ scalar-tensor theory solution
is therefore given by equations \eref{newphi} and \eref{finds} in the
coordinates $(t,X,Y,Z)$.  Consequently, all Bianchi type VI$_h$ models in
the scalar-tensor theory satisfying equations \eref{omega} with
$(\alpha-2)^2 > 2\omega_0 + 3$ asymptote towards the exact anisotropic
solution given by equations \eref{newphi} and \eref{finds}.

{\bf 3) \ \ } An open set of scalar field models of Bianchi type
VII$_h$ with an exponential potential with $k^2>2$ asymptote towards
the isotropic (but non-inflationary) negative-curvature FRW model
\cite{Coley1997a} with metric
\be
\up{sf}ds^2 = -dt_*^2 + t_*^2 d\sigma^2, \label{ex3}
\ee
where $d\sigma^2$ is the three-metric of a space of constant negative 
curvature.  Again, $\Phi$ is given by \eref{Jphi} and the metric is
given by \eref{conform}, which becomes after the time
recoordinatization \eref{newT}
\be
\up{st}ds^2 = -dt^2 +C^2t^2 d\sigma^2, \label{ex4}
\ee
where $C^2\equiv \Phi_0^{-1}c^{-2} = \left[\frac{1+k\bar{\omega}}
{k\bar{\omega}}\right]^2$.  This negatively-curved FRW metric is
equivalent to that given by \eref{ex3}.  Finally, the scalar field is
given by equation \eref{newphi}.

Therefore, when $(\alpha-2)^2> 2\omega_0+3$, there is an open set of
(BVII$_h$) scalar-tensor theory solutions satisfying equations
\eref{omega} which asymptote towards the exact isotropic solution
given by equations \eref{newphi} and \eref{ex4}.

Equations \eref{Ephi} and \eref{Jphi} and the resulting analysis are
only valid for scalar-tensor theories satisfying \eref{omega}.
However, the asymptotic analysis will also apply to generalized
theories of the forms discussed in subsection 2.1.2.  Finally, a
similar analysis can be applied in Brans-Dicke theory with $V=0$
\cite{Coley1998a}.  In \cite{Billyard1999a} it was shown that all
solutions derived above are self--similar; i.e. for each example,
there exists a vector $\bv{X}$ which satisfies
\be
{\cal L}_{\bv{X}_*}\up{sf}g^{\alpha\beta}={\cal C}\ \up{sf}g^*_{\alpha\beta},
\ee
where ${\cal L}_{\bv{X}_*}$ denotes the Lie derivative along $\bv{X}$
and ${\cal C}$ is a constant.

\section{Reverse Transformation: String Theory in the Einstein Frame\label{StringtoEinstein}}

The transformations (\ref{SFtoSTtrans}) can be equally applied in
reverse: given a ST theory with a power-law potential, then a GR
scalar field theory with an exponential potential can be derived.
This is particularly useful in the context of string cosmologies where
the dynamics are analysed in the Jordan frame for $\omega=-1$.  In
particular, it shall be shown that certain string theories in the
Jordan frame are conformally equivalent to a GR scalar field theory containing an
exponential potential and matter terms.
 
Taking the string action defined by (\ref{sigmaaction_i}) and apply the transformations (\ref{SFtoSTtrans}) as
well as (\ref{powerphi}) (for $\omega=-1$, $\bar\omega=\pm1/\sqrt2$),
then the following action in the Einstein frame can be written 
\begin{eqnarray}
\up{sf}S&=&\int d^4 x \sqrt{-\up{sf}g} \left\{
	\up{sf}R - \left( \nabla \phi \right)^2  
	-6 \left( \nabla \beta_m \right)^2 
	-\half \left( \nabla \bar\sigma \right)^2 e^{\mp2\sqrt2\phi} 
	-2\bar\Lambda e^{\mp\sqrt 2\phi} \right. \nonumber \\
	&& \qquad \qquad \qquad -\left.\half\bar{Q}^2e^{-6\beta_m\mp2\sqrt2\phi} 
	-\bar{\Lambda}_{\rm M}e^{\mp2\sqrt2\phi}\right\} \nonumber \\
  &\equiv& \int d^4 x \sqrt{-\up{sf}g} \left\{\up{sf}R -
	\left( \nabla \phi \right)^2 - 2V 
	-6 \left( \nabla \beta_m \right)^2 
	-\frac{A^2}{2} \left( \nabla \bar\sigma \right)^2 e^{\mp2\sqrt2\phi}
	-{2\cal U} \right\}, \nonumber \\
\label{string_Einstein}
\end{eqnarray}
where 
\be
V +{\cal U} \equiv \bar\Lambda e^{\mp\sqrt 2\phi} 
	+ \half \bar{\Lambda}_{\rm M}e^{\mp2\sqrt2\phi}
	+\quart  \bar{Q}^2e^{-6\beta_m\mp2\sqrt2\phi},\\
\ee
$A=\Phi_0^{-1} e^{\pm \sqrt2\phi_0}$, $\bar\sigma=A^2\sigma$,
$\bar\Lambda=A\Lambda$, $\bar{\Lambda}_{\rm M} =A^2 \Lambda_{\rm M}$,
$\bar{Q}^2=A^2 Q^2$ and $V=V(\phi)$ (i.e. either the $\Lambda$ term or
the $\Lambda_{\rm M}$ term).  This thesis explicitly assumes (without
loss of generality) that $\Phi_0=e^{\pm \sqrt2\phi_0}$ so that $A=1$
and $\{\bar\sigma, \bar{\Lambda}, \bar{\Lambda}_{\rm M}, \bar Q\}
\rightarrow \{\sigma, {\Lambda}, {\Lambda}_{\rm M}, Q\}$.  In the Einstein
frame, the field equations are
\begin{eqnarray}
\up{sf}T_{\alpha\beta} & = & \grad_\alpha\phi\grad_\beta\phi
	+ 6\grad_\alpha\beta\grad_\beta \beta_m
	+\half\grad_\alpha\sigma\grad_\beta \sigma e^{\mp2\sqrt2\phi}
\nonumber \\ &&
		-\up{sf}g_{\alpha\beta}\left[\half\left( \grad\phi\right)^2
				+ 3\left( \grad\beta_m\right)^2 
				+\quart\left(\grad\sigma\right)^2
				+V +\half {\cal U}\right] \label{s_E_EFE},
\end{eqnarray}
with the constraint
\begin{eqnarray}
0 &= &
\grad_\beta\phi\left[\Box\phi -\frac{dV}{d\phi}\pm \frac{\sqrt2}{2}
	\left(\grad\sigma\right)^2 e^{\mp2\sqrt2\phi}
	-\frac{\di{\cal U}}{\di\phi} \right]  \nonumber\\&&
+\half\grad_\beta\sigma\left[\Box\sigma \mp 2\sqrt2
	\grad^\alpha\sigma\grad_\alpha \right]e^{\mp2\sqrt2\phi} 
	\nonumber\\&&
+6\grad_\beta\beta_m\left[ \Box\beta_m -\frac{1}{6}\frac{\di{\cal U}}
	{\di\beta_m}\right].\label{string_conserve}
\end{eqnarray}
In general, each line in (\ref{string_conserve}) will {\em not} be
separately conserved, although the equivalent equations in the Jordan
frame will be; in fact, if the conservation equations are separately
conserved in one frame they will not generally be separately conserved
in the other frame.  Therefore, if some of the terms of action
(\ref{string_Einstein}) can be re-written in terms of a perfect fluid,
it would be expected that interaction terms in the Einstein frame
between the scalar field and the matter terms will become apparent.
Such terms provides a motivation for forms to choose
when considering interaction terms, which is the focus of Chapter
\ref{reheat}.

The models in chapters \ref{string} - \ref{Qsection} which will be
examined can contain both curvature terms (parameterized by a constant
$k$) and shear terms ($\beta_s$) and represent any one of the three
Bianchi models: type I for $k=0$, type V for $k<0$ and type IX for
$k>0$.  When the shear terms are absent, the models are curved FRW
models.  The relevant field equations for such models in the Einstein
frame are
\beqn
\mainlabel{fields_Einstein}
\up{sf}\dot H +\up{sf}H^2 - \third\left(V -\dot\phi^2+{\cal U}
	-3\left[\dot\beta_m^2+\dot\beta_s^2\right] -\half\dot{\sigma}^2
	e^{\mp 2\sqrt2\phi} \right) &=&0, \\
\dot\phi\left(\ddot\phi+3\up{sf}H\dot\phi +\frac{dV}{d\phi}\pm 
	\frac{\sqrt2}{2}\dot{\sigma}^2e^{\mp2\sqrt2\phi}+\frac{\di{\cal U}}
	{\di\phi}  \right) & = & 0,\\
\half\dot{\sigma}^2e^{\mp 2\sqrt2\phi} \left(\ddot{\sigma} 
	+3\up{sf}H\dot{\sigma} \mp 2\sqrt2\dot{\sigma}
	\dot\phi\right) & = & 0,\\
\label{modulus_beta}
6\dot\beta_m\left(\ddot \beta_m + 3\up{sf}H\dot\beta_m +\frac{1}{6}
	\frac{\di{\cal U}}{\di\beta_m}\right) & = & 0, \\
\label{shear_beta}
6\dot\beta_s\left(\ddot \beta_s + 3\up{sf}H\dot\beta_s \right) 
	& = & 0, \\
3\up{sf}H^2 +3ke^{-2\alpha^*} -\half\dot\phi^2-V
	-3\left[ \dot\beta_m^2+\dot\beta_s^2\right] -\quart\dot{\sigma}^2
	e^{\mp 2\sqrt2\phi}- {\cal U} & = & 0,
\eeqn
where $a^*\equiv e^{\alpha^*}$ is the scale factor in the Einstein frame,
related to the scale factor in the Jordan frame by
$\alpha^*=\alpha\pm\sqrt2\phi/2 -\ln A$, and
$\up{sf}H=\dot{\alpha^*}$.  Equation (\ref{shear_beta}) arises when
ensuring that $G^1_1=G^2_2=G^3_3$ (in either frame) since
$T^1_1=T^2_2=T^3_3$.  When considering the string action
(\ref{sigmaaction_i}) in this thesis, $Q$ and $\Lambda_{\rm M}$ are
never both non-zero, and so the two cases are separately treated here.

When $Q=0$ then $\di{\cal U}/\di\beta_m=0$ (hence equations
(\ref{modulus_beta}) and (\ref{shear_beta}) become identical) and
either $V=\Lambda e^{\mp \sqrt2\phi}$ ($k^2=2$), ${\cal
U}=\half\Lambda_{\rm M}e^{\mp 2\sqrt2\phi}$ or $V=\Lambda e^{\mp
2\sqrt2\phi}$ ($k^2=8$), ${\cal U}=\half\Lambda_{\rm M}e^{\mp
\sqrt2\phi}$.  In either instance, one can write ${\cal U}= {\cal
U}_0\exp(\mp a\sqrt2\phi)$ where $a=1$ or $a=2$.  Both $\beta_m$ and
$\beta_s$ may be combined via $\dot\beta^2\equiv
\dot\beta_m^2+\dot\beta_s^2$, and so a Bianchi string model with or without a modulus field, or indeed a curved FRW
string model with a modulus field, is equivalent to a GR scalar field Bianchi
model.  Equations (\ref{fields_Einstein}) can be rewritten in terms of
$\dot\beta$ in a straight forward manner (equations
(\ref{modulus_beta}) and (\ref{shear_beta}) will be replaced by a
single equation of the same form).  Now, the following identifications
are made
\beqn
\mainlabel{matter1}
\mu & \equiv & \quart \dot{\sigma}^2 e^{\mp2\sqrt2\phi} +{\cal U}, \\
p & \equiv & \quart \dot{\sigma}^2 e^{\mp2\sqrt2\phi} -{\cal U},
\eeqn
then (\ref{fields_Einstein}) may be written
\beqn
\mainlabel{fields_Einstein_1}
\up{sf}\dot H +\up{sf}H^2 & = & \third\left(V -\dot\phi^2
	-3\dot\beta^2 -\half \left(\mu+3p\right) \right), \\
\dot\phi\left(\ddot\phi+3\up{sf}H\dot\phi +\frac{dV}{d\phi}\right) 
	 &=& \mp\sqrt2\dot\phi\left[\left(1-\frac{a}{2}\right)\mu+
	\left(1+\frac{a}{2}\right)p\right]\equiv-\delta,\\
\dot\mu + 3\up{sf}H\left(\mu+p\right) &=& \pm\sqrt2\dot\phi\left
	[\left(1-\frac{a}{2}\right)\mu+\left(1+\frac{a}{2}\right)p\right]
	=+\delta,\\
\ddot \beta &=& -3\up{sf}H\dot\beta, \\
3\up{sf}H^2 +3ke^{-2\alpha^*} -3\dot\beta^2&=& \half\dot\phi^2+V+\mu,
\eeqn
which are the appropriate equations for a GR scalar field theory containing an
exponential potential and a matter field.  By choosing
$V=\Lambda e^{\mp \sqrt2\phi}$ ($k^2=2$) and ${\cal
U}=\half\Lambda_{\rm M}e^{\mp 2\sqrt2\phi}$ then the interaction
term is $\delta = \pm2\sqrt2\dot\phi p$, whilst choosing
$V=\Lambda e^{\mp 2\sqrt2\phi}$ ($k^2=8$) and ${\cal
U}=\half\Lambda_{\rm M}e^{\mp \sqrt2\phi}$ yields the interaction
term $\delta=\pm\frac{\sqrt2}{2}\dot\phi(\mu+3p)$.  Note that the
fluid here is not linear ($p\neq[\gamma-1]\mu$) in general.

For the $Q\neq0$ case, $\Lambda_{\rm M}=0$ and the shear terms and the modulus terms cannot be combined as in the previous case.  For
this case, the choice ${\cal U} = \quart
{Q}^2e^{-6\beta_m\mp2\sqrt2\phi}$ and $V=\Lambda
e^{\mp\sqrt2\phi}$ ($k^2=2$) is made.  Choosing the
identifications
\beqn
\mainlabel{matter2}
\mu_1 & \equiv &  3\dot\beta_m^2 +{\cal U}, \\
p_1 & \equiv & 3\dot\beta_m^2 -{\cal U},\\
\mu_2&=&p_2=\quart \dot{\sigma}^2 e^{\mp2\sqrt2\phi},
\eeqn
then (\ref{fields_Einstein}) may be written
\beqn
\mainlabel{fields_Einstein_2}
\up{sf}\dot H +\up{sf}H^2 & = & \third\left(V -\dot\phi^2
	-3\dot\beta_s^2 -\half \left(\mu_1+3p_1\right) 
	-2 \mu_2\right), \\
\dot\phi\left(\ddot\phi+3\up{sf}H\dot\phi +\frac{dV}{d\phi}\right) 
	 &=& \mp\sqrt2\dot\phi\left[\left(\mu_2-\mu_1\right)+
	\left(p_2+p_1\right)\right],\\
\dot\mu_1 + 3\up{sf}H\left(\mu_1+p_1\right) &=& 0\\
\dot\mu_2 + 3\up{sf}H\left(\mu_2+p_2\right) &=& \mp\sqrt2\dot\phi
	\left[\left(\mu_2-\mu_1\right)+	\left(p_2+p_1\right)\right],\\
\ddot \beta_s &=& -3\up{sf}H\dot\beta_m, \\
3\up{sf}H^2 +3ke^{-2\alpha^*} -3\dot\beta_m^2&=& \half\dot\phi^2+V
	+\mu_1+\mu_2,
\eeqn
which are the field equations of a GR scalar field theory with an exponential
potential and with two matter fields, one of which is a stiff perfect
fluid ($\gamma=2$) interacting with the scalar field, and the other
representing a non-interacting fluid that, in general,
has a non-linear equation of state ($p\neq[\gamma-1]\mu$).

This formalism will be used in Chapters \ref{string} when discussing
the qualitative analysis of the string cosmologies, primarily using
either (\ref{matter1}) or (\ref{matter2}) to comment on the equivalent
solutions in the GR scalar field theory.

\section{Discussion}

In this chapter, the asymptotic behaviour of a special subclass of
spatially homogeneous cosmological models in scalar-tensor theories,
which are conformally equivalent to general relativistic Bianchi
models containing a scalar field with an exponential potential, has
been studied by exploiting results found in previous work
\cite{Coley1997a}.

The method of studying the particular example of Brans-Dicke theory
with a power-law potential and various generalizations thereof has
been illustrated, paying particular attention to the possible
isotropization and inflation of such models.  In addition, physical
constraints on possible late-time behaviour have been discussed and,
in particular, whether the scalar-tensor theories under consideration
have a general relativistic limit at late times.  Similarly, the
reverse transformation is applied to string cosmologies in the Jordan
frame to produce in the Einstein frame a GR scalar field theory with exponential
potential and matter terms.

In particular, several exact scalar-tensor theory cosmological models
(both inflationary and non-inflationary, isotropic and anisotropic)
which act as attractors were discussed, and all such exact
scalar-tensor solutions were shown to be self-similar in Billyard {\em et al.} \cite{Billyard1999a}.

This analysis need not be limited to spatially anisotropic solutions.
For instance, in Billyard {\em et al.} \cite{Billyard1999g}, the
relationships (\ref{SFtoSTtrans}) and (\ref{potentialtransform}) were
used to generate exact solutions in the Jordan frame from known exact
spatially inhomogeneous $G_2$ models\footnote{\gtwo \ models are those
models which admit two commuting Killing vectors} in the Einstein
frame.  These models, whose metric has the form
\be
\up{sf}ds^2 = e^F\left(-dt_*^2+dz^2\right)+G\left(e^pdx^2 + e^{-p}dy^2\right)
\ee
(where all metric functions depend upon $t^*$ and $z$) contains a
scalar field $\phi=\phi(t^*,z)$ with an exponential potential.  Since
the transformations (\ref{SFtoSTtrans}) here depend in general on both $z$
and $t^*$, these transformations will typically be singular for a
particular value of $z$. However, the transformation
is well defined for $z>0$ (for example) and scalar-tensor \gtwo\ \ 
solutions can be obtained formally by analytic continuation.

The formal relations (\ref{SFtoSTtrans}) will be applied in subsequent
chapters to extend the results developed there to the alternative
frame.  For instance, the work in Chapters \ref{scaling} and
\ref{BianchiB} is confined to the Einstein frame, and the solutions to
the equilibrium points found there will be transformed to the Jordan
frame and summarized at the end of Chapter \ref{BianchiB}.  In
Chapters \ref{string} - \ref{Qsection}, string theory in the Jordan
frame is examined and so the
solutions to the equilibrium points found there will be transformed to
the Einstein frame and summarized at the end of each chapter.

\chapter{Matter Scaling Solutions: Perturbations to Shear and Curvature\label{scaling}}
Scalar field cosmological models are of great importance in the study
of the early universe, particularly in the investigation of inflation
\cite{Guth1981a,Olive1990a} .  Models with a variety of
self-interaction potentials have been studied, and one potential that
is commonly investigated and which arises in a number of physical
situations has an exponential dependence on the scalar field
\cite{Halliwell1987a,Burd1988a,Kitada1993a,Coley1997a,Wetterich1988a,Wands1993a,Copeland1997a,Ferreira1997b,Ferreira1998a,Wetterich1995a,Copeland1994b}.

A number of authors have studied scalar field cosmological models with
an exponential potential within GR.  Homogeneous and isotropic
Friedmann-Robertson-Walker (FRW) models were studied by Halliwell
\cite{Halliwell1987a} using phase-plane methods (see also \cite{Olive1990a}).
Homogeneous but anisotropic models of Bianchi types I and III (and
Kantowski-Sachs models) have been studied by Burd and Barrow
\cite{Burd1988a} in which they found exact solutions and discussed
their stability.  Lidsey \cite{Lidsey1992a} and Aguirregabiria {\em et
al.} \cite{Aguirregabiria1993b} found exact solutions for Bianchi type
I models, and in the latter paper a qualitative analysis of these
models was also presented.  Bianchi models of types III and VI were
studied by Feinstein and Ib\'{a}\~{n}ez \cite{Feinstein1993a}, in
which exact solutions were found.  A qualitative analysis of Bianchi
models with $k^2<2$, including standard matter satisfying standard
energy conditions, was completed by Kitada and Maeda
\cite{Kitada1992a,Kitada1993a}; they found that the well-known
power-law inflationary solution is an attractor for all initially
expanding Bianchi models (except a subclass of the Bianchi type IX
models which will recollapse).

The governing differential equations in spatially homogeneous Bianchi
cosmologies containing a scalar field with an exponential potential
exhibit a symmetry \cite{Bluman1989a,Coley1994d}, and when appropriate
expansion- normalized variables are defined, the governing equations
reduce to a dynamical system, which was studied qualitatively in
detail in \cite{Coley1997a} (where matter terms were not considered).  In particular, the question of
whether the spatially homogeneous models inflate and/or isotropize,
thereby determining the applicability of the so-called cosmic no-hair
conjecture in homogeneous scalar field cosmologies with an exponential
potential, was addressed. The relevance of the exact solutions (of
Bianchi types III and VI) found by Feinstein and Ib\'{a}\~{n}ez
\cite{Feinstein1993a}, which neither inflate nor isotropize, was also
considered.
In a follow up paper \cite{vandenHoogen1997b} the isotropization of
the Bianchi VII$_h$ cosmological models possessing a scalar field with
an exponential potential and no matter was further investigated; in the case
$k^2>2$, it was shown that there is an open set of initial conditions
in the set of anisotropic Bianchi VII$_h$ initial data such that the
corresponding cosmological models isotropize asymptotically.  Hence,
scalar field spatially homogeneous cosmological models having an
exponential potential with $k^2>2$ can isotropize to the future.
However, in the case of the Bianchi type IX models having an
exponential potential with $k^2>2$ the result is different in that
typically expanding Bianchi type IX models do not 
isotropize to the future; the analysis of \cite{vandenHoogen1999a}
indicates that if $k^2>2$, then the model recollapses.

Recently cosmological models which contain both a perfect fluid
description of matter and a scalar field with an exponential potential
have come under heavy analysis.  One of the exact solutions found for
these models has the property that the energy density due to the
scalar field is proportional to the energy density of the perfect
fluid, hence these models have been labelled matter scaling
cosmologies
\cite{Wetterich1988a,Wands1993a,Copeland1997a,Ferreira1997b,Ferreira1998a,Wetterich1995a}.  With the discovery of these matter scaling
solutions, it has become imperative to study spatially homogeneous
Bianchi cosmologies containing a scalar field with an exponential
potential and an additional matter field consisting of a barotropic
perfect fluid.  The matter scaling solutions studied in
\cite{Wetterich1988a,Wands1993a,Copeland1997a,Ferreira1997b,Ferreira1998a,Wetterich1995a},
which are spatially flat isotropic models in which the scalar field
energy density tracks that of the perfect fluid, are of particular
physical interest (e.g., dark matter candidate).  It is known that the matter scaling
solutions are late-time attractors (i.e., stable) in the subclass of
flat isotropic models \cite{Wetterich1988a,Wands1997a,Copeland1997a,Ferreira1997b,Ferreira1998a,Wetterich1995a,Copeland1997a}.

In addition to the scaling solutions described above, curvature
scaling solutions and anisotropic scaling solutions are also possible.
In \cite{vandenHoogen1999b} homogeneous and isotropic spacetimes with
non-zero spatial curvature were studied in detail and three possible
asymptotic future attractors in an ever-expanding universe were
found. In addition to the zero-curvature power-law inflationary
solution and the zero-curvature matter scaling solution alluded to
above, there is a solution with negative spatial curvature where the
scalar field energy density remains proportional to the curvature
which also acts as a possible future asymptotic attractor.  In
\cite{Coley1998b} spatially homogeneous models with a perfect fluid
and a scalar field with an exponential potential were also studied and
the existence of anisotropic scaling solutions was discovered;
the stability of these anisotropic scaling solutions within a
particular class of Bianchi type models was discussed.  These
anisotropic scaling solutions are indeed matter scaling solutions, but
the name is used to indicate the anisotropic nature of the solution,
whereas the term matter scaling solution is used specifically for the isotropic
solution.

Clearly these matter scaling models are of potential cosmological
significance.  It is consequently of prime importance to
determine the genericity of such models by studying their stability in
the context of more general spatially homogeneous models.  It is this
question that shall be addressed in this chapter.  This chapter, a
precursor to the next, performs a perturbation analysis of the matter
scaling solutions to determine whether or not they are stable to shear
and curvature perturbations.  The results herein will be generalized
in the next chapter where scalar field with exponential potentials are
examined in Bianchi class B models.  Since this chapter and the next
primarily works in the Einstein frame, the index ``(sf)'' shall be
suppressed for ease in notation.

\section{The Matter Scaling Solutions} 

The governing equations for a scalar field with an 
exponential potential
$V = V_0 e^{k \phi}, $
where $V_0$ and $k$ are positive constants, evolving in a flat
FRW model containing a separately conserved perfect which satisfies
the linear equation of state 
$p_\gamma = (\gamma -1) \mu_\gamma,$
where the constant $\gamma$ satisfies $0\le\gamma\le 2$ (although we
shall only be interested in the range $0<\gamma < 2$ here), are given
by
\beqn 
\mainlabel{governing_eqns}
\label{Hdot}
\dot{H} &= -\frac{1}{2} (\gamma \mu_\gamma + \dot{\phi}^2), \\
\label{the_mu}
\dot{\mu}_\gamma &= -3 \gamma H \mu_\gamma, \\
\label{the_p}
\ddot{\phi}&= -3H\dot{\phi} - k V, 
\eeqn
subject to the Friedmann constraint
\be
\label{Fried_1}
H^2 = \frac{1}{3} (\mu_\gamma + \frac{1}{2} \dot{\phi}^2 + V), 
\ee
where $H$ is the Hubble parameter and an overdot denotes ordinary
differentiation with respect to time $t$.  Note that the
total energy density of the scalar field is given by
$\mu_\phi = \frac{1}{2}\dot{\phi}^2 +V$. 

Defining
\be
\label{def_eqns}
x \equiv \frac{\dot{\phi}}{\sqrt{6}H} \q , 
	\q y \equiv \frac{\sqrt{V}}{\sqrt{3}H}, 
\ee
and the new logarithmic time variable $\tau$ by
\be
\label{new_time}
\frac{d \tau}{dt} \equiv H,
\ee
equations (\ref{governing_eqns}) can be written as the
plane-autonomous system \cite{Copeland1997a}:
\beqn
\mainlabel{xy_1}
\label{x_1}
x' & = -3x - \sqrt{\frac{3}{2}} k y^2 + \frac{3}{2} x [2x^2 + \gamma (1-x^2 -y^2)], \\
\label{y_1}
y' &= \frac{3}{2}y \left[-\sqrt{\frac{2}{3}} k x + 2x^2 + \gamma (1-x^2 -y^2)\right], 
\eeqn
where a prime denotes differentiation with respect to $\tau$, and 
equation (\ref{Fried_1}) becomes 
\benonumber \Omega + \Omega_\phi =1, \eenonumber
where
\be
\label{O_1}
\Omega \equiv \frac{\mu_\gamma}{3 H^2}, \q \Omega_\phi \equiv 
	\frac{\mu_\phi}{3H^2} = x^2 + y^2, 
\ee
which implies that $0 \leq x^2 +y^2 \leq 1$ for $\Omega \geq 0$ so that the
phase-space is bounded.

A qualitative analysis of this plane-autonomous system
is given in \cite{Copeland1997a}.  The well-known power-law inflationary
solution for $k^2<2$ \cite{Halliwell1987a,Burd1988a,Kitada1993a,Coley1997a} corresponds to the equilibrium point
$x = k /\sqrt{6}$, $y = (1 - k^2/6)^{1/2}$ ($\Omega_\phi =1$, $\Omega =0$) of the system (\ref{xy_1}), which is shown to be
stable (i.e., attracting) for $k^2 < 3 \gamma$ in the presence of a
barotropic fluid.  Previous analysis has shown that when $k^2<2$
this power-law inflationary solution is a global attractor in 
spatially homogeneous models in the absence of a perfect 
fluid (except for a subclass of Bianchi type IX models 
which recollapse).

In addition, for $\gamma > 0$ there exists a matter scaling solution
corresponding to the equilibrium point \index{equilibrium sets!${\cal F}_S(I)$, ${\cal F}_S$}
\be
\label{eqm_pts_1}
x = x_0 = -\sqrt{\frac{3}{2}} \frac{\gamma}{k}, \q 
y = y_0 = [3 (2 - \gamma) \gamma/2
k^2]^{\frac{1}{2}}, 
\ee
 whenever $k^2 > 3\gamma$.  The
linearization of system (\ref{xy_1}) about the equilibrium point 
(\ref{eqm_pts_1})
yields the two eigenvalues with negative real parts
\be
\label{eigen_1}
-\frac{3}{4}\left( 2-\gamma\right) \pm \frac{3}{4k}
\sqrt{(2-\gamma) [24\gamma^2-k^2(9\gamma-2)]}
\ee 
when $\gamma<2$.
The equilibrium point is con\-se\-quently sta\-ble (a spi\-ral for $k^2 >
\frac{24 \gamma^2}{(9\gamma-2)}$, else a node) so that the corresponding
cosmological solution is a late-time attractor in the class of flat
FRW models in which neither the scalar-field nor the perfect fluid
dominates the evolution.  The effective equation of state parameter for the
scalar field is given by 
\benonumber
\gamma_\phi \equiv \frac{(\mu_\phi +p_\phi)}{ \mu_\phi} = 
	\frac{2x^2_0}{x^2_0 + y^2_0} = \gamma, 
\eenonumber 
which is the same as the equation of state parameter for the perfect fluid.
The solution is referred to as a matter scaling solution since the energy
density of the scalar field remains proportional to that of the
barotropic perfect fluid according to $\Omega/\Omega_\phi =
k^2/3\gamma -1$ \cite{Wetterich1988a,Wands1993a}.  Since the
matter scaling solution corresponds to an equilibrium point of the system
(\ref{xy_1}) it is a self-similar cosmological model
\cite{Wainwright1997a}.

\section{Stability of the Matter Scaling Solution}

The stability of the matter scaling solution shall be studied here 
with respect to anisotropic and curvature perturbations within
the class of spatially homogeneous models.

\subsection{Bianchi I models\label{BI}}  

In order to study the stability of the matter scaling solution with respect
to shear perturbations the class of anisotropic Bianchi I models shall
be investigated first, which are the simplest spatially homogeneous
generalizations of the flat FRW models which have non-zero shear but
zero three-curvature.  The governing equations in the Bianchi I models
are equations (\ref{the_mu}) and (\ref{the_p}), and equation
(\ref{Fried_1}) becomes 
\be
\label{Fried_2}
H^2 = \frac{1}{3} \left(\mu_\gamma +
\frac{1}{2} \dot{\phi}^2 + V\right) + \Sigma^2,
\ee  where $\Sigma^2 =\frac{1}{3} \Sigma^2_0 a^{-6}$
is the contribution due to the shear,
where $\Sigma_0$ is a constant and $a$ is the scale factor.  Equation
(\ref{Hdot}) is replaced by the time derivative of equation (\ref{Fried_2}).

Using the definitions (\ref{def_eqns}), (\ref{new_time}) and
(\ref{O_1}) the governing ordinary differential
equations can be deduced. Due to the $\Sigma^2$ term in (\ref{Fried_2}) this equation
can no longer be used to substitute for $\mu_\gamma$ in the remaining
equations, and consequently the three-dimensional autonomous system is
obtained:
\beqn
\mainlabel{xy_2}
\label{x_2}
x' &=& -3x - \sqrt{\frac{3}{2}} k y^2 + \frac{3}{2} x [2 + (\gamma -2) \Omega -2y^2], \\
\label{y_2}
y' &=& \frac{3}{2}y \left\{  \sqrt{\frac{2}{3}} k x +2 + (\gamma -2) \Omega
-2y^2\right\}, \\
\label{O_2}
\Omega' &=& 3 \Omega \{(\gamma -2)(\Omega -1) -2y^2\}, 
\eeqn
where equation (\ref{Fried_2}) yields 
$1 - \Omega -x^2 -y^2 = \Sigma^2 H^{-2} \geq 0, $
so that again the phase-space is bounded.

The matter scaling solution, corresponding to the flat FRW 
solution, is now represented by the equilibrium point
\be
\label{eqm_pts_2} 
x = x_0, ~~y = y_0, ~~\Omega =1 - \frac{3 \gamma}{k^2}. 
\ee
The linearization of system (\ref{xy_2}) about the equilibrium point
(\ref{eqm_pts_2}) yields three eigenvalues, two of which are given by
(\ref{eigen_1}) and the third has the value $-3(2-\gamma)$, all with
negative real parts when $\gamma<2$.  Consequently the matter scaling
solution is stable to Bianchi type I shear perturbations.

\subsection{Curved FRW models\label{Curved_FRW}}  

In order to study the stability of the matter scaling solution with respect
to curvature perturbations the class of FRW models which have
curvature but no shear shall first be studied. Again equations
(\ref{the_mu}) and (\ref{the_p}) are valid, but in this case equation
(\ref{Fried_1}) becomes
\be
\label{Fried_3}
H^2 = \frac{1}{3} (\mu_\gamma +
\frac{1}{2} \dot{\phi}^2 +V) + K,
\ee 
where $K = -k a^{-2}$ and $k$ is a
constant that can be scaled to $0$, $\pm 1$.  Equation (\ref{Hdot}) is
again replaced by the time derivative of equation (\ref{Fried_3}).

As in the previous case equation (\ref{Fried_3}) cannot be used to
replace $\mu_\gamma$, and using the definitions (\ref{def_eqns}),
(\ref{new_time}) and (\ref{O_1}) the three-dimensional
autonomous system is obtained:
\beqn
\mainlabel{xy_3}
\label{x_3}
x' &= -3x - \sqrt{\frac{3}{2}}k y^2 + \frac{3}{2}x \left[ \left( 
	\gamma - \frac{2}{3} \right) \Omega + \frac{2}{3} (1 + 2x^2 -y^2)  
	\right], \\
\label{y_3}
y' &= \frac{3}{2} y \left\{ \sqrt{\frac{2}{3}} k x + \left( \gamma 
	-\frac{2}{3}  \right) \Omega + \frac{2}{3} (1+2x^2 -y^2) \right\}, \\
\label{O_3}
\Omega' &= 3 \Omega \left\{\left(\gamma - \frac{2}{3}\right)(\Omega -1) + 
	\frac{2}{3} (2x^2 -y^2)  \right\}, 
\eeqn
where 
\benonumber
1 - \Omega - x^2 -y^2 = KH^{-2}.
\eenonumber
The phase-space is bounded for $k =0$ or $k=-1$, but not for $k = +1$.

The matter scaling solution again corresponds to the equilibrium point
(\ref{eqm_pts_2}).  The linearization of system (\ref{xy_3}) about
this equilibrium point yields the two eigenvalues with negative real
parts given by (\ref{eigen_1}) and the eigenvalue $(3 \gamma
-2)$. Hence the matter scaling solution is only stable for $\gamma <
\frac{2}{3}$.  For $\gamma > \frac{2}{3}$ the equilibrium point
(\ref{eqm_pts_2}) is a saddle with a two-dimensional stable manifold
and a one-dimensional unstable manifold.

Consequently the matter scaling solution is unstable
to curvature perturbations in the case of realistic
matter $(\gamma \geq 1)$; i.e., the matter scaling solution is no longer
a late-time attractor in this case.  However, the matter scaling solution
does correspond to an equilibrium point of the governing 
autonomous system of ordinary differential equations
and hence there are cosmological models that can spend
an arbitrarily long time `close' to this solution.  Moreover,
since the curvature of the universe is presently constrained
to be small by cosmological observations, it is possible
that the matter scaling solution could be important in 
the description of the actual universe.  That is, not enough
time has yet elapsed for the curvature instability to
have effected an appreciable deviation from the flat
FRW model (as in the case of the standard perfect fluid
FRW model).

Hence the matter scaling solution may still be of physical interest.  To
further study its significance it is important to determine its
stability in a general class of spatially homogeneous models.  We
shall therefore study the stability of the matter scaling solution in the
(general) class of Bianchi type VII$_h$ models, which are perhaps the most
physically relevant models since they can be regarded as
generalizations of the open (negative-curvature) FRW models.

\subsection{Bianchi VII$_h$ models}  

The Bianchi VII$_h$ models are sufficiently complicated that a simple
coordinate approach (similar to that given above) is not desirable.
To study Bianchi VII$_h$ spatially homogeneous models with a minimally
coupled scalar field with an exponential potential and a barotropic
perfect fluid it is best to employ a group-invariant orthonormal frame
approach with expansion-normalized state variables governed by a set
of dimensionless evolution equations (constituting a `reduced'
dynamical system) with respect to a dimensionless time subject to a
non-linear constraint \cite{Wainwright1997a}, generalizing previous
work in which there is no scalar field \cite{Hewitt1993a} and in which
there is no matter
\cite{vandenHoogen1997a}.

The reduced dynamical system is seven-dimensional (subject to a
constraint) \cite{Billyard1998a}.  The matter scaling solution is again an
equilibrium point of this seven-dimensional system. This equilibrium
point, which only exists for $k^2>3\gamma$, has two eigenvalues given
by (\ref{eigen_1}) which have negative real parts for $\gamma<2$, two
eigenvalues (corresponding to the shear modes) proportional to
$(\gamma -2)$ which are also negative for $\gamma <2$, and two
eigenvalues (essentially corresponding to curvature modes)
proportional to $(3 \gamma -2)$ which are negative for $\gamma
<\frac{2}{3}$ and positive for $\gamma >\frac{2}{3}$
\cite{Billyard1998a}.  The remaining eigenvalue (which also
corresponds to a curvature mode) is equal to $3\gamma-4$.  Hence for
$\gamma <
\frac{2}{3}$ ($k^2 > 3 \gamma$) the matter scaling solution is again
stable.  However, for realistic matter ($\gamma \geq 1$) the
corresponding equilibrium point is a saddle with a (lower) four- or
five-dimensional stable manifold (depending upon whether $\gamma>4/3$
or $\gamma<4/3$, respectively).

\section{Discussion} 

Perhaps these stability results can be understood heuristically as
follows.  From the conservation law the barotropic matter redshifts as
$a^{-3\gamma}$.  In subsection \ref{BI} the shear $\Sigma^2$ redshifts
as $a^{-6}$ and so always redshifts faster than the matter, resulting
in the stability of the matter scaling solution.  Note that the bifurcation
that occurs at $\gamma=2/3$ in subsection \ref{Curved_FRW} corresponds
to the case in which the curvature $K$ is formally equivalent to a
barotropic fluid with $\gamma=2/3$, and in which both the matter and
the curvature redshift as $a^{-2}$.  For $\gamma>2/3$, the barotropic
matter redshifts faster than $a^{-2}$ and the curvature eventually
dominates.  A complete qualitative analysis of cosmological models
with a perfect fluid and a scalar field with an exponential potential
will be performed in the next chapter.

\chapter{Matter Scaling Solutions in Bianchi Class B Models\label{BianchiB}}

The purpose of this chapter is to comprehensively study the qualitative
properties of spatially homogeneous models with a barotropic fluid and
a non-interacting scalar field with an exponential potential in the
class of Bianchi type B models (except for the exceptional case
Bianchi VI$_{-1/9}$), using the Hewitt and Wainwright formalism
\cite{Wainwright1997a,Hewitt1993a}. In particular, the generality of the 
scaling solutions shall be studied.  The chapter is organized as
follows.  In section \ref{I} the governing equations are defined, which are
modified from those developed in \cite{Wainwright1997a}, and the invariant 
sets and the existence of monotonic functions are discussed.  In section
\ref{III}, all of the equilibrium points are classified and listed, and their
local stability is discussed in section \ref{IV}.  Section
\ref{BianchiB_Jordan} lists the equilibrium points in the Jordan frame
and discussed values of $\omega_0$ for which the models can inflate.
Conclusions and discussion are reserved for section \ref{VI}.

\section{ The Equations\label{I}}

It shall be assumed that the matter content is composed of two
non-interacting components. The first component is a separately
conserved barotropic fluid with a gamma-law equation of state, i.e.,
$p=(\gamma-1)\mu$, where $\gamma$ is a constant with $0\leq\gamma\leq
2$, while the second is a noninteracting scalar field $\phi$ with an
exponential potential $V(\phi)=V_0 e^{k\phi}$, where $V_0$ and $k$ are
positive constants (units in which $8\pi G=c=1$ are used). The term
``non-interacting'' means that the energy-momentum of the two matter
components will be separately conserved.

The state of any Bianchi type B model with the above matter content
can be described in a tetrad formalism requiring no specific form for the
metric and can be described by the evolution of the variables
\begin{equation}
  \left( H, \sigma_+, \tilde \sigma, \delta, \tilde a, n_+,\dot
	\phi,\phi\right)\in \mR^8,
\end{equation} 
where $H$ is the expansion rate of the fluid(s), $\tilde\sigma\equiv
\frac{3}{2} \tilde\sigma^{ab}\tilde\sigma_{ab}$ is the ``magnitude''
of the trace-free component of the fluid congruence's shear rate,
$\sigma_{ab}$
($\tilde\sigma_{ab}=\sigma_{ab}-\half\sigma^c_c\delta_{ab}$),
$\sigma_+$ is proportional to the shear rate's trace,
$\sigma_+=\frac{3}{2}\sigma^a_a$, the scalars $\tilde a$ and
$n_+\equiv \frac{3}{2}n^a_a$ describe the curvature of the spacelike
hypersurfaces orthogonal to the fluid congruence and
$\delta\equiv\frac{3}{2}\tilde\sigma_{ab}\tilde n^{ab}$ where ($\tilde
n_{ab}$ is the trace-free component of $n_{ab}$).  The evolution of the
state variables are given as \cite{Hewitt1993a}
\beqn
3\dot H +3H^2 & = & -\frac{2}{3}\left(\sigma_+^2+\tilde\sigma\right) +V
	-\dot\phi^2 -\half\left(3\gamma - 2\right)\mu,  \\
\dot\sigma_+ & = & -3H\sigma_+ -\frac{2}{3}\tilde n, \\
\dot{\tilde\sigma} & = & -6H\sigma -\frac{4}{3}n_+\delta -\frac{4}{3}
	\tilde a\sigma_+, \\
\dot\delta & = & \frac{2}{3}\left(\sigma_+-6H \right) +\frac{2}{3}n_+
	\left(\tilde\sigma-\tilde n\right), \\
\dot{\tilde a} & = & \frac{2}{3}\left(2\sigma_+-3H \right)\tilde a, \\
\dot{n_+} & = & \third\left(2\sigma_+-3H \right)n_+ +2\delta,
\eeqn
where $\tilde n\equiv \third(n_+^2-l\tilde a)$, 
with the addition
of the Klein-Gordon equation for the scalar field,
\begin{equation}
\ddot \phi+3H\dot\phi+kV(\phi)=0
\end{equation}
(in \cite{Hewitt1993a} scalar fields are not included, but can be 
in a straight forward manner).

By introducing dimensionless variables, the evolution equation for
$H$ decouples and the resulting reduced system has one less dimension 
\cite{Wainwright1997a}.  Defining
\cite{Wainwright1997a,Coley1997a} 
\begin{eqnarray}
\Sigma_+&=&\frac{\sigma_+}{3H}, \quad \tilde{\Sigma}=\frac{\tilde\sigma}{9H^2},
\quad \Delta= \frac{\delta}{9H^2}, \quad \tilde A = \frac{\tilde a}{9H^2},
\nonumber \\ N_+&=&\frac{n_+}{3H},\quad \Psi=\frac{\dot \phi}{\sqrt{6}H}, \quad
\Upsilon=\frac{\sqrt{V(\phi)}}{\sqrt 3H},\quad \Omega=\frac{\mu}{3H^2}, \label{variable_def}
\end{eqnarray} the differential equations for the quantities 
\begin{equation}{\bf X}=
\left( \Sigma_+, \tilde{\Sigma}, \Delta, \tilde A,  N_+,\Psi,\Upsilon\right)\in
\mR^7  \label{TheSet}\end{equation}  \ are as follows:

\beqn
\mainlabel{DE_all}
\Sigma_+'   	&=& (q - 2)\,{\Sigma _{+}} - 2\,{\tilde N},\label{DE_1}\\
\tilde{\Sigma}'	&=& 2\,(q - 2)\,{\tilde{\Sigma}} - 4\,\Delta \,N_+ 
			- 4\,{\Sigma _{+}}\,\tilde A, \\
\Delta'    	&=& 2\,(q + {\Sigma _{+}} - 1)\,\Delta  + 2\,({\tilde{\Sigma}} 
			-{\tilde N})\,N_+, \\
\tilde{A}'   	&=& 2\,(q + 2\,\Sigma_+)\,\tilde A, \\
N_+'        	&=& (q + 2\,\Sigma_+)\,N_+ + 6\,\Delta, \\
\label{PsiDE}
\Psi'       &=& (q - 2)\,\Psi  - {\displaystyle \frac {1}{2}} \,\sqrt{6}\,k\,\Upsilon ^{2},\\
\label{PhiDE}
\Upsilon'       &=& (q+1  + {\displaystyle \frac {1}{2}} \,\sqrt{6}\,k\,\Psi 
)\,\Upsilon, \label{DE_7}
\eeqn
where a prime denotes differentiation with respect to the time $\tau$, where
$dt/d\tau=H$.  
The deceleration parameter $q$ is defined by $q\equiv-(1+H'/H)$, and both
$\tilde N$ (a curvature term) and $\Omega$ (a matter term) are obtained from first integrals:
 \beqn
  q  &=&2\,\Sigma_+^{2} + 2\,{\tilde{\Sigma}} + 
{\displaystyle \frac 
     {1}{2}}\,(3\,\gamma  - 2)\,\Omega  + 2\,\Psi ^{2} - \Upsilon ^{2},\\
\tilde N &=&{\displaystyle \frac {1}{3}} \,N_+^{2} - 
   {\displaystyle \frac {1}{3}} \,l\,\tilde A,\label{tilden}\\
\Omega &=& 1 - \Psi ^{2} - \Upsilon ^{2} - \Sigma_+^{2} - {\tilde{\Sigma}} - {\tilde N} - \tilde A .\label{constraint2}
\eeqn
The evolution of $\Omega$ is given by the auxiliary equation
\begin{equation}
\Omega'           = \Omega \,(2\,q - 3\,\gamma  + 2). \label{auxiliary}
\end{equation}
The parameter $l= 1/h$ where h is the group parameter is equivalent to
Wainwright's $\tilde h$ in \cite{Wainwright1997a}.  If $l<0$ and $\tilde A>0$ then the model 
is of Bianchi
type VI$_h$.  If $l>0$ and $\tilde A>0$ and $N_+\not= 0$ then the model is of
Bianchi type VII$_h$. If $l =0$ then the model is either Bianchi IV or V.  
If $\tilde A =0$ then the model is either a Bianchi type I or a Bianchi type II
model.

There is one constraint equation that must also be satisfied:
\begin{equation}  
G({\bf X})={\tilde{\Sigma}}\,{\tilde N} - \Delta ^{2} - \tilde A\,
   \Sigma_+^{2}= 0 \label{constraint1},   
   \end{equation}
Therefore the state space is six-dimensional; the seven evolution equations
(\ref{DE_all}) are subject to the constraint equation 
(\ref{constraint2}).  The seven-dimensional state space
(\ref{TheSet}) shall be referred to as the {\em extended} state space.

By definition $\tilde A$ is
non-negative, which implies from equations (\ref{constraint1}) and (\ref{tilden}) that $
\tilde{\Sigma}$ and $\tilde N$ are also non-negative.   Thus 
\begin{equation}
\tilde A\geq0,\qquad\qquad \tilde{\Sigma}\geq0,\qquad\qquad \tilde N \geq 
0.  
\end{equation} 
In addition, from the physical constraint $\Omega\geq0$ together with
equation (\ref{constraint2}), it is easily verified that the state space is
compact.  Indeed, the variables are bounded by
\begin{equation}
0\leq\left\{\Sigma_+^2, \tilde\Sigma, \Delta^2, \tilde A,\tilde N, \Psi^2,
\Upsilon \right\}\leq 1.
\end{equation}
Since both $\tilde A$ and $\tilde N$ are bounded, then from equation
(\ref{tilden}) it is apparent $N_+$ is bounded.  In equation
(\ref{variable_def}) the ``positive square root'' shall be assumed.
In principle, there exists negative and positive values for
$\Upsilon$, but from the definition (\ref{variable_def}) a negative
$\Upsilon$ implies a negative $H$ and hence $H<0$ for all time; i.e.,
the models are contracting.  Since the system is invariant under
$\Upsilon\rightarrow -\Upsilon$, without loss of generality, only
$\Upsilon\ge0$ shall be considered.

\subsection{Invariant Sets\label{IIA}}

There are a number of important invariant sets.  Recall that the state
space is constrained by equation (\ref{constraint1}) to be a
six-dimensional surface in the seven-dimensional {\em extended} space.  
Taking the constraint equation (\ref{constraint1}) into account the dimension of each invariant set shall be counted.
These invariant sets can be classified into various classes according
to Bianchi type and/or according to their matter content. Some invariant
sets (notably the Bianchi invariant sets) have lower-dimensional
invariant subsets. Equilibrium points and orbits occuring in each
Bianchi invariant set correspond to cosmological models of that
Bianchi type.  The notation used here has been adapted from
\cite{Wainwright1997a}.  Various lower-dimensional invariant sets can
be constructed by taking the intersection of any Bianchi invariant set
with the various Matter invariant sets. For example, $B(I)\cap {\cal M}$ is
a 3-dimensional invariant set describing Bianchi type I models with a
massless scalar field.
 
\begin{table}
\begin{center}
\begin{tabular}{|llcl|}\hline
Bianchi &&&\\ Type& Notation & Dim.  & Restrictions \\
\hline \hline
{\it I}  &      $B(I)$ &    4 & $\tilde{A}=\Delta=N_+=0$\\
                 &      $S(I)$ &    2 & $\tilde{A}=\Sigma_+=\tilde\Sigma=\Delta=
                                        N_+=0$ \\
\hline
{\it II} & $B^\pm(II)$ &    5 & $\tilde{A}=0, \q N_+>0 \mbox{ or } 
                                         N_+<0$\\
                 & $S^\pm(II)$ &    4 & $\tilde{A}=0, \q \tilde\Sigma=
                                          3{\Sigma_+}^2, \q \Delta={\Sigma_+}
                                          {N_+}$ \\
\hline
{\it IV} & $B^\pm(IV)$ &    6 & $l=0,\q \tilde{A}>0, \q N_+>0 
                                          \mbox{ or } N_+<0$\\
\hline
{\it V}  &      $B(V)$ &    4 & $l=0,\q \tilde{A}>0, \q 
                                         \Sigma_+=\Delta=N_+=0$\\
                 &      $S(V)$ &    3 & $l=0,\q \tilde{A}>0, \q \Sigma_+=\tilde\Sigma=
                                        \Delta=N_+=0$\\
\hline
{\it VI$_h$}&$B(VI_h)$ &    6 & $l<0, \q \tilde{A}>0$\\
                 &   $S(VI_h)$ &    4 & $l<0,\q \tilde{A}>0, \q 3\Sigma_+^2+l\tilde
					\Sigma = 0, \q 	N_+=\Delta=0$\\ 
                 &$S^\pm(III)$ &    5 & $l=-1,\q \tilde{A}>0, \q 3\Sigma_+^2-
					\tilde\Sigma = 0,\q {\Delta}=
                                        {\Sigma_+}{N_+}$\\
\hline 
{\it VII$_h$}&$B^\pm(VII_h)$&6& $l>0,\q \tilde{A}>0, \q N_+>0 
                                        \mbox{ or } N_+<0$\\
                 & $S^\pm(VII_h)$ & 3 & $l>0,\q \tilde{A}>0,\q \Sigma_+=\tilde\Sigma=\Delta=0, 
					\q N_+^2=l\tilde{A}>0$\\
\hline
\end{tabular}
\end{center}
\caption[{\em Bianchi Invariant Sets}]{{\em Bianchi Invariant Sets.
The third column represents the dimension of the phase space.  Note
that $B(I)$ and 
$B^\pm(II)$ are class A Bianchi invariant sets which occur in the
closure of the appropriate higher-dimensional Bianchi type B invariant
set (see Fig. 1).  In addition, if $l$ is non-negative, $N_+>0$ and
$N_+<0$ define disjoint invariant sets (indicated by a superscript
$\pm$ in the table).  Due to the discrete symmetry $\Delta\rightarrow
-\Delta$, $N_+\rightarrow -N_+$, these pairs of invariant sets are
equivalent.}\label{Table1}}
\end{table}

\begin{table}[htp]
\begin{center}
\begin{tabular}{|llcl|}
\hline
Matter Content  & Notation & Dimension & Restrictions \\
\hline \hline
Scalar Field            & ${\cal S}$   & 5 & $\Omega =0$; 
	\q $\Psi\not=0,\q \Upsilon\not =0$ \\ \hline
Massless Scalar Field   & ${\cal M}$   & 4 & $\Omega =0$; 
	\q $\Psi\not =0,\q \Upsilon =0$ \\ \hline
Vacuum                  & ${\cal V}$     & 3 & $\Omega =0$; 
	\q $\Psi=0,\q \Upsilon =0$ \\
\hline \hline
Perfect Fluid &&& \\
+ Scalar Field        & ${\cal FS}$ & 6 & $\Omega\neq 0$;
	\q $\Psi\not =0,\q \Upsilon \not =0$  \\ \hline
Perfect Fluid &&& \\
+ Massless Scalar Field     & ${\cal FM}$ & 5 & $\Omega\neq 0$;
	\q $\Psi\not=0,\q \Upsilon =0$  \\ \hline
Perfect Fluid                       & ${\cal F}$     & 4 & $\Omega\neq 0$;
	\q $\Psi=0,\q \Upsilon =0$  \\ \hline
\end{tabular}
\end{center}
\caption{{\em Matter Invariant Sets.}\label{Table2}}
\end{table} 

An analysis of the dynamics in the invariant sets ${\cal V}$ and ${\cal F}$ has
 been presented by Wainwright and Hewitt \cite{Hewitt1993a}.
 Equilibrium points and orbits in the invariant set ${\cal M}$
 correspond to models with a massless scalar field; i.e., scalar field
 models with zero potential.  These models are equivalent to models
 with a stiff perfect fluid (i.e., $\gamma=2$) equation of state; see
 \cite{Hewitt1993a}.  Equilibrium points and orbits in the
 invariant set ${\cal FM}$ can be interpreted as representing a
 two--perfect--fluid model with $\gamma_2=2$ \cite{Coley1992c}. A
 partial analysis of the isotropic equilibrium points in the invariant set
 ${\cal S}$ was completed by van den Hoogen et
 al. \cite{vandenHoogen1997a}.  Note that the so-called
 scaling solutions \cite{Copeland1997a,Wetterich1988a,Wands1993a} are in
 the invariant set ${\cal FS}$.

The isotropic and spatially homogeneous models are found in the
 invariant sets $S^\pm(VII_h) \cup S(I)$ if $l\not =0$, and $S(V) \cup
 S(I)$ if $l=0$. In particular the zero curvature isotropic models are
 found in the two dimensional set $S(I)$, while the negative curvature
 models are found in the three-dimensional sets $S^\pm(VII_h)$ or
 $S(V)$ depending upon the value of $l$.  See van den Hoogen {\em et
 al.} for a comprehensive analysis of the isotropic scaling models
 \cite{vandenHoogen1999b}.

Note that in the invariant set $B(I)$ there exists the invariant set 
$\tilde{\Sigma}+\Sigma_+^2+\Psi^2<1$, $\Delta=\tilde{A}=N_+=\Upsilon=0$, 
which may be directly integrated to yield
\begin{equation}
\tilde{\Sigma}+\Sigma_+^2+\Psi^2 = \left[1+\zeta e^{3(2-\gamma)\tau} 
\right]^{-1}, \q \zeta=\mbox{constant},
\end{equation}
where $\tau$ is the time parameter.  This solution asymptotes into the
past towards the paraboloid ${\cal K}$ (section \ref{IIIB}), and
asymptotes to the future towards the point P$(I)$. This
solution belongs to the matter invariant set ${\cal FM}$, asymptoting into the
past towards the set ${\cal M}$.

The existence of strictly monotone functions, $W({\bf X}):\mR^n\to\mR$,
on any invariant set, $S$, proves the non-existence of periodic or
recurrent orbits in $S$ and can be used to provide information about the global
behaviour of the dynamical system in $S$. (See Theorem 4.12 in
\cite{Wainwright1997a} for details.)

\begin{table}[h]
\begin{center}
\begin{scriptsize}
\begin{tabular}{|lll|} 
\hline
Function: $W_i({\bf X})$  & Derivative: $W_i'({\bf X})$ & Region of 
	Monotonicity  \\ \hline\hline
	& & Monotonically approaches zero \\
$\displaystyle W_1\equiv \left(1+\Sigma_+\right)^2-\tilde A$ &
	$\displaystyle W_1' = -2\left(2-q\right) 	
	W_1+$ & in the invariant set $\calm \cup \calv $.  \\
	& $\ \ \ \ \ \ \ \ +3\left(1+\Sigma_+\right)
	\left(2\Upsilon^2+\left(2-\gamma\right) \Omega\right)$ & 
		\\ & & \\ \hline
	& & Monotonically decreasing to zero \\
$\displaystyle W_2\equiv \frac{1-\Omega -\Upsilon^2-\Psi^2}{\Omega}$ & 
	$\displaystyle W_2' = -W_2(2-3\gamma) - \frac{1}{\Omega}\left
	(\Sigma_+^2+\tilde\Sigma\right)$ & 
	in the set $({\cal FS} \cup {\cal FM} \cup {\cal F})\backslash S(I)$ \\
	& & when $0\leq \gamma\leq 2/3$\\
\hline
	& & Monotonically decreasing to zero \\
$\displaystyle W_3\equiv \tilde{\Sigma}$ &
	$\displaystyle W_3' = -2\left(2-q\right) W_3-4(\Delta N_+ 
	+\Sigma_+\tilde{A})$ & 	in the invariant sets $B(I)\backslash S(I)$ \\
	& & and $B(V)\backslash S(V)$. \\ && 
 	\\
\hline
	& & Monotonically approaches zero \\
$\displaystyle W_4\equiv \frac{\tilde{A}^2}{N_+} $ &
	$\displaystyle W_4' = 3W_4\left(q+2\frac{\Sigma_+N_+-\Delta}{N_+}\right)$ &
 	in the invariant set $S^{\pm}(III)\backslash({\cal S}\cup{\cal FS})$,\\
 	& & when $\gamma>2/3$.\\ \hline
\end{tabular}
\end{scriptsize}
\caption{{\em Functions, their derivatives and the sets in which they are monotonic.}\label{Table3}}
\end{center}
\end{table} 

Hewitt and Wainwright found a number of monotone functions in the
invariant sets of dimension less than four in the perfect fluid case
(i.e., in lower-dimensional subsets of the perfect fluid invariant
set) and these are summarized in an Appendix in Hewitt and Wainwright
\cite{Wainwright1997a,Hewitt1993a}.  However, they were not able
to find a monotonic function in the full perfect fluid invariant set
for $2/3<\gamma<2$.

\subsection{The Constraint Surface\label{IIC}}

The constraint equation $G({\bf X})=0$ and the Implicit Function
Theorem can generally be used to eliminate one of the variables at any
point in the {\em extended} state-space provided the constraint
equation is not singular there, i.e., $\grad(G({\bf X}))\not = {\bf
0}$.  The constraint surface is singular for all points in the
invariant sets $S(I)$, $B(V)$ and $S(VII_h)$ and therefore cannot be
used to eliminate one of the variables (and hence reduce the dimension
of the dynamical system to six).

Therefore, the local stability of equilibrium
points cannot be determined in the sets $S(I)$, $B(V)$ or $S(VII_h)$ within the
six-dimensional state-space, and hence it is required to determine
the local stability of these equilibrium points in the {\em extended}
space, due to the singular nature of the constraint surface. This
leads to further complications because of the limited use of the
Stable Manifold Theorem.  If these equilibrium points are stable in
the {\em extended} state space, then they are stable in the
six-dimensional constrained surface.  However, if these equilibrium
points are saddles in the {\em extended} state-space, then one
cannot easily determine the dimension of the stable manifold within the
constraint surface.
   
\section{Classification of the Equilibrium Points\label{III}} 

The evolution equations for the matter variables will be analysed,
namely equations (\ref{PsiDE}) and (\ref{PhiDE}) and the auxiliary
equation (\ref{auxiliary}).  From equation (\ref{auxiliary}) it can be
shown that at the equilibrium points either
\begin{equation}
(A) \quad \Omega  =  0, \label{caseA}
\end{equation} 
or 
\begin{equation}
(B) \quad q  =  \frac{3}{2}\gamma-1. \label{caseB}
\end{equation}
In the scalar field case $(A)$ there is no perfect fluid present.
This is the scalar field invariant set ${\cal S}$.  The equilibrium
points and their stability will be studied in subsection \ref{IIIA}.  These
models include the massless scalar field case in which $\Upsilon=0$
($V=0$), but not the vacuum case $\Upsilon=\Psi=0$ which will be dealt with
as a subcase of the perfect fluid case (see below).  The equilibrium
points of case $(A)$ include the isotropic Bianchi VII$_h$ models
studied in \cite{vandenHoogen1997a}.

If, on the other hand, equation (\ref{caseB}) is satisfied, assuming that
$\gamma<2$ so that $q\neq 2$, then equations (\ref{PsiDE}) and (\ref{PhiDE}) 
yield
\begin{equation}
(B1) \quad \Psi=0, \Upsilon=0 \label{caseB1}\end{equation}
or \begin{equation}
(B2) \quad \Psi=\frac{-\sqrt 3 \gamma}{\sqrt 2 k}, \quad
     \Upsilon^2=\frac{3\gamma(2-\gamma)}{2k^2} \label{caseB2}.
\end{equation}
In case $(B1)$, in which both equations (\ref{caseA}) and (\ref{caseB}) are 
valid, there is no scalar field present.  The perfect fluid subcase, which
was studied by Hewitt and Wainwright \cite{Hewitt1993a}, will be dealt
with in subsection \ref{IIIB}.  Note that from equation (\ref{PhiDE}) $\Upsilon=0$
is an invariant set, denoted by ${\cal M}$.

The final case $(B2)$, in which equation (\ref{caseB}) is valid and neither
the scalar field nor the perfect fluid is absent, corresponds to the scaling
solutions when $\gamma>0$.  By using the definition 
\begin{equation}
\mu_\phi \equiv \frac{1}{2}\dot\phi^2 +V(\phi), \quad 
	  p_\phi \equiv \frac{1}{2} \dot\phi^2-V(\phi), \label{phiFluid}
\end{equation}
then from equation (\ref{caseB2}) it is clear that
\begin{equation}
\gamma_\phi \equiv \frac{\mu_\phi+p_\phi}{p_\phi} = 
	\frac{2\Psi^2}{\Psi^2+\Upsilon^2} = \gamma,
\end{equation}
so that the scalar field ``inherits'' the equation of state of the fluid.  It
can be shown that there are exactly three equilibrium points corresponding to
scaling solutions; the flat isotropic scaling solution described in 
\cite{Copeland1997a}, and whose stability was discussed within Bianchi 
type 
VII$_h$ models in \cite{Billyard1998a}, and two anisotropic scaling 
solutions \cite{Coley1998a}.  This will be further discussed in 
subsection \ref{IIIC}.

Hereafter, $0<\gamma<2$ shall be assumed.  The value $\gamma=0$
corresponds to a cosmological constant and the model can be analyzed
as a scalar field model with the potential $V=\Lambda +V_0 e^{k\phi}$
\cite{Billyard1999a}.  The value $\gamma=2$, corresponding to the
stiff fluid case, is a bifurcation value (bifurcation is defined on page \pageref{bifurcation} of appendix \ref{OD}) and will not be considered further.

\subsection{Scalar Field Case\label{IIIA}} 

There are seven equilibrium points and one equilibrium set 
in the scalar field invariant set ${\cal S}$ in which
$\Omega=0$.  The first four equilibrium points were given in
\cite{Coley1997a} (wherein matter terms were not included); they
represent isotropic models ($\Sigma_+=\tilde \Sigma=\tilde N=\Delta=0$):

\

\noindent{\bf 1)} P$_{\cal S}(I)$: $\Sigma_+=\tilde{\Sigma}=\Delta=\tilde{A}=N_+=0, \Psi=-k/\sqrt{6}, \Upsilon=\sqrt{1-k^2/6}$ \index{equilibrium sets!P$_{\cal S}(I)$}
    
This equilibrium point, for which $q=-1+k^2/2$ and which exists only
for $k^2\leq 6$, is in the Bianchi I invariant set $B(I)$.  This point
represents a flat FRW model which is inflationary for $k^2<2$
\cite{Halliwell1987a,Coley1997a}.  The corresponding eigenvalues in
the extended state space are (throughout this chapter, the corresponding
eigenvectors will not be explicitly displayed):
\begin{equation}
-\half (6-k^2), \q-\half (6-k^2), \q  -(6-k^2), \q -(4-k^2), \q -(2-k^2),
\q -\half(2-k^2), \q k^2-3\gamma.
\end{equation}

\

\noindent{\bf 2)} P$^{\pm}_{\cal S}(VII_h)$: $\Sigma_+=\tilde{\Sigma}=\Delta=0, \tilde{A}=\frac{(k^2-2)}{k^2}$, $N_+=\pm\frac{\sqrt{l(k^2-2)}}{k}$, $\Psi=-\frac{\sqrt{2}}{\sqrt{3}k}$, $\Upsilon=\frac{2}{\sqrt{3}k}$
\label{BVIIh_end} \index{equilibrium sets!P$^{\pm}_{\cal S}(VII_h)$}

These two equilibrium points (the indices ``$\pm$'' correspond to the $\pm$
values for $N_+$), which occur in the Bianchi VII$_h$ invariant set $S(VII_h)$
(since $\tilde{A}\geq0$, then $k^2\geq 2$ and therefore $l>0$), have $q=0$.  These
equilibrium points
represent an open FRW model \cite{vandenHoogen1997a}.  The
corresponding eigenvalues in the extended state space are:
\begin{eqnarray}
\nonumber
& 2-3\gamma, \q -1\pm\frac{\sqrt{3}i}{k}\sqrt{k^2-8/3}, &\\ 
& -2\pm\frac{\sqrt 2}{k}
	\sqrt{k^2-4(k^2-2)l\pm\sqrt{\left[k^2-4(k^2-2)l\right]^2+16l
	\left(k^2-2\right)^2+k^4}}. &
\end{eqnarray}

\

{\bf 2a)} P$_{\cal S}(V)$: $\Sigma_+=\tilde{\Sigma}=\Delta=0, 
\tilde{A}=\frac{(k^2-2)}{k^2},
     N_+=0,\Psi=-\frac{\sqrt{2}}{\sqrt{3}k}, 
     \Upsilon=\frac{2}{\sqrt{3}k}$ \label{BVIIh_end2}

This case corresponds to points 2) for $l=0$ and belongs to the set $S(V)$. 
The corresponding eigenvalues in the extended state space are:
\begin{equation}
2-3\gamma, \q -1\pm\frac{\sqrt{3}i}{k}\sqrt{k^2-8/3}, \q -2,\q -2, \q
	0, \q -4.
\end{equation}

\noindent{\bf 3)} P$^{\pm}_{\cal S}(II)$: $\Sigma_+=-\frac{k^2-2}{k^2+16}, 
   \tilde{\Sigma}=3\Sigma_+^2$, $\Delta=\Sigma_+N_+$, $\tilde{A}=0$, 
   $N_+=\pm 3\frac{\sqrt{-(k^2-2)(k^2-8)}}{k^2+16}$, \\ \hspace*{25mm}
    $\Psi=-\frac{3\sqrt6 k}{k^2+16}$, $\Upsilon=6\frac{\sqrt{8-k^2}}{k^2+16}$
\index{equilibrium sets!P$^{\pm}_{\cal S}(II)$}

These two equilibrium points, for which $q=8(k^2-2)/(k^2+16)>0$, exist
only for $2\leq k^2 \leq 8$.  These two points represent Bianchi type
II models analogous to those found in \cite{Hewitt1993a}.  The
corresponding eigenvalues are:
\begin{eqnarray}
\nonumber
& 12\frac{k^2-2}{k^2+16}, \q 6\frac{k^2-8}{k^2+16},
   \q 6\frac{k^2-8}{k^2+16},&\\& \q 3\frac{(k^2-8)\pm \sqrt{(13k^2-32)(k^2-8)}}
   {k^2+16}, \q -3\gamma+18\frac{k^2}{k^2+16}. &
\end{eqnarray}
\

\noindent{\bf 4)} P$^\pm_{\cal S}(VI_h)$: $\Sigma_+=\frac{-l(k^2-2)}{n}, 
    \tilde{\Sigma}=-3\Sigma_+^2/l,\Delta=0,\tilde{A}=\frac{9(k^2-2l)(k^2-2)}
     {n^2}, N_+=0$, \\
	\hspace*{25mm} $\Psi=\frac{\sqrt 6k(1-l)}{n},
    \Upsilon=\frac{2\sqrt 3 \sqrt{(k^2-2l)(1-l)}}{n},$\label{BVIh_end2}
where $n\equiv k^2(l-3)+4l$.  \index{equilibrium sets!P$^\pm_{\cal S}(VI_h)$}

Since $\tilde{\Sigma}>0$, then
$l<0$ and hence this equilibrium point occurs in the Bianchi
VI$_h$ invariant sets.  The deceleration parameter is given by
$q=2l(k^2-2)/[k^2(l-3)+4l]\geq0$, where $k^2\geq2$, and this point corresponds
 to a Collins Bianchi type VI$_h$ solution \cite{Collins1971a}.
The corresponding eigenvalues are:
\begin{eqnarray} \nonumber
&  3\frac{(k^2-2l)\pm\sqrt{(k^2-2l)^2+8l(1-l)(k^2-2)}}{[k^2(l-3)+4l]}, &\\
& 6\frac{k^2-2l}{[k^2(l-3)+4l]}, \q
  -3\gamma-6\frac{k^2(1-l)}{[k^2(l-3)+4l]}, \q
   3\frac{(k^2-2l)\pm\sqrt{(k^2-2l)[(k^2-2l)-4(1-l)(k^2-2)]}}{[k^2(l-3)+4l]}.&
   \label{PSVIh}
\end{eqnarray}

Next, consider {\em the Massless Scalar Field Invariant Set
${\cal M}$}: there is one equilibrium set here, which generalizes the work in
\cite{Hewitt1993a} to include scalar fields:

\

\noindent{\bf 5)} ${\cal K}_{\cal M}$: $\tilde{\Sigma}+\Sigma_+^2+\Psi^2=1, 
\Delta=\tilde{A}=N_+=\Upsilon=0, \Psi\neq 0$ \index{equilibrium sets!${\cal K}_{\cal M}$, ${\cal K}$, ${\cal K}^\pm$}

This paraboloid, for which $q=2$, generalizes the parabola ${\cal K}$
in \cite{Hewitt1993a} defined by $\tilde{\Sigma}+\Sigma_+^2=1$ to
include a massless scalar field, and represents Jacobs' Bianchi type I
non-vacuum solutions \cite{Collins1971a}.  However, the eigenvalues
are considerably different from those found in \cite{Hewitt1993a}, and
so all are listed here (the variables which define the subspaces in which the corresponding
eigendirections reside are included below in curly braces):
\begin{eqnarray}
\nonumber
&{2[(1+\Sigma_+)\pm\sqrt{3\tilde\Sigma}], \atop \{\Delta, 
N_+\}} \q
{0, \atop \{\Sigma_+, \tilde{\Sigma}\}} \q {0, \atop \{\Sigma_+, 
\tilde{\Sigma},\Psi\}} \q
{3(2-\gamma), \atop \{\Sigma_+,\tilde{\Sigma},\Psi\}} & \\ \nonumber \\ &
{4(1+\Sigma_+), \atop \{\Sigma_+,\tilde{\Sigma},\tilde{A},\Psi\}} \q
{\frac{\sqrt 6}{2}\left( \sqrt{6} +
k\Psi\right). \atop \{\tilde{\Sigma},\Upsilon\}} &
\end{eqnarray}

\subsection{Perfect Fluid Case, $\Psi=\Upsilon=0$\label{IIIB}} 

As mentioned earlier, the perfect fluid invariant set ${\cal F}$ in
which $\Psi=\Upsilon=0$ was studied by Hewitt and Wainwright
\cite{Hewitt1993a}; hence this subsection generalizes their
results by including a scalar field with an exponential potential.  We
shall use their notation to label the equilibrium points/sets.  There
are five such invariant points/sets.  In all of these cases the extra
two eigenvalues associated with $\Psi$ and $\Upsilon$ are (respectively)
\begin{equation} -\frac{3}{2}(2-\gamma)<0, \q \frac{3}{2}\gamma>0. 
\end{equation}

\

\noindent{\bf 1)} P(I): 
$\Sigma_+=\tilde{\Sigma}=\Delta=\tilde{A}=N_+=\Psi=\Upsilon=0$ \index{equilibrium sets!$P(I)$}

This equilibrium point, for which $\Omega=1$, is a saddle for
$2/3<\gamma<2$ in ${\cal F}$ \cite{Hewitt1993a} (and is a sink
for $0\le\gamma<2/3$), and corresponds to a flat FRW model.

\

\noindent{\bf 2)} P$^\pm$(II): $\Sigma_+=-\frac{1}{16}(3\gamma-2)$, 
  $\tilde{\Sigma}=3\Sigma_+^2$, $\Delta=\Sigma_+N_+, \tilde{A}=0$, 
$N_+=\pm\frac{3}{2}
   \sqrt{-\Sigma_+(2-\gamma)}$, \\\hspace*{25mm}$\Psi=\Upsilon=0$
\index{equilibrium sets!P$^\pm$(II)}

This equilibrium point, for which $\Omega=\frac{3}{16}(6-\gamma)$, is a saddle
in the perfect fluid invariant set \cite{Hewitt1993a}. 

\

\noindent{\bf 3)} P(VI$_h$): $\Sigma_+=-\frac{1}{4}(3\gamma-2),
\tilde{\Sigma}=-  3\Sigma_+^2/l, \Delta=0, 
\tilde{A}=-\frac{9}{16l}(3\gamma-2)(2-\gamma)$,
 \\\hspace*{25mm}$N_+= \Psi=\Upsilon=0$\index{equilibrium sets!P(VI$_h$))}

Since $\tilde{\Sigma}>0$ and $\tilde{A}>0$, this equilibrium point
occurs in the Bianchi VI$_h$ invariant set and corresponds to the
Collins solution \cite{Collins1971a}, where $\Omega=\frac{3}{4}(2-\gamma)
+\frac{3}{4l}(3\gamma-2)$ (and therefore $2/3\leq\gamma\leq 2(-l-1)/(3-l)$
and so $l\leq -1$). In \cite{Hewitt1993a} this was a sink in
${\cal F}$, but is a saddle in the extended state space due to the
fact that the two new eigenvalues have values of different sign.

\

There are also two equilibrium sets, which generalize the work in 
\cite{Hewitt1993a} to include scalar fields:

\

\noindent{\bf 4)} ${\cal L}^\pm_l$: $\tilde{\Sigma}=-\Sigma_+(1+\Sigma_+)$, 
$\Delta=0$,  $\tilde{A}=(1+\Sigma_+)^2$, \label{BVIh_end1}
     $N_+=\pm\sqrt{[l\tilde A-3\Sigma_+(1+\Sigma_+)]}$, \\
	\hspace*{25mm}$\Psi=\Upsilon=0$
\index{equilibrium sets!${\cal L}^\pm_l$}

For this set $\Omega=0$. The local sinks in this set 
occur when \cite{Hewitt1993a} 

(a) $l<0$ (Bianchi type VI$_h$) for $-\frac{1}{4}(3\gamma-2)<\Sigma_+ 
<l/(3-l)$ and \\ \hspace*{67mm}$l>-(3\gamma-2)/(2-\gamma)<0$,

(b) $l=0$ (Bianchi type IV) for $-\frac{1}{4}(3\gamma-2)<\Sigma_+ < 0$,

(c) $l>0$ (Bianchi type VII$_h$) for $-\frac{1}{4}(3\gamma-2)<\Sigma_+ < 0$.
  
\noindent The additional two eigenvalues for the full system are:
\begin{equation}
1-2\Sigma_+,\q -2(1+\Sigma_+).
\end{equation}

Finally, consider {\em the Massless Scalar Field Invariant Set
${\cal FM}$}:

\noindent{\bf 5)} ${\cal K}$: $\tilde{\Sigma}+\Sigma_+^2=1, 
\Delta=\tilde{A}=N_+=\Upsilon=\Psi=0$\label{general_end}\index{equilibrium sets!${\cal K}_{\cal M}$, ${\cal K}$, ${\cal K}^\pm$}

This parabola, for which $q=2$, is the special case of ${\cal K}_{\cal
M}$ for which $\Psi=0$ and corresponds to the parabola ${\cal K}$ in
\cite{Hewitt1993a}.  However, the eigenvalues are considerably
different from those found in \cite{Hewitt1993a}, and so all are
listed here (the variables which define the subspaces in which the corresponding eigendirections
reside are included below in curly braces):
\begin{equation}
{2[(1+\Sigma_+)\pm\sqrt{3\tilde{\Sigma}}], \atop \{\Delta, 
N_+\}} \q
{0, \atop \{\Sigma_+, \tilde{\Sigma}\}} \q {0, \atop \{\Psi\}} \q
{3(2-\gamma), \atop \{\Sigma_+,\tilde{\Sigma}\}} \q
{4(1+\Sigma_+), \atop \{\Sigma_+,\tilde{\Sigma},\tilde{A}\}} \q
{3. \atop \{\Upsilon\}} 
\end{equation}

Table \ref{TableIII} is included, listing the equilibrium sets
and corresponding eigenvalues as listed in \cite{Hewitt1993a}.
\begin{table}[htp]
\begin{center}
\begin{tabular}{|lcl|}
\hline
Eqm. set & Eigenvalues & Comment \\
\hline \hline
$P(I)$ & $-\frac{3}{2}(2-\gamma) \q -3(2-\gamma) \q (3\gamma -4)$  &  \\ & 
$\frac{1}{2}(3\gamma-2) \q \frac{1}{2}(3\gamma-2)$ & \\
\hline
$P^\pm(II)$ & $\frac{3}{4}(3\gamma-2) \q -\frac{3}{2}(2-\gamma)$ & Constraint eqn. used \\
& $-\frac{3}{4}(2-\gamma)\left\{1\pm\sqrt{1-\frac{(3\gamma-2)(6-\gamma)}{2(2-\gamma)}}\right\}$ & to eliminate $\tilde\Sigma$ \\
\hline
$P(VI_h) ^\dagger$ & $-\frac{3}{4}(2-\gamma)(1\pm\sqrt{1-r^2})$ 
& Constraint eqn. used \\ &
$-\frac{3}{4}(2-\gamma) (1\pm \sqrt{1-q^2})$  & to eliminate $\tilde\Sigma$\\
\hline
${\cal K}$ & $0 \q 2(1+\Sigma_+) \q 2(2-\gamma)$ & 1-D invariant set \\
&  $2\left[1+\Sigma_+ \pm \sqrt{3(1-\Sigma_+^2)} \right]$ & \\
\hline
${\cal D}$ & $ 0 \q 0 $ & 2-D invariant set, $\gamma=2$\\
& $ 2
\left[1+\Sigma_+\pm\sqrt{3\tilde\Sigma}\right] \q
2\left(1+\Sigma_+\right) $  & \\
\hline
${\cal L}_l$ & $0 \q -4\Sigma_+-(3\gamma-2)$ & Constraint eqn. used \\ 
& $-2\left[ (1+\Sigma_+) \pm 2iN_+ \right] $ & to eliminate $\tilde\Sigma$ \\
\hline
${\cal F}_l$ & $-2 \q 4 \q -2 \q 0 \q 0$ & $l\geq 0$, non-hyperbolic \\
	&& $\gamma=2/3$ \\ \hline
\end{tabular}
\end{center}
\caption[{\em Equilibrium sets found by Hewitt and Wainwright}]{{\em Equilibrium sets found by Hewitt and Wainwright,
and the corresponding eigenvalues in the extended space.  In the table
$r^2\equiv 2(3\gamma-2)(1-l_c/l)$, $q^2\equiv 2r^2/(2-\gamma)$ and
$l_c\equiv -(3\gamma-2)/(2-\gamma)$.\label{TableIII}}}
\end{table}

\subsection{Scaling Solutions\label{IIIC}}

Defining 
\begin{equation}
\Psi_S\equiv -\sqrt{\frac{3}{2}} \frac{\gamma}{k}, \quad 
    \Upsilon^2_S\equiv \frac{3\gamma(2-\gamma)}{2k^2},
\end{equation}
and recalling that $0<\gamma<2$, there are three equilibrium points
corresponding to scaling solutions.  Because the scalar field mimics
the perfect fluid with the exact same equation of state
($\gamma_\phi=\gamma$) at these equilibrium points, 
these two ``fluids'' can be combined via $p_{tot}=p_\phi+p$,
$\mu_{tot}=\mu_\phi+\mu$, $p_{tot}=(\gamma-1)\mu_{tot}$; therefore,
all of these equilibrium points will correspond to exact perfect fluid
models analogous to the equilibrium points found in
\cite{Hewitt1993a}.

\

The flat isotropic FRW scaling solution \cite{Wetterich1988a,Wands1993a}:

\noindent{\bf 1)} ${\cal F}_S(I)$: $\Sigma_+=\tilde{\Sigma}=\Delta=A=N_+=0,
\Psi=\Psi_S, \Upsilon=\Upsilon_S$\index{equilibrium sets!${\cal F}_S(I)$, ${\cal F}_S$}

The eigenvalues for these points in the extended space, for which 
$\Omega=1-3\gamma/k^2$ (and therefore $k^2\geq 3\gamma$) are:
\begin{eqnarray}
\nonumber
&\frac{1}{2}(3\gamma-2), \q -3(2-\gamma), \q 3\gamma-4, \q -\frac{3}{2}(2-\gamma), \q
3\gamma-2,& \\ & -\frac{3}{4}(2-\gamma)\pm
\frac{3}{4}\sqrt{(2-\gamma)(2-9\gamma +24\gamma/k^2)}  & \label{FSI}
\end{eqnarray}

There are two anisotropic scaling solutions:

\

\noindent{\bf 2)} ${\cal A}_S(II)$: $\Sigma_+=-\frac{1}{16}{(3\gamma-2)}$, 
 $\tilde\Sigma=3\Sigma_+^2$,
$\Delta=\Sigma_+N_+$, $\tilde{A}=0$, $N_+=\pm\frac{3}{2}\sqrt{-\Sigma_+
(2\!-\!\gamma)}$, \\\hspace*{25mm}
$\Psi=\Psi_S$, $\Upsilon=\Upsilon_S$\label{BVIh_end3}\index{equilibrium sets!${\cal A}_S(II)$}

The eigenvalues for these points, for which
$\Omega=\frac{3}{16}(6-\gamma)-3\gamma/k^2$ (and therefore
$k^2\geq16\gamma/[6-\gamma]$), are:
\begin{eqnarray}\nonumber &
\frac{3}{4}\left(3\gamma-2\right), \q -\frac{3}{2}(2-\gamma),&\\
	&-\frac{3}{4}\left[r_\gamma
	\pm\sqrt{ r_\gamma^2 -\frac{3}{4}r_\gamma 
          \left\{
	  2\left(3\gamma-2\right) +\frac{\gamma(6-\gamma)}{k^2}\left(
	  k^2 -\frac{16\gamma}{6-\gamma}\right) \pm \sqrt{E_1} 
			\right\}}\right],& 
\end{eqnarray}
where $E_1\equiv \left[
	  2\left(3\gamma-2\right) -\frac{\gamma(6-\gamma)}{k^2}\left(
	  k^2 -\frac{16\gamma}{6-\gamma}\right)\right]^2 
	+\frac{8}{9}\left(3\gamma-2\right)\frac{\gamma(6-\gamma)}{k^2}\left(
	  k^2 -\frac{16\gamma}{6-\gamma}\right)$ and $r_\gamma=2-\gamma$.

\

\noindent{\bf 3)}${\cal A}_S(VI_h)$:
\label{an_attractor}$\Sigma_+=-\frac{1}{4}\left(3\gamma- 2\right),
 \tilde\Sigma=-3\Sigma_+^2/l,\Delta=0,\tilde{A}=-\frac{9}{16l}(2-\gamma)
\left(3\gamma-2\right)$, \\\hspace*{25mm}
$N_+=0,\Psi=\Psi_S,\Upsilon=\Upsilon_S$\index{equilibrium sets!${\cal A}_S(VI_h)$}

These points occur in the Bianchi VI$_h$ invariant set ($l<0$ since
$\tilde{\Sigma}>0$) for which
$\Omega=\frac{3}{4}(2-\gamma)+\frac{3}{4l}(3\gamma-2) -3\gamma/k^2$
(and therefore $-l^{-1}\leq (2-\gamma)/(3\gamma-2)$ and $k^2\geq 4\gamma/[(2-\gamma)+(3\gamma-2)/l]$ ) and
correspond to the Collins Bianchi VI$_h$ perfect fluid solutions
\cite{Collins1971a}.  The eigenvalues for these equilibrium points are:
\begin{eqnarray}
\nonumber
&-\frac{3}{4}\left[r_\gamma\pm
\sqrt{r_\gamma^2-4(3\gamma-2)^2\left(\frac{r_\gamma}{3\gamma-2}+\frac{1}{l}
	\right)}\right], &
\\
\label{bugger}
& -\frac{3}{4}\left[r_\gamma\pm
	\sqrt{r_\gamma^2-r_\gamma\left[
	4\gamma\left(1-\frac{3\gamma}{k^2}\right)+\left(3\gamma-2\right)
	\left(\frac{r_\gamma}{3\gamma-2}+\frac{1}{l}\right) \pm \sqrt{E_2} 
        \right]}\right],&
\end{eqnarray}
where $E_2\equiv
	\left[4\gamma\left(1-\frac{3\gamma}{k^2}\right)-\left(3\gamma-2\right)
	\left(\frac{2-\gamma}{3\gamma-2}+\frac{1}{l}\right)\right]^2
	-128\frac{\gamma^2}{k^2}$.

All equilibrium sets are tabulated in table \ref{table_sum} along with
the deceleration parameter and the allowed values of $k$ for each set.
\begin{table}[htb]
\begin{center}
\begin{tabular}{|lccc|} \hline 
Eqm.  & $q$ & $k$ & Inflation \\
Point &&& for \\ \hline\hline
P$_{\cal S}(I) $  & $\half k^2 -1$ & $k^2\leq 6$ & $k^2<2$ \\ \hline
P$_{\cal S}(II) $ & $\frac{8\left[k^2-2\right]}{\left[k^2+16\right]}$ & 
	$2\leq k^2\leq 8$ & $-$ \\ \hline
P$_{\cal S}(V) $ & $0$ & $k^2\geq 2$ &	$-$ \\ \hline
P$^\pm_{\cal S}(VI_h)$ & $\frac{2l(k^2-2)}{(k^2(l-3)+4l)}$ 
	& $k^2\geq 2$ & $-$ \\ \hline
P$_{\cal S}(VII_h)$   & 0  & $k^2\geq 2$ &$-$ \\ \hline
${\cal K}_{\cal M}$ & $2$ & $-$ & $-$ \\ \hline
P(I) &&& \\
P$^\pm$(II) &\raisebox{0ex}[0pt]{{\Huge\}}} $\frac{3}{2}\gamma-1$&$-$& $-$\\
P(VI$_h$) &&& \\ \hline
${\cal L}_l^\pm$ &$-2\Sigma_+$& $-$ & $\Sigma_+>0$  \\
${\cal K}$ &2&$-$& $-$\\ \hline 
${\cal F}_{\cal S}(II)$ & &$k^2\geq 3\gamma$ &$-$ \\
${\cal A}_{\cal S}(II)$ &\raisebox{0ex}[0pt]{{\Huge\}}}$\frac{3}{2}\gamma-1$
	 & $k^2\geq \frac{16\gamma}{6-\gamma}$& $-$ \\
${\cal A}_{\cal S}(VI_h)$ & &$k^2\geq \frac{4\gamma}{(2-\gamma)+(3\gamma-2)/l}$
 &$-$  \\ \hline
\end{tabular}
\end{center}
\caption[{\em Summary of Equilibrium Sets}]
{{\em This table lists all of the equilibrium sets as well as the value of
$q$  and allowed values of $k$.  Note that $q<0$ represent inflationary models.
The last column gives the values of $k$ for which the models inflate.
\label{table_sum}}}
\end{table}

\section{Stability of the Equilibrium Points and Some Global Results\label{IV}}

The stability of the equilibrium points listed in the previous section
can be easily determined from the eigenvalues displayed.  Often the
stability can be determined by the eigenvalues in the extended state
space, otherwise the constraint must be utilized to determine the
stability in the six-dimensional state space (i.e., within the
constraint surface).  In the cases in which this is not possible, the
eigenvalues in the extended seven-dimensional state-space must be
analysed and the conclusions that can be drawn are consequently
limited.  Employing local stability results and utilizing the monotone
functions found in Table \ref{Table3}, some
global results can be proven.  In the absence of monotone functions, and in the same
spirit as Refs. \cite{Wainwright1997a} and
\cite{Hewitt1993a}, plausible results can be conjectured which are
consistent with the local results and the dynamical behaviour on the
boundaries and which are substantiated by numerical experiments
\cite{Billyard1999g}.

\subsection{The Case  $\Omega=0$}

If $\Omega=0$ and $\Phi=0$, then the function $W_1$ in Table
\ref{Table3} monotonically approaches zero.  The existence of the
monotone function $W_1$ implies that the global behaviour of models in
the set ${\cal M}\cup{\cal V}$ can be determined by the local
behaviour of the equilibrium points in ${\cal M}\cup{\cal V}$.
Consequently, a portion of the equilibrium sets ${\cal K}$ and ${\cal
K}_{\cal M}$ (corresponding to local sources) represent the past
asymptotic states while the future asymptotic state is represented by
${\cal L}_l$, or in the case of Bianchi types I and II, by a point on
${\cal K}$.

Therefore, all vacuum models and all massless scalar field models are
asymptotic to the past to a Kasner state and are asymptotic to the
future either to a plane wave solution (Bianchi types IV, VI$_h$ and
VII$_h$), or to a Kasner state (Bianchi types I and II), or to a Milne
state (Bianchi type V).

If $\Omega=0$ and $\Upsilon\neq0$, then the models only contain a
scalar field.  It was proven in \cite{Kitada1992a,Kitada1993a} that
all Bianchi models evolve to a power-law inflationary state
(represented by P$_{\cal S}(I)$) when $k^2<2$.  If $k^2>2$, then it was
shown in
\cite{vandenHoogen1997a} that a subset of Bianchi models of
types V and VII$_h$ evolve towards negatively curved isotropic models
represented by points P$_{\cal S}(V)$ and P$^\pm_{\cal S}(VII_h)$.  In
\cite{vandenHoogen1997b} it was shown that when $k^2>2$ the future
state of a subset of Bianchi type VI$_h$ solutions is represented by
the point P$_{\cal S}(VI_h)$.  It can be seen here that the future
state of a subset of Bianchi type II models is represented by the
point P$^\pm_{\cal S}(II)$.  

Therefore, all scalar field models with $\Omega=0$ evolve to a
power-law inflationary state if $k^2<2$.  If $k^2>2$, then the future
asymptotic state for all Bianchi types IV, V and VII$_h$ is
conjectured to be a negatively--curved, isotropic model and the future
asymptotic state for all Bianchi type VI$_h$ is conjectured to be the
Feinstein-Ib\'{a}\~{n}ez anisotropic scalar field model
\cite{Feinstein1993a}.  If $2<k^2<8$, then the future asymptotic state
for all Bianchi type II models is the anisotropic Bianchi type II
scalar field model, and if $k^2\geq8$ then the future asymptotic state
is that of a Kasner model.  If $2<k^2<6$, then the Bianchi type I
models approach a non-inflationary, isotropic (i.e., the point
P$_{\cal S}(I))$; if $k^2\geq 6$, then they evolve to a Kasner state in
the future.

\subsection{The Case $\Omega\neq0$, $0\leq\gamma\leq2/3$}

If $\Omega\neq0$ and $0\leq\gamma\leq2/3$ then the function $W_2$ in
Table \ref{Table3} is monotonically decreasing to zero.  Therefore, we
conclude that the omega-limit set of all non-exceptional orbits (i.e.,
those orbits which are not equilibrium points, heteroclinic orbits,
etc.) of the dynamical system (\ref{DE_all}) is a subset of
$S(I)$.  This implies that all non-exceptional models with
$\Omega\neq0$ evolve towards the zero--curvature,
spatially--homogeneous and isotropic models in $S(I)$ and hence
isotropize to the future.  In \cite{vandenHoogen1999b}, it was shown
that the zero--curvature spatially--homogeneous and isotropic models
evolve towards the power-law inflationary model, represented by the
point P$_{\cal S}(I)$ when $k^2<3\gamma$ or towards the isotropic
scaling solution, represented by the point ${\cal F}_{\cal S}(I)$, when
$k^2>3\gamma$.  Using $W_1$, it is concluded that the past asymptotic
state(s) of all non-exceptional models (including models in $S(I)$) is
characterized by $\Omega=0$.  In other words, matter is dynamically
{\em unimportant} as these models evolve to the past.  It was shown in
\cite{vandenHoogen1999b} that all models evolve in the past to some
portion of ${\cal K}$ or ${\cal K}_{\cal M}$ (the Kasner models) which
are local sources.

\subsection{The Case $\Omega\neq0$, $\frac{2}{3}<\gamma<2$}

The following table lists the local sinks for $\frac{2}{3}<\gamma<2$.

\begin{table}[htp]
\begin{center}
\begin{tabular}{|lccc|} \hline 
Sink   & Bianchi& k & Other constraints\\&Type && \\ \hline\hline
P$_{\cal S}(I) $              & I       & $k^2\leq 2$ & \\
P$_{\cal S}(VII_h)^\dagger$   & I       & $k^2=2$ & \\ \hline
P$_{\cal S}^\pm(VI_{-1}) $    & III     & $k^2\geq 2$ & $\gamma>k^2/(k^2+1),\q l=-1$  \\
${\cal L}^\pm_l(VI_{-1}) $    & III     & all &  $\gamma>1,\q \Sigma_+=-1/4$\\ \hline
P$_{\cal S}(V)$               & V       & $k^2\geq2$ & \\ \hline
P$_{\cal S}^\pm(VI_h) $       & VI$_h$  & $k^2\geq 2$ & $\ \ \gamma>2k^2(1-l)/[k^2(l-3)+4l]$ \\
${\cal L}^\pm_l(VI_h) $       & VI$_h$  & all &$\gamma>4/3,\q \Sigma_+<-1/2$\\ 
${\cal A_S}(VI_h)$	      & VI$_h$  & $k^2 \geq \frac{ 4\gamma} 
	{[(2-\gamma)+(3\gamma-2)/l]}$ & $l \leq \frac{-(3\gamma-2)}{2-\gamma}$
 \\ \hline
P$^\pm_{\cal S}(VII_h)$ & VII$_h$ & $k^2\geq 2$ &
	$\ \ \ \ l>\frac{k^2}{4(k^2-2)(4-k^2)}$ for $2<k^2\leq4$ \\ 
		    &&&$l<\frac{k^2}{4(k^2-2)(k^2-4)}$ for $k^2>4$  \\ \hline
\end{tabular}
\end{center}
\caption[{\em Sinks in the various Bianchi invariant sets for $2/3<\gamma<2$}]{{\em This table lists all of the sinks in the
various Bianchi invariant sets for $2/3<\gamma<2$.  A subset of ${\cal
K}_{\cal M}$ acts as a source for all Bianchi class B models.
$^\dagger$Note: in this case $N_+=0$ (i.e., P$_{\cal S}$=P$_{\cal
S}^\pm$) and in fact corresponds to a Bianchi I model.\label{TableIV}}}
\end{table}

The function $W_3$ mono\-ton\-ically de\-crea\-ses to zero in
$B(I)\backslash S(I)$ and $B(V)\backslash S(V)$.  This implies that
there do not exist any periodic or recurrent orbits in these sets and,
furthermore, the global behaviour of the Bianchi type I and V models
can be determined from the local behaviour of the equilibrium points
in these sets.  It is conjectured that there do not exist any periodic or
recurrent orbits in the entire phase space for $\gamma>2/3$, whence it
follows that all global behaviour can be determined from Table
\ref{TableIV}.

\begin{figure}[htp]
  \centering
   \epsfbox{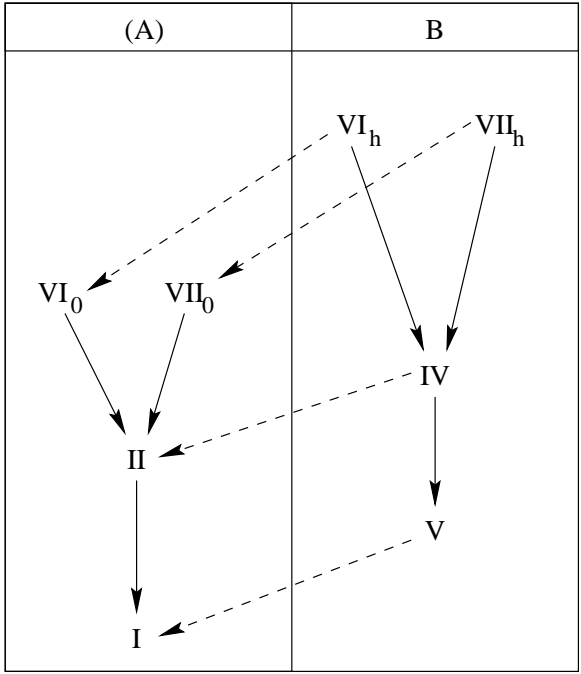}
  \caption[{\em Specialization diagram for Bianchi class B models}]{{\em Specialization diagram for Bianchi class B models
obtained by letting a non-zero parameter go to zero.  A broken arrow
indicates the group class changes (from B to A).}\label{Spec}}
 \end{figure}

Note that a subset of ${\cal K}_{\cal M}$ with
$(1+\Sigma_+)^2>3\tilde\Sigma$, $\Psi>-\sqrt{6}/k$ acts as a source
for all Bianchi class B models.  For $k^2<2$, P$_{\cal S}(I)$ is the
global attractor (sink).  Note that from Table \ref{TableIV} there are unique
global attractors (both past and future) in all invariant sets and
hence the asymptotic properties are simple to determine.  The sinks
and sources for a particular Bianchi invariant set, which may appear
in that invariant set or on the boundary corresponding to a
(lower-dimensional) specialization of that Bianchi type, can be easily
determined from table \ref{TableIV} and figure \ref{Spec} which lists the specializations of
the Bianchi class B models \cite{MacCallum1971b}.

The most general models are those of Bianchi types VI$_h$ and VII$_h$.
The Bianchi type VII$_h$ models are of particular physical interest
since they contain open FRW models as special cases.  From Table \ref{TableIV}
and figure \ref{Spec} it can be argued that generically these models (with a
scalar field) isotropize to the future, a result which is of great
significance.  The Bianchi type VI$_h$ models are also of interest
since they contain a class of anisotropic scaling solutions that act
as attractors for an open set of Bianchi type B models.  Note that
generically Bianchi type VI$_h$ models do not isotropize for $k^2\geq
2$.

It is also of interest to determine the intermediate behaviour of the
models.  In order to do this, the saddles need to be investigated,
determine the dimension of their stable submanifolds, and construct
possible heteroclinic sequences.  This could then be used, in
conjunction with numerical work, to establish the physical properties
of the models.  For example, an investigation could be made to determine whether {\em
intermediate isotropization} can occur in Bianchi type VII$_h$ models
\cite{Wainwright1998a}.  There are many different cases to
consider depending upon the various bifurcation values and the
particular Bianchi invariant set under investigation.  For example, in
\cite{Billyard1999f} the heteroclinic sequences in the
four-dimensional invariant set $S(VI_h)$ were studied, because it
illustrated the method and because such a study emphasizes the
importance of anisotropic scaling solutions.

\section{Asymptotic Analysis in the Jordan Frame\label{BianchiB_Jordan}}

Since the conformal transformation (\ref{trans2}) is well defined in
all cases of interest, the Bianchi type is invariant under the
transformation and the asymptotic properties of the
scalar-tensor theories can be deduced from the corresponding behaviour in the
Einstein frame.  In particular, isotropization can be readily
determined; if a model isotropizes in one frame, it will isotropize in
the other frame.  However, inflation must be deduced using equations
(\ref{SFtoSTtrans}).  In particular, this section lists for each
equilibrium point discussed in this chapter, the deceleration and
Hubble parameter.  

Recalling the time transformation (\ref{jesus2})
\be
	\frac{dt^*}{dt} = \pm \sqrt\Phi = \pm e^{\half\phi/\bar\omega},
	\label{t_trans}
\ee
where $t^*$ is the time in the Einstein frame and $t$ is the time in
the Jordan frame, and using a similar transformation for the scale
factors
\be
a^* = \pm \sqrt\Phi ~ a \qquad \Leftrightarrow \qquad a = 
	e^{-\half\phi/\bar\omega} a^*,
	\label{a_trans}
\ee
where $a^*$ is the scale factor in the Einstein frame and $a$ is the
scale factor in the Jordan frame, the Hubble parameter of each
frame can be related to one another by
\be
\up{st}H = \pm e^{\half\phi/\bar\omega} \left[\up{sf}H 
	-\frac{\dot\phi}{2\bar\omega} \right].
	\label{H_trans}
\ee
For the cases in which $\phi=\phi_0$ (and hence $\Phi=\Phi_0$) both
frames are equivalent (up to a constant positive conformal factor);
hence, it is clear that the upper sign should be chosen so that such
models are expanding (or contracting) in both frames.  Henceforward, it
will be assumed that the ``$+$'' transformation in (\ref{t_trans}) -
(\ref{H_trans}) shall be used.  By making use of the normalized
variables (\ref{variable_def}), the deceleration parameter in the
Jordan frame is written
\be
\up{st}q = \bar\omega \frac{\left[\bar\omega\up{sf}q -\sqrt6\Psi -\frac{3}{2}k
	\Upsilon^2\right]}{\left[\bar\omega-\half\sqrt6\Psi\right]^2},
\ee
and the Hubble parameter is written
\be
\up{st}H = \frac{e^{\half\phi/\bar\omega}}{\bar\omega}~ H 
	\left[\bar\omega-\half\sqrt6\Psi\right].
\ee
In the Einstein frame $\up{sf}H=0$ is an invariant set and therefore
all models therein are either ever-expanding or ever-contracting.
Therefore it is assumed that $\up{sf}H>0$ since this is physically
relevant for the current epoch of the universe.  Models in the Jordan
frame will be inflationary if $\up{st}q<0$ and $\up{st}H>0$ and table
\ref{Jordan_isotropize} lists the values of $\up{st}q$ and
$\up{st}H$, indicating for which values of $\bar\omega$ these models
are inflationary.  For compact notation, $\up{sf}\bar H\equiv
e^{\half\phi/\bar\omega} H >0$ is used.

\begin{table}[htb]
\begin{center}
\begin{tabular}{|lcccc|} \hline 
Eqm.  & $\up{st}q$ & $\up{st}\bar H$ & Inflationary & Type \\
Point &&& for & \\ \hline\hline
P$_{\cal S}(I) $  & $\frac{\left[k^2-2\right]}{\left[2+k/\bar\omega\right]}$ & 
	$\bar H \left[1+\half k/\bar\omega \right]$ &
	$k^2 < 2$  & (A) \\ &&& \& $k/\bar\omega >-2$ & \\ \hline
P$_{\cal S}(II) $ & $\frac{8\left[k^2-2\right]}
	{\left[(k^2+16)+9k/\bar\omega\right]}$ & 
	$\bar H \left[1+\frac {k/\bar\omega}{(k^2+16)} \right]$ &
	$-$ & (S) \\ \hline
P$_{\cal S}(V) $ & $0$ & $\bar H \left[1+\frac{1}{k/\bar\omega}\right]$ &
	$-$ & (A) \\ \hline
P$^\pm_{\cal S}(VI_h)^\dagger$ & $\frac{2l(k^2-2)}{n}\left[1-\frac{3k}{\bar\omega}
	\frac{(1-l)}{n}\right]^{-1}$ & $\bar H \left[1-\frac{3k}{\bar\omega}
	\frac{(1-l)}{n}\right]$ & $-$ & (A) \\ \hline
P$_{\cal S}(VII_h)$   & 0  & $\bar H \left[1+\frac{1}{k/\bar\omega}\right]$ 
	&$-$& (A)\\ \hline
${\cal K}_{\cal M}$ & $2\left[1-\half\sqrt6\Psi/\bar\omega \right]^{-1}$ &
	$\bar H\left[1-\half\sqrt6\Psi/\bar\omega \right]$ & $-$ & (R) 
	\\ \hline
$\in {\cal F}$ & $\up{sf}q$ & $\bar H\geq 0$ & 
	\multicolumn{2}{c|}{see table \ref{table_sum}} \\ \hline
${\cal F}_{\cal S}(II)$ & & &$-$ & (S)\\
${\cal A}_{\cal S}(II)$ &\raisebox{0ex}[0pt]{{\Huge\}}}
	$\half\frac{(3\gamma-2)}{\left[1+
	\frac{ 3\gamma}{2k\bar\omega}\right]}$
	 & $\bar H\left[1+\frac{ 3\gamma}{2k\bar\omega}\right]$& $-$& (S)
	 \\
${\cal A}_{\cal S}(VI_h)$ & & &$-$ & (A) \\ \hline
\end{tabular}
\end{center}
\caption[{\em Values of $q$ and $H$ in the Jordan frame}]
{{\em This table lists all of the equilibrium points and, for each corresponding model, the value of
$\up{st}q$ and $\up{st}H$ in the Jordan frame, and gives the
conditions for inflation.  The notation $\up{sf}\bar H\equiv
e^{\half\phi/\bar\omega} H >0$ is used.  The last column denotes
whether the equilibrium set is a source (R), saddle (S) or sink (A)
for the values of $k$ for which the model inflates.  For the
equilibrium sets $\{\mbox{P(I)}, \mbox{P$^\pm$(II)},
\mbox{P(VI$_h$)},{\cal L}_l^\pm ,{\cal K}\}\in{\cal F}$ the respective
deceleration parameters are identical to those in the Einstein frame
and are listed in table \ref{table_sum}.  $^\dagger$For the point
P$^\pm_{\cal S}(VI_h)$, $n\equiv k^2(l-3)+4l<0$.\label{Jordan_isotropize}}}
\end{table}

In table \ref{Jordan_isotropize}, the only possible inflationary
equilibrium point is P$_{\cal S}(I)$ which is an attractor for
$k^2\leq 2$.  However, unlike in the Einstein frame, it is possible for
P$_{\cal S}(I)$ to represent a non-inflationary model if
$k/\bar\omega<-2$.  For instance, if $k^2=1$ and
$\omega_0<\frac{5}{4}$ then this equilibrium point represents a
non-inflating model if, say, $k=-1$ and $\bar\omega<
+1/\sqrt{2}$ or $k=+1$ and $\bar\omega<-1/\sqrt{2}$.

Although it is possible to have $\up{st}q<0$ at other equilibrium
points, the conditions requiring $\up{st}q<0$ will also require
$\up{st}\bar H<0$ which is not an inflating solution but rather a
contracting solution whose contraction is decreasing to the future.

\section{Discussion\label{VI}}

In this chapter the qualitative properties of Bianchi type B
cosmological models containing a barotropic fluid and a scalar field
with an exponential potential has been discussed.  The most general
models are those of type VI$_h$, which include the anisotropic scaling
solutions, and those of type VII$_h$, which include the open FRW
models.  

In cases in which monotone functions can be found, global results can
be proven.  Otherwise, based on the local analysis of the stability of
equilibrium points and the dynamics on the boundaries of the
appropriate state space, plausible global results have been presented 
(this is similar to the analysis of perfect fluid models in
\cite{Wainwright1997a} and \cite{Hewitt1993a} in which no monotone
functions were found in the Bianchi type VI and VII invariant sets).

The following is a summary list of the main results:

\begin{itemize}
\item All models with $k^2<2$ asymptote toward the flat FRW power-law
inflationary model \cite{Halliwell1987a,Coley1997a},
corresponding to the global attractor $P_{\cal S}(I)$, at late times;
i.e., all such models isotropize and inflate to the future.

\item $F_{\cal S}(I)$ is a saddle and hence the flat FRW scaling
solutions \cite{Wetterich1988a,Wands1993a} do not act as late-time
attractors in general \cite{Billyard1998a}.

\item A subset of ${\cal K}_{\cal M}$ acts as a source for all Bianchi
type B models; hence all models are asymptotic in the past to a massless
scalar field analogue of the Jacobs anisotropic Bianchi I solutions

\item For $k^2\geq 2$, Bianchi type VII$_h$ models generically
asymptote towards an open FRW scalar field model, represented by one
of the local sinks $P_{\cal S}(V\!)$ or $P_{\cal S}^\pm\!(VII_h)$, and
hence isotropize to the future.

\item For $k^2\geq 2$, Bianchi type VI$_h$ models generically
asymptote towards either an anisotropic scalar field analogue of the Collins
solution \cite{Collins1971a}, an anisotropic vacuum solution (with no
scalar field) or an anisotropic scaling solution
\cite{Coley1998a}, corresponding to the local sinks
$P_{\cal S}^\pm(VI_h)$, ${\cal L}_l(VI_h)$ or ${\cal A}_{\cal
S}(VI_h)$, respectively (depending on the values of a given model's
parameters - see table IV for details).  These models do not generally
isotropize.

\item In particular, the equilibrium point ${\cal A}_{\cal S}(VI_h)$
is a local attractor in the Bianchi VI$_h$ invariant set and hence
there is an open set of Bianchi type B models containing a perfect
fluid and a scalar field with exponential potential which asymptote
toward a corresponding anisotropic scaling solution at late times.
\end{itemize}

It should be stressed that the analysis and results herein are applicable to a
variety of other cosmological models in, for example, scalar-tensor
theories of gravity (which has been demonstrated in the preceding section),
theories with multiple scalar fields with exponential potentials
\cite{Liddle1998a} and string theory \cite{Billyard1999b,Kaloper1998b}.

\chapter{Interaction Terms\label{reheat}}

Although the exponential models are interesting models because they
lead to late-time isotropization, they also have some shortcomings.
For $k^2<2$, it has been shown in chapter
\ref{BianchiB} that {\em all} Bianchi class B models asymptote towards
a power-law inflationary model in which the matter terms are driven to
zero, and there is no graceful exit from this inflationary phase.
Furthermore, the scalar field cannot oscillate in this inflationary
model and so reheating cannot occur by the conventional scenario.
Clearly, these models need to be augmented in an attempt to alleviate
these problems.  For example, exponential potentials are believed to
be an approximation, and so the theory could include more complicated
potentials.  However, it is not clear what these other potentials should be,
and so another way to augment the models is to introduce interaction
terms where the energy of the scalar field is transferred to the
matter field.  This chapter examines how some interaction terms can
affect the qualitative behaviour of models containing matter and a
scalar field with an exponential potential.  Such an interaction term,
represented by $\delta$ in this thesis, arises in the conservation
equations, via
\beqn
\mainlabel{conserve_interact}
\label{phi_con}
&& \dot\phi\left(\ddot\phi +3H\dot\phi +kV\right) = -\delta \\
\label{mu_con}
&& \dot\mu + 3\gamma H\mu = \delta.
\eeqn

The form of such a term is often discussed in the literature within
the context of inflation and reheating.  In many inflationary
scenarios, inflation drives the matter content to zero
\cite{Linde1987a}.  In order to ``reheat'' the universe after
inflation, the energy of the scalar field is converted into photons.
In models where the scalar field's potential is quadratic or quartic
\cite{Linde1983a,Linde1983b,Abbott1982a,Amendola1996a,Morikawa1984a,Berera1996a,Calzetta1995b,Kamenshchik1998b,Graziani1989a,Liddle1998b},
inflation occurs during a ``slow roll'' regime where $\dot\phi\ll V$
and ends when the scalar field oscillates about the potential's
minimum.  During this latter phase, the scalar field is coupled
 to the other fields (fermionic and bosonic) in which the
scalar field's energy is transferred to the matter content, thereby
reheating the universe.  An alternative model is the warm
inflationary model \cite{Berera1995a,Bellini1999a,Ferreira1998a}, whereby
an interaction term is significant throughout the inflationary
regime (not just after slow-roll) and so the energy of the scalar
field is continually transferred to the matter content throughout inflation.  In
this model, the matter content is {\em not} driven to zero and so
reheating is not necessarily required.

Several examples of interaction terms appear in the literature for
models with a variety of self-interaction potentials.  In particular,
potentials which have a global minimum have attracted much attention.
Albrecht {\em et al.} \cite{Albrecht1982b} considered
$\delta=a\dot\phi^d\phi^{5-2d}$ (where $a$ is a constant), derived
from dimensional arguments, in the reheating context after inflation
(with potentials derived from Georgi--Glashow SU(5) models).  In a
similar context, Berera
\cite{Berera1995b} considered interaction terms of the form $\delta =
a\dot\phi^2$.  Quadratic potentials and interaction terms of the form
$\delta= a\dot\phi^2$ and $\delta= a\phi^2\dot\phi^2$ were considered
by de Oliveira and Ramos \cite{Oliveira1998a} and a graceful exit from
inflation was demonstrated numerically. Similarly, Yokoyama {\em et
al.}
\cite{Yokoyama1987a}
showed that an interaction term, which is negligible during the
slow-roll inflationary phase, dominates at the end of inflation when
the scalar field is oscillating about its minimum; during this
reheating phase it is assumed that the energy transferred from the
scalar field is solely converted into particles.

Within the context of exponential potentials, Yokoyama and Maeda
\cite{Yokoyama1988a} as well as Wands {\em et al.}
\cite{Wands1993a} considered interactions of the form
$\delta=a\sqrt{V}$.  The main goal in both these papers was to show
that power-law inflation can occur for $k^2>2$, thereby showing that
inflation can exist for steeper potentials.  The main motivation for
this work is the fact that exponential potentials which arise
naturally from other theories, such as supergravity or superstring
models, typically have $k^2>2$.  Wetterich
\cite{Wetterich1995a} considered interaction terms containing a matter
dependence, namely $\delta = a \mu$, in which perturbation analysis
showed that the matter scaling solutions were stable solutions when
such interaction terms are included.  In \cite{Wetterich1995a}, it was
shown that the age of the universe is older when $\delta$ is included
and that the scalar field can still significantly contribute to the
energy density of models at late times.  As discussed in section
\ref{StringtoEinstein} of chapter \ref{SFtoST}, certain string
theories in which the energy sources are separately conserved in the
Jordan frame naturally lead to interaction terms in the Einstein
frame, although this is {\em not} specific to string cosmologies; any
scalar-tensor theory with matter terms and a power-law potential will
yield the same results \cite{Wetterich1995a}.  If some of these energy
sources can be modeled as matter sources, then the corresponding
interaction term is of the form $a\dot\phi\mu$ (in chapters
\ref{string} - \ref{Qsection} it will be shown that in the string
models in the Einstein frame the interaction terms are linear in {\em
both} $\mu$ and $p$ and the matter does not have a linear equation of
state, $p\not\propto\mu$, except at the equilibrium points; however,
this chapter explicitly considers linear equations of state,
$p=(\gamma-1)\mu$, and the coefficient involving $\gamma$ will be
absorbed into the constant $a$ in the interaction term).  Finally, a
term of the form $\delta=a\mu H$ might be motivated by analogy with
dissipation.  For example, a fluid with bulk viscosity may give rise
to a term of this form in the conservation equation
\cite{Eckart1940a,Coley1990a}

If a matter term is absent from $\delta$ then unphysical situations
may arise.  For conventional interaction terms without $\mu$,
numerical studies show that $\mu$ becomes zero.  Heuristically,
suppose that at some instant of time $\mu=0$, then (\ref{mu_con})
becomes
\begin{displaymath}
\dot\mu = \delta,
\end{displaymath}
and if $\delta<0$ then the matter energy density will become negative
(such a negative interaction term is possible, for instance, in
transforming from the Jordan frame to the Einstein frame, the sign of
$\pm\sqrt{\omega +3/2}$ determines the sign of $\delta$ which can thus
be negative; similarly, the term $\delta\propto
\dot\phi\phi$ in \cite{Albrecht1982a} can also lead to $\delta<0$).
Thus, $\mu$ shall be included in the interaction terms to follow in
order to ensure that $\mu\geq 0$.  In \cite{Wetterich1995a} it was
shown that $\delta<0$ will {\em not} admit static solutions.
Furthermore, the sign of $\delta$ will should be positive at the
equilibrium points representing inflationary models, otherwise the
matter field will be ``feeding'' the scalar field and will redshift to
zero even faster than in the absence of the interaction terms.

This chapter examines interaction terms of the general form
$\delta=\bar\delta\mu H$ (where
$\bar\delta=\bar\delta(\dot\phi,V,H)$) in the context
of flat FRW models within the physically relevant range
$2/3<\gamma<2$, in order to determine the asymptotic properties of
these models.  In particular, it will be determined whether these
models can asymptote towards inflationary models in which the matter
terms are not driven to zero.  This would partially alleviate the need
for reheating; since the matter content tracks that of scalar field
(in this context) it is never driven to zero (unless both are driven
to zero in which case the solution is not inflationary).  However, a
more comprehensive reheating model would still be necessary.  The
structure of the chapter is as follows.  In section
\ref{i_govern}, the governing equations are defined and the case $\delta=0$  studied in \cite{Copeland1997a} is reviewed.  In section \ref{i_I}, the case
$\delta=a\dot\phi\mu$ is studied, motivated by the conformal relationships
between the Jordan and Einstein frame in string theory, and extends
the work of \cite{Wetterich1995a}.  In section \ref{i_II} the case
$\delta=a\mu H$ is examined as another example of an interaction term.
The chapter ends with a discussion in section \ref{i_discuss}.  This
chapter works entirely within the Einstein frame and so the ``(sf)''
notation is suppressed.

\section{Governing Equations\label{i_govern}}

The governing field equations are given by the equations (\ref{conserve_interact}) and
\be 
\label{i_Hdot}
\dot{H} = -\frac{1}{2} (\gamma \mu + \dot{\phi}^2), 
\ee
subject to the Friedmann constraint
\be
\label{i_Fried_1}
H^2 = \frac{1}{3} (\mu + \frac{1}{2} \dot{\phi}^2 + V), 
\ee
(an overdot denotes ordinary
differentiation with respect to time $t$).  Note that the
total energy density of the scalar field is given by
$\mu_\phi = \frac{1}{2}\dot{\phi}^2 +V$. 
The deceleration parameter for this system is given by
\be
q = \third H^{-2} \left[\dot\phi^2-V+\half\left(3\gamma-2\right)\mu \right],
\ee
and is {\em independent} of the interaction term.

Defining
\be
\label{i_def_eqns}
x \equiv \frac{\dot{\phi}}{\sqrt{6}H} \q , \q y \equiv 
	\frac{\sqrt{V}}{\sqrt{3}H},
\ee
and the new logarithmic time variable $\tau$ by
\be
\label{i_new_time}
\frac{d \tau}{dt} \equiv H,
\ee
the governing differential equations can be written as the
plane-autonomous system:
\beqn
\mainlabel{i_xy_1}
\label{i_x_1}
x' & =& -3x - \sqrt{\frac{3}{2}} k y^2 + \frac{3}{2} x [2x^2 + \gamma (1-x^2 -y^2)] - \frac{\bar\delta}{6x}(1-x^2 -y^2), \\
\label{i_y_1}
y' &=& \frac{3}{2}y \left[\sqrt{\frac{2}{3}} k x + 2x^2 + \gamma (1-x^2 -y^2)\right], 
\eeqn
where a prime denotes differentiation with respect to $\tau$.
Note that $y=0$ is an invariant set, corresponding to $V=0$.  The
equations are invariant under $y\rightarrow -y$ and $t\rightarrow -t$
and so the region $y<0$ is a time-reversed mirror to the region $y>0$;
therefore, only $y>0$ will be considered.  Similarly, $k>0$ will be
considered since the equations are invariant under $k\rightarrow -k$ and
$x\rightarrow -x$.

Equation (\ref{i_Fried_1}) can be written as 
\benonumber \Omega + \Omega_\phi =1, \eenonumber
where
\be
\label{i_O_1}
\Omega \equiv \frac{\mu_\gamma}{3 H^2}, \q \Omega_\phi \equiv 
	\frac{\mu_\phi}{3H^2} = x^2 + y^2, 
\ee
which implies that $0 \leq x^2 +y^2 \leq 1$ for $\Omega \geq 0$ so that the
phase-space is bounded.  The deceleration parameter is now written
\be
q = -1 +3 x^2+\frac{3}{2}\gamma\left(1-x^2-y^2\right).
\ee

\subsection{Comments on arbitrary $\bar\delta$}

The fact that $q$ is independent of the interaction term implies that
the regions of phase space which represent inflationary models is the
same for all models considered.  Namely, $q=0$ occurs along the
ellipse $\gamma y^2 = (2-\gamma)x^2+\third(3\gamma-2)$.  For any value
of $\gamma$, the lines intersect the boundary of the phase space at
$x^2=\third$.

It is possible to make some qualitative comments about the system
(\ref{i_xy_1}) for arbitrary $\bar\delta$.  First, the dynamics on the
boundary $x^2+y^2=1$ ($\Omega=0$) are independent of the choice of
such interaction terms;  the three equilibrium points (and their
associated eigenvalues) which exist on the boundary for any $\bar\delta$ are:
\beqn
\mainlabel{boundary_points}
{\cal K}^+:&& (x,y)=(+1,0) \\ \nonumber\index{equilibrium sets!${\cal K}_{\cal M}$, ${\cal K}$, ${\cal K}^\pm$}
&&	\left(\lambda_1,\lambda_2\right) = \left(3(2-\gamma)+\third 
	\left.\bar\delta\right|_{{\cal K}^+},\sqrt{\frac{3}{2}}
	\left[\sqrt6 + k\right]\right), \\
{\cal K}^-:&& (x,y)=(-1,0) \\ \nonumber
&&	\left(\lambda_1,\lambda_2\right) = \left(3(2-\gamma)+\third 
	\left.\bar\delta\right|_{{\cal K}^-}, \sqrt{\frac{3}{2}}
	\left[\sqrt6 - k\right]\right), \\
P_{\cal S}:&& (x,y)=\left(-\frac{k}{\sqrt6},\sqrt{1-k^2/6} \right)\index{equilibrium sets!P$_{\cal S}(I)$} \\ \nonumber
&&	\left(\lambda_1,\lambda_2\right) = \left(-\half\left[6-k^2\right],
	-\left[3\gamma-k^2-\third\left.\bar\delta\right|_{P_{\cal S}}\right]
	\right).
\eeqn
The points ${\cal K}^\pm$ represent the isotropic subcases of Jacobs'
Bianchi type I solutions \cite{Collins1971a} (subcases of the Kasner
models), generalized to include a massless scalar field.  These
solutions are non-inflationary ($q=2$).  The point $P_{\cal S}$, which
exists only for $k^2<6$, represents the FRW power-law model
\cite{Halliwell1987a,Coley1997a}, which is inflationary for $k^2<2$
($q=\half\left[k^2-2\right]$).  Although these three points exists for
any $\bar\delta$, the interaction term {\em does} affect the stability
of these solutions, as is evident from the eigenvalues in \eref{boundary_points}.  In
particular, for $k^2<3\gamma$ the point $P_{\cal S}$ can become a
saddle point if
\benonumber
\left.\bar\delta\right|_{P_{\cal S}} > 3\left(3\gamma-k^2 \right).
\eenonumber
Hence, if the interaction term is significant, solutions will spend an
indefinite period of time near this power-law inflationary model, but
will then evolve away and typically be attracted to another
equilibrium point in some other region of the phase space.

Matter scaling solutions (i.e. those solutions in which
$\gamma_\phi=\gamma$), denoted by ${\cal F}_{\cal S}$ in chapter
\ref{BianchiB}, exist only in special circumstances when such
interaction terms are present, and occur at the point
\be
x_{{\cal F}_{\cal S}} = -\sqrt\frac{3}{2}\frac{\gamma}{k}, \qquad 
	y_{{\cal F}_{\cal S}} = \sqrt{\frac{3\gamma(2-\gamma)}{2k^2}}.
\ee
Substituting these solutions into equations (\ref{i_xy_1}) yields
\beqn
x' &=& \frac{(3\gamma-k^2)}{3\sqrt6k\gamma}
	\left.\bar\delta\right|_{{\cal F}_{\cal S}}, \\
y' &=& 0.
\eeqn
Hence, the matter scaling solutions will be represented by an
equilibrium point only if $\left.\bar\delta\right|_{{\cal F}_{\cal
S}}=0$ (or in the special case $3\gamma=k^2$, which is typically a
bifurcation value).  For the simple forms for $\delta$ given in the literature
and those used in this chapter, this condition will not be satisfied
and so the matter scaling solutions cannot be asymptotic attracting
solutions.  

However, an analogous situation does arise.  In
particular, any equilibrium point within the boundary of the phase
space will satisfy
\be
y_0^2 = \frac{(2-\gamma_\phi)}{\gamma_\phi} x_0^2,
\ee
in which the scalar field is equivalent to a perfect fluid of the form
$p_\phi=(\gamma_\phi -1)\mu_\phi$, but where $\gamma_\phi\neq\gamma$.
Consequently, any attracting equilibrium point within the phase space
will represent models in which neither the matter field nor the scalar
field is negligible and the scalar field mimics a barotropic fluid
different from the matter field and therefore could still constitute a
possible dark matter candidate.

Finally, it can be shown that any equilibrium point within (but not
on) the boundary will occur for $x_0<0$.  For $y\neq0$ and $\Omega\neq 0$ equation (\ref{i_y_1}), which does not depend on $\delta$, yields 
\be
\label{yp_zero}
\gamma y^2 = \gamma +\sqrt\frac{2}{3}kx +(2-\gamma)x^2,
\ee
a relationship any such equilibrium point must satisfy.

Now, $\gamma(y^2+x^2)<\gamma$ since $y^2+x^2<1$ inside the boundary, and hence  (\ref{yp_zero}) yields
\be
x\left(x+\frac{k}{\sqrt 6}\right) <0,
\ee
which cannot be satisfied for $x>0$ (since $k>0$).

\subsection{Review of the case $\delta=0$\label{i_d0}}

Copeland {\em et al.} \cite{Copeland1997a} performed a phase-plane
analysis of the system (\ref{i_xy_1}) for $\delta=0$, and found five
equilibrium points.  One of the equilibrium points (denoted here by
$P$) represents a flat, non-inflating FRW model
\cite{Copeland1997a}, for which $\Omega=1$.   For $2/3<\gamma<2$ this point is a
saddle in the phase space.  The flat FRW matter scaling solution
 ($F_{\cal S}$) was found to exist for $k^2>3\gamma$ and was shown to
 be a sink.  The equilibrium point ${\cal K}_{\cal M}^+$ was shown to
 be a source for all $k$ and ${\cal K}^-$ a source for $k^2<6$.  The
 FRW power-law model ($P_{\cal S}$) was shown to be a sink for
 $k^2<3\gamma$, and was shown to represent an inflationary model for
 $k^2<2$.  The results found in \cite{Copeland1997a} are summarized in
 table \ref{i_table1}.
\begin{table}[ht]
\begin{center}
\begin{tabular}{|l||c|c|c|c|}
\hline
	& $0\leq k^2 \leq 2$ 	& \multicolumn{2}{|c|}{$2<k^2<6$} 		& $k^2>6$ \\ \hline 
	&  	& $k^2<3\gamma$ & $k^2>3\gamma$ &  \\ \hline 
\hline
$P$ 	& {\scriptsize saddle}	& \multicolumn{2}{|c|}{\scriptsize saddle}	& {\scriptsize saddle} \\
	& {\tiny (NI)}		& \multicolumn{2}{|c|}{\tiny (NI)}		& {\tiny (NI)} \\
\hline
${\cal K}^+$ & source 		& \multicolumn{2}{|c|}{source} 	& source \\
	& {\tiny (NI)}		& \multicolumn{2}{|c|}{\tiny (NI)}		& {\tiny (NI)} \\
\hline
${\cal K}^-$ & source 		& \multicolumn{2}{|c|}{source} 	& {\scriptsize saddle} \\
	& {\tiny (NI)}		& \multicolumn{2}{|c|}{\tiny (NI)}		& {\tiny (NI)} \\
\hline
$P_{\cal S}$ & sink 	&  sink &  {\scriptsize saddle} & $-$ \\
	& {\tiny (I) (for $k^2<2$)} & {\tiny (NI)} & {\tiny (NI)}&  \\
\hline
${\cal F}_{\cal S}$ & $-$ &$-$	&  sink & sink \\
	& && {\tiny (NI)}	&  {\tiny (NI)} \\
\hline
\end{tabular}
\end{center}
\caption[Equilibrium points for $\delta=0$]{{\em The equilibrium
points for $\delta=0$ and their stability for various values of $k$.
The label ``(NI)'' denotes non-inflationary models whereas ``(I)''
represents inflationary models.}\label{i_table1}}
\end{table}

\section{Case I: $\delta=-a\dot\phi\mu$\label{i_I}}

In this section, an interaction term of the form
$\delta=-a\dot\phi\mu$ (and hence $\bar\delta=-a\dot\phi/H$) shall be considered.  Again, it will be assumed that $k>0$.  The explicit sign choice for $\delta$, with the
assumption that $a>0$, is to guarantee that all equilibrium points
within the phase space will represent models in which energy is being
transferred from the scalar field to the perfect fluid, since it was
shown that all equilibrium points within the phase space occur for
$x<0$ ($\dot\phi<0$).  Indeed, this is even true for the equilibrium
point $P_{\cal S}$ on the boundary $\Omega=0$, since it is located at
$x<0$.  With this particular choice for $\delta$, equations
(\ref{i_xy_1}) become
\beqn
\mainlabel{i_xy_2}
\label{i_x_2}
x' & =& -3x(1-x^2) - \sqrt{\frac{3}{2}} k y^2 + \left(\frac{3}{2}\gamma x 
	+ \sqrt{\frac{3}{2}}a\right)\left(1-x^2-y^2\right), \\
\label{i_y_2}
y' &=& \frac{3}{2}y \left[\sqrt{\frac{2}{3}} k x + 2x^2 + 
	\gamma (1-x^2 -y^2)\right].
\eeqn

\noindent There are five equilibrium points for this system:
\begin{enumerate}
\item ${\cal K}^+:\qquad (x,y)=(+1,0), \qquad \Omega=0, \qquad q=2.$ \\
The eigenvalues for this equilibrium point are 
\be
\left(\lambda_1,\lambda_2\right) = \left(3(2-\gamma)-\sqrt6 a,
		\sqrt{\frac{3}{2}}\left[\sqrt6 + k\right]\right)
\ee
This equilibrium point is a source for $a<\sqrt\frac{3}{2}(2-\gamma)$
and a saddle otherwise.

\item ${\cal K}^-:\qquad (x,y)=(-1,0), \qquad \Omega=0, \qquad q=2.$ \\
The eigenvalues for this equilibrium point are 
\be
\left(\lambda_1,\lambda_2\right) = \left(3(2-\gamma)+\sqrt6 a, 
	\sqrt{\frac{3}{2}} \left[\sqrt6 - k\right]\right),
\ee
and so ${\cal K}^-$ is a source for $k^2<6$ and a saddle otherwise.

\item $\displaystyle P_{\cal S}:\qquad
(x,y)=\left(-\frac{k}{\sqrt6},\sqrt{1-\frac{k^2}{6}} \right), \qquad \Omega=0,
\qquad q=\half(k^2-2).$  \\ 
The eigenvalues for this equilibrium point are 
\be
\left(\lambda_1,\lambda_2\right) = \left(-\half\left[6-k^2\right],
	-\left[3\gamma-k^2-ka\right]\right).
\ee
This point exists only for $k^2<6$ (when $k^2=6$ then $P_{\cal S}$
merges with the equilibrium point ${\cal K}^-$).  Here, it is evident
that $P_{\cal S}$ is a sink for $a<(3\gamma-k^2)/k$ and a saddle
otherwise.

\index{equilibrium sets!${\cal N}$}
\item $\displaystyle {\cal N}:\qquad (x,y)=\left(-\sqrt\frac{3}{2}
	\frac{\gamma}{\Delta},\sqrt{\frac{3\gamma(2-\gamma)+2a\Delta}
	{2\Delta^2}} \right), \qquad
	\Omega=\frac{k\Delta-3\gamma}{\Delta^2},$ \\ 
	\hspace*{\fill} $\displaystyle q=\frac{3\gamma k-2\Delta}{\Delta},$ 
	\hspace*{4em}\\
where $\Delta\equiv k+a>0$.  Note that this solution is physical
(i.e., $\Omega\geq 0)$ either for $k^2>3\gamma$ or for $k^2<3\gamma$  and
\be
a\geq (3\gamma-k^2)/k\label{exist_1}.
\ee  
These solutions were discussed in \cite{Wetterich1995a} for $a<k$ and
are related to similar power-law solutions discussed in
\cite{Wands1993a}.  This model inflates if
\be 
a\geq(\frac{3}{2}\gamma-1)k\label{inflate_1}.
\ee
(note that in \cite{Wetterich1995a} only $a<k$ was considered and
hence the solutions therein were {\em not} inflationary).  For
$k^2<2$, if condition (\ref{exist_1}) is satisfied then
(\ref{inflate_1}) is automatically satisfied and so these models
inflate for $k^2<2$.  For $2<k^2<3\gamma$, if condition
(\ref{inflate_1}) is satisfied then (\ref{exist_1}) is automatically
satisfied and therefore models can inflate for $k^2>2$ if
$a\geq(\frac{3}{2}\gamma-1)k$.  For $k^2>3\gamma$ there is no constraint on
$a$ for the point to exist and therefore whether this models inflates
is solely determined by (\ref{inflate_1}).  The eigenvalues for this
equilibrium point are 
\begin{eqnarray}
\label{newsink_1}
\lambda_\pm &=& \frac{-3\left[(2-\gamma)k+2a\right]}{4\Delta}\nonumber \\
	&& \pm \frac{\sqrt{9\left[(2-\gamma)k+2a\right]^2 
	-24\left[3\gamma(2-\gamma)+2a\Delta\right] \left[k\Delta-3\gamma\right]
		}}{4\Delta},
\end{eqnarray}
and so ${\cal N}$ is always a sink when it exists.  Note that the scalar
field acts as a perfect fluid with an equation of state parameter given by
\be
\gamma_\phi = \frac{\gamma}{1+\frac{a\Delta}{3\gamma}} < \gamma.
\ee

\index{equilibrium sets!${\cal N}_2$}
\item $\displaystyle {\cal N}_2:\qquad (x,y)=\left(\sqrt\frac{2}{3}
	\frac{a}{(2-\gamma)},0\right), \qquad
	\Omega=1-\frac{2a^2}{3(2-\gamma^2)},$ \\ 
	\hspace*{\fill} $\displaystyle q=\half(3\gamma-2)
	+\frac{a^2}{(2-\gamma)}>0.$\hspace*{5em}\\
This equilibrium exists for $a < \sqrt\frac{3}{2}(2-\gamma)$ and is a saddle, as determined from its eigenvalues:
\be
\left(\lambda_1,\lambda_2\right) = \left(-\frac{3}{2}[2-\gamma]
	\left[1-\frac{2a^2}{3(2-\gamma)^2}\right], \frac{3}{2}\gamma+
	\frac{a(k+a)}{(2-\gamma)}\right).
\ee
\end{enumerate}

Table \ref{i_table2} lists the equilibrium points and their stability
for the ranges of $k$ and $a$.
\begin{table}[ht]
\begin{center}
\begin{scriptsize}
\begin{tabular}{|l||c|c||c|c|c||c|c||c|c|}
\hline
& \multicolumn{2}{|c||}{$0< k^2 < 2$} & \multicolumn{3}{|c||}{$2<k^2<3\gamma$}
& \multicolumn{2}{|c||}{$3\gamma<k^2<6$} &\multicolumn{2}{|c|}{$k^2>6$}\\ 
\hline 
& $a<G_1$ & $a>G_1$ & $a<G_1$ &$G_1<a<G_2$ & $a>G_2$&$a<G_2$&$a>G_2$&$a<G_2$ 
& $a>G_2$
\\ \hline 
\hline
${\cal K}^+$ 	& \multicolumn{9}{|c|}{\rule[0em]{0em}{1.2em}{\bf R} 
			for $2a^2<3(2-\gamma)^2$} \\ 
   {\tiny(NI)}	& \multicolumn{9}{|c|}{s for $2a^2>3(2-\gamma)^2$ } \\ \hline 
${\cal K}^-$ 	& \multicolumn{7}{|c||}{\rule[0em]{0em}{1.2em}
			\raisebox{-.7em}[0pt]{\bf R}}&\multicolumn{2}{|c|}
			{\raisebox{-.7em}[0pt]{s}} \\ 
   {\tiny(NI)}	&  \multicolumn{7}{|c||}{ } &\multicolumn{2}{|c|}{ }\\ \hline 
$P_{\cal S}$ 	&  \rule[0em]{0em}{1.2em} {\bf A} & s & {\bf A} 
			& \multicolumn{4}{|c||}{s} &\multicolumn{2}{|c|}
			{\raisebox{-.7em}[0pt]{$-$}} \\
	& {\tiny (I)} & {\tiny (I)}& {\tiny (NI)} 
		 &\multicolumn{4}{|c||}{{\tiny(NI)}}&\multicolumn{2}{|c|}{ } \\
		\hline
${\cal N}$ 		 & \rule[0em]{0em}{1.2em}\raisebox{-.7em}[0pt]{$-$} & {\bf A} 
			& \raisebox{-.7em}[0pt]{$-$} & {\bf A}& {\bf A}
			&{\bf A}&{\bf A}&{\bf A}&{\bf A}\\
   	&     	 & {\tiny (I)} &   & {{\tiny (NI)}} & {{\tiny (I)}} 
			&{{\tiny (NI)}} &{{\tiny (I)}} & {{\tiny (NI)}}
			&{{\tiny (I)}}\\\hline
${\cal N}_2$ 		 & \multicolumn{9}{|c|}{\rule[0em]{0em}{1.2em} s for 
			$2a^2<3(2-\gamma)^2$} \\ 
     {\tiny(NI)} & \multicolumn{9}{|c|}{$-$ for $2a^2>3(2-\gamma)^2$}\\ \hline 
\end{tabular}
\end{scriptsize}
\end{center}
\caption[Equilibrium points for $\delta=-a\dot\phi\mu$]{{\em The equilibrium
points for the model with $\delta=-a\dot\phi\mu$ and their stability
for various values of $k$ and $a$.  Note that
$G_1\equiv(3\gamma-k^2)/k$ and $G_2\equiv(3\gamma-2)k/2$.  The symbol
``{\bf R}'' denotes when the equilibrium point is a source (repellor),
``s'' for when it is a saddle, ``{\bf A}'' for when it is a sink
(attractor), and ``$-$'' when it does not exist within the particular
parameter space.  The label ``(NI)'' denotes non-inflationary models
whereas ``(I)'' represents inflationary models.}\label{i_table2}}
\end{table}

As is evident, the presence of the interaction term can substantially
change the dynamics of these models.  Of particular interest, when $k^2<2$ and $a>(3\gamma-k^2)/2$, then the power-law model
$P_{\cal S}$ is no longer a sink.  Therefore, trajectories approach
this equilibrium point (therefore the models inflate for an indefinite
period of time), and eventually asymptote towards the inflating model
${\cal N}$.  For $a\approx (3\gamma-k^2)/2$, the eigenvalues for ${\cal N}$
(\ref{newsink_1}) are real and negative and so the attracting
solution is represented by an attracting node.  However, for
$a\gg(3\gamma-k^2)/2$ equation (\ref{newsink_1}) becomes
\be
\lambda_\pm \approx -\frac{3}{2} \pm \sqrt{-3ak}.
\ee
Hence for large $a$ this equilibrium point is a {\em spiral node}; 
trajectories exhibit a {\em decaying oscillatory behaviour} as they asymptote
towards ${\cal N}$.

\subsection{An Example}

\begin{figure}[htp]
  \centering
   \includegraphics*[width=5in]{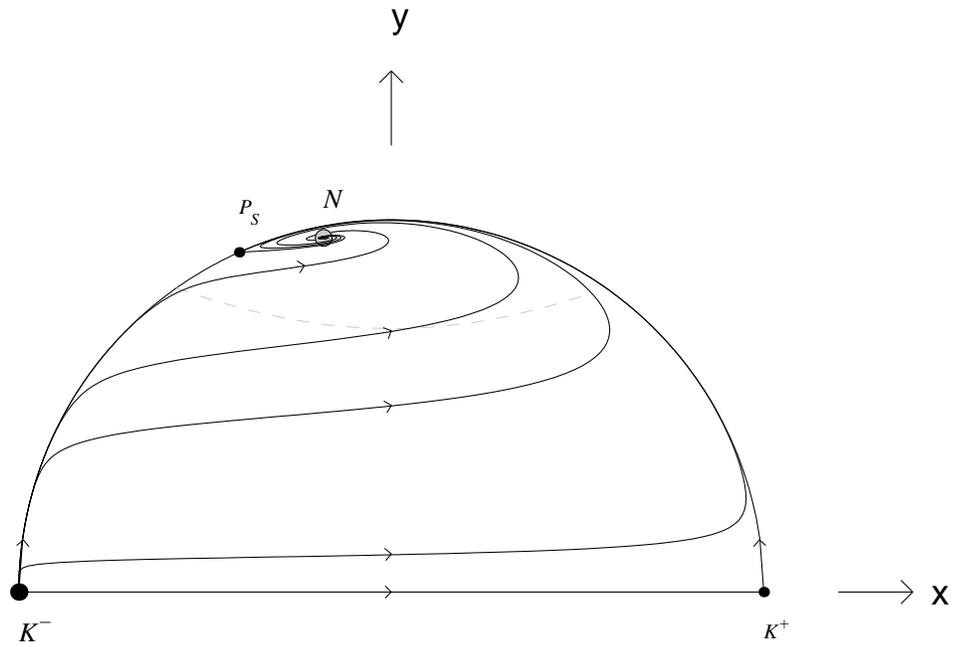}
  \caption[{\em Phase diagram for the case
$\delta=-a\dot\phi\mu$}]{{\em Phase diagram of the system
(\ref{i_xy_1}) when $\delta=-a\dot\phi\mu$ for the choice of parameters
$k=1$, $\gamma=4/3$ and $a=8$.  In this figure, the black dot
represents the source (i.e., the point ${\cal K}^-$), the large grey dot
represents the sink (i.e., the point ${\cal N}$) and small black dots
represent saddle points.  The region above the grey dashed line
represents the inflationary portion of the phase space. Arrows on the
trajectory indicate the direction of time. } \label{freheat1}}
\end{figure}

\begin{figure}[htp]
  \centering
  \framebox{ \includegraphics*[width=5in]{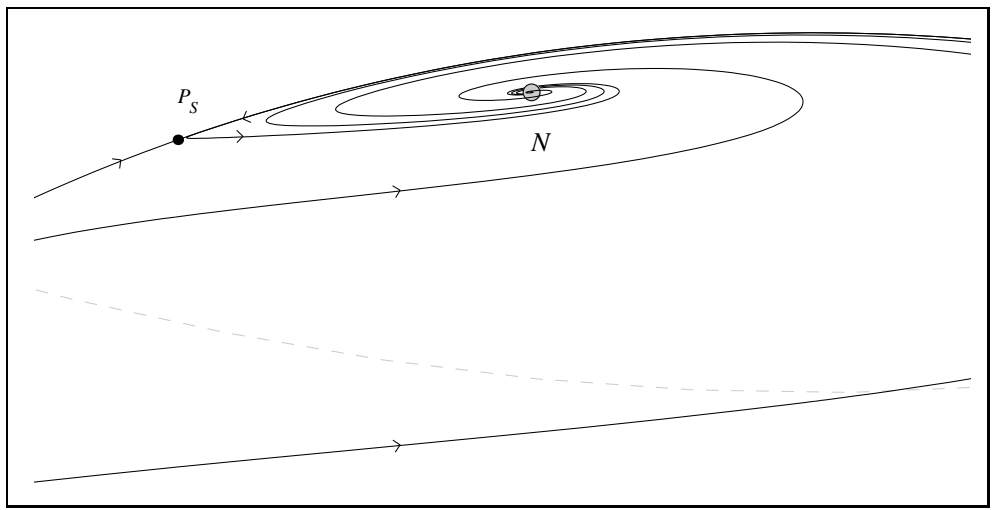}}
 \caption[{\em
  A magnification of the attracting region of figure \ref{freheat1}}]{{\em A
  magnification of the attracting region of the phase space depicted
  in figure \ref{freheat1}.  See also caption to figure
  \ref{freheat1}. } \label{freheat2}} \end{figure}

To illustrate this oscillatory nature, an explicit example is chosen
with $k=1$ and $\gamma=4/3$ (radiation).  The equilibrium points
and their respective eigenvalues are:
\beqn
\nonumber
{\cal K}^\pm: && (x,y)=(+1,0) \\
	&& \left(\lambda_1,\lambda_2\right) = \left(\pm\sqrt6 a+2,
	\sqrt\frac{3}{2}\right),\\\nonumber\\
\nonumber
P_{\cal S}: && (x,y)=\left(-\frac{k}{\sqrt6},\sqrt{1-\frac{k^2}{6}} \right) \\ 
	&& \left(\lambda_1,\lambda_2\right) = \left(a-3,-\frac{5}{2}\right),\\
\nonumber\\\nonumber
{\cal N}: && (x,y)=\left(\frac{-2\sqrt6}{3(a+1)}, \frac{\sqrt{4+3a(a+1)}}
		{\sqrt3(a+1)}\right)\\
	&&\lambda_\pm =\frac{-(3a+1)}{2(a+1)}\pm \frac{\sqrt{49+26a
	+33a^2-12a^3}}{2(a+1)},
\eeqn
where $a>3$ in order for ${\cal N}$ to exist and for it to be a sink,
as well as for $P_{\cal S}$ to be a saddle.  Note that for $a>3$
${\cal N}_2$ does not exist, ${\cal K}^+$ is a saddle and ${\cal K}^-$
is a source.  Numerical analysis shows that ${\cal N}$ is a spiral
sink for $a\gtrsim 3.65$.  Figure
\ref{freheat1} depicts this phase space for $a=8$, and the attracting region therein is magnified in figure \ref{freheat2}.  These figures are typical for
other values of $\gamma$ (this comment is important since we note
that in the context of conformally transformed scalar-tensor theories,
strictly speaking $\delta=0$ for $\gamma=4/3$).

\section{Case II: $\delta=a\mu H$: Perturbation Analysis\label{i_II}}

This section provides a second example to demonstrate that other types
of interaction terms can lead to a similar behaviour; i.e., that there
will be a range of parameters for which the inflationary models which
drive the matter fields to zero are {\em not} late-time attractors,
and for which the trajectories exhibit an oscillatory behaviour as
they asymptote to the late-time attracting solution.  Specifically, the
interaction term $\delta=a\mu H$ is chosen, where $a>0$.  

With this choice, equations (\ref{i_xy_1}) become
\beqn
\mainlabel{i_xy_3}
\label{i_x_3}
x' & =& -3x(1-x^2) - \sqrt{\frac{3}{2}} k y^2 + \left(\frac{3}{2}\gamma x 
	- \frac{a}{2x}\right)\left(1-x^2-y^2\right), \\
\label{i_y_3}
y' &=& \frac{3}{2}y \left[\sqrt{\frac{2}{3}} k x + 2x^2 + 
	\gamma (1-x^2 -y^2)\right].
\eeqn
For physical reasons, the line $x=0$ will not be considered.
Consequently, a full phase-plane analysis is not possible using these
variables.  However, it is still possible to determine the equilibrium
points with $x\neq0$ for the system and determine their local
stability.  Therefore, the analysis provided will constitute a
perturbation analysis whereby the stability of equilibrium points
cannot determine the global behaviour of the system.

\noindent There are four equilibrium points for this system for $x\neq0$:
\begin{enumerate}
\item ${\cal K}^+:\qquad (x,y)=(+1,0), \qquad \Omega=0, \qquad q=2,$ \\
The eigenvalues for this equilibrium point are 
\be
\left(\lambda_1,\lambda_2\right) = \left(3[2-\gamma] +a,
		\sqrt{\frac{3}{2}}\sqrt6 + 3\right)
\ee
This equilibrium point is a source.

\item ${\cal K}^-:\qquad (x,y)=(-1,0), \qquad \Omega=0, \qquad q=2.$ \\
The eigenvalues for this equilibrium point are 
\be
\left(\lambda_1,\lambda_2\right) = \left(3[2-\gamma]+ a, 
	-\sqrt{\frac{3}{2}} +3 \right),
\ee
and so ${\cal K}^-$ is a source for $k^2<6$ and a saddle otherwise.

\item $\displaystyle P_{\cal S}:\qquad
(x,y)=\left(-\frac{k}{\sqrt6},\sqrt{1-\frac{k^2}{6}} \right), \qquad \Omega=0,
\qquad q=\half(k^2-2).$  \\ 
The eigenvalues for this equilibrium point are 
\be
\left(\lambda_1,\lambda_2\right) = \left(-\half\left[6-k^2\right],
	-\left[\Xi-k^2\gamma-k^2\right]\right),
\ee
where $\Xi\equiv 3\gamma-a$.  This point exists only for $k^2<6$, and is a sink for $k^2<3\gamma$ and $a<3\gamma-k^2$ (saddle otherwise).

\index{equilibrium sets!${\cal N}_1$}
\item $\displaystyle {\cal N}_1:\qquad (x,y)=\left(-\frac{\Xi}{\sqrt 6k},
	\sqrt{\frac{a}{3\gamma}+\frac{(2-\gamma)}{\gamma}\frac{\Xi^2}{6k^2}} 
	\right), \qquad \Omega=\frac{\Xi(k^2-\Xi)}{3\gamma k^2},$ \\ 
	\hspace*{\fill} $\displaystyle q=-1+\half\Xi,$ 
	\hspace*{4em}\\
Note that this solution is physical ($\Omega\geq 0)$ for
$[3\gamma-k^2]<a<3\gamma$ and $k^2<3\gamma$, and represents an
inflationary model for $a>(3\gamma-2)$ (consequently, this model will
always inflate for $k^2<2$ and can inflate for $2<k^2<3\gamma$). The
eigenvalues for this equilibrium point are
\begin{eqnarray}
\label{newsink_2}
\lambda_\pm &=& -\frac{[2a(2\Xi-k^2)+(2-\gamma)\Xi^2]}{4\gamma\Xi}
	\nonumber \\ \nonumber
	&& \pm \frac{\sqrt{\left[2a(2\Xi-k^2)+(2-\gamma)\Xi^2\right]^2k^2 
	-8\gamma\Xi^2(k^2-\Xi)\left[2ak^2+(2-\gamma)\Xi\right]}}
	{4\gamma\Delta k},\\
\end{eqnarray}
and so ${\cal N}_1$ is a sink for $a<3\gamma-\half k^2$.  Note that the scalar
field acts as a perfect fluid with an equation of state parameter given by
\be
\gamma_\phi = \frac{\gamma}{1+\frac{ak^2}{\Xi^2}} < \gamma.
\ee
\end{enumerate}

Table \ref{i_table3} lists the equilibrium points and their stability
for the ranges of $k$ and $a$.
\begin{table}[ht]
\begin{center}
\begin{tabular}{|l||c|c|c||c|c|c||c|}
\hline
& \multicolumn{3}{|c||}{$0< k^2 < 2$} & \multicolumn{3}{|c||}{$2<k^2<3\gamma$}
& $3\gamma < k^2 < 6$\\ 
\hline 
& $a<G_3$ & \multicolumn{2}{|c||}{$a>G_3$} & $a<G_3$ & \multicolumn{2}{|c||}{$a>G_3$} &  \\ \hline
& & $a<G_4$ & $a>G_4$ &  & $a<G_4$ & $a>G_4$ & \\ \hline 
\hline
${\cal K}^\pm$ 	& \multicolumn{7}{|c|}{\rule[0em]{0em}{1.2em}{\bf R}} \\ 
   		& \multicolumn{7}{|c|}{{\tiny(NI)} } \\ \hline 
$P_{\cal S}$ 	&  \rule[0em]{0em}{1.2em} {\bf A} & \multicolumn{2}{|c||}{s} 
		& {\bf A}  & \multicolumn{3}{|c|}{s}\\
		& {\tiny (I)} & \multicolumn{2}{|c||}{{\tiny (I)}} 
		& {\tiny (NI)} &\multicolumn{3}{|c|}{{\tiny(NI)}} \\
		\hline
${\cal N}_1$ 		& \rule[0em]{0em}{1.2em} & {\bf A} 
		& s &  & {\bf A}& s  & 	 \\
   		&  {$-$} & {\tiny (I)}& {\tiny (I)} 
			& {$-$}  & {\tiny (I) for}
			& {\tiny (I) for}  & {$-$} \\
   		&     	 &   &  &   & \raisebox{.7em}[0pt]{\tiny $a>3\gamma-2$}
			& \raisebox{.7em}[0pt]{\tiny $a>3\gamma-2$}  &  \\
\hline
\end{tabular}
\end{center}
\caption[Equilibrium points for $\delta=a\mu H$]{{\em The equilibrium
points for the model with $\delta=a\mu H$ and their stability for various values of
$k$ and $a$.  Note that $G_3\equiv3\gamma-k^2$ and
$G_4\equiv3\gamma-\half k^2$.  The symbol ``{\bf R}'' denotes when the
equilibrium point is a source (repellor), ``s'' for when it is a
saddle, ``{\bf A}'' for when it is a sink (attractor), and ``$-$''
when it does not exist within the particular parameter space.  The
label ``(NI)'' denotes non-inflationary models whereas ``(I)''
represents inflationary models.  For $k^2>6$ the only equilibrium
points to exist are the points ${\cal K}^\pm$, for which ${\cal K}^+$
is a source and ${\cal K}^-$ is a saddle.}\label{i_table3}}
\end{table}

Again, for the range $k^2<2$ and $(3\gamma-\half k^2)<a<(3\gamma-\half
k^2)$, the power-law model $P_{\cal S}$ is no longer a sink and ${\cal N}_1$
is a sink.  Therefore, solutions approach the equilibrium point
which is represented by $P_{\cal S}$ (thereby inflating) for an
indefinite period of time, but eventually evolve away.  It can be
shown numerically that within this range for $k$ and $a$, the
equilibrium point ${\cal N}_1$ is a spiral node (for instance, for $k=1$ and
$\gamma=4/3$, ${\cal N}_1$ is a sink for $3<a<3.5$ and a spiral sink for
$3.24 \lesssim a < 3.5$); therefore the scalar field exhibits
oscillatory motion as the solutions asymptotes to ${\cal N}$.

\section{Discussion\label{i_discuss}}

Without an interaction term, it is known that for $k^2<2$ the
late-time attractor for the system (\ref{i_xy_1}) is a power-law
inflationary model in which the matter is driven to zero
\cite{Copeland1997a}.  The purpose of this chapter was to determine
whether this behaviour could be altered qualitatively with the
introduction of an interaction term.  Two examples given have show
that this is indeed possible.  In particular, there are values in the
parameter space for which this particular power-law inflationary model
becomes a saddle point, and so while the models may spend an
indefinite period of time inflating with $\Omega\rightarrow 0$, they
eventually evolve away from this solution.  The late-time attractors
within the same parameter space are also inflationary, however
the matter field's energy density remains a fixed fraction of the
scalar field's energy density and indeed to the total energy density.

In the absence of an interaction term, matter scaling solutions are
represented by equilibrium points of the dynamical system.  It was
shown here that for simple models found in the literature, they {\em
cannot} be represented by equilibrium points when an interaction term
is introduced.  However, new equilibrium points can arise which
represent models in which the energy densities of the matter and
scalar field remain a fixed proportion to one another.  Indeed, these
models are equivalent to a two-perfect-fluid model (for an example of
two-perfect-fluid model see \cite{Coley1986a}); for the examples
given, the constant $\gamma_\phi$ for the scalar field was given,
obeying $\gamma_\phi < \gamma$.  The solutions corresponding to these
equilibrium points are analogues of the matter scaling solutions in
which $\gamma_\phi=\gamma$.

In the first example ($\delta=-a\dot\phi\mu$) a phase-plane analysis
was performed to determine the qualitative properties and detailed
calculations for particular values $k=1$ and $\gamma=4/3$ (radiation)
were used to describe the qualitative behaviour described in more
detail (see figure \ref{freheat1}).  In the second example
($\delta=a\mu H$), the system as described is not well behaved for
$x=0$ and so only the local behaviour of the relevent equilibrium points is 
discussed. Again, the power-law inflationary model, represented by
$P_{\cal S}$, becomes a saddle point and the only attracting
equilibrium point is ${\cal N}_1$.

For an appropriate parameter range, the equilibrium point ${\cal N}_1$ is an
attracting focus, and hence as solutions approach this late-time
attractor the scalar field oscillates.  Although the late-time
behaviour is still inflationary, the
oscillatory behaviour provides a possible mechanism for inflation to
stop and for conventional reheating to ensue (indeed this is similar
to the mechanism for reheating in scalar field models with a potential
containing a global minimum 
\cite{Linde1987a,Linde1983a,Linde1983b,Amendola1996a,Berera1996a}). To
study reheating properly more complicated physics in which the
oscillating scalar field is coupled to both fermionic and bosonic
fields needs to be included.  This contrasts with the situation for
exponential models for $k^2<2$ with no interaction term which have no
graceful exit from inflation and in which there is no conventional
reheating mechanism.

Therefore, it has been shown that there are general relativistic
scalar field models with an exponential potential which evolve towards
an inflationary state in which the matter is not driven to zero and
which exhibit late-time oscillatory behaviour; these models may
constitute a first step towards a more realistic model.  There is the
question of how physical these models are, since they correspond to
relatively large values of $a$.  In the context that the interaction
term represents energy transfer, for physical reasons it might be
expected that $a$ must be small; i.e., $a<0.1$ \cite{Wetterich1995a}
(see also \cite{Amendola1999b}).  On the other hand, in the context of
scalar-tensor theories $a$ is of order unity and can certainly attain
values large enough to produce the behaviour described above; this is
also the situation in the context of string theories.

It is also of interest to study the cosmological consequences of the
`decaying cosmological constant' or `quintessential' cosmological
models, since they may be consistent with the observations of
accelerated expansion \cite{Perlmutter1999a,Riess1998a} and may lead
to a physically interesting residual scalar field energy-density which
is still present today.  These issues have recently been addressed by
Amendola \cite{Amendola1999a,Amendola1999c} in the context of the conformally
related scalar-tensor theories of gravity.

Finally, it is worth noting that inflationary solutions can be
obtained when $k^2>2$, unlike the case in which there is no
interaction term. Moreover, from Table \ref{i_table2} it can be seen that these
inflationary solutions occur for $a>\half(3\gamma-2)k$ and are sinks
(attractors). This result complements the results of
\cite{Wands1993a,Yokoyama1988a} who looked for inflationary solutions
for steeper potentials.  When $k^2<2$ and $0<a<(3\gamma-k^2)/2$ the
power-law inflationary solution corresponding to the equilibrium point
$P_{\cal S}$ is again a global attracting solution.  The late-time
behaviour of these models, both inflationary and non-inflationary, may
also be of cosmological interest.  Due to recent observations of
accelerated expansion \cite{Perlmutter1999a,Riess1998a}, models that
are presently inflating are also of interest.

\chapter{String Models I: Non-Zero Central Charge Deficit\label{string}}

The next four chapters consider string models defined by action
(\ref{sigmaaction_i}) in the Introduction.  The general solutions to
the field equations of action (\ref{sigmaaction_i}) are known
analytically when $\Lambda=\Lambda_{\rm M} =Q=0$ for both the spatially flat
and isotropic Friedmann--Robertson--Walker (FRW) universes and the
anisotropic Bianchi type I models
\cite{Copeland1994b,Copeland1995b,Meissner1991a}.  It can be shown
that the action (\ref{sigmaaction_i}) is invariant under a global
${\rm SL}(2, R)$ transformation on the dilaton and axion fields
\cite{Shapere1991a,Sen1993a}, when $\Lambda=\Lambda_{\rm M}=Q=0$.  The
general Friedmann--Robertson--Walker (FRW) cosmologies derived from
equation (\ref{sigmaaction_i}) (with $\Lambda=\Lambda_{\rm M}=Q=0$)
have been found by employing this symmetry
\cite{Copeland1994b,Copeland1995b}.  Specifically, the general
solution where only $\Lambda =\Lambda_{\rm M}=Q=0$ is the
`dilaton--moduli--axion' solution
\cite{Copeland1994b,Copeland1995b} (see equation (\ref{dma}) for $d=0$).  This cosmology asymptotically approaches one of the
dilaton--moduli--vacuum solutions (see (\ref{dmv}) for $h_0^2=\third$) in
the limits of high and low spacetime curvature.  The axion field
induces a smooth transition between these two power-law solutions and
causes a bounce to occur.  It is only dynamically important for a
short time interval when $s \approx s_0$
\cite{Copeland1994b,Copeland1995b}.  However, the ${\rm SL}(2, R)$ symmetry is broken when a cosmological constant is present \cite{Kar1996a} (either $\Lambda$,
$\Lambda_{\rm M}$ or $Q$) and the general FRW solution is not known in
this case.  There are two known solutions, which are the ``rolling
radii'' solutions found by Mueller
\cite{Mueller1990a,Tseytlin1992a,Tseytlin1992b,Tseytlin1992c,Tseytlin1994a}
(see equation (\ref{roll})), and the ``linear dilaton-vacuum''
solution
\cite{Myers1987b,Antoniadis1988a,Antoniadis1989a} (see
equation (\ref{static})).  Previously, analytical solutions had not been
found for this model when the axion field is trivial and $\Lambda_{\rm
M} >0$
\cite{Barrow1990a,Lidsey1997a}.    Moreover, the combined effects of 
$\Lambda_{\rm M}$ and axion field had not been considered previously.
The work herein extends previous qualitative analyses where one or
more of the terms in action (\ref{sigmaaction_i}) was neglected
\cite{Goldwirth1994a,Kaloper1995a,Easther1996a,Kaloper1996a,Behrndt1994a,Behrndt1994b}.

As will be demonstrated, most of the analysis to follow is equally
applicable to curved FRW models with a modulus field {\em and} to
certain Bianchi type I, V and IX models with or without a modulus
field, and therefore extends the work by Easter {\em et al.}
\cite{Easther1996a}, who performed a perturbation analysis to a static $k=+1$
FRW solution to show that it was a late-time attractor.  The only section
where this equivalence does not hold is in chapters \ref{NSNSRR} and \ref{Qsection}
wherein only curved FRW models are considered (the shear terms have
been neglected).

In this chapter, $\Lambda_{\rm M}=0$ and $Q=0$ in order to determine
the r\^{o}le of $\Lambda$ in the dynamics of the system (\ref{rr}),
although several comments here are equally applicable to the following
chapters.  The chapter is organized as follows.  In section
\ref{Governing_Equations}, the field equations with all parameters present
are derived from the string action.  All of the terms of action
(\ref{sigmaaction_i}) will be kept arbitrary for this section so that
they can be used in subsequent chapters, but $\Lambda_{\rm M}=Q=0$
will be used for the actual analysis of this chapter.  All the known
corresponding exact solutions are listed in section
\ref{Exact_Solutions} for $\Lambda_{\rm M}=Q=0$.  Section
\ref{Asymptotic_Behaviour} discusses the asymptotic behaviour of the axion
and curvature terms.  There it will be shown that for most cases
either one of the two will asymptotically dominate at early or late
times, and hence the four-dimensional dynamical system can be reduced
to a three-dimensional system.  The arguments in this section are
equally applicable to the analysis of chapter \ref{RR}.  For the
remainder of the chapter, the case $\Lambda_{\rm M}=Q=0$ is explicitly
studied, examining the combined effects of the axion field, modulus
field and central charge, thereby extending previous qualitative
analyses
\cite{Goldwirth1994a,Kaloper1995a,Easther1996a,Kaloper1996a,Behrndt1994a,Behrndt1994b}.
The chapter ends with a summary section and a section which
discusses the corresponding solutions and asymptotic behaviour in the
Einstein frame.  Since the work in this chapter is primarily confined
to the Jordan frame, the index ``(st)'' shall be omitted to save
notation, but must be introduced again in the final section when both
frames are discussed.

\section{Governing Equations\label{Governing_Equations}}

A class of spatially-homogeneous
spacetimes with non-trivial curvature shall be considered here.  It shall assumed that the
three-dimensional Ricci curvature tensor is isotropic; i.e.,
$^3R_{\mu\nu}$ is proportional to $k\delta_{\mu\nu}$ on the spatial
hypersurfaces (and hence the spatial hypersurfaces have constant
curvature $k$) \cite{MacCallum1973a}.  This class of models, more commonly
known as the isotropic curvature models, contain the Bianchi type I ($k=0$)
and V ($k<0$) models and a special case of the Bianchi type IX ($k>0$)
model.  In essence, these isotropic curvature models can be regarded
as the simplest anisotropic generalizations of FRW (flat, open,
closed, respectively) models.

In particular, the four--dimensional, anisotropic, space--time with
constant curvature $k$ shall be assumed, with a line element of the
form
\begin{equation}
ds^2 =-dt^2 +e^{2\alpha (t)} \left[ \left(\mathbf{\omega}^1\right)
	+e^{-2\sqrt3\beta_s(t)}\left(\mathbf{\omega}^2\right)^2 +
	e^{2 \sqrt3\beta_s(t)} \left(\mathbf{\omega}^3\right)^2\right], 
	\label{BI_V_IX}
\end{equation}
where 
\beqn
\mbox{for $k=0$ (Bianchi I)} && [\mathbf{\omega}^1,\mathbf{\omega}^2,
	\mathbf{\omega}^3]=\left[dx,dy,dz\right]\\
\mbox{for $k<0$ (Bianchi V)} && [\mathbf{\omega}^1,\mathbf{\omega}^2,
	\mathbf{\omega}^3]=\left[dx,e^xdy,e^xdz\right]\\
\mbox{for $k>0$ (Bianchi IX)} && [\mathbf{\omega}^1,\mathbf{\omega}^2,
	\mathbf{\omega}^3]=\left[dx-\sin y \,dz, 
	\right. \nonumber \\
	&& \qquad \qquad \qquad \cos x\,dy +\cos y\,\sin x\,dz, \nonumber \\
	&& \left. \qquad \qquad \qquad -\sin x\,dy + \cos y\,\cos x\,dz\right].
\eeqn
Since, the modulus field ($\beta_m$) has
the same functional form in the field equations as the shear terms
($\beta_s$) when $Q=0$, both terms can be combined via
\begin{equation}
{\dot\beta}^2\equiv{\dot\beta}_s^2+{\dot\beta}_m^2, \label{Betadefine}
\end{equation}
to yield the same field equations.  Indeed, several modulus fields can
be included via 
\begin{equation}
\label{sumup}
{\dot\beta}^2 = {\dot\beta}_s^2 +\Sigma {\dot\beta}_m^2.
\end{equation}
It shall be implicit throughout the text that the appearance of the
``$\beta$'' term is the combination of the shear and modulus field  
for the $Q=0$ cases, but will only represent
$\beta_m$ when $Q\neq0$ (when $Q\neq0$, only flat FRW models will be
considered and hence there will be no shear terms).

The field equations derived from the action (\ref{sigmaaction_i}) are 
then given by 
\beqn
\mainlabel{rr}
\label{rr1}
\ddot{\alpha} -\dot{\alpha}\dot{\varphi} -\frac{1}{2} 
     \rho +\tilde K +\frac{1}{2}\Lambda_{\rm M}
     e^{\varphi+3\alpha} +\quart Q^2 e^{-6\beta +\varphi+3\alpha}=0,\\
\label{rr2}
2\ddot{\varphi} -\dot{\varphi}^2 -3\dot{\alpha}^2 - 6 \dot{\beta}^2 
     +\frac{1}{2}\rho - 3 \tilde K +2 \Lambda =0, \\
\label{rr3}
\ddot{\beta} -\dot{\beta} \dot{\varphi} -\quart Q^2 e^{-6\beta +\varphi
	+3\alpha}=0, \\
\label{rr4}
\dot{\tilde K} +2\dot\alpha \tilde K = 0, \\
\label{rr5}
\dot{\rho} +6\dot{\alpha} \rho =0,
\eeqn
together with the generalized Friedmann constraint equation
\begin{equation}
\label{rrfriedmann}
3\dot{\alpha}^2 -\dot{\varphi}^2 +6 \dot{\beta}^2 +\frac{1}{2} \rho
-3\tilde K +2\Lambda + \Lambda_{\rm M} e^{\varphi+3\alpha} 
	+\half Q^2 e^{-6\beta +\varphi +3\alpha}=0,
\end{equation}
where $\varphi \equiv \hat{\Phi} -3\alpha$ defines the `shifted' dilaton
field, $\tilde K\equiv 2k \exp(-2\alpha)$ represents the curvature
term, $\rho \equiv \dot\sigma^2\exp(2\varphi+6\alpha)>0$ may be
interpreted as the effective energy density of the pseudo--scalar
axion field, $\beta$ is defined by (\ref{Betadefine}), and a dot
denotes differentiation with respect to cosmic time, $t$.

\section{Exact Solutions\label{Exact_Solutions}}

The exact solutions that were either quoted or defined in
\cite{Billyard1999b} and \cite{Billyard1999c} are here listed.

\index{equilibrium sets!$L^\pm$, $L^{\pm}_{(\pm)}$}
\index{exact solutions!dilaton--moduli--vacuum}
The dilaton--moduli--vacuum solutions, which are found in all sectors
(and therefore $\Lambda=\Lambda_{\rm M}=Q=0$), are given by
\begin{eqnarray}
\nonumber
a & = & a_0\left| t \right|^{\pm h_0}, \\
\nonumber
e^{\hat{\Phi}} & = &e^{\hat{\Phi}_0}\left|t\right|^{\pm3h_0-1}, \nonumber \\
\nonumber
e^\beta & =& e^{\beta_0}\left|t\right|^{\pm\epsilon\sqrt{(1-3h_0^2)/6}} ,\\
\nonumber
\sigma&=&\sigma_0,\\
\label{dmv}
k&=&0,
\end{eqnarray}
where $\{ \hat\Phi_0,\sigma_0, \beta_0 ,h_0\} $ are constants, $a$ is
the scale factor ($a\equiv e^\alpha$), the $\pm$ sign corresponds
to the sign of $t$ (note that $\dot\varphi >0$ for $t<0$ and
$\dot\varphi<0$ for $t>0$) and $\epsilon=\pm1$.  The ``$+$'' solution for $h_0=\pm\third$
corresponds to the $\dot\beta=0$ equilibrium points $L^-_{(\pm)}$
throughout this chapter and the ``$-$'' solution  for $h_0=\pm\third$ correspond to the
equilibrium points $L^+_{(\pm)}$.  Because the modulus field is static
for these four end-points, they are referred to only as
\index{exact solutions!dilaton--moduli--vacuum!dilaton--vacuum}
dilaton--vacuum solutions.  For $t<0$ (i.e. $L^+$) these solutions
are inflationary (i.e., $H>0$, $q<0$) for $h_0>0$.

\index{exact solutions!dilaton--moduli--axion}
These exact solutions are a subset to the dilaton-moduli-axion solutions:
\begin{eqnarray}
a&=&a_0 \left| \frac{s}{s_0} \right|^{\half}
\left[ \left| \frac{s}{s_0} \right|^c +\left| 
\frac{s}{s_0} \right|^{-c} \right]^{\half}, \nonumber \\
e^{\hat{\Phi}} &=&\frac{e^{\hat{\Phi}_0}}{2} \left[ 
\left| \frac{s}{s_0} \right|^c + 
\left| \frac{s}{s_0} \right|^{-c} \right], \nonumber \\
e^{\beta} &=&e^{\beta_0} \left| \frac{s}{s_0} \right|^d   ,\nonumber \\
\nonumber
\sigma &=& \sigma_0 \pm e^{-\hat{\Phi}_0} 
\left[ \frac{ |\frac{s}{s_0}|^{-c} -|\frac{s}{s_0}|^c}{|\frac{s}{s_0}|^{-c} + 
|\frac{s}{s_0}|^c} \right], \\
\label{dma}
k&=&0,
\end{eqnarray}
where $\{a_0,\hat{\Phi}_0,\beta_0,\sigma_0\}$ are an arbitrary
constants, $c \equiv |\sqrt{3-12d^2} |$, and
$s\!\equiv\!\int^tdt'/a(t')$.

\index{equilibrium sets!$S^\pm$}
\index{exact solutions!Milne}
Another solution which appears in both the NS-NS and matter sectors
($\Lambda=\Lambda_{\rm M}=0$) but not in the R--R sector
($Q\neq0$) is the saddle-point Milne solution
\cite[p. 205]{Rindler1977a} (i.e. Minkowski space):
\begin{eqnarray}
\nonumber
a&=&a_0\left(\pm t\right), \\
\nonumber
\hat{\Phi}&=&\hat{\Phi}_0,\\
\nonumber
\beta&=&\beta_0,\\
\nonumber
\sigma&=&\sigma_0,\\
\label{curv_drive}
k&=&-a_0^2,
\end{eqnarray}
where $\{a_0,\hat{\Phi}_0,\beta_0,\sigma_0\}$ are constants.  The
``$\pm$'' sign corresponds to the sign of $t$.  These solutions are
represented by the equilibrium points denoted by $S^\mp$ throughout
this chapter.

\index{exact solutions!10D isotropic, frozen axion}
The ten-dimensional isotropic, frozen axion solution is given by
\begin{eqnarray}
\nonumber
a&=&a_0 \left| {\rm tanh} (\half A t) 
\right|^{\pm\third}, \\
\nonumber
e^{-\hat{\Phi}} &=&e^{-\hat{\Phi}_0} \left| 
\cosh (\half At) \right|^{(1\pm3)} 
\left| \sinh (\half At)  \right|^{(1\mp3)}, \\
\nonumber
e^{\beta} &=&e^{\beta_0} \left| {\rm tanh} (\half At) \right|^{\pm\third},\\
\nonumber
\sigma&=&\sigma_0, \\
\label{ifa}
k&=&0,
\end{eqnarray}
where $A \equiv \sqrt{2\Lambda}$ and $\{a_0,\hat{\Phi}_0,\beta_0,\sigma_0\}$
are constants.  The generalization of this solution is the ``rolling radii'' solutions:
\index{exact solutions!rolling radii}
\begin{eqnarray}
\nonumber
a&=&a_0 \left| {\rm tanh} (\half A t) 
\right|^m, \\
\nonumber
e^{-\hat{\Phi}} &=&e^{-\hat{\Phi}_0} \left| 
\cosh (\half At) \right|^{2p-6n} 
\left| \sinh (\half At)  \right|^{2l+6n}, \\
\nonumber
e^{\beta} &=&e^{\beta_0} \left| {\rm tanh} (\half At) \right|^{6n},\\
\nonumber
\sigma&=&\sigma_0, \\
\label{roll}
k&=&0,
\end{eqnarray}
where $A\equiv \sqrt{2\Lambda}$ and the real numbers $\{l,m,n,p\}$ satisfy
\begin{displaymath}
3m^2+6n^2=1, \qquad 3m+6n=p-l, \qquad p+l=1.
\end{displaymath}
The corresponding solutions for $\Lambda<0$ are related to
(\ref{roll}) by redefining $A\equiv-i\sqrt{2\Lambda}$.  In this case,
the range of $t$ is bounded such that $0<t<\pi/A$.

\

\index{equilibrium sets!$C^\pm$}
\index{exact solutions!linear dilaton--vacuum}
The static `linear dilaton--vacuum' solution \cite{Myers1987b,Antoniadis1988a,Antoniadis1989a} is given by
\begin{eqnarray}
\nonumber
a&=&a_0, \\
\nonumber
\hat{\Phi}&=&\hat{\Phi}_0\pm\sqrt{2\Lambda} t,\\
\nonumber
\beta&=&\beta_0, \\
\nonumber
\sigma&=&\sigma_0,\\
\label{static}
k&=&0,
\end{eqnarray}
where $\{a_0, \hat{\Phi}_0, \beta_0,\sigma_0\}$ are constants.  This solution
is represented by the equilibrium points $C^\pm$, where
``$\pm$'' corresponds to the ``$\pm$'' sign in (\ref{static}).

\index{equilibrium sets!$L_1$}
\index{exact solutions!linear dilaton--vacuum!generalized linear dilaton--vacuum}
This solution had been generalized in \cite{Billyard1999d} to
the `generalized linear dilaton--vacuum', to include the axion field
and a curvature term:
\begin{eqnarray}
\nonumber
a&=&a_0, \\
\nonumber
\hat{\Phi}&=&\hat{\Phi}_0 +n\sqrt{\frac{6\Lambda}{2+n^2}} t, \\
\nonumber
\beta&=&\beta_0, \\
\nonumber
\sigma&=&\sigma_0 \pm \sqrt{\frac{2\left(1-n^2\right)}{3n^2}} \exp
\left(-\hat{\Phi}_0 -n\sqrt{\frac{6\Lambda}{2+n^2}} t \right), \\
\label{static_general}
k&=&\frac{1-n^2}{2+n^2}\Lambda a_0^2, 
\end{eqnarray}
where $n\in [-1,1]$.  The solutions are represented by the line of
equilibrium points denoted throughout this chapter by $L_1$.  Note
that $C^\pm=L_1$ when $n^2=1$ ($\rho=k=0$).

\section{Asymptotic Behaviour\label{Asymptotic_Behaviour}}

Returning to the field equations (\ref{rr}) and (\ref{rrfriedmann}),
the variables $\rho$ and $\tilde K$ may be combined to define the new
variable
\begin{equation}
\Xi\equiv \frac{\rho-\tilde K}{\rho+\tilde K},
\end{equation}
and hence (\ref{rr4}) and (\ref{rr5}) can be combined to yield the
evolution equation
\begin{equation}
\dot\Xi = -2\dot\alpha\left(1-\Xi^2\right). \label{dchi}
\end{equation}
Hence, all of the equilibrium sets occur either for $\dot\alpha=0$ or
for $\Xi^2=1$, and therefore if $\dot\alpha\neq0$ then either $\rho=0$
($\Xi=-1$) or $\tilde K =0$ ($\Xi=+1$) asymptotically.  Equations
(\ref{rr}) and (\ref{rrfriedmann}) define a four-dimensional dynamical
system (although there are five ODE's, equation (\ref{rrfriedmann})
can be used to globally reduce the system by one dimension). In the
cases in which all of the equilibrium sets lie on $\Xi^2=1$, the
asymptotic properties of the string cosmologies can be determined from
the dynamics in the three-dimensional sets $\rho=0$ or $\tilde K=0$.  

In what follows, the three-dimensional $\rho=0$ case and $\tilde K=0$
case will be consequently explicitly examined separately.  The only
case in which there exist equilibrium sets with $\dot\alpha=0$ but
$\Xi^2\neq 1$ occurs in the NS-NS case ($\Lambda_{\rm M}=Q=0$) in
which $\tilde K>0$ and $\Lambda>0$, and therefore the
full four-dimensional system shall be examined in this case, although the
three-dimensional subset $\rho=0$ still plays a principal r\^{o}le in
the asymptotic analysis.

\section{Analysis}
In this chapter, $\Lambda_{\rm M}=0$ and $Q=0$ in order to determine
the r\^{o}le of $\Lambda$ in the dynamics of the system (\ref{rr}).
Through equation (\ref{rrfriedmann}), the variable $\rho$ is eliminated
from the field equations, and the following definitions are made:
\begin{equation}
X\equiv \frac{\sqrt3\dot\alpha}{\xi}, \quad
Y\equiv \frac{\dot\varphi}{\xi}, \quad
Z\equiv \frac{6\dot{\beta}^2}{\xi^2}, \quad
U\equiv \frac{\pm 3 \tilde K}{\xi^2}, \quad
V\equiv \frac{\pm 2\Lambda}{\xi^2}, \quad
\frac{d}{dt}\equiv \xi \frac{d}{dT},\label{TheDefs}
\end{equation}
where the $\pm$ sign in the definitions for $U$ and $V$ are to ensure
$U>0$ and $V>0$, where necessary.  With these definitions, all
variables are bounded: $0\leq \{X^2,Y^2,Z,U,V\}\leq 1$.  Equation
(\ref{rrfriedmann}) now reads
\begin{equation}
\kappa = Y^2 \pm U \mp V -X^2-Z \geq 0, \label{newFried1}
\end{equation}
where 
\be
\kappa \equiv \half\frac{\rho}{\xi^2}
\ee
The
variable $\xi$ is defined in each of the following six cases by:
\begin{itemize}
\item $\Lambda>0$\begin{itemize} \item $\tilde K>0$: $\xi^2 \equiv 3\tilde K
	+\dot\varphi^2$ (subsection \ref{NSpKp}),
\item $\tilde K\leq0$: $\xi^2 \equiv \dot\varphi^2$ (subsections \ref{NSpK0} and \ref{NSpKn}),\end{itemize}
\item $\Lambda<0$ \begin{itemize} \item $\tilde K>0$: 
	$\xi^2 \equiv 3\tilde K
	+\dot\varphi^2-2\Lambda$ (subsection \ref{NSpKp}),
\item $\tilde K\leq0$: $\xi^2 \equiv \dot\varphi^2-2\Lambda$ (subsections \ref{NSnK0} and \ref{NSnKn}).
\end{itemize}\end{itemize}
For example, consider $\Lambda>0$ with $\tilde K>0$; for this case $Y^2+U=1$ and equation (\ref{newFried1}) reads
\be
\kappa =1-V-X^2-Z \geq 0.
\ee
Hence the variables $\{X,V,Z\}$ will be used as the phase space variables (see section
\ref{NSpKp} for details).  The only exception to this approach is the
$\tilde K=0$ cases in sections \ref{NSpK0} and \ref{NSnK0}
in which $\dot\beta$ is
removed from the field equations in lieu of $\rho$. Although this
deviation from the orthodox of the other sections was performed merely
to give an example where the axion field explicitly remains in the dynamics, it
leads to the exact qualitative behaviour if $\dot\beta$ had been kept and not $\rho$.

Each of six cases will now be considered.  Subscripts will be added to the
variables $\{X,Y,Z,U,V\}$ to indicate each case, although the $\tilde
K=0$ cases will have the same subscript as the $\tilde K>0$ (for
either $\tilde K>0$ or $\tilde K<0$, the phase space variables reduce
to the same form when $\tilde K=0$ and so the notation is consistent
with either case; however, since the $\tilde K>0$ case follows the
$\tilde K=0$ in what follows, the above subscript choice will be
made).  In each of the following subsections, the four-dimensional
dynamical systems for $\tilde K\neq0$ will be established, and the
$\rho=0$ invariant set will be examined.  As discussed in section
\ref{Asymptotic_Behaviour}, all equilibrium sets discussed below
which have $\dot\alpha=0$ also have $\Xi\neq 1$, and hence nearly all
orbits asymptote towards the equilibrium sets in one of the
invariant sets $\rho=0$ or $\tilde K=0$, where the three-dimensional
$\tilde K=0$ case will be studied in its own subsection.  The
qualitative behaviour of the full four-dimensional phase space in each
$\tilde K\neq 0$ case will also be examined.

\subsection{The Case $\Lambda>0$, $\tilde K=0$ \label{NSpK0}}

For this case, it proves convenient to employ the generalized Friedmann
constraint equation (\ref{rrfriedmann}) to eliminate the modulus
field rather than the axion field, and the definition $\xi^2 =
\dot\varphi^2$ is used.  From the generalized Friedmann equation one
has that
\begin{equation}
0 \leq X_1^2 + \kappa + Z_1 \leq 1,
\end{equation}
and equations (\ref{rr}) and (\ref{rrfriedmann}) transform to the
three--dimensional autonomous system:
\beqn
\mainlabel{s_main}
\label{s3}
\frac{d X_1}{dT} &=& \kappa +\frac{X_1}{\sqrt{3}}  
\left[ 1-X_1^2 -Z_1 \right], \label{dmu} \\
\label{s4}
\frac{d \kappa}{dT} &=& -2 \kappa \left[ X_1 + 
\frac{1}{\sqrt{3}} 
\left( Z_1 + X_1^2 \right) \right], \label{dnu}\\
\frac{d Z_1}{d T} &=& \frac{2}{\sqrt{3}} 
Z_1 \left( 1 - X_1^2 -Z_1 \right) \label{dlambda}  .
\eeqn
The sets $\kappa=0$ and $Z_1=0$ are invariant sets
corresponding to $\rho=0$ and $\dot{\beta}=0$, respectively.  In addition,
$X_1^2+\kappa+Z_1=1$ is an invariant set corresponding to
$\Lambda=0$.  Note that the right-hand side of equation (\ref{dlambda}) is
positive-definite and this simplifies the
dynamics considerably.

\subsubsection{The Frozen Modulus ($\dot\beta=0$) Invariant Set}

\begin{figure}[htp]
  \centering
   \includegraphics*[width=5in]{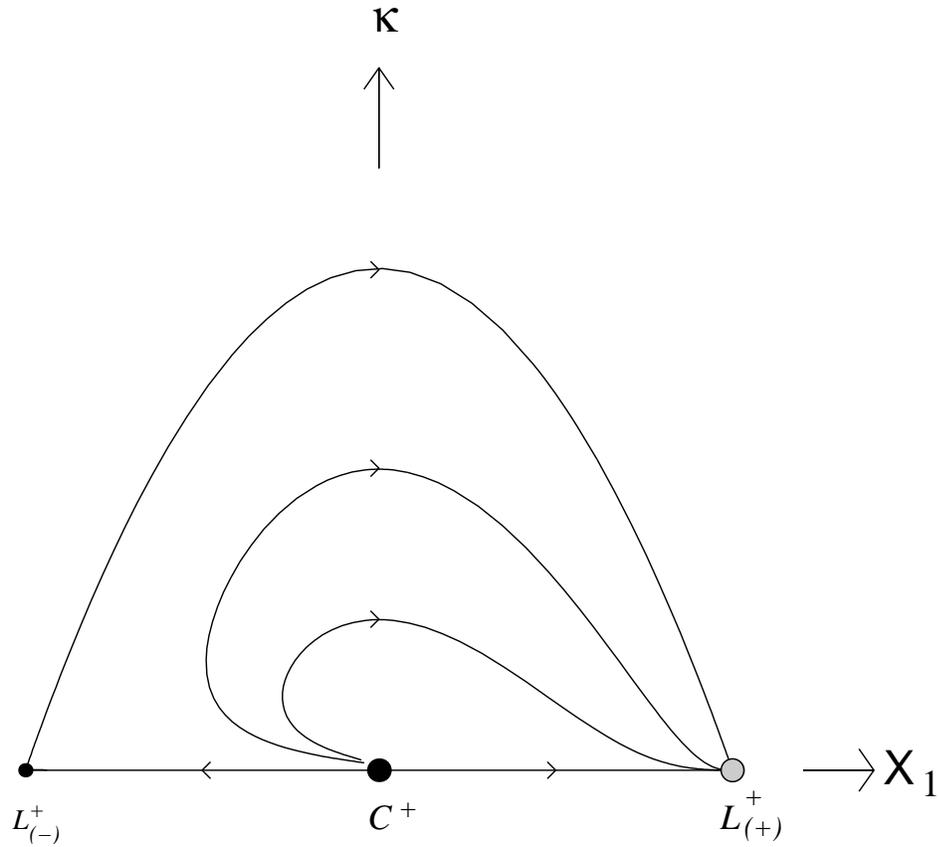}
  \caption[{\em Phase diagram for $\Lambda>0$, $\tilde
K=\dot\beta=0$}]{{\em Phase diagram of the system (\ref{s_main}) in
the NS-NS ($\Lambda>0$) sector with $\tilde K=\dot\beta=0$. In this
phase space, $\dot\varphi>0$ is assumed.  Equilibrium sets are denoted
by dots and the labels in all figures correspond to those equilibrium
sets (and hence the exact solutions they represent) discussed in the
text.  The convention adopted throughout will be that large black dots
represent sources (i.e., repellors), large grey-filled dots represent
sinks (i.e., attractors), and small black dots represent saddles.
Arrows on the trajectories denote the direction of increasing
time.}\label{fNSpK0}}
 \end{figure}

It is also instructive to first consider the dynamics on the boundary
corresponding to $Z_1=0$, since the case $\dot\beta=0$ is of physical
interest in its own right as a four--dimensional model.  Here the
dynamical system becomes equations (\ref{dmu}) and (\ref{dnu}) with
$Z_1=0$, and the equilibrium points (and their eigenvalues) are given by
\beqn
C^+: && X_1=\kappa=0; \nonumber \\
	&& (\lambda_1,\lambda_2)= \left(\frac{1}{\sqrt{3}}, 0\right), \\
L^+_{(-)}:&& X_1=-1, \kappa=0; \nonumber\\
	&& (\lambda_1,\lambda_2)= \left(-2\left[ -1+\frac{1}
	{\sqrt{3}} \right], -\frac{2}{\sqrt{3}}\right), \\
L^+_{(+)}:&& X_1=+1, \kappa=0; \nonumber\\
	&& (\lambda_1,\lambda_2)= \left(-2\left[ 1+\frac{1}
	{\sqrt{3}} \right], -\frac{2}{\sqrt{3}}\right)
\eeqn
(throughout these chapters, the corresponding eigenvectors will not be
explicitly given).  Point $C^+$ is a non-hyperbolic equilibrium point;
however, by changing to polar coordinates $C^+$ can be shown to be a
repeller with an invariant ray $\theta=\tan^{-1}(-\sqrt{3})$.  The
phase portrait is given in figure \ref{fNSpK0}.

\subsubsection{Three--Dimensional System ($\dot\beta\neq0$)}

\begin{figure}[htp]
  \centering
   \epsfxsize=5in
   \includegraphics*[width=5in]{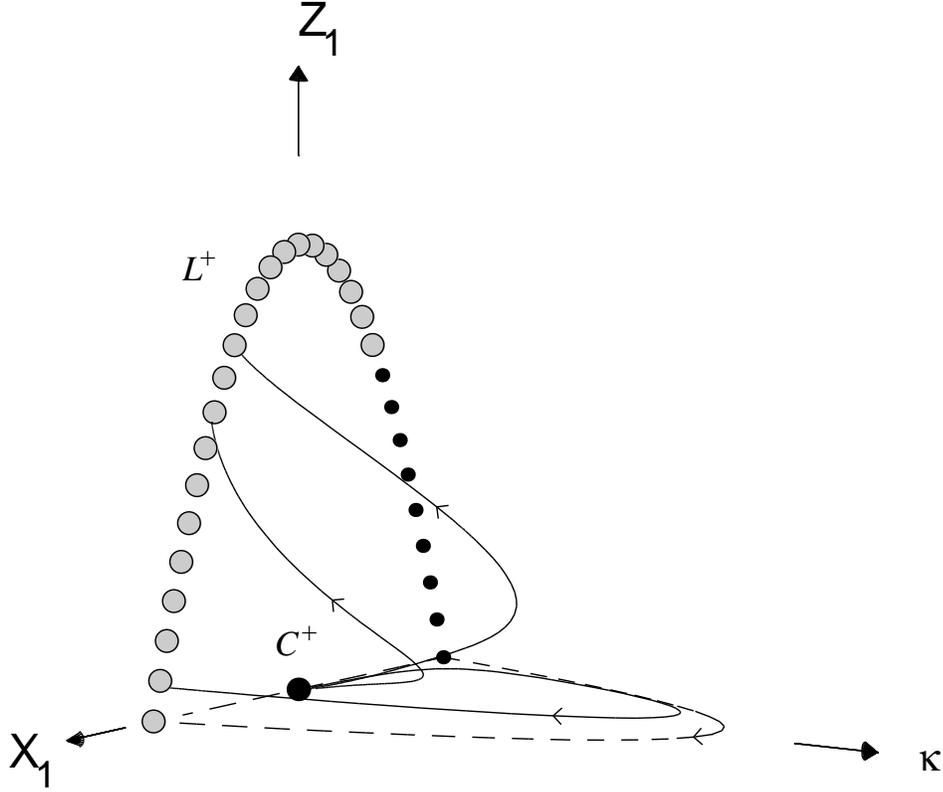}
  \caption[{\em Phase diagram for $\Lambda>0$, $\rho\neq0$ and $\tilde
K=0$}]{{\em Phase diagram of the system (\ref{s_main}) in the NS-NS
($\Lambda>0$) sector with $\rho\neq0$ and $\tilde K=0$.  Note that
$L^+$ represents a {\it line} of equilibrium points.  In this phase
space, $\dot\varphi>0$ is assumed.  The invariant set $Z_1=0$
corresponds to the phase portrait \ref{fNSpK0}.  Dashed black lines are those trajectories along the intersection of
the invariant sets, and solid black lines are typical trajectories
within the full three-dimensional phase space.  See also caption to
figure
\ref{fNSpK0} on page \pageref{fNSpK0}.} \label{fNSpK0b0}} \end{figure}

Returning to the full three-dimensional dynamical system
(\ref{s_main}), the equilibrium sets of the system (and their
corresponding eigenvalues) are 
\beqn
C^+: && X_1=\kappa=Z_1=0; \nonumber \\
	&& (\lambda_1,\lambda_2,\lambda_3)= \left(\frac{1}{\sqrt{3}}, 
	\frac{2}{\sqrt{3}}, 0\right), \\
L^+:&& X_1^2+Z_1=1, \kappa=0; \nonumber\\
	&& (\lambda_1,\lambda_2,\lambda_3)= \left(-2\left[ X_1+\frac{1}
	{\sqrt{3}} \right], -\frac{2}{\sqrt{3}}, 0\right).
\eeqn
Although $C^+$ is non-hyperbolic, a simple analysis shows that it is a
global source.  Equilibrium points on $L^+$ are saddles for $X_1<-\frac{1}{\sqrt{3}}$ and local sinks for $X_1>-\frac{1}{\sqrt{3}}$.  Note
that the points $L^+_{(-)}$ and $L^+_{(+)}$ in the two--dimensional
invariant set $\dot\beta=0$ are the endpoints to the line $L^+$.  The
phase space is depicted in figure \ref{fNSpK0b0}.

The set $C^+$ represents the linear dilaton--vacuum solution
(\ref{static}) from which all solutions asymptote away.  The line
$L^+$ represents the dilaton--moduli--vacuum solutions (\ref{dmv})
(the ``$-$'' branch) towards which all solutions asymptote into the future.

\subsection{The Case $\Lambda>0$, $\tilde K>0$\label{NSpKp}}

In this case the choice $\xi^2 = \dot\varphi^2+3\tilde K$ is made and the
positive signs for $U_1$ and $V_1$ has been utilized, as defined by 
(\ref{TheDefs}).  The generalized Friedmann equation yields 
\begin{equation}
0 \leq X_1^2 + Z_1 + V_1 \leq 1,\qquad Y_1^2+U_1=1,
\end{equation}
and therefore $U_1$ (which is proportional to  $\tilde K$) may be eliminated, 
yielding the
four-dimensional system of ODEs for $0\leq \{X_1^2, Y_1^2, Z_1, V_1\}\leq1$:
\beqn
\mainlabel{dX_1_main}
\label{dX_1}
\frac{dX_1}{dT} & = &\frac{1}{\sqrt3}\left(1-X_1^2\right)\left(2+Y_1^2\right)-\sqrt{3}\left(Z_1+V_1 \right) +X_1Y_1\left(1\!-\!X_1^2\!-\!Z_1\right)\!\!, \\
\label{dY_1}
\frac{dY_1}{dT} &=& \left(1-Y_1^2\right) \left(X_1^2+Z_1
        +\frac{1}{\sqrt3}X_1Y_1\right),\\
\label{dZ_1}
\frac{dZ_1}{dT} &=& 2Z_1\left[ Y_1\left( 1-X_1^2-Z_1\right) 
        + \frac{1}{\sqrt3} X_1 \left( 1-Y_1^2 \right) \right], \\
\label{dV_1}
\frac{dV_1}{dT} &=& -2V_1 \left[ Y_1\left(X_1^2+Z_1\right) 
        -\frac{1}{\sqrt{3}} X_1	 \left( 1 - Y_1^2 \right) \right].
\eeqn
The invariant sets $Y_1^2=1$, $X_1^2+Z_1+V_1=1$,
$Z_1=0$ and $V_1=0$ define the boundary of the phase space.  The
equilibrium sets and their corresponding eigenvalues (denoted by
$\lambda$) are
\beqn
L^\pm: & & Y_1=\pm1, Z_1=1-X_1^2, V_1=0; \nonumber \\
   &&
   \left(\lambda_1,\lambda_2, \lambda_3, \lambda_4\right) =  \left(
    \mp\frac{2}{\sqrt3} \left[\sqrt3 \pm X_1 \right], \mp 2, 0,
	-2\sqrt3\left[X_1\pm\frac{1}{\sqrt3}\right]\right), \\
L_1: & & X_1=0,Z_1=0,V_1=\third(2+Y_1^2); \nonumber \\
   && 
   \left(\lambda_\pm, \lambda_2, \lambda_3\right) =  \left(
   \frac{1}{2} \left[ Y_1\pm\frac{1}{\sqrt3}\sqrt{19Y_1^2-16} \right ], 
   2Y_1, 0    \right). \label{mother}
\eeqn
The zero eigenvalue in each arise because these are all {\em lines} of
equilibrium points.  Here, the global sources are the lines $L_1$ (for
$Y_1>0$ or $\dot\varphi>0$) and $L^-$ (for $X_1<\frac{1}{\sqrt3}$).
The global sinks are the lines $L_1$ (for $Y_1<0$) and $L^+$ (for
$X_1>-\frac{1}{\sqrt3}$). 

This case is {\em different} from the other cases to be
considered, since it is the only one with the line of equilibrium
points, $L_1$, {\em inside} the phase space and this line acts as both sink
and source (all the other cases, as will be shown, have 
equilibrium points only on the boundaries of the phase
space).  The line $L_1$ corresponds
to exact static solutions (\ref{static_general}) which generalize the static `linear dilaton--vacuum'
solution (\ref{static}).
This solution was examined in \cite{Easther1996a} for $\dot\varphi>0$ and was shown
to be an attractor in the perturbation analysis found there.

The line $L^\pm$ corresponds to the dilaton--moduli--vacuum
solutions, given by equation (\ref{dmv}).  Note that in
equation (\ref{dmv}) $L^+$ corresponds to the ``$-$'' branch (the
corresponding attracting solutions are represented by the range of the
integration constant $h_0>-\frac{1}{3}$), whereas
$L^-$ corresponds to the ``$+$'' branch (the corresponding attracting
solutions are represented by the range of the integration constant
$h_0<\third$).  

\subsubsection{The Invariant Set $\rho=0$ for $\Lambda>0$, $\tilde K>0$}

\begin{figure}[htp]
  \centering
   \epsfxsize=5in
   \includegraphics*[width=5in]{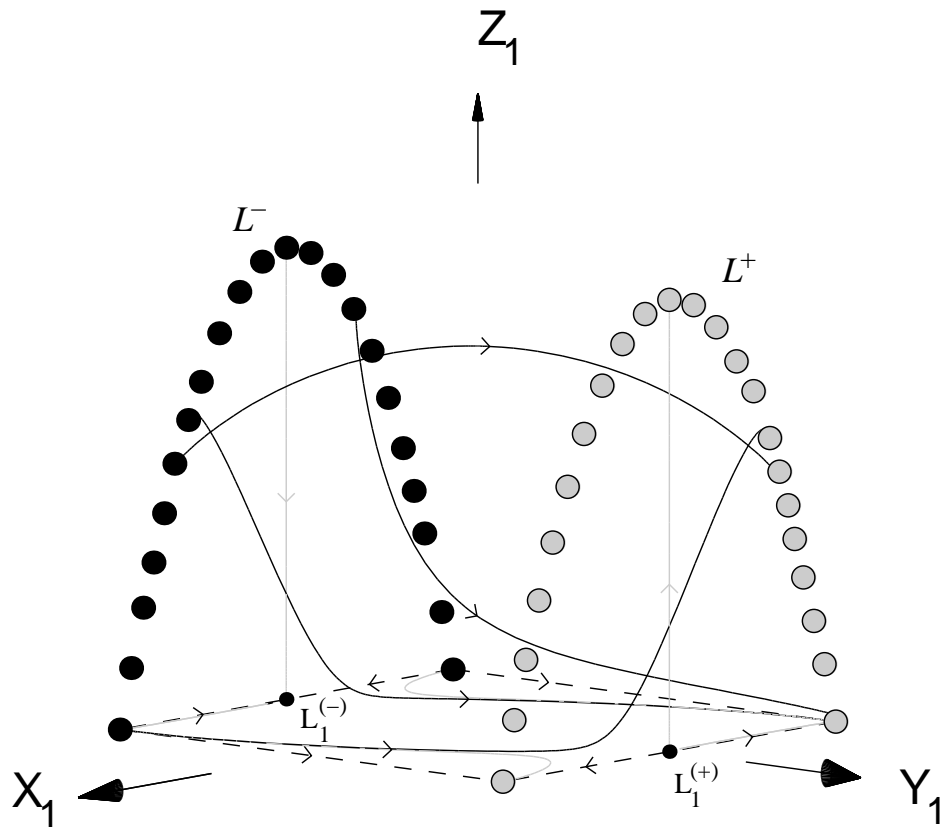}
  \caption[{\em Phase diagram for $\Lambda>0$, $\rho=0$ and $\tilde
K>0$}]{{\em Phase diagram of the system (\ref{dX_1_main}) in the NS-NS
($\Lambda>0$) sector with $\rho=0$ and $\tilde K>0$. Note that $L^+$
and $L^-$ represent {\it lines} of equilibrium points, and the labels
$L_1^{(+)}$ and $L_1^{(-)}$ represent the equilibrium points which are
the endpoints of the line $L_1$ (for which $X_1=+1$ and $X_1=-1$,
respectively).  Grey lines represent typical trajectories found within
the two-dimensional invariant sets, d.  See also caption to figure
\ref{fNSpK0} on page \pageref{fNSpK0}.}\label{fNSpKp}}
 \end{figure}

In the $\rho=0$ case, the system reduces to the three dimensions of
$\{X_1,Y_1,Z_1\}$ ($V_1=1-X_1^2-Z_1$).  The equilibrium sets are the
lines $L^\pm$ with eigenvalues $\lambda_1$, $\lambda_2$ and
$\lambda_3$ from subsection \ref{NSpKp}, and the two endpoints of $L_1$
(i.e. $L_1$ for $Y_1^2=1$; these endpoints are denoted by $L_1^{(\pm)}$,
where the ``$\pm$'' in the superscript reflects the sign of $Y_1$)
with eigenvalues $\lambda_\pm$ and $\lambda_3$ from subsection \ref{NSpKp}.  Note that for this invariant set the entire line $L^+$ acts as a
global sink and the entire line $L_-$ acts as a global source.
Furthermore, $\lambda_-=0$ for $L_1$, and so these two
points are non-hyperbolic.  However, the eigenvectors associated with
these zero eigenvalues are both $[-\frac{2}{\sqrt3},1,0]$ which lies
completely in the $X_1-Y_1$ plane.  Hence, if $Z_1=0$ is chosen and
the $X_1-Y_1$ axes are rotated via
\begin{displaymath}
\tilde x \equiv (Y_1\mp 1)-\frac{\sqrt3}{2}X_1, \qquad 
\tilde y \equiv (Y_1\mp 1)+\frac{2}{\sqrt3}X_1,
\end{displaymath}
then it can be shown that  trajectories along $\tilde x$ for $\tilde y=0$ are along
these eigenvectors (close the equilibrium point).  Hence
for $\tilde y=0$ and for small $\tilde x$,
\begin{displaymath}
\frac{d\tilde x}{dT} \approx \mp \frac{\tilde x}{7},
\end{displaymath}
and for $Y_1=+1$ the trajectory along $\tilde x$
asymptotes towards the equilibrium point, whereas the trajectory along
$\tilde x$ for $Y_1=-1$ asymptotes away from the equilibrium point,
and therefore the points $L_1$ are saddle points.  Figure \ref{fNSpKp}
depicts this phase space.

The quantity $X_1/\sqrt{Z_1}$ is {\em monotonically decreasing}.  Such
a monotonic function excludes the possibility of periodic or recurrent
orbits in this three-dimensional space.  Therefore, solutions
generically asymptote into the past towards the $\dot\varphi<0$
dilaton--moduli--vacuum solutions (\ref{dmv}), and into the future
towards the $\dot\varphi>0$ dilaton--moduli--vacuum solutions
(\ref{dmv}).  In this three-dimensional set, curvature is dynamically
important only at intermediate times.

\subsubsection{Qualitative Analysis of the Four-Dimensional System}

The qualitative dynamics in the full four-dimensional phase space is
as follows.  In the four-dimensional phase space, the past attractors
are the line $L^-$ for $X_1<\frac{1}{\sqrt3}$ and the line $L_1$ for
$Y_1>0$.  The future-attractor sets are the lines $L^+$ (for
$X_1>-\frac{1}{\sqrt3}$) and $L_1$ (for $Y_1<0$).  Note that
$Y_1/\sqrt{V_1}$ is {\em monotonically increasing}.  The existence of
such a monotonic function excludes the possibility of periodic or
recurrent orbits in the full four--dimensional phase space.
Therefore, solutions are generically asymptotic to the past to either
the $\dot\varphi<0$ dilaton--moduli--vacuum solutions (\ref{dmv}) for
$h_0<\third$ or the generalized $\dot\varphi<0$ linear dilaton--vacuum
solution (\ref{static_general}) for $n>0$.  Similarly, solutions are
generically asymptotic to the future to either the $\dot\varphi>0$
dilaton--moduli--vacuum solutions (\ref{dmv}) for $h_0>-\frac{1}{3}$
or the generalized $\dot\varphi<0$ linear dilaton--vacuum solution
(\ref{static_general}) for $n<0$.

Since $Y_1/\sqrt{V_1}$ is monotonically increasing, $Y_1\rightarrow
+1$ or $V_1\rightarrow 0$ asymptotically to the future, corresponding
to the global sinks on the line $L^+$.  Similarly, $Y_1\rightarrow -1$
or $V_1\rightarrow 0$ asymptotically to the past, corresponding to the
global sources on the line $L^-$.  There are also equilibrium sets
for the points $Y_1=Y_0$ inside the phase space.  Again, since
$Y_1/\sqrt{V_1}$ is monotonically increasing, the points $Y_0<0$ are
global sinks and $Y_0>0$ are global sources.  All of this is
consistent with the above discussion concerning asymptotics.

Due to the existence of the monotonic function and the continuity of
orbits in the four-dimensional phase space, solutions cannot start and
finish on $L_1$.  Investigation of this invariant set also indicates
which sources and sinks are connected; not all orbits from $L^-$ can
evolve towards $L^+$.

Although both lines $L^\pm$ lie in both invariant sets $\rho=0$ and
$\tilde K=0$, the line $L_1$ does not, and so both the axion field and
curvature term can be dynamically significant at all times (early,
intermediate and late) for the corresponding solutions.  On the line
$L_1$, $X_1=0$ corresponds to $\dot\alpha=0$, and represents static
solutions.  Similarly, both curvature and axion field are non-zero
(and indeed the variables $\rho$, $\tilde K$ and $\Lambda$ are
proportional to one another).  The fact that $\dot\alpha=0$ in this
case means that neither $\tilde K$ nor $\rho$ need be zero, consistent
with the analysis of equation (\ref{dchi}).

\subsection{The Case $\Lambda>0$, $\tilde K<0$\label{NSpKn}}

In this case, the negative sign
in (\ref{TheDefs}) for $U_2$ has been chosen as has been the positive sign for $V_2$, and the definition $\xi^2 = \dot\varphi^2$ is used.  The generalized
Friedmann constraint equation is now written to read
\begin{equation}
0 \leq X_2^2 + Z_2 + U_2 + V_2\leq 1.
\end{equation}
For this system $Y_2^2=1$ and so the
four-dimensional system consists of the variables $0\leq \{X_2^2, Z_2, U_2, 
V_2\} \leq 1$:
\beqn
\mainlabel{dX_2_main}
\label{dX_2}
\frac{dX_2}{dT} & = & \sqrt 3 \left( 1-X_2^2-Z_2-V_2-\frac{2}{3}U_2  \right) 
		     + X_2 \left( 1-X_2^2-Z_2 \right), \\
\label{dZ_2}
\frac{dZ_2}{dT} &=& 2Z_2 \left( 1-X_2^2-Z_2 \right) > 0, \\
\label{dU_2}
\frac{dU_2}{dT} &=& -2U_2 \left(X_2^2+Z_2+\frac{1}{\sqrt3}X_2 \right),\\
\label{dV_2}
\frac{dV_2}{dT} &=& -2V_2 \left( X_2^2 + Z_2\right) < 0.
\eeqn
The invariant sets $X_2^2+Z_2+U_2+V_2=1$, $Z_2=0$, $V_2=0$, $U_2=0$ define the
boundary of the phase space.  The equilibrium sets and their
corresponding eigenvalues (denoted by $\lambda$) are
\beqn
S^+: & & X_2=-\frac{1}{\sqrt3}, Z_2=0, U_2=\frac{2}{3}, V_2=0; \nonumber \\ &&
   \left(\lambda_1, \lambda_2, \lambda_3, \lambda_4\right) =  \left(
   -\frac{2}{3}, \frac{2}{3}, \frac{4}{3}, \frac{4}{3} \right), \\
C^+: & & X_2=0,Z_2=0, U_2=0, V_2=1; \nonumber \\&&
   \left(\lambda_1,\lambda_2, \lambda_3, \lambda_4\right) =  \left(
   1,0, 2, 0 \right), \\
L^+: & & Z_2=1-X_2^2, U_2=0, V_2=0; \nonumber \\ &&
   \left(\lambda_1, \lambda_2, \lambda_3, \lambda_4\right) =  \left(
    -\frac{2}{\sqrt3} \left[ X_2+\sqrt3 \right],-2,0,-2\sqrt3 \left[ X_2+\frac{1}{\sqrt3} \right]    \right).
\eeqn

The point $C^+$ represents the static `linear dilaton--vacuum'
solution (\ref{static}).  The saddle $S^+$ represents the Milne
solution (\ref{curv_drive}), where only the curvature term and scale
factor are dynamic.

\subsubsection{The Invariant Set $\rho=0$ for $\Lambda>0$, $\tilde K<0$}

\begin{figure}[htp]
  \centering
   \epsfxsize=5in
   \includegraphics*[width=5in]{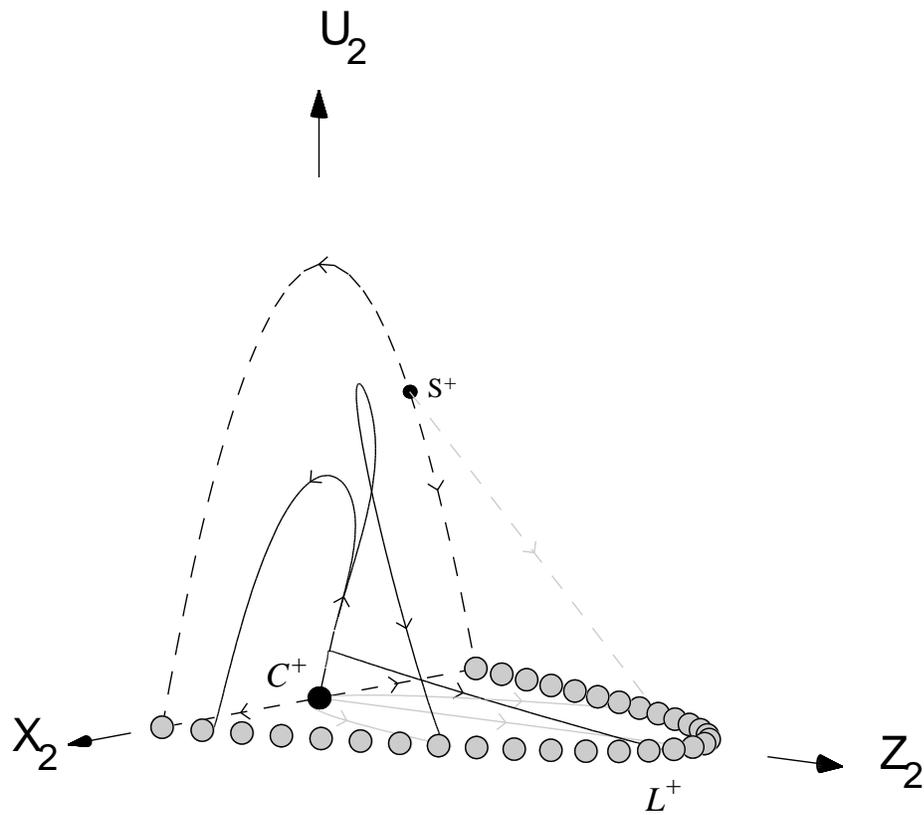}
  \caption[{\em Phase diagram for $\Lambda>0$, $\rho=0$ and $\tilde K<0$}]{{\em Phase diagram of the system
(\ref{dX_2_main}) in the NS-NS ($\Lambda>0$) sector with
$\rho=0$ and $\tilde K<0$. Note that $L^+$ represents a {\em line} of
equilibrium points.  See also caption to figure \ref{fNSpK0} on page \pageref{fNSpK0}.}\label{fNSpKn}}
 \end{figure}

In the $\rho=0$ case, the system reduces to three dimensions of
$\{X_2,Z_2,U_2\}$ ($V_2=1-X_2^2-Z_2-U_2$).  The equilibrium sets are
the same as above with eigenvalues $\lambda_1$, $\lambda_2$ and
$\lambda_3$.  Note that for the invariant set $\rho=0$ the entire
line $L^+$ acts as a global sink, and $C^+$ acts as a source (although
one of the eigenvalues is zero, this point will be shown to be a
source in the full four--dimensional in the next subsubsection and the
arguments there equally apply here).  Figure \ref{fNSpKn} depicts this three-dimensional phase space.

The variable $Z_2$ is {\em monotonically increasing}, and as such it
is apparent that the modulus field is negligible at early times, but
becomes dynamically significant at late times.  The existence of such
a monotonic function excludes the possibility of periodic or recurrent
orbits in this three--dimensional phase space.  Generically, solutions
asymptote into the past towards the static linear dilaton--vacuum
solution (\ref{static}), represented by the point $C^+$.  Into the
future, solutions generically asymptote towards the $\dot\varphi>0$
dilaton--moduli--vacuum solution (\ref{dmv}), represented by the line
$L^+$.

\subsubsection{Qualitative Analysis of the Four-Dimensional System}

The qualitative behaviour for the invariant set $\tilde K=0$ is
discussed in subsection \ref{NSpK0} (see figure \ref{fNSpK0} on page
\pageref{fNSpK0}).

In the four-dimensional set, the point $C^+$ is non-hyperbolic because
of the two zero eigenvalues.  However, it can be shown that the point
$C^+$ is a source in the four-dimensional set by the following
argument.  Note that the variable $Z_2$ is {\em monotonically
increasing} and hence orbits asymptote into the past towards the
invariant set $Z_2=0$.  Similarly, $V_2$ is a {\em monotonically decreasing}
function, and so orbits asymptote into the past towards large
$V_2$, which will be shown to be $V_2=1$, and therefore towards the point
$C^+$.

Since $Z_2=0$ asymptotically, consider the invariant set $Z_2=0$.
Equations (\ref{dX_2_main}) in this set become
\beqn
\label{dX}
\frac{dX_2}{dT} & = & \sqrt 3 \left( 1-X_2^2-V_2-\frac{2}{3}\tilde U_2 \right)
	+ X_2 \left( 1-X_2^2\right), \\
\label{dU}
\frac{dU_2}{dT} &=& -2U_2 \left(X_2^2+\frac{1}{\sqrt3}X_2 \right),\\
\label{dV}
\frac{dV_2}{dT} &=& -2 V_2 X_2^2.
\eeqn
It is clear from (\ref{dV}) that $V_2$ increases monotonically into
the past (this is also true for $Z_2\neq0$).  Now, this
three-dimensional phase space is bounded by the surface
$X_2^2+U_2+V_2=1$, the ``apex'' of which lies at $V_2=1$ (and
$X_2=U_2=0$).  Therefore, all orbits on or inside of the boundary of
this phase space lie below $V_2=1$, and therefore asymptote into the
past towards $V_2=1$.
To help illustrate that this point is indeed a source, consider the
invariant set $Z_2=U_2=0$, $X_2^2+V_2=1$ in the neighborhood of $C^+$.
Here equation (\ref{dX}) becomes $dX_2/dT = X_2(1-X_2^2)$,
indicating that orbits are repelled from $X_2=0$.  Hence, the point
$C^+$ is the past attractor to the full four-dimensional set.  The
future attractor for the four-dimensional set is the line $L^+$ (for
$X_2>-\frac{1}{\sqrt3}$).  Both $C^+$ and $L^+$ lie in both the
invariant sets $\rho=0$ and $\tilde K=0$, which is consistent with the
analysis of equation (\ref{dchi}).  Generically, solutions are asymptotic
in the past to the static, linear dilaton--vacuum solutions (\ref{static}),
 and to the future to the $\dot\varphi>0$
dilaton--moduli--vacuum solutions (\ref{dmv}), and the
curvature terms as well as the axion field are dynamically important only at
intermediate times, and are negligible at both early and  late
times.  The variables $Z_2$ increases monotonically so that the effect
of the modulus field becomes increasingly important dynamically,
whilst the variable $V_2$ decreases monotonically and so that the
effect of the central charge deficit, $\Lambda$, becomes increasing
negligible, dynamically.  In addition, the existence of monotone
functions in the four--dimensional space prohibits closed orbits and
serves as proof of the evolution described above.

\subsection{The Case $\Lambda<0$, $\tilde K=0$ \label{NSnK0}}

In this case, it proves convenient 
to employ the generalized Friedmann constraint  equation 
(\ref{rrfriedmann}) to eliminate the  
modulus field rather than the axion field. 
This equation can be written as
\begin{equation} 1-X_3^2-\kappa=Z_3.
\end{equation} 
The resulting field equations are
\beqn
\mainlabel{dXK0}
\frac{dX_3}{dT} &=& \kappa \left( \sqrt{3} +Y_3 X_3 \right), \label{dxi}\\
\frac{dY_3}{dT} &=& (1-\kappa ) \left( 1- Y_3^2 \right), \label{deta}\\
\frac{d\kappa}{dT} &=& -2\kappa \left[ \sqrt{3} X_3 + Y_3 ( 1- \kappa )
		\right]. \label{dkappa}
\eeqn
Note that the invariant set $\dot\beta=0$ corresponds to $\kappa=1-X_3^2$.

\subsubsection{The Frozen Modulus ($\dot\beta=0$) Invariant Set}

\begin{figure}[htp]
  \centering
   \epsfxsize=3in
   \includegraphics*[width=5in]{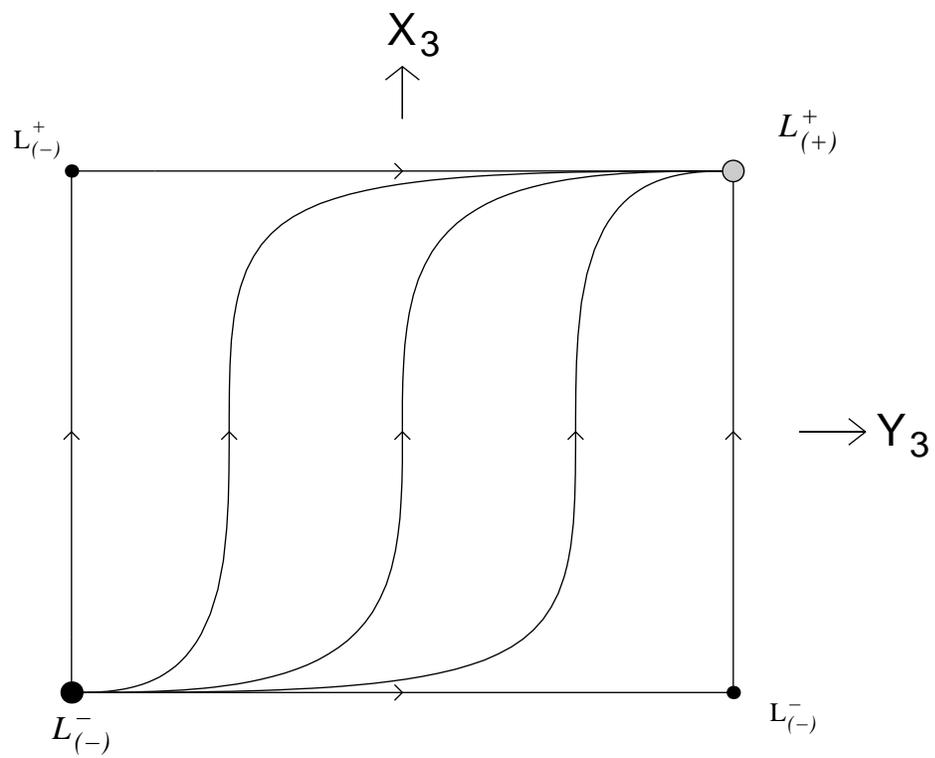}
  \caption[{\em Phase diagram for $\Lambda<0$, $\tilde
K=\dot\beta=0$}]{{\em Phase portrait of the system (\ref{dXK0}) in the
NS-NS ($\Lambda<0$) sector for $K=0$ and $\dot\beta=0$.  See also
caption to figure \ref{fNSpK0} on page
\pageref{fNSpK0}.}\label{fNSnK0b0}} \end{figure}

Again, the invariant set $\dot\beta=0$, corresponding to
$\kappa=1-X_3^2$, is first examined.  For this case, the dynamical
system becomes equations (\ref{dxi}) and (\ref{deta}) with
$\kappa=1-X_3^2$, and the equilibrium points (and their eigenvalues)
are given by
\beqn
L^+_{(+)}: && X_3=1, Y_3=1; \nonumber \\
    && (\lambda_1, \lambda_2) =  \left(-2, -2[\sqrt{3} +1[ \right),\\
L^+_{(-)}: && X_3=-1, Y_3=1; \nonumber \\
    && (\lambda_1, \lambda_2) =  \left(-2, 2[\sqrt{3}X_3-1] \right),\\
L^-_{(+)}: && X_3=1, Y_3=-1; \nonumber \\
    && (\lambda_1, \lambda_2) =  \left(2, -2[\sqrt{3} -1] \right),\\
L^-_{(-)}: && X_3=1, Y_3=1; \nonumber \\
    && (\lambda_1, \lambda_2) =  \left(2, 2[\sqrt{3} +1] \right).\\
\eeqn
Note that $L^+_{(-)}$ and $L^-_{(+)}$ are saddles, $L^+_{(+)}$ is a
sink, and $L^-_{(-)}$ is a source.  The phase portrait is given in
figure \ref{fNSnK0b0}.

\subsubsection{Three--Dimensional System ($\dot\beta\neq0$)}

\begin{figure}[htp]
  \centering
   \epsfxsize=5in
   \includegraphics*[width=5in]{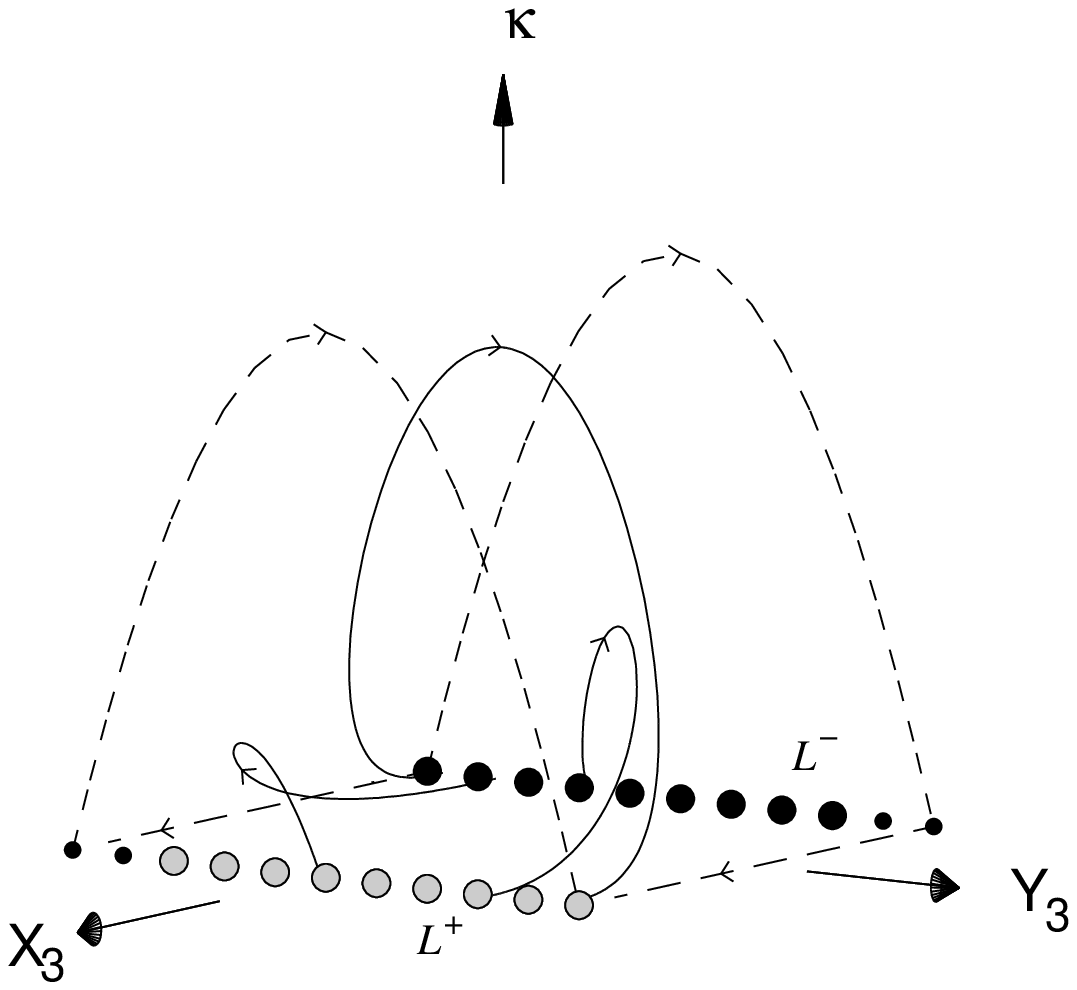}
  \caption[{\em Phase diagram for $\Lambda<0$, $\rho\neq0$ and $\tilde
K=0$}]{{\em Phase diagram of the system (\ref{dXK0}) in the NS-NS
($\Lambda<0$) sector for $K=0$ and $\rho\neq 0$.  Note that $L^+$ and
$L^-$ represent {\em lines} of equilibrium points.  The invariant set
$\kappa=1-X_3^2$ portion (upper parabola sheet) is depicted in
figure \ref{fNSnK0b0} on page \pageref{fNSnK0b0}.  See also caption to
figure \ref{fNSpK0} on page \pageref{fNSpK0}.}\label{fNSnK0}}
\end{figure}

Returning to the three--dimensional system (\ref{dXK0}), the
equilibrium points of this system of ODEs all lie on one of the two
lines of non-isolated equilibrium points (or one-dimensional
equilibrium sets)
\begin{eqnarray}
L^\pm: && Y_3=\pm 1,\kappa=0; \nonumber \\
    && (\lambda_1, \lambda_2, \lambda_3) =  \left(-2Y_3, -2(\sqrt{3}X_3 +Y_3),
	0\right),
\end{eqnarray}
where $X_3$ is arbitrary.  These equilibrium sets are normally
hyperbolic.  The third eigenvalue is zero since this is a set of
equilibrium points.  Thus, on the line $L^+$ the
equilibrium points are saddles for $X_3<-\frac{1}{\sqrt{3}}$ and
local sinks for $X_3>-\frac{1}{\sqrt{3}}$. On the line
$L^-$ the equilibrium points are local sources for
$X_3<\frac{1}{\sqrt{3}}$ and saddles for $X_3>\frac{1}{\sqrt{3}}$.
In the two--dimensional invariant set $\dot\beta=0$, the
points $L^+_{(\pm)}$ are the endpoints to the line $L^+$ and the
points $L^-_{(\pm)}$ are the endpoints to the line $L^-$.  
The dynamics is very simple due to the fact that the right-hand sides
of equations (\ref{dxi}) and (\ref{deta}) are positive--definite and
hence $Y_3$ and $X_3$ are always monotonically increasing
functions. The curved upper boundary $\kappa=1-X_3^2$ denotes the
invariant set $\dot\beta=0$.  The phase diagram is given in figure
\ref{fNSnK0}.

The lines $L^\pm$ represent the ``$\mp$'' dilaton--moduli--vacuum
solutions (\ref{dmv}); solutions asymptote into the past towards the
``$-$'' solutions of (\ref{dmv}) whilst solutions asymptote into the
future towards the ``$+$'' solutions of (\ref{dmv}). 

\subsection{The Case $\Lambda<0$, $\tilde K>0$\label{NSnKp}}

For this case, $\xi^2 = \dot\varphi^2+3\tilde K-2\Lambda$ has been chosen
and from (\ref{TheDefs}) the positive sign for $U_3$ is chosen as is the negative
sign for $V_3$ in order for  these variables to be positive definite.  The
generalized Friedmann constraint equation can be rewritten as
\begin{equation}
0 \leq X_3^2 + Z_3 \leq 1,\qquad Y_3^2+U_3+V_3=1,
\end{equation}
and again $U_3$ (which is proportional to $\tilde K$) may be
eliminated, yielding the four-dimensional system of ODEs for $0\leq
\{X_3^2, Y_3^2, Z_3, V_3\} \leq1$:
\beqn
\mainlabel{dX_3_main}
\label{dX_3}
\frac{dX_3}{dT} & = & \left(1-X_3^2-Z_3\right)\!\! \left(\sqrt 3\!+\!
	X_3Y_3\right)
-\!\frac{1}{\sqrt 3}\left(1\!-\!Y_3^2\!-\!V_3 \right)\!\! \left(1\!-\!X_3^2\!-\!Z_3\right), \\
\label{dY_3}
\frac{dY_3}{dT} &=& \frac{1}{\sqrt3}X_3Y_3 \left( 1-Y_3^2 -V_3 
	\right) +\left(1-Y_3^2 \right)\left(X_3^2+Z_3\right) ,\\
\label{dZ_3}
\frac{dZ_3}{dT} &=& 2Z_3\left[ \frac{1}{\sqrt3}X_3 
	\left( 1-Y_3^2-V_3 \right) +Y_3\left( 1-X_3^2-Z_3 
	\right) \right], \\
\label{dV_3}
\frac{dV_3}{dT} &=& 2V_3 \left[ \frac{1}{\sqrt3} X_3
	\left(1-Y_3^2-V_3 \right) - Y_3\left( X_3^2 + Z_3
	\right) \right].
\eeqn
The invariant sets $Y_3^2+V_3=1$, $X_3^2+Z_3=1$, $Z_3=0$ define the
boundary of the phase space.  The only equilibrium set and its
corresponding eigenvalues (denoted by $\lambda$) are
\beqn
L^\pm: & & Y_3=\pm 1, Z_3=1-X_3^2, V_3=0; \nonumber \\ &&
   \left(\lambda_1, \lambda_2, \lambda_3, \lambda_4\right) \!=\! \left(
   \mp\frac{2}{\sqrt3} \left[ \sqrt3\! \pm\! X_3\right], \mp2, 0, \mp2\sqrt3
	\left[ \frac{1}   {\sqrt3} \!\pm\! X_3 \right]    \right), 
\eeqn
where again the zero eigenvalues arise because these are all {\em
lines} of equilibrium points.  Here, the global sink is the line $L^+$
for $X_3>-\frac{1}{\sqrt3}$, and the global source
is the line $L^-$ for $X_3<\frac{1}{\sqrt3}$.  The
exact solutions corresponding to these two lines are the
dilaton--moduli--vacuum solutions given by equation (\ref{dmv}).

\subsubsection{The Invariant Set $\rho=0$ for $\Lambda<0$, $\tilde K>0$}

\begin{figure}[htp]
  \centering
   \epsfxsize=5in
   \includegraphics*[width=5in]{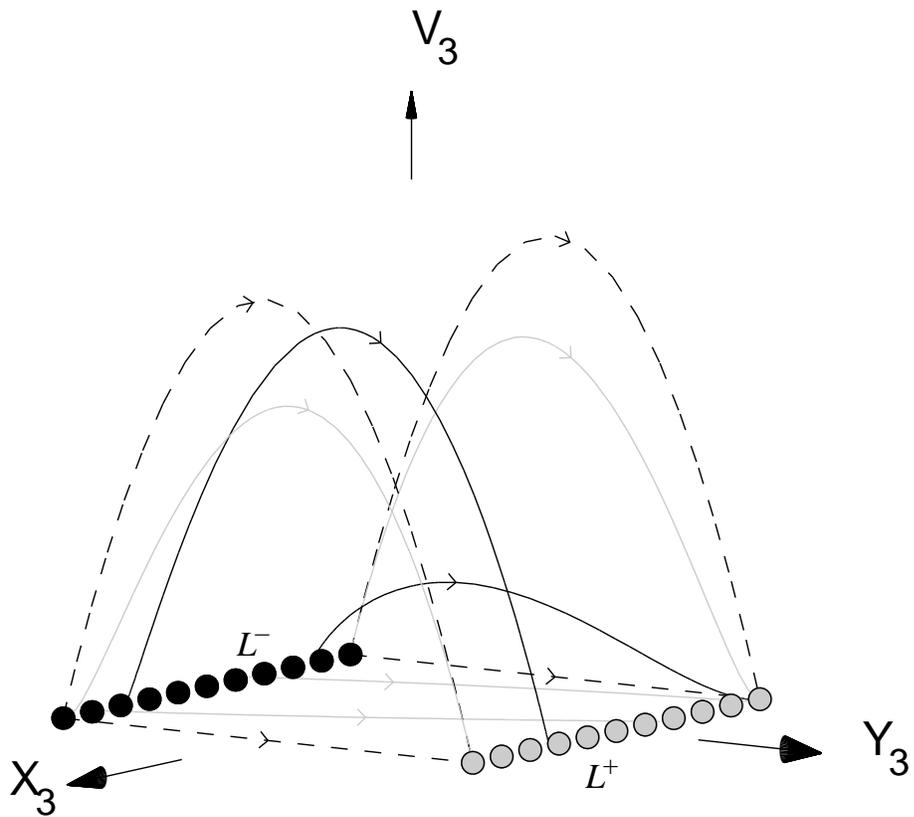}
  \caption[{\em Phase diagram for $\Lambda<0$, $\rho=0$ and $\tilde K>0$}]{{\em Phase diagram of the system
(\ref{dX_3_main}) in the NS-NS ($\Lambda<0$) sector for
$\rho=0$ and $\tilde K>0$.  Note that $L^+$ and $L^-$  represent {\em lines} of
equilibrium points.  See also caption to figure \ref{fNSpK0} on page \pageref{fNSpK0}.}\label{fNSnKp}}
 \end{figure}

Because $\dot\alpha\neq0$ at the equilibrium sets, the
$\rho=0$ case is examined, in which the system reduces to three dimensions
$\{X_3,Y_3,V_3\}$ ($Z_3=1-X_3^2$).  The equilibrium sets are the
same as above with eigenvalues $\lambda_1$, $\lambda_2$, $\lambda_3$.
Note that the entire line $L^+$ acts as a global sink and that the
entire line $L^-$ acts as a global source in this three--dimensional
invariant set.  Figure \ref{fNSnKp} depicts this three-dimensional
phase space.

The function $Y_3/\sqrt{V_3}$ is {\em monotonically increasing},
eliminating the possibility of periodic orbits.  Therefore, solutions
generically asymptote into the past towards the $\dot\varphi<0$
dilaton--moduli--vacuum solutions (\ref{dmv}), and into the future
towards the $\dot\varphi>0$ dilaton--moduli--vacuum solutions
(\ref{dmv}).  The curvature term and central charge deficit are
dynamically significant only at intermediate times.

\subsubsection{Qualitative Analysis of the Four-Dimensional System}

The qualitative dynamics in the full four-dimensional phase space is
as follows.  The global repellors and attractors of this phase space
are the lines $L^-$ and $L^+$, respectively, corresponding to the
dilaton--moduli--vacuum solutions.  Hence orbits generically asymptote
into the past towards the line $L^-$ (for $X_3<\frac{1}{\sqrt3}$), and
generically asymptote into the future towards the line $L^+$ (for
$X_3>-\frac{1}{\sqrt3}$); both lines lie in both of the invariant sets
$\rho=0$ and $\tilde K=0$, consistent with the analysis of equation
(\ref{dchi}).  Hence, solutions generically asymptote into the past
towards the $\dot\varphi<0$ dilaton--moduli--vacuum solutions
(\ref{dmv}) for $h_0<\third$, and into the future towards the
$\dot\varphi>0$ dilaton--moduli--vacuum solutions (\ref{dmv}) for
$h_0>-\frac{1}{3}$.  Again, the curvature term, the axion field and
the central charge deficit are dynamically important only at
intermediate times, and are negligible at early and late times.  The
existence of the {\em monotonically increasing} function
$Y_3/\sqrt{V_3}$ excludes the possibility of periodic orbits and
serves to verify that the evolution of solutions in the
four--dimensional set discussed above.

\subsection{The Case $\Lambda<0$, $\tilde K<0$\label{NSnKn}}

The negative signs are chosen for both $U_4$ and $V_4$ from
(\ref{TheDefs}) in this case, and the definition $\xi^2 =
\dot\varphi^2-2\Lambda$ is chosen.  The generalized Friedmann constraint
equation is now written to read
\begin{equation}
0 \leq X_4^2 + Z_4 + U_4\leq 1, \qquad Y_4^2+V_4=1,
\end{equation}
where now $V_4$ is considered to be the extraneous variable, resulting is
the four-dimensional system consisting of the variables $0\leq \{X_4^2, 
Y_4^2, Z_4, U_4\} \leq 1$:
\beqn
\mainlabel{dX_4_main}
\label{dX_4}
\frac{dX_4}{dT} & = & \left(1-X_4^2-Z_4\right)\left(\sqrt 3 +X_4Y_4\right) 
-\frac{2}{\sqrt 3}U_4, \\
\label{dY_4}
\frac{dY_4}{dT} &=& \left(1-Y_4^2\right) \left( X_4^2 + Z_4\right) > 0, \\
\label{dZ_4}
\frac{dZ_4}{dT} &=& 2Z_4Y_4 \left( 1-X_4^2-Z_4 \right), \\
\label{dU_4}
\frac{dU_4}{dT} &=& -2U_4 \left[ Y_4\left(X_4^2+Z_4\right)
	+\frac{1}{\sqrt3}X_4 \right].
\eeqn
The invariant sets $X_4^2+Z_4+U_4=1$, $Z_4=0$, $Y_4^2=1$, $U_4=0$ define the
boundary of the phase space.  The equilibrium sets and their
corresponding eigenvalues (denoted by $\lambda$) are
\beqn
S^\pm: & & X_4=\mp\frac{1}{\sqrt3}, Y_4=\pm1, Z_4=0, U_4=\frac{2}{3}; \nonumber \\ &&
   \left(\lambda_1, \lambda_2, \lambda_3, \lambda_4\right) =  \left(
   \pm\frac{2}{3}, \mp\frac{2}{3}, \pm\frac{4}{3}, \pm\frac{4}{3} \right), \\
L^\pm: & & Y_4=\pm1, Z_4=1-X_4^2, U_4=0; \nonumber \\ &&
   \left(\lambda_1, \lambda_2, \lambda_3, \lambda_4\right) =  \left(
   \mp\frac{2}{\sqrt3} \left[ \sqrt3 \!\pm\! X_4\right], \mp 2 , 0,
  \mp 2\sqrt3 \left[ \frac{1}{\sqrt3} \!\pm\! X_4 \right] \right)\!.
\eeqn
The global source for this system is the line $L^-$ (for
$X_4<\frac{1}{\sqrt3}$), and the global sink is the line $L^+$ (for
$X_4>-\frac{1}{\sqrt3}$).  These lines correspond to the
dilaton--moduli--vacuum solutions, given by equation (\ref{dmv}).  The
saddle points $S^\pm$ represent the Milne models (\ref{curv_drive}).

\subsubsection{The Invariant Set $\rho=0$ for $\Lambda <0$, $\tilde K<0$}

\begin{figure}[htp]
  \centering
   \epsfxsize=5in
   \includegraphics*[width=5in]{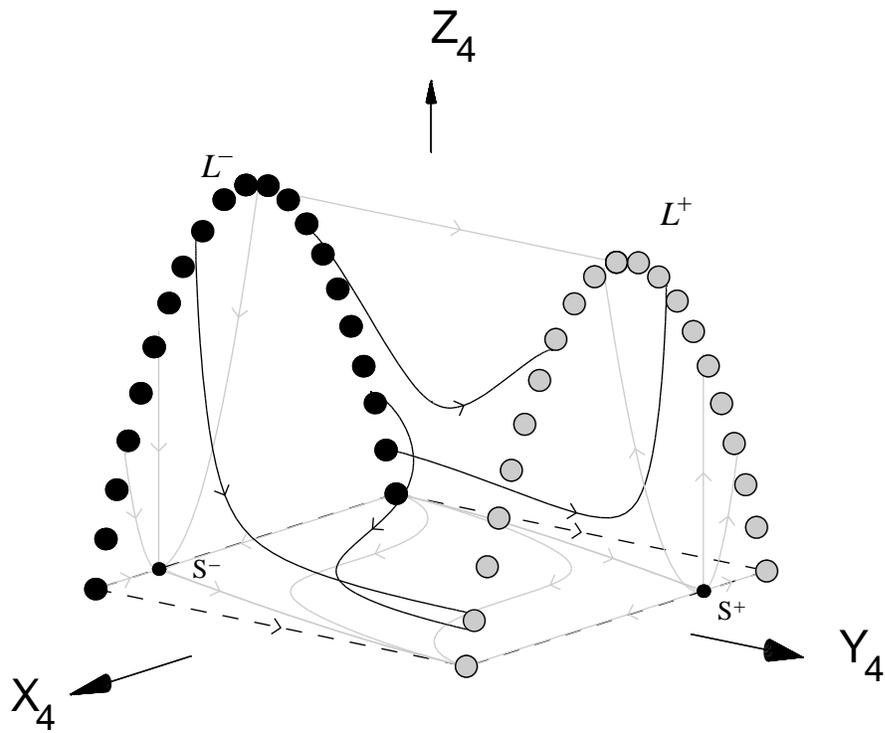}
  \caption[{\em Phase diagram for $\Lambda<0$, $\rho=0$ and $\tilde
K<0$}]{{\em Phase diagram of the system (\ref{dX_4_main}) in the
NS-NS ($\Lambda<0$) sector with $\rho= 0$ and $\tilde K<0$.  Note that
$L^+$ and $L^-$ represent {\em lines} of equilibrium points.  See also
caption to figure \ref{fNSpK0} on page
\pageref{fNSpK0}.}\label{fNSnKn}}
 \end{figure}

In the $\rho=0$ case, the system reduces to three dimensions of
$\{X_4,Y_4,Z_4\}$ ($U_4=1-X_4^2-Z_4$).  The equilibrium sets are the
same as above with eigenvalues $\lambda_1$, $\lambda_2$, $\lambda_3$.
Note that for the invariant set $\rho=0$ the entire line $L^+$ acts as
a global sink and the entire line $L^-$ acts as a global source.  The
variable $Y_4$ is {\em monotonically increasing}, the existence of
which excludes the possibility or recurrent orbits.  Hence, solutions
generically asymptote into the past towards the $\dot\varphi<0$
dilaton--moduli--vacuum solutions (\ref{dmv}), and into the future
towards the $\dot\varphi>0$ dilaton--moduli--vacuum solutions
(\ref{dmv}).  The curvature and central charge deficit are dynamically
significant only at intermediate times.  Figure \ref{fNSnKn} depicts
this three-dimensional phase space.

\subsubsection{Qualitative Analysis of the Four-Dimensional System}

The dynamical behaviour in the invariant set $\tilde K=0$ is identical
to that described in subsection \ref{NSnKn} (see figure \ref{NSnK0}).

The qualitative dynamics in the full four-dimensional phase space is
as follows.  Note that $Y_4$ is {\em monotonically increasing} and so
orbits asymptote into the past towards $Y_4=-1$ and into the future
towards $Y_4=+1$.  Again, the existence of such monotonic functions
exclude the possibility of periodic orbits in the four--dimensional
phase space.  Most orbits asymptote into the past towards the line
$L^-$ (for $X_4<\frac{1}{\sqrt3}$), and into the future towards the
line $L^+$ (for $X_4>-\frac{1}{\sqrt3}$).  Both lines $L^\pm$ lie in both
invariant sets $\rho=0$ and $\tilde K=0$, which is consistent with the
analysis of equation (\ref{dchi}).  Generically, solutions asymptote
to the past to the $\dot\varphi<0$ dilaton--moduli--vacuum solutions
(\ref{dmv}) for $h_0<\third$, and to the future to the $\dot\varphi>0$
dilaton--moduli--vacuum solutions (\ref{dmv}) for $h_0>-\frac{1}{3}$.  The
variable $Y_4$ increases monotonically along orbits and so
$\dot\varphi$ is dynamically significant at early and late times, but
dynamically negligible at intermediate times.  Conversely, it can
be seen that the curvature, the central charge deficit and the
axion field are dynamically important only at intermediate times, and
are negligible at both early and late times.

\section{Summary of Analysis in the Jordan Frame\label{summary_I}}

This discussion begins with two tables; Table \ref{term_asymp_I} lists
which terms are the dominant variables for each equilibrium set as
well as the deceleration parameter, $q$, for the corresponding model
and Table \ref{points_asymp_I} lists the attracting behaviour of the
equilibrium sets.
\begin{table}[htb]
\begin{center}
\begin{tabular}{|c|cccccccc|c|c|}\hline
Set & \multicolumn{8}{c|}{Dominant Variables}&$q$&H \\
\hline \hline
$L^+$ {\tiny($t<0$)} & $\alpha$ &$\hat\Phi$ & $\dot\beta$ &&&&&
	&$-(1+h_0)h_0^{-1}$ &$h_0(-t)^{-1}$\\
$L^-$ {\tiny($t>0$)} & $\alpha$ &$\hat\Phi$ & $\dot\beta$ &&&&&
	&$(1-h_0)h_0^{-1}$ &$h_0t^{-1}$\\ \hline
$C^\pm$     & & $\hat\Phi$ &&&& $\Lambda>0$ &&&$0$&$0$\\
$L_1$       & &$\hat\Phi$ && $\dot\sigma$ & $k\geq 0$ & $\Lambda>0$&&&$0$&$0$\\
\hline
$S^\pm$     & $\alpha$ &&&& $k<0$ &&&&$0$& $t^{-1}$\\ \hline
\end{tabular}
\end{center}
\caption[{\em Dominant Variables and $q$ for Equilibrium Sets}]{{\em The
dominant variables for each equilibrium set as well as the equilibrium
set's deceleration parameter, $q$, and Hubble parameter $H$.
Inflation occurs when $q<0$ and $H>0$, whereas ``deflation'' occurs
for $q>0$ and $H<0$.  Note that the only ``anisotropic'' solutions are
represented by the lines $L^\pm$, except when
$h_0^2=\third$.}\label{term_asymp_I}}
\end{table}
Note that the only inflationary models are those represented by $L^+$
(for $h_0>0$).

\begin{table}[ht]
\begin{center}
\begin{tabular}{|ccccc||c|c|c|}\hline
\multicolumn{5}{|c||}{Terms Present} & Early & Intermediate & Late \\
\hline \hline
$\dot\beta$ &     & $\Lambda>0$ &             &      &$C^+$&$L^+$ & $L^+$\\
$\dot\beta$ &$k>0$& $\Lambda>0$ &  &    &$L_1$, $L^-$ &$L^\pm$ &$L_1$, $L^+$ \\
$\dot\beta$ &$k<0$& $\Lambda>0$ &        &      &$C^+$ & $S^+$, $L^+$&$L^+$ \\
$\dot\beta$ & $k\geq0$& $\Lambda<0$ &         &      &$L^-$ &$L^\pm$ &$L^+$ \\
$\dot\beta$ &$k<0$& $\Lambda<0$ &         &  &$L^-$ &$S^\pm$, $L^\pm$ &$L^+$ \\
\hline
\end{tabular}
\end{center}
\caption[{\em Summary of Equilibrium Sets for Models
considered}]{{\em Summary of the early-time, intermediate, and
late-time attractors for the various models examined.  Note that
$\dot\alpha$, $\hat\Phi$ and $\dot\sigma$ are present in every
model.}\label{points_asymp_I}}
\end{table}
For every case studied, monotonic functions have been established
which precludes the existence of recurrent or periodic orbits, thereby
allowing the early-time and late-time behaviour conclusions to be made
based upon the equilibrium sets of the system.  In all cases, the
dilaton--moduli--vacuum solutions act as either early-time or
late-time attractors (and, in many of the cases, both).  Because these
solutions lie in both the $\rho=0$ and $\tilde K=0$ invariant sets and
contain no central charge deficit, it is conclusive that the modulus
and dilaton fields are dynamically important asymptotically.
Furthermore, with the exception of the $\Lambda>0$, $\tilde K>0$ case,
all early-time and late-time attracting sets lie in either the
$\rho=0$ invariant set, or the $\tilde K=0$ invariant set, and a
majority of these sets lie in both; thus, there seems to be a generic
feature in which the curvature and the axion field are
dynamically significant at intermediate times and are asymptotically
negligible at early and at late times.  The exception to this generic
behaviour is the $\Lambda>0$, $\tilde K>0$ case, where the generalized
linear dilaton--vacuum solution (\ref{static_general}), in which
neither $\rho=0$ nor $\tilde K=0$, acts both as a repellor (for
$\dot\varphi>0$) and as an attractor (for $\dot\varphi<0$).  In these
solutions, the variables $\rho$ and $\tilde K$ are proportional to the
central charge deficit, $\Lambda$.

When $\Lambda<0$, the central charge deficit is dynamically
significant only at intermediate times.  The only repelling and
attracting sets in this instance are the dilaton--moduli--vacuum
solutions.  When $\tilde K>0$, the central charge deficit can be
dynamically significant at early and at late times, and the corresponding
solutions are the generalized dilaton--vacuum solutions
(\ref{static_general}), represented by the line $L_1$.  When $\tilde
K>0$, these solutions can be repelling ($\dot\varphi>0$) and
attracting ($\dot\varphi<0$).  When $\tilde K<0$, then the endpoint of
this line, $C^+$ (representing equation (\ref{static})) is a repellor.

Typically, the curvature term is found to be be dynamically
significant only at intermediate times, and is asymptotically
negligible.  However, the only exceptions to this is the case
$\Lambda>0$, $\tilde K>0$ where the generalized linear dilaton--vacuum
attractors and repellors have a non-negligible curvature.  

The field equations derived from action (\ref{sigmaaction_i}) for this
background for $\Lambda_{\rm M}=Q=0$ are formally {\em identical} to
those for either FRW models with a modulus field or certain Bianchi
models with or without a modulus field.  With an anisotropic model,
however, the variables $\alpha$ and $\beta$ now parameterize the
averaged scale factor and the shear parameter of the universe,
respectively.  More generally, a finite number of modulus fields and
shear modes can be introduced by defining the variable $6\dot\beta^2$
via equation (\ref{sumup}).  If isotropization is defined by
$\dot\beta\rightarrow 0$, then the only anisotropic models are the
dilaton--moduli-vacuum solutions (\ref{dmv}).  These solutions are
attracting solutions (either into the past or future) for most models.
However, these are not the only asymptotic solutions and all other
solutions mentioned in this chapter are isotropic ($\dot\beta=0$).  In
particular, the models in which $\Lambda>0$ and $k>0$ do have
isotropic late-time attractors.

The only at\-tracting so\-lu\-tions which are in\-fla\-tion\-ary are
the di\-la\-ton--mo\-du\-li--va\-cuum so\-lu\-tions (the ``$-$''
branch of (\ref{dmv}) for $h_0>0$) which occur in the pre-big bang
portion of the theory.  Note that the time reverse dynamics of
these models correspond to $\dot\alpha\rightarrow -\dot\alpha$ (i.e.,
contracting $\rightarrow$ expanding and vice versa) as well as
$\dot\varphi \rightarrow -\dot\varphi$ and correspond to a post-big
bang era. Hence, all late--time attracting solutions in the post-big bang regime are
{\em not} inflationary.

\section{Exact Solutions in the Einstein Frame\label{Einstein_Frame_I}}

The analysis of this chapter can equally be applied in the Einstein
frame, resulting in the same equilibrium points.  This section
describes the solutions of the equilibrium point in the Einstein frame
and then relates the quantities $\{\dot\beta, \dot\sigma\, \Lambda,
\Lambda_{\rm M}, Q\}$ in terms of matter fields, as described in
Chapter \ref{SFtoST}, thereby mathematically casting the analysis into
a theory of general relativity containing a matter source and a scalar
field with exponential potential.

First, the transformation (\ref{jesus2}) on page \pageref{jesus2} 
between the scalar fields
$\phi$ and $\Phi$ for $\omega_0=-1$ is rewritten as
\be
\frac{dt^*}{dt} = \pm e^{\pm\phi/\sqrt2} \equiv e^{-\varepsilon\phi/\sqrt2};
	\label{t2_trans}
\ee
i.e., the ``$\pm$'' sign before the exponential term is explicitly
taken to be ``$+$'' to ensure that $t^*$ (time in the Einstein frame)
and $t$ (time in the Jordan frame) grow in the same direction which
ensures that $\up{sf}H$ and $\up{st}H$ are of the same sign when
$\phi=\phi_0$ ($\Phi=\Phi_0$), and the ``$\pm$'' sign inside the
exponential term is replaced with the constant $\varepsilon=\mp1$ to save
confusion with the ``$\pm$'' signs to follow in the exact solutions.

To refresh, the scale factors between the two frames are related by
\be
a^* = e^{-\half\hat\Phi} a \qquad \Leftrightarrow \qquad 
	a = e^{\varepsilon\phi/\sqrt2} a^*, \label{a2_trans}
\ee 
where $a^*$ is the scale factor in the Einstein frame and $a$ is the
scale factor in the Jordan frame.

The dilaton--moduli--vacuum solutions transform to the Einstein frame as
\index{equilibrium sets!$L^\pm$, $L^{\pm}_{(\pm)}$}
\index{exact solutions!dilaton--moduli--vacuum}
\begin{eqnarray}
\nonumber
L^\mp: & & a^* = a^*_0\left| t_* \right|^{\third} 
	\qquad \Rightarrow \qquad \up{sf}H = \third t_*^{-1} 
	\qquad \Rightarrow \qquad \up{sf}q=2, \\
\nonumber
& & e^{\phi} = e^{\bar{\phi}^*_0}
	\left|t_*\right|^{\frac{\varepsilon\sqrt2(\pm3h_0-1)}
			{3(1 \mp h_0)}}, \nonumber \\
\nonumber
& & e^\beta = e^{\beta^*_0}\left|t_*\right|^{\frac{\pm\sqrt{2(1-3h_0^2)}}
			{3\sqrt3(1 \mp h_0)}} ,\\
\nonumber
&& \sigma = \sigma_0,\\
&& k=0,
\label{E_dmv}
\end{eqnarray}
where $\{ a^*_0, \bar{\phi}^*_0, \beta^*_0\}$ are suitable redefined
integration constants to absorb positive definite constants 
which arise from (\ref{t2_trans}) and (\ref{a2_trans}), and again the
$\pm$ sign corresponds to the sign of $t_*$.  Note that in the Einstein
frame, the scale factor does {\em not} depend on the constant $h_0$.
From $\up{sf}H$ and $\up{sf}q$, it is apparent that $L^+$ represents
deflationary models, whereas $L^-$ represents expanding,
non-inflationary models.  This is substantially different than from
the Jordan frame, where inflationary models exist for $t_*<0$.  From the
analysis of the previous sections, $L^-$ is an early-time attractor
for most models considered and $L^+$ is a late-time attractor for most
models.  In all models considered $L^\pm$ are also saddle points.

In the Einstein frame, the saddle-point Milne solution becomes
\index{equilibrium sets!$S^\pm$}
\index{exact solutions!Milne}
\begin{eqnarray}
\nonumber
S^\mp: && a^*=a^*_0\left(\pm t_*\right) \qquad \Rightarrow \qquad 
	\up{sf}H=t_*^{-1} \qquad \Rightarrow \qquad  \up{sf}q=0, \\
\nonumber
&& \phi=\phi_0,\\
\nonumber
&& \beta=\beta_0,\\
\nonumber
&& \sigma=\sigma_0,\\
\label{E_curv_drive}
&& k=-a_0^2,
\end{eqnarray}
where $a^*_0=a_0$. Again, the ``$\pm$'' sign corresponds to the sign
of $t_*$.  These solutions are a saddle solutions in all models where
$\Lambda\neq0$ and $k\neq0$.

\index{equilibrium sets!$C^\pm$}
\index{exact solutions!linear dilaton--vacuum}
The static linear dilaton--vacuum solution in the Einstein frame becomes
\begin{eqnarray}
\nonumber
C^\mp: && a^*=a^*_0 \left| t_*\right| \qquad \Rightarrow \qquad 
	\up{sf}H=t_*^{-1} \qquad \Rightarrow \qquad \up{sf}q=0, \\
\nonumber
 && e^{\varepsilon\sqrt2\phi} = \frac{2}{\Lambda t_*^2},\\
\nonumber
&& \beta =\beta_0, \\
\nonumber
&& \sigma = \sigma_0,\\
\label{E_static}
&& k=0,
\end{eqnarray}
where $a^*_0=\sqrt{\half \Lambda}a_0$.  The solution $C^+$ is an
early-time attractor for all $\Lambda>0$ models considered (except
when $k>0$ in which case this solution is replaced by its
generalization $L_1$ discussed below), and $C^-$ arises as a late-time
attractor in the model where $\Lambda>0$ and $\Lambda_{\rm M}<0$
(chapter \ref{NSNSRR}).  Unlike in the Jordan frame in which the
universe is static, in the Einstein frame the universe is linearly
contracting for $t_*<0$ and linearly expanding for $t_*>0$.

The generalized linear dilaton--vacuum solutions in the Einstein frame become
\index{equilibrium sets!$L_1$}
\index{exact solutions!linear dilaton--vacuum!generalized linear dilaton--vacuum}
\begin{eqnarray}
\nonumber
L_1: && a = a^*_0 \left|t_*\right| \qquad \Rightarrow \qquad \up{sf}H=t_*^{-1} 
	\qquad \Rightarrow \qquad \up{sf}q=0, 
\\
\nonumber
&& e^{\varepsilon\sqrt2\phi} =\frac{2\left(2+n^2\right)}{3n^2\Lambda t_*^2}, \\
\nonumber
&& \beta =\beta_0, \\
\nonumber
&& \sigma = \sigma_0 \pm \frac{\sqrt{6n^2\left(1-n^2\right)}}
	{2\left(2+n^2\right)}\Lambda t_*^2, \\
&& k = \frac{2(1-n^2)}{3n^2} a^{*2}_0, 
\label{E_static_general}
\end{eqnarray}
where $n\in [-1,1]$.  Here the $n<0$ solutions correspond to $t_*>0$,
representing positively-curved, expanding models whereas $n>0$
(corresponding to $t_*<0$) represent contracting models.
Again, this is a different behaviour than that in the Jordan frame
wherein the models are static.  As determined in the analysis in
sections \ref{NSpKp}, for $\Lambda>0$ and $k>0$, $L_1$ represents
early-time attracting solutions for $n>0$ and late-time attracting
solutions for $n<0$.

In the Einstein frame, there are no inflationary late-time attracting
solutions.

\subsection{Mathematical Equivalence to Matter Terms in the Einstein Frame\label{math_I}}

As mentioned in Chapter \ref{SFtoST}, these string models are 
mathematically equivalent (in the Einstein frame) to a theory of general
relativity containing a matter source and a scalar field with an 
exponential potential; i.e., $V=V_0e^{k\phi}$, where either
$k^2=2$ or $V_0=0$.  This subsection is devoted to explicitly
determining the form of the matter field for each of the equilibrium sets
discussed in this chapter.  Before each case is examined, it is useful
to restate that upon transformation from the Jordan frame into the
Einstein frame, interaction terms arise so that each source (matter
field and scalar field) will {\em not} be separately conserved.  In
fact, the conservation equations in the Einstein frame can be
typically written as
\beqn
\mainlabel{E_conserves_I}
\dot\phi\left(\ddot\phi+3\up{sf}H\dot\phi +\frac{dV}{d\phi}\right) 
	 &=& -\delta,\label{E_phi_conserves_I}\\
\dot\mu + 3\up{sf}H\left(\mu+p\right) &=& +\delta.\label{E_mu_conserves_I}
\eeqn

Furthermore, in each instance, the total energy
density and pressure of the scalar field may be calculated, namely,
\beqn
\mu_\phi & = & \half\dot\phi^2+V,\\
  p_\phi & = & \half\dot\phi^2-V,
\eeqn
and hence $\gamma_\phi\equiv(\mu_\phi+p_\phi)/\mu_\phi$ can be compared with
the $\gamma$ of the matter field.

For these models there are two scenarios from which to choose:
\renewcommand{\labelenumi}{\Alph{enumi})}
\begin{enumerate}
\item $V=\Lambda e^{\sqrt2\varepsilon\phi}$  $(k^2=2)$, ${\cal U}=0$ \\
The interaction term for this case is $\delta=-2\sqrt2\varepsilon\dot\phi\ p$.
\item $V=0$, ${\cal U}=\Lambda e^{\sqrt2\varepsilon\phi}$ \\
The interaction term for this case is $\delta=-\frac{\sqrt2}{2}
\varepsilon\dot\phi(\mu+3p)$.
\end{enumerate}
For these two scenarios, the matter field is defined by
\beqn
\mu &\equiv & \quart \dot\sigma^2e^{2\sqrt2\varepsilon\phi} + {\cal U} \\
p &\equiv & \quart \dot\sigma^2e^{2\sqrt2\varepsilon\phi} - {\cal U}
\eeqn
and do not in general represent barotropic matter with a linear
equation of state [$p=(\gamma-1)\mu$], although the equations of state
are linear at the equilibrium points.  Tables
\ref{model_A_I} and \ref{model_B_I} list
$\{\mu, p, \gamma, \mu_\phi, p_\phi, \gamma_\phi, \delta\}$ for each
equilibrium set in each of the two scenarios discussed above.
\begin{table}[htb]
\begin{center}
\begin{tabular}{|c|ccccccc|}\hline
\multicolumn{8}{|c|}{\bf Scenario A: \qquad 
	$V=\Lambda e^{\sqrt2\varepsilon\phi}$, \qquad
	${\cal U}=0$, \qquad
	$\delta=-2\sqrt2\varepsilon\dot\phi\ p$} \\ \hline
\hline
Set & $\mu$ & $p$ & $\gamma$ & $\mu_\phi$ & $p_\phi$ & $\gamma_\phi$ 
		& $\delta$ \\
\hline 
$L^\pm$ & $0$ & $0$ & $-$ & $\frac{(\pm3h_0-1)^2}{9(1\mp h_0)^2}t_*^{-2}$ & 
	$\mu_\phi$ & $2$ & $0$ \\
$S^\pm$ & $0$ & $0$ & $-$ & $0$ & $0$ & $-$ & $0$ \\
$C^\pm$ & $0$ & $0$ & $-$ & $3t_*^{-2}$ & $-t_*^{-2}$ & $\frac{2}{3}$ & $0$ \\
$L_1$   & $\frac{2(1-n^2)}{3n^2}t_*^{-2}$& $\mu$ & $2$ & $\frac{4+5n^2}{3n^2}
	t_*^{-2}$ & $\frac{n^2-4}{3n^2}	t_*^{-2}$ & $\frac{6n^2}{4+5n^2}
	\in[0,\frac{2}{3}]$ & $\frac{8(1-n^2)}{3n^2}t_*^{-3}$\\
\hline
\end{tabular}
\end{center}
\caption[{\em Matter Terms for Model A for $\Lambda_{\rm
M}=Q=0$}]{{\em The matter terms ($\mu$,
$p$, $\gamma$) as well as $\mu_\phi$, $p_\phi$, $\gamma_\phi$ and
$\delta$ for each of the equilibrium sets derived in scenario A for
$\Lambda_{\rm M}=Q=0$.}\label{model_A_I}}
\end{table}

From section \ref{summary_I}, the asymptotic behaviour of the
models in the Einstein frame is known and comments on which solutions 
represent asymptotic states in the Einstein frame are equally applicable here.

In scenario A, it is evident that asymptotically the matter field
either vanish (represented by the sets $L^\pm$ or $C^\pm$) or
asymptote towards a stiff equation of state (represented by the set
$L_1$), the latter arising only in the case of negative curvature.
There are no matter scaling solutions at the equilibrium sets except
the trivial case of the saddle Milne model, in which all sources are
zero.

\begin{table}[ht]
\begin{center}
\begin{tabular}{|c|ccccccc|}\hline
\multicolumn{8}{|c|}{\bf Scenario B: \qquad 
	$V=0$,\qquad 
	${\cal U}=\Lambda e^{\sqrt2\varepsilon\phi}$, \qquad
	$\delta=-\frac{\sqrt2}{2}\varepsilon\dot\phi(\mu+3p)$}\\\hline\hline
Set & $\mu$ & $p$ & $\gamma$ & $\mu_\phi$ & $p_\phi$ & $\gamma_\phi$ 
		& $\delta$ \\
\hline 
$L^\pm$ & $0$ & $0$ & $-$ & $\frac{(\pm3h_0-1)^2}{9(1\mp h_0)^2}t_*^{-2}$ & 
	$\mu_\phi$ & $2$ & $0$ \\
$S^\pm$ & $0$ & $0$ & $-$ & $0$ & $0$ & $-$ & $0$ \\
$C^\pm$ & $2t_*^{-2}$ & $-\mu$ & $0$ & $t_*^{-2}$ & $\mu_\phi$ 
	& $2$ & $-4t_*^{-3}$ \\
$L_1$   & $\frac{2}{n^2}t_*^{-2}$& $\frac{-2(1+2n^2)}{3n^2}t_*^{-2}$ & 
	$\frac{2}{3}(1-n^2)\in[0,\frac{2}{3}]$  & $t_*^{-2}$ 
	& $\mu_\phi$ & $2$ & $-4t_*^{-3}$\\
\hline
\end{tabular}
\end{center}
\caption[{\em Matter Terms for Model B for $\Lambda_{\rm M}=Q=0$}]{{\em The matter terms ($\mu$,
$p$, $\gamma$) as well as $\mu_\phi$, $p_\phi$, $\gamma_\phi$ and
$\delta$ for each of the equilibrium sets derived in scenario
B for $\Lambda_{\rm M}=Q=0$.}\label{model_B_I}}
\end{table}

In scenario B, it is evident that asymptotically the matter field
vanishes as trajectories asymptote to $L^\pm$ or $C^\pm$ (which is the
same behaviour for scenario A).  However, for trajectories asymptoting
towards the line $L_1$ (for negatively curved models), the matter
field asymptotes to a linear equation of state with $0\leq \gamma <
\frac{2}{3}$.  Again, there are no matter scaling solutions at any of the
equilibrium sets.

\chapter{String Models II: Non-Zero Cosmological Constant ($\Lambda=Q=0$)\label{RR}}
 
In this chapter, $\Lambda=0$ and $Q=0$ in order to determine the
r\^{o}le of $\Lambda_{\rm M}$ in the dynamics of the system
(\ref{rr}).  As demonstrated in the last chapter, most of the analysis
to follow is equally applicable to curved FRW models with a modulus
field {\em and} to certain Bianchi type I, V and IX models with or
without a modulus field.  The chapter is organized as follows.  In
section
\ref{Governing_Equations_II}, the field equations (\ref{rr}) and 
(\ref{rrfriedmann}) are reexamined with $\Lambda=0$ and $Q=0$, and all
the known corresponding exact solutions are listed in section
\ref{Exact_Solutions_II} for $\Lambda_{\rm M}\neq 0$.  It was shown in 
section \ref{Asymptotic_Behaviour} that for most cases either the
curvature or the axion field will asymptotically dominate at early or
late times, and hence the four-dimensional dynamical system can be
reduced to a three-dimensional system.  There was one exception to
this comment which existed only in the NS-NS case and hence the
analysis there is completely applicable to this chapter.  Section
\ref{Analysis_II} proceeds with the analysis of the equations.  The
chapter ends with a summary section and a section which discusses the
corresponding solutions and asymptotic behaviour in the Einstein
frame.  Again, this chapter is primarily confined to the Jordan frame
(except the final section), and so the index ``(st)'' shall be omitted
to save notation, (but must be introduced again in the final section
when both frames are discussed).

\section{Exact Solutions in the Matter Sector\label{Exact_Solutions_II}}

There are several exact solutions which exist for $\Lambda_{\rm
M}\neq0$ and $\Lambda=Q=0$ which are represented by equilibrium points
found in the analysis of this chapter.

\index{equilibrium sets!$S^\pm_1$}
\index{exact solutions!dilaton--axion}
The solution found in Billyard {\em et al.} \cite{Billyard1999c} is given by
\begin{eqnarray}
\nonumber
a & = &  a_0 \left[\frac{\sqrt{3\Lambda_{\rm M}}}{4}
	\left|t\right|\right]^{\third}, \\
\nonumber
\hat{\Phi} & = &  -\ln \left[\frac{3\Lambda_{\rm M}}{16} t^2 \right], \\
\nonumber
\beta& = & \beta_0,\\
\nonumber
\sigma & = & \sigma_0\pm \frac{\sqrt{15}}{16}\Lambda_{\rm M} t^2, \\
\label{newsol}
k&=&0,
\end{eqnarray}
where $\{ a_0,\beta_0 , \sigma_0\}$ are arbitrary constants
and time is defined over the interval $t<0$ ($\dot\varphi>0$) for the
equilibrium point $S^+_1$ and $t>0$ $(\dot\varphi<0$) for the
equilibrium point $S^-_1$.  Note that $S^+_1$ represents a {\em deflationary} model (i.e. $q>0$ and $H<0$).

\index{equilibrium sets!$N$}
\index{exact solutions!negative--curvature driven}
The negative-curvature cosmological model is given by
\begin{eqnarray}
\nonumber
a &=& \frac{1}{2}a_0\sqrt \Lambda_{\rm M}\left| t\right|, \\
\nonumber
\hat{\Phi} &=& - \ln \left[ \frac{1}{4}\Lambda_{\rm M} t^2 \right], \\
\nonumber
\beta &=& \beta_0, \\
\nonumber 
\sigma&=&\sigma_0, \\
\label{open_new} \nonumber 
k &=& -\frac{3}{4}\Lambda_{\rm M} a_0^2 \\
\end{eqnarray}
where $\{ a_0,\beta_0, \sigma_0\}$ are integration constants and the
time is defined for $t<0$.  This solution is represented by the
equilibrium point $N$ throughout this chapter.

\

For the $\tilde K=0$, $\Lambda_{\rm M}>0$ case (section \ref{RRpK0}),
there exists an invariant line which connects the point ``$S_1^+$''
(i.e. equation (\ref{newsol})) to the line $L^+$ for
$\mu_1=-\frac{1}{\sqrt{27}}$ (i.e. the ``$-$'' solution of equation
(\ref{dmv}) for $h_0=-\frac{1}{9}$). In terms of cosmic time, $t$,
this exact solution satisfies
\begin{equation}
\Lambda_{\rm M}e^{\varphi+3\alpha} -\frac{16}{27} \dot\varphi^2
	+\frac{288}{13}\dot\beta^2=0, \label{B.1}
\end{equation}
and 
\begin{equation}
\dot{\alpha} =-\frac{1}{9}\dot{\varphi}, \label{B.2}
\end{equation}
whence equations (\ref{rr}) and (\ref{rrfriedmann}) yield
\begin{equation}
 \ddot{\varphi}=\dot{\varphi}^2 +k_\varphi^2e^{2\varphi},\label{B.3}
\end{equation}
which is a second-order ODE for $\varphi$, where $k_\varphi$ is an
integration constant.  Defining 
\begin{equation}
\varrho \equiv \frac{e^{-\varphi}}{k_\varphi} ,
\end{equation} 
this equation simplifies to 
\begin{equation}
\varrho\ddot{\varrho} = -1,
\end{equation}
which can be integrated exactly to obtain $\dot \varphi$
\cite{Lidsey1995a} and a second integration then yields $\varphi$ in
terms of the Inverse Error function, so that in principle the scale
factor as a function of time $t$ can be obtained.
 
\index{equilibrium sets!$R/A$}
\index{exact solutions!dilaton--cosmological constant}
For $\Lambda_{\rm M}<0$, there are the solutions
\begin{eqnarray}
\nonumber
a & = & \frac{a_0}{\sqrt{\pm2 t}}, \\
\nonumber
\hat{\Phi} & = & -\ln\left(-2\Lambda_{\rm M}t^2 \right), \\
\nonumber
\beta&=&\beta_0, \\
\nonumber
\sigma&=&\sigma_0,\\
\label{at_half}
k&=&0,
\end{eqnarray}
where $\{a_0,\beta_0,\sigma_0\}$ are integration constants.  The $\pm$
sign corresponds to the sign of $t$ and the ``$+$'' solution
(represented in this thesis by the equilibrium point $R$) is a
repelling solution while the ``$-$'' solution (represented by the
equilibrium point $A$) is an attracting solution.  Note that the $-$
solution (i.e. $A$) is inflationary.

The dilaton--moduli--vacuum solutions (\ref{dmv}) and the saddle Milne
model (\ref{curv_drive}) exist for $\Lambda=\Lambda_{\rm M}=Q=0$ and
so will appear as well in this chapter, denoted by $L^\pm$ and
$S^\pm$, respectively.

\section{The Governing Equations\label{Governing_Equations_II}
}

Equations (\ref{rr}) and (\ref{rrfriedmann}) are reexamined for
$\Lambda=Q=0$ by defining the new variables
\begin{equation}
\label{beta}
\frac{d}{dt}\equiv e^{\frac{1}{2}(\varphi+3\alpha)}\frac{d}{dT}, \quad
N \equiv 6{\beta'}^2 , 
\qquad \psi \equiv \varphi' , \qquad h \equiv \alpha', \qquad K \equiv
\tilde K e^{-(\varphi+3\alpha)},
\end{equation}
where a prime denotes differentiation with respect to the new time
coordinate, $T$. Equations (\ref{rr}) and (\ref{rrfriedmann}) may then 
be written as the set of ODEs: 
\beqn
\mainlabel{rra}
\label{rr1a}
h ' &=& -\frac{3}{2} h^2+\frac{1}{2}h\psi - K -\frac{1}{2} \Lambda_{\rm M} +
		\frac{1}{2}\rho e^{-(\varphi+3\alpha)}, \\
\label{rr2a}
\psi' &=& \frac{3}{2}h^2 -\frac{3}{2}h\psi +\frac{1}{2} N+ \frac{3}{2} K
		- \frac{1}{4} \rho e^{-(\varphi+3\alpha)}, \\
\label{rr3a}
N' &=& (\psi-3h) N, \\
\label{rr4a}
K' &=& -(\psi + 5h) K, \\ 
\label{rr5a}
\rho' &=& -6h \rho,
\eeqn
and
\be
\label{rrfriedmanna}
3h^2 -\psi^2 +N -3K +\Lambda_{\rm M} +\frac{1}{2}\rho e^{-(\varphi+3\alpha)}=0.
\ee

Through equation (\ref{rrfriedmanna}), the variable $\rho$ is eliminated 
from the field equations, and the following definitions are made:
\begin{equation}
\mu\equiv \frac{\sqrt3h}{\xi}, \quad
\chi\equiv \frac{\psi}{\xi}, \quad
\nu\equiv \frac{N}{\xi^2}, \quad
\zeta\equiv \frac{\pm 3 K}{\xi^2}, \quad
\lambda\equiv \frac{\pm \Lambda_{\rm M}}{\xi^2}, \quad
\frac{d}{dT}\equiv \xi \frac{d}{d\tau},\label{TheDefs2}
\end{equation}
where the $\pm$ sign in the definitions for $\zeta$ and $\lambda$ are
to ensure $\zeta>0$ and $\lambda>0$, where necessary.  With these
definitions, all variables are bounded: $0\leq \{\mu^2, \chi^2, \nu, \zeta,
\lambda\} \leq 1$.  Equation (\ref{rrfriedmanna}) now reads
\begin{equation}
\tilde\kappa = \chi^2 \pm \zeta \mp \lambda 
  -\mu^2-\nu >0.\label{newFried2}
\end{equation}
where
\be
\tilde \kappa \equiv \frac{1}{2}\rho\xi^{-2} e^{-(\varphi+3\alpha)}.
\ee
The variable $\xi$ is defined in each of the following
six cases by:
\begin{itemize}
\item $\Lambda_{\rm M}>0$\begin{itemize} \item $K>0$: $\xi^2 \equiv 3K
	+\psi^2$  (subsection \ref{RRpKp}),
\item $K\leq0$: $\xi^2 \equiv \psi^2$  (subsections \ref{RRpK0} and 
	\ref{RRpKn}),	\end{itemize}
\item $\Lambda_{\rm M}<0$ \begin{itemize} \item $ K>0$: $\xi^2 \equiv 3 K
	+\psi^2-\Lambda_{\rm M}$  (subsection \ref{RRnKp}),
\item $K\leq0$: $\xi^2 \equiv \psi^2-\Lambda_{\rm M}$ (subsections 
	\ref{RRnK0} and \ref{RRnKn}).
\end{itemize}\end{itemize}
For example, consider $\Lambda_{\rm M}>0$ with $K>0$.  For this case,
$\chi^2+\zeta=1$ and equation (\ref{newFried2}) will read
\beqn
\tilde\kappa=1-\mu^2-\nu-\lambda \geq 0.
\eeqn
Hence, the variables $\{\mu,\nu,\lambda\}$ will be used for the phase
space (see subsection \ref{RRpKp} for details).  Each of these cases
will now be considered.  Subscripts will be added to the variables
$\{\mu,\chi,\nu,\zeta,\lambda\}$ to indicate separate cases, although
the $K=0$ cases will have the same subscript as the $K>0$.  As
discussed in section \ref{Asymptotic_Behaviour}, all equilibrium
points discussed in this chapter
will have $\dot\alpha\neq0$ and $\Xi^2= 1$, and hence nearly all
orbits asymptote towards the equilibrium points in one of the
invariant sets $\rho=0$ or $K=0$, where the three-dimensional $K=0$
case will be studied in its own subsection.  The qualitative behaviour
of the four-dimensional phase space in each $K\neq 0$ case will also
be examined.

\section{Analysis\label{Analysis_II}}
\subsection{The Case $\Lambda_{\rm M}>0$, $K=0$ \label{RRpK0}}

For $K=0$, equation (\ref{rrfriedmanna}) is written in the new variables as
\begin{equation}
0\leq \mu_1^2+\nu_1+\lambda_1\leq 1, \qquad \chi_1^2=1,
\end{equation}
where the ``$+$'' sign is chosen for both $\lambda$ and $\zeta$ in
(\ref{TheDefs2}).  For this case, the variable $\chi_1=+1$ will
be considered.  The system (\ref{rra}) then
reduces to the four-dimensional system:
\beqn
\mainlabel{xp_main}
\frac{d\mu_1}{d\tau} &= & (\mu_1+\sqrt{3})[1-\mu_1^2-\nu_1-\lambda_1] 
	+\frac{1}{2} \lambda_1[\mu_1-\sqrt{3}], 
\label{xp} \\
\frac{d\nu_1}{d\tau} &= & 2\nu_1\left\{[1-\mu_1^2-\nu_1-\lambda_1]
	+\frac{1}{2}\lambda_1\right\}, 
\label{yp} \\
\frac{d\lambda_1}{d\tau} &= & 2\lambda_1\left\{[1-\mu_1^2-\nu_1-\lambda_1] 
	-\frac{1}{2}(1-\lambda_1-\sqrt{3}\mu_1)\right\} \label{zp}   .
\eeqn
The invariant sets $\mu_1^2+\nu_1+\lambda_1=1$ ($\rho=0$), $\nu_1=0$
($N=0$) and $\lambda_1=0$ ($\Lambda_{\rm M}=0$) define the boundaries
to the phase space.  The dynamics is also determined by the fact that
the right-hand side of equation (\ref{yp}) is positive--definite so that
$\nu_1$ is a monotonically increasing function. This guarantees that
there are no closed or recurrent orbits in the three-dimensional phase
space.

In the invariant set $1-\mu_1^2-\lambda_1=0$, corresponding to the case of a
zero axion field, equations (\ref{xp}) and (\ref{yp}) reduce to the
single ordinary differential equation: 
\begin{equation}
\label{invariantset}
\frac{d\mu_1}{d\tau} = \frac{1}{2}\left(1-\mu_1^2\right)
	\left(\mu_1-\sqrt 3\right),
\end{equation}
which can be integrated to yield an exact solution in terms of 
$\tau$--time.

\subsubsection{The Frozen Modulus ($\dot\beta=0$) Invariant Set}

\begin{figure}[htp]
  \centering
   \epsfxsize=3in
   \includegraphics*[width=5in]{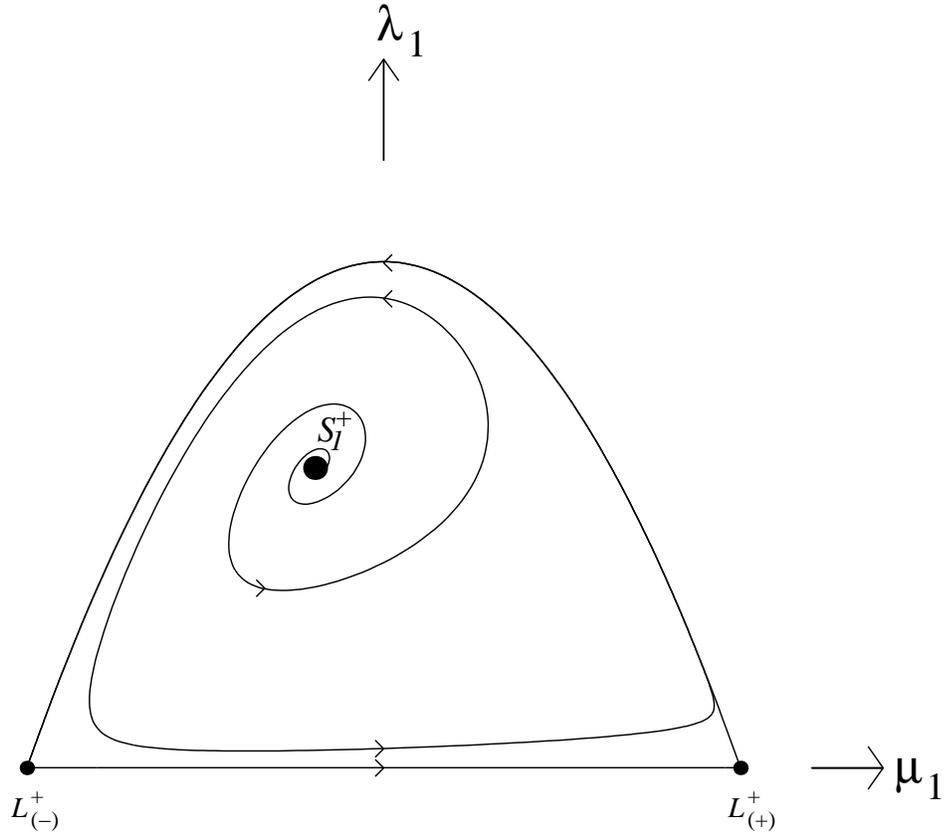}
  \caption[{\em Phase diagram for $\Lambda_{\rm M}>0$,
$K=\dot\beta=0$}]{{\em Phase diagram of the system (\ref{xp_main}) in
the matter sector ($\Lambda_{\rm M}>0$) with $\rho\neq0$ and
$K=\dot\beta=0$.  In this phase space, $\psi>0$ is assumed.
See also caption to figure
\ref{fNSpK0} on page \pageref{fNSpK0}.}\label{fRRpK0b0}}
 \end{figure}

The invariant set $\nu_1=0$, corresponding to
$\dot\beta=0$, is first examined.  For this case, the dynamical
system becomes equations (\ref{xp}) and (\ref{zp}) with
$\nu_1=0$, and the equilibrium points (and their eigenvalues)
are given by
\beqn
S^+_1: && \mu_1=-\frac{1}{3\sqrt{3}}, \lambda_1=\frac{16}{27}; 
	\nonumber \\
	&& (\lambda_1,\lambda_2)=\left(\third +\frac{i\sqrt{231}}{9},
		\third -\frac{i\sqrt{231}}{9}\right), \\
L^+_{(+)}: && \mu_1=1, \lambda_1=0;  \nonumber \\
	&& (\lambda_1,\lambda_2)=\left( -2\sqrt 3\left[1
	+\frac{1}{\sqrt{3}}\right], \sqrt 3\left[1-\frac{1}{\sqrt{3}} 
	\right]\right), \\
L^+_{(-)}: && \mu_1=-1, \lambda_1=0;  \nonumber \\
	&& (\lambda_1,\lambda_2)=\left( 2\sqrt 3\left[\mu_1
	-\frac{1}{\sqrt{3}}\right], -\sqrt 3\left[\mu_1+\frac{1}{\sqrt{3}} 
	\right]\right).
\eeqn
Note that there are no sinks in this two--dimensional phase space,
only a source ($S^+_1$) and two saddles ($L^+_{(\pm)}$).  Figure
\ref{fRRpK0b0} depicts this phase space.

For trajectories confined to the invariant set $\nu_1=0$, most
trajectories asymptote from the equilibrium point $S^+_1$
(corresponding to the solution (\ref{newsol})) and are future
asymptotic to a heteroclinic orbit, consisting of two saddle
equilibrium points (which are the points $L^+_{(+)}$ and $L^+_{(-)}$;
i.e., the ``$-$'' solutions given by equation (\ref{dmv}) with
$h_0=\pm\frac{1}{\sqrt3}$) and the single boundary orbits in the
invariant sets $\lambda_1=0$ and $\lambda_1+\mu_1^2=1$.  The former
set corresponds to a $\Lambda_{\rm M}=0$ solution, whilst the
latter to a solution with a constant axion field.  The dynamics of
this two-dimensional system is of interest from a mathematical point
of view due to the existence of the quasi--periodic behaviour. In a
given `cycle' an orbit spends a long time close to $L^+_{(-)}$ and
then moves quickly to $L^+_{(+)}$ shadowing the orbit in the invariant
set $\lambda_1=0$.  It is then again quasi-stationary and remains
close to the equilibrium point $L^+_{(+)}$ before quickly moving back
to $L^+_{(-)}$ shadowing the orbit in the invariant set
$1-\mu_1^2-\lambda_1=0$.  It must be stressed that the motion is {\em
not} periodic, and on each successive cycle a given orbit spends more
and more time in the neighbourhood of the equilibrium points
$L^+_{(+)}$ and $L^+_{(-)}$.  This qualitative behaviour is quite
generic in the following sections whenever $\Lambda_{\rm M}>0$ and
$K=0$ and will reappear in sections
\ref{NSpRRp} and \ref{NSnRRp}.

\subsubsection{Three--Dimensional System ($\dot\beta\neq0$)}

Returning to the three--dimensional system (\ref{xp_main}), the equilibrium
sets and their corresponding eigenvalues are
\beqn
S^+_1: && \mu_1=-\frac{1}{3\sqrt{3}}, \nu_1=0, \lambda_1=\frac{16}{27}; 
	\nonumber \\
	&& (\lambda_1,\lambda_2,\lambda_3)=\left(\third +\frac{i\sqrt{231}}{9},
		\third -\frac{i\sqrt{231}}{9},\frac{4}{3}\right), \\
L^+: && \nu_1=1-\mu_1^2, \lambda_1=0;  \nonumber \\
	&& (\lambda_1,\lambda_2,\lambda_3)=\left( -2\sqrt 3\left[\mu_1
	+\frac{1}{\sqrt{3}}\right], \sqrt 3\left[\mu_1-\frac{1}{\sqrt{3}} 
	\right],0\right).
\eeqn
Points on $L^+$ with $\mu_1^2< \frac{1}{3}$ are local
sinks, while the equilibrium point $S^+_1$ is a spiral source.  Note that the
points $L^+_{(\pm)}$ in the two--dimensional system ($\dot\beta=0$)
are the endpoints to the line $L^+$.

Note that in the three--dimensional phase space the variable $\nu_1$
is {\em monotonically increasing}.  When a modulus field is included
$(\nu_1 \ne 0)$, $S^+_1$ still represents the {\em only} source in the
system.  The orbits follow cyclical trajectories in the neighbourhood
of the invariant set $\nu_1=0$ and they spiral outwards monotonically,
since equation (\ref{dnu_1}) implies that $d\nu_1/d\tau >0$.  After a
finite (but arbitrarily large) number of cycles the kinetic energy of
the modulus field becomes more important until a critical point is
reached where it dominates the axion and cosmological constant. The
orbits then asymptote to the dilaton--moduli--vacuum solutions
(\ref{dmv}).  The general behaviour for most trajectories
in this phase space is to asymptote away from the equilibrium point
$S^+_1$, spiraling about the line $\mu_1=\frac{1}{\sqrt{27}}$,
$\nu_1=\frac{13}{8}(\lambda_1-\frac{16}{27})$ (see (\ref{B.1}) for this
particular line), asymptoting towards the line $L^+$ for
$\mu_1^2<\third$.

\begin{figure}[htp]
  \centering
   \epsfxsize=5in
   \includegraphics*[width=5in]{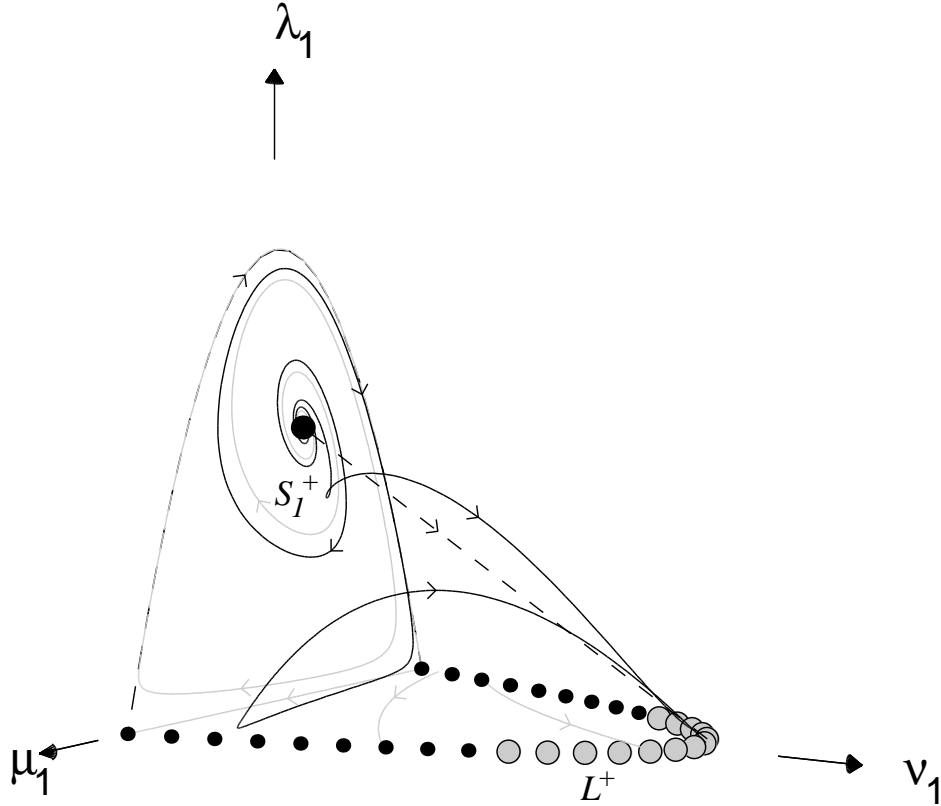}
  \caption[{\em Phase diagram for $\Lambda_{\rm M}>0$, $\rho\neq0$ and
$K=0$}]{{\em Phase diagram of the system (\ref{xp_main}) in the matter
sector ($\Lambda_{\rm M}>0$) with $\rho\neq0$ and $K=0$.  Note that
$L^+$ represents a {\em line} of equilibrium points.  The invariant
set $\dot\beta=0$ is depicted in \ref{fRRpK0b0} on page
\pageref{fRRpK0b0}.  In this phase space, $\psi>0$ is assumed.  See
also caption to figure
\ref{fNSpK0} on page \pageref{fNSpK0}.}\label{fRRpK0}}
 \end{figure}

Figure \ref{fRRpK0} depicts the full three-dimensional space.  Note
that the phase diagram depicted in the figure \ref{fRRpK0} is similar
to that of figure 1(e) in \cite{Nilsson1997a} that describes the locally
rotationally symmetric submanifold of the stationary Bianchi type I
perfect fluid models in general relativity, although in \cite{Nilsson1997a}
the independent variable is space-like.  Orbits in the full phase
space of figure \ref{fRRpK0} with non-trivial modulus field or shear
term (represented by the variable $\nu_1$) are repelled from the
source $S^+_1$, where $\nu_1$ increases monotonically, spiraling around
the exact solution given by $\mu_1=\frac{1}{\sqrt{27}}$ and
$\nu_1=\frac{13}{8}(\lambda_1-\frac{16}{27})$ (see also
equations (\ref{B.1})-(\ref{B.3})) represented by the dashed line in
figure \ref{fRRpK0}.  This implies that solutions are asymptotic in
the past to the solution given by equation (\ref{newsol}).  At early times
the orbits `shadow' the orbits in the invariant set $\nu_1 = 0$ and
undertake cycles between the saddles (in three-dimensional phase
space) on the equilibrium set $L^+$ close to $L^+_{(-)}$ and
$L^+_{(+)}$. These saddles on $L^+$ may be interpreted as Kasner--like
solutions \cite{Wainwright1997a,Nilsson1997a}.  Note that $\nu_1 = 0$
at $L^+_{(+)}$ and $L^+_{(-)}$, however, and there is no modulus field or shear
term in these cases.  The orbits thus experience a finite number of
cycles in which the orbits interpolate between different Kasner-like
states.  The orbits eventually asymptote toward a source on the line
$L^+$.

\subsection{The Case $\Lambda_{\rm M}>0$, $K>0$\label{RRpKp}}

For $K>0$, equation (\ref{rrfriedmanna}) is written in the new variables as
\begin{equation}
0\leq \mu_1^2+\nu_1+\lambda_1\leq 1, \qquad \zeta_1+\chi_1^2=1,
\end{equation}
where the ``$+$'' sign is chosen for both $\lambda$ and $\zeta$ in
(\ref{TheDefs2}).  For this case, the variable $\zeta_1\equiv 1-\chi_1^2$ will
be considered extraneous.  The system (\ref{rra}) then
reduces to the four-dimensional system:
\beqn
\mainlabel{dmu_1_main}
\nonumber
\frac{d\mu_1}{d\tau} &=& \left(1-\!\mu_1^2-\!\nu_1-\!\frac{1}{2}\lambda_1
\right)\!
\left(\sqrt 3+\!\mu_1\chi_1\right)-\frac{1}{\sqrt3}\left(1-\!\mu_1^2\right)\!
\left(1-\!\chi_1^2\right)-\sqrt3\lambda_1, \\  \label{dmu_1} \\
\label{dchi_1}
\frac{d\chi_1}{d\tau} & = & \left(1-\chi_1^2\right)\left[\mu_1^2+\nu_1
	+\frac{1}{2}\lambda_1+\frac{1}{\sqrt 3}\mu_1\chi_1\right], \\
\label{dnu_1}
\frac{d\nu_1}{d\tau} &=& \nu_1\left[\frac{2}{\sqrt 3} \mu_1\left(1
	-\chi_1^2\right)+2\chi_1\left(1-\mu_1^2-\nu_1
	-\frac{1}{2}\lambda_1\right)\right], \\ 
\label{dlambda_1}
\frac{d\lambda_1}{d\tau} &=& \lambda_1\left[\frac{1}{\sqrt 3}\mu_1
	\left(5-2\chi_1^2\right) +\chi_1\left(1-2\mu_1^2-2\nu_1
	-\lambda_1\right)\right].
\eeqn
The invariant sets $\mu_1^2+\nu_1+\lambda_1=1$ ($\rho=0$),
$\chi_1^2=1$ ($K=0$), $\nu_1=0$ ($N=0$) and $\lambda_1=0$
($\Lambda_{\rm M}=0$) define the boundaries to the phase space.  
The equilibrium points and their respective eigenvalues
(denoted by $\lambda$) are given by
\beqn
L^\pm: && \chi_1=\pm1, \mu_1^2+\nu_1=1, \lambda_1=0; \nonumber \\ \nonumber &&
  (\lambda_1,\lambda_2,\lambda_3,\lambda_4) = \left (0,
   \mp\frac{2}{\sqrt 3}\left[\sqrt 3\pm \mu_1\right], \sqrt 3 \left[\mu_1\mp 
     \frac{1}{\sqrt 3}\right],\mp2\sqrt{3}\left[\frac{1}{\sqrt 3}\pm\mu\right]\right).\\ \label{Lpm}\\
\nonumber
S^\pm_1: & & \mu_1=\mp\frac{1}{\sqrt{27}}, \chi_1=\pm1, \nu_1=0, \lambda_1=\frac{16}{27}; \nonumber \\ &&
   \left(\lambda_1,\lambda_2, \lambda_3, \lambda_4\right) =  \left(
   \mp\third\left[1+i\frac{\sqrt{231}}{3}\right],\mp\third\left[1-i\frac{\sqrt{231}}{3}\right], \mp\frac{4}{3}, \pm\frac{4}{9} \right).
\eeqn
From the eigenvalues, it is clear that $L^+$ is a late-time attractor
for $\mu_1^2<\third$, and that $L^-$ is an early-time repellor for
$\mu_1^2<\third$.  In both cases, $\chi_1^2=1$ and therefore $\zeta_1=0$
($K=0$).  Both lines represent the dilaton--moduli--vacuum
solutions (\ref{dmv}), and the points $S^\pm_1$ are
saddle points on the boundary of the phase space, corresponding to the
exact solutions (\ref{newsol}).

\subsubsection{The Invariant Set $\rho=0$ for $\Lambda_{\rm M}>0$, $K>0$}

\begin{figure}[htp]
  \centering
   \epsfxsize=5in
   \includegraphics*[width=5in]{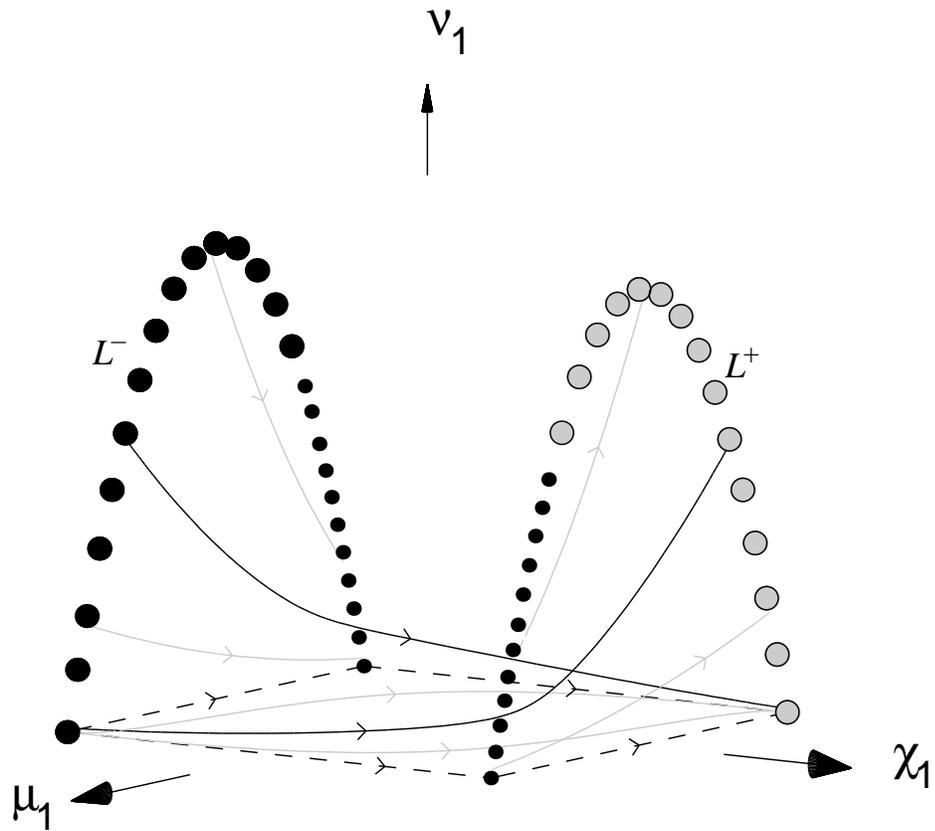}
  \caption[{\em Phase diagram for $\Lambda_{\rm M}>0$, $\rho=0$ and
$K>0$}]{{\em Phase diagram of the system (\ref{dmu_1_main}) in the
matter ($\Lambda_{\rm M}>0$) sector with $\rho=0$ field and $K>0$.
Note that $L^+$ and $L^-$ represent {\em lines} of equilibrium points.
See also caption to figure \ref{fNSpK0} on page
\pageref{fNSpK0}.}\label{fRRpKp}}
 \end{figure}

For the invariant set $\rho=0$, the system (\ref{dmu_1_main}) reduces
to three dimensions $\{\mu_1,
\chi_1, \nu_1 \}$ ($\lambda_1=1-\mu_1^2-\nu_1$).  The only
equilibrium points in this set are the lines $L^\pm$ with the first
three eigenvalues in (\ref{Lpm}), and therefore the early and time
attractors are the lines $L^-$ (for $\mu_1>-\frac{1}{\sqrt3}$) and
$L^+$ (for $\mu_1<\frac{1}{\sqrt3}$), respectively.  In this invariant
set, $\mu$ is a {\em monotonically decreasing} function, and so the
possibility of periodic orbits is excluded.  Therefore, solutions
generically asymptote into the past towards the $\dot\varphi<0$
dilaton--moduli--vacuum solutions (\ref{dmv}) for $h_0>-\frac{1}{3}$,
and asymptote into the future towards the $\dot\varphi>0$
dilaton--moduli--vacuum solutions (\ref{dmv}) for $h_0 <\third$.  The
curvature and cosmological constant are only dynamically significant
at intermediate times.  The phase space is depicted in figure
\ref{fRRpKp}.

\subsubsection{Qualitative Analysis of the Four-Dimensional System}

The qualitative dynamics in the full four-dimensional phase space is
as follows.  The only past--attractors belong to the line $L^-$ for
$\mu_1^2<\third$, and the only future attractors belong to the line
$L^+$ for $\mu_1^2<\third$.  Both lines $L^\pm$ lie in both invariant
sets $\rho=0$ and $\tilde K=0$, which is consistent with the analysis
of equation (\ref{dchi}).  The function
$\chi_1\nu_1^{\frac{1}{2}}\left(1-\mu_1^2-\nu_1-\lambda_1
\right)^{\frac{1}{6}}\lambda_1^{-\third}$ {\em
monotonically increases} and so there can be no periodic orbits.
Hence, solutions generically asymptote into the past towards the
$\dot\varphi<0$ dilaton--moduli--vacuum solutions (\ref{dmv}) for
$h_0^2<\frac{1}{9}$, and into the future towards the $\dot\varphi>0$
dilaton--moduli--vacuum solutions (\ref{dmv}) for $h_0^2<\frac{1}{9}$.
The curvature, cosmological constant term and the axion field are
only dynamically significant at intermediate times.

For orbits in the four-dimensional phase space, the complex
eigenvalues of the saddle point $S^\pm_1$ suggest that the
heteroclinic sequences may exist in the four--dimensional set.
Indeed, those orbits which asymptote to the $\tilde K=0$ invariant set
{\em do} generically spend time in a heteroclinic sequence, interpolating
between two equilibrium points representing two
dilaton--vacuum solutions, as discussed in subsection \ref{RRpK0}.
However, for those orbits which asymptote towards the $\rho=0$
invariant set, there are no heteroclinic sequences (as is evident from
figure \ref{fRRpKp}), and asymptote to equilibrium points representing
dilaton--moduli--vacuum solutions.

\subsection{The Case $\Lambda_{\rm M}>0$, $K<0$\label{RRpKn}}

For $K<0$, equation \eref{rrfriedmanna} is written in the new variables as
\begin{equation}
0\leq \mu_2^2+\nu_2+\zeta_2+\lambda_2\leq 1, \qquad \chi_2^2=1,
\end{equation}
where the ``$+$'' sign for $\lambda$ and the ``$-$'' sign for $\zeta$
has been chosen in (\ref{TheDefs2}).  For this case, $\chi_2=+1$ is
explicitly chosen, as $\chi_2=-1$ corresponds to a time reversal of
(\ref{rra}).  The system (\ref{rra}) then reduces to the
four-dimensional system:
\beqn
\mainlabel{dmu_2_main}
\label{dmu_2}
\frac{d\mu_2}{d\tau} &=& \left( 1-\mu_2^2-\nu_2-\frac{1}{2}\lambda_2\right)
	\left(\sqrt3+\mu_2\right)-\sqrt3\left(\lambda_2
	+\frac{2}{3}\zeta_2\right) , \\
\label{dnu_2}
\frac{d\nu_2}{d\tau} &=& 2\nu_2\left(1-\mu_2^2-\nu_2
	-\frac{1}{2}\lambda_2\right), \\ 
\label{dzeta_2}
\frac{d\zeta_2}{d\tau} & = & -2\zeta_2\left(\mu_2^2+\nu_2
	+\frac{1}{2}\lambda_2+\frac{1}{\sqrt 3}\mu_2\right), \\
\label{dlambda_2}
\frac{d\lambda_2}{d\tau} &=& \lambda_2\left(1-2\mu_2^2-2\nu_2-\lambda_2
	+\sqrt {3}\mu_2\right).
\eeqn
The invariant sets $\mu_2^2+\nu_2+\zeta_2+\lambda_2=1$ ($\rho=0$),
$\zeta_2=0$ ($K=0$), $\nu_1=0$ ($N=0$) and $\lambda_1=0$
($\Lambda_{\rm M}=0$) define the boundaries to the phase space.  The
equilibrium points and their respective eigenvalues (denoted by
$\lambda$) are given by
\beqn
S^+: && \mu_2 =-\frac{1}{\sqrt3},\zeta_2=\frac{2}{3} ,\lambda_2=0,\nu_2=0 ; 
	\nonumber \\ 
	&& (\lambda_1,	\lambda_2, \lambda_3,\lambda_4) = 
		\third(2,4,-2,4), \\
N: && \mu_2=-\frac{\sqrt 3}{5}, \nu_2=0,\zeta_2=\frac{18}{25},
	\lambda_2=\frac{4}{25};  \nonumber \\ 
	&& (\lambda_1, \lambda_2, \lambda_3,\lambda_4) = 
	\frac{2}{5}(1+i\sqrt2,1-i\sqrt2, 4, 2), \\
L^+: && \mu_2^2+\nu_2=1, \zeta_2= 0, \lambda_2=0; \nonumber \\ \nonumber
	&&   (\lambda_1, \lambda_2, \lambda_3,\lambda_4) = 
	\left( 0,-\frac{2}{\sqrt3}
	\left[\mu_2+\sqrt3\right],\sqrt3\left[\mu_2-\frac{1}{\sqrt3} 
	\right], -2\sqrt3\left[\mu_2+\frac{1}{\sqrt3}\right]\right). \\
\eeqn
From the eigenvalues, it is clear that $L^+$ is a late-time attractor
for $\mu_1^2<\third$.  This line represents the
dilaton--moduli--vacuum solutions (\ref{dmv}).  The point $N$ inside
the phase space is the early-time attractor for the system, and
represents the curvature--driven, static--modulus, static--axion
solution (\ref{open_new}).  The saddle point $S^+$ corresponds to the
Milne model (\ref{curv_drive}).

\subsubsection{The Invariant Set $\rho=0$ for $\Lambda_{\rm M}>0$, $K<0$}

\begin{figure}[htp]
  \centering
   \epsfxsize=5in
   \includegraphics*[width=5in]{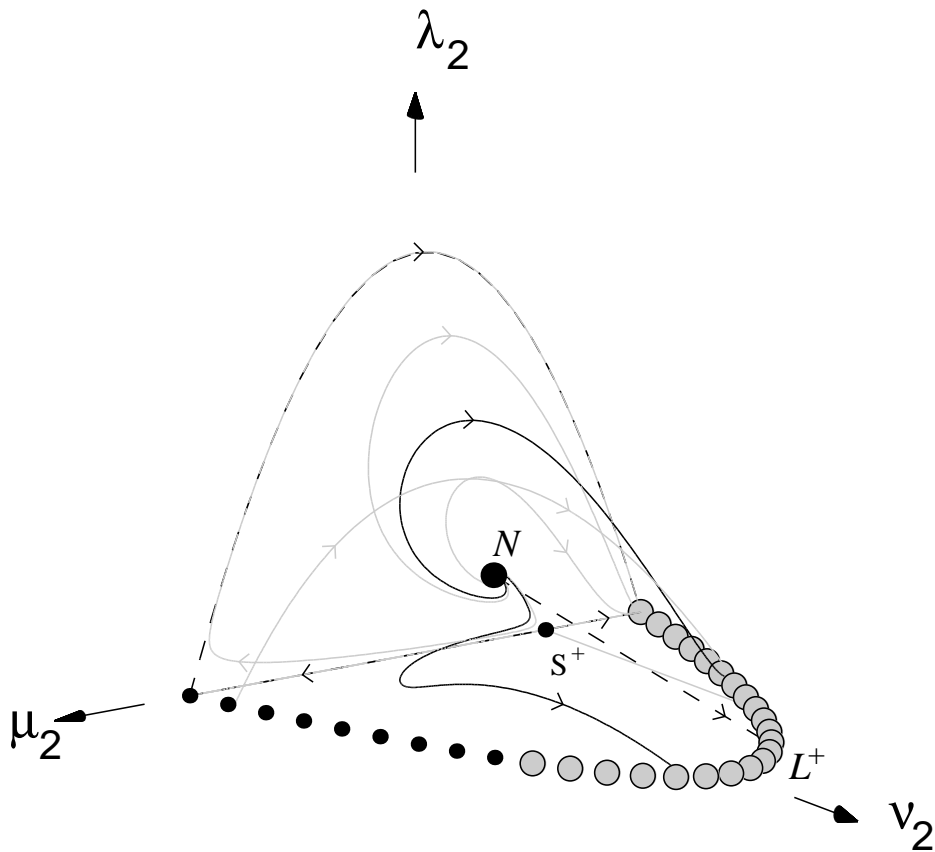}
  \caption[{\em Phase diagram for $\Lambda_{\rm M}>0$, $\rho=0$ and $K<0$}]{{\em Phase diagram of the system
(\ref{dmu_2_main}) in the matter ($\Lambda_{\rm M}>0$) sector
with $\rho=0$ and $K<0$.  Note that $L^+$ represents a {\em line} of
equilibrium points.  In this phase space, $\psi>0$ is assumed.
See also caption to figure \ref{fNSpK0} on page \pageref{fNSpK0}.}\label{fRRpKn}}
 \end{figure}

For this invariant set, the four--dimen\-sion\-al system
(\ref{dmu_2_main}) reduces to a three--dimen\-sion\-al system
involving the coordinates $\{\mu_2,\nu_2,\lambda_2\}$
($\zeta_2=1-\mu_2^2-\nu_2-\lambda_2$).  The equilibrium points are the
same as the full four-dimensional set, but with eigenvalues
($\lambda_1,\lambda_2,\lambda_3$), and so the line $L^+$ is a sink for
$\mu_2<\frac{1}{\sqrt3}$, and $N$ is a source.  The variable $\nu$ is
a {\em monotonically increasing} function, the existence of which
eliminates the possibility for recurrent orbits to occur.  Therefore,
the generic behaviour of this model is for solutions to asymptote into
the past towards the curvature--dominated, static--modulus,
static--axion solution (\ref{open_new}), and to the future towards the
$\dot\varphi>0$ dilaton--moduli--vacuum solutions (\ref{dmv}) for
$h_0<\third$.  Figure
\ref{fRRpKn} depicts this phase space.

\subsubsection{Qualitative Analysis of the Four-Dimensional System}

In the full four-dimensional set, the point $N$ is the early-time
attractor, and $L^+$ is the late-time attractor for $\mu_2^2<\third$.
The point $N$ lies in the invariant set $\rho=0$ and $L^+$ lies in
both the invariant sets $\rho=0$ and $\tilde K=0$, which is consistent
with the analysis of equation (\ref{dchi}).  It is apparent that
$\nu_2$ is a {\em monotonically increasing} function, and so the
possibility of recurrent or periodic orbits is not allowed.
Therefore, generically, solutions asymptote into the past towards the
curvature--dominated, static--modulus, static--axion solution
(\ref{open_new}), and into the future towards the $\dot\varphi>0$
dilaton--moduli--vacuum solutions (\ref{dmv}) for $h_0^2<\frac{1}{9}$.
Since, $\nu$ monotonically increases, the modulus field is initially
dynamically trivial, but becomes dynamically significant
asymptotically into the future.  The axion field and the cosmological
constant are dynamically significant only at intermediate times and do
not play a r\^{o}le in the early- and late-time dynamical evolution.
The curvature term is dynamically significant at early and
intermediate times, but becomes dynamically trivial at late times.

Orbits which asymptote into the future towards the $K=0$ invariant
set, generically end in a heteroclinic sequence as described in
section 4.1.1 and depicted in figure \ref{fRRpK0}.  However, such a
sequence does not occur for orbits which asymptote into the future
towards the $\rho=0$ invariant set.  Indeed, by examining the
eigenvalues of the equilibrium points of the four-dimensional system,
there do not seem to be heteroclinic sequences outside of the $\tilde
K=0$ invariant set.

For mathematical completeness, the $\Lambda_{\rm M}<0$ cases will now be
studied.

\subsection{The Case $\Lambda_{\rm M}<0$, $K=0$ \label{RRnK0}}

For this case, equation (\ref{rrfriedmanna}) is written in the new variables as
\begin{equation}
0\leq \mu_3^2+\nu_3\leq 1, \qquad \chi_3^2+\lambda_3=1,
\end{equation}
where the ``$-$'' sign for $\lambda$ has been chosen in (\ref{TheDefs2}).  The variable $\lambda_3$ is
chosen as the extraneous variable and the system
(\ref{rra}) then reduces to the four-dimensional system:
\beqn
\mainlabel{dudXi_main}
\frac{d\mu_3}{d\tau} & = & \frac{\sqrt 3}{2}\left(1-\mu_3^2\right)
	\left(1-\chi_3^2\right) + \left(1-\mu_3^2-\nu_3\right)
	\left(\sqrt 3 +\mu_3\chi_3\right), \label{dudXi} \\
\frac{d\chi_3}{d\tau} & = & -\frac{1}{2}\left(1-\chi_3^2\right)
	\left(1-2\mu_3^2-2\nu_3	+\sqrt{3}\mu_3\chi_3\right),  \label{dvdXi}\\
\frac{d\nu_3}{d\tau} & = & \nu_3\left[ 2\chi_3\left(1-\mu_3^2-\nu_3\right)
	-\sqrt 3 \mu_3\left(1-\chi_3^2\right)\right]. \label{dwdXi}
\eeqn

The phase space is bounded by the sets $\chi_3=\pm1$ and
$\nu_3=1-\mu_3^2$, where the latter corresponds to a zero axion field.
The dynamics is determined by the fact that the right-hand side of
equation (\ref{dudXi}) is positive definite so that $\mu_3$ is a
monotonically increasing function.

\subsubsection{The Frozen Modulus ($\dot\beta=0$) Invariant Set}

\begin{figure}[htp]
  \centering
   \epsfxsize=3in
   \includegraphics*[width=5in]{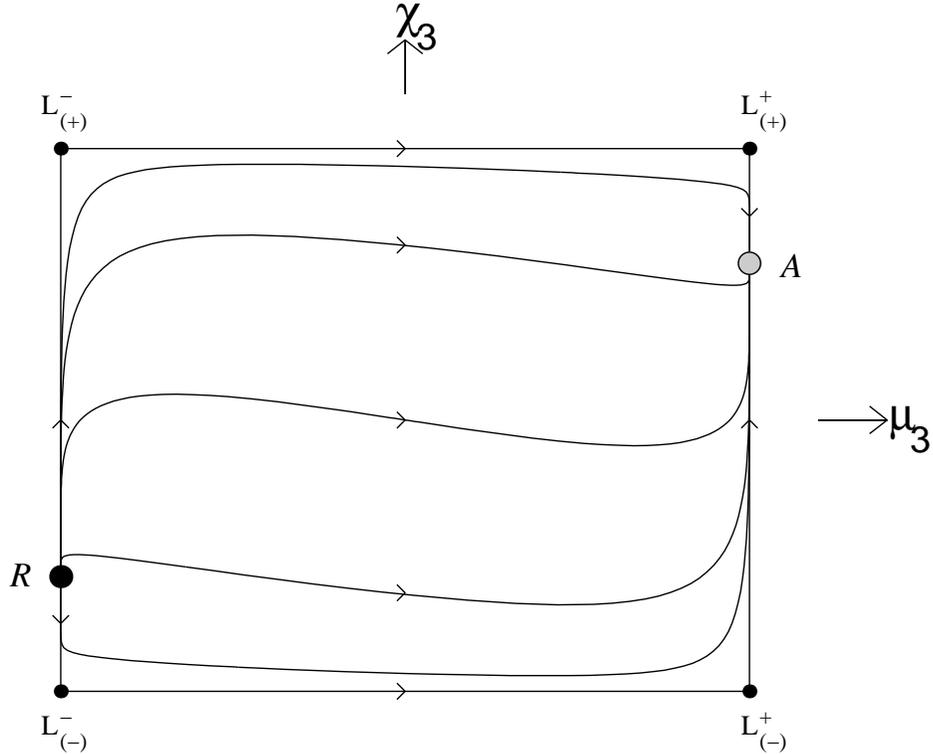}
  \caption[{\em Phase diagram for $\Lambda_{\rm M}<0$, $K=\dot\beta=0$}]{{\em Phase diagram of the system
(\ref{dudXi_main}) in the matter ($\Lambda_{\rm M}<0$)
sector with $K=0$ and $\dot\beta=0$.  See also caption to figure \ref{fNSpK0} on page \pageref{fNSpK0}.}\label{fRRnK0b0}}
 \end{figure}

The invariant set $\dot\beta=0$, corresponding to
$\nu_3$, is first examined.  For this case, the dynamical
system becomes equations (\ref{dudXi}) and (\ref{dvdXi}) with
$\nu_3=0$, and the equilibrium points (and their eigenvalues)
are given by
\beqn
\nonumber
L^+_{(+)} : && \chi_3=1, \mu_3=1; \\ 
        && (\lambda_1,\lambda_2) = \left( \sqrt{3}\left[1 - \frac{1}
        {\sqrt3}\right], -2\sqrt 3\left[1 + \frac{1}{\sqrt3}
        \right]\right), \\
\nonumber
L^+_{(-)} : && \chi_3=1, \mu_3=-1; \\ 
        && (\lambda_1,\lambda_2) = \left( -\sqrt{3}\left[1 + \frac{1}
        {\sqrt3}\right], 2\sqrt 3\left[1 + \frac{1}{\sqrt3}
        \right]\right), \\
\nonumber
L^-_{(+)} : && \chi_3=-1, \mu_3=1; \\ 
        && (\lambda_1,\lambda_2) = \left( \sqrt{3}\left[1 + \frac{1}
        {\sqrt3}\right], -2\sqrt 3\left[1 - \frac{1}{\sqrt3}
        \right]\right), \\
\nonumber
L^-_{(-)} : && \chi_3= -1, \mu_3=-1; \\ 
        && (\lambda_1,\lambda_2) = \left( -\sqrt{3}\left[1- \frac{1}
        {\sqrt3}\right], 2\sqrt 3\left[1+ \frac{1}{\sqrt3}
        \right]\right), \\
\nonumber
R: && \chi_3=-\frac{1}{\sqrt3}, \mu_3=-1, \nu_3=0; \\
   && (\lambda_1,\lambda_2) = \frac{1}{\sqrt3}
        \left(1, 10\right), \\
\nonumber
A: && \chi_3=\frac{1}{\sqrt3}, \mu_3=1, \nu_3=0; \\
   && (\lambda_1,\lambda_2) = -\frac{1}{\sqrt3}
        \left(1, 10\right).
\eeqn
The four points $L^\pm_{(\pm)}$ are saddles, whereas the point $R$ is a
source and the point $A$ is a sink.  Figure \ref{fRRnK0b0} depicts this
phase space.

\subsubsection{Three--Dimensional System ($\dot\beta\neq0$)}

\begin{figure}[htp]
  \centering
   \epsfxsize=5in
   \includegraphics*[width=5in]{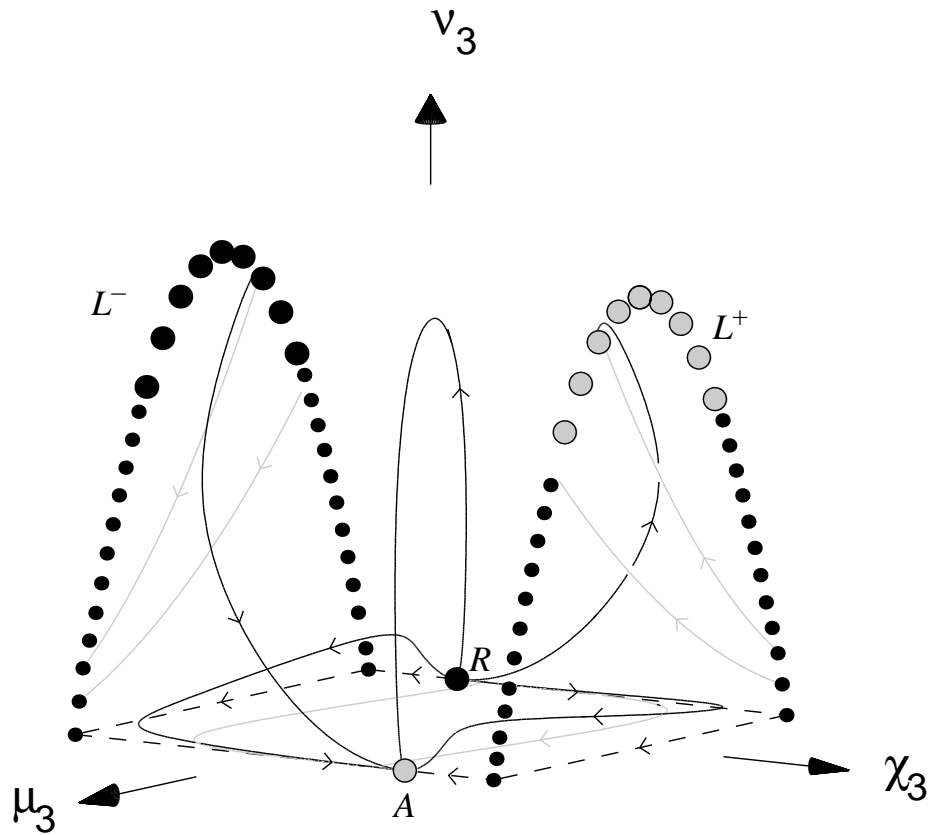}
  \caption[{\em Phase diagram for $\Lambda_{\rm M}<0$, $\rho\neq0$ and
$K=0$}]{{\em Phase diagram of the system (\ref{dudXi_main}) in the
matter ($\Lambda_{\rm M}<0$) sector with $\rho\neq0$ and $K=0$.  Note
that $L^+$ and $L^-$ represent {\em lines} of equilibrium points.  The
invariant set $\nu_3=0$ is depicted in figure \ref{fRRnK0b0} on page
\pageref{fRRnK0b0}.  See also caption to figure \ref{fNSpK0} on page
\pageref{fNSpK0}.}\label{fRRnK0}} \end{figure}

Returning to the three--dimensional system (\ref{dudXi_main}), the equilibrium points and their respective eigenvalues
are:
\beqn
\nonumber
L^\pm : && \chi_3=\pm 1, \mu_3^2+\nu_3=1; \\ 
        && (\lambda_1,\lambda_2,\lambda_3) = \left( 0,
        \sqrt{3}\left[\mu_3\mp\frac{1}
        {\sqrt3}\right], -2\sqrt 3\left[\mu_3\pm\frac{1}{\sqrt3}
        \right]\right), \\
\nonumber
R: && \chi_3=-\frac{1}{\sqrt3}, \mu_3=-1, \nu_3=0; \\
   && (\lambda_1,\lambda_2,\lambda_3) = \frac{1}{\sqrt3}
        \left(1, 2, 10\right), \\
\nonumber
A: && \chi_3=\frac{1}{\sqrt3}, \mu_3=1, \nu_3=0; \\
   && (\lambda_1,\lambda_2,\lambda_3) = -\frac{1}{\sqrt3}
        \left(1, 2, 10\right).
\eeqn
Note that the points $L^+_{(+)}$ and $L^+_{(-)}$ from the
two--dimensional system ($\dot\beta=0$) are the endpoints to the line
$L^+$ whereas the points $L^-_{(+)}$ and $L^-_{(-)}$ are the endpoints
to the line $L^-$.  The line $L^-$ represents early--time attracting
solutions for $\mu_3^2<\third$ and saddles otherwise.  The saddle points
The line $L^+$ corresponds to late--time attracting solutions for
$\mu_3^2<\third$ and saddles otherwise.  Hence, there are
two early--time attractors given by the point $R$ and the line $L^-$
for $\mu_3^2<\third$. There are also two late--time
attractors corresponding to the point $A$ and the line $L^+$ for
$\mu_3^2 <\third$.  Figure \ref{fRRnK0} depicts this phase
space.

Because $\mu_3$ {\em monotonically increases}, most trajectories in
this phase space represent bouncing cosmologies which are initially
contracting and end up expanding.  Most trajectories asymptote into
the past towards either $L^-$, representing the $\dot\varphi<0$
dilaton--moduli--vacuum solutions (\ref{dmv}), or to $R$, representing the 
static-moduli solution (\ref{at_half}).  To the future,
most orbits asymptote towards either $L^+$, representing the
$\dot\varphi>0$ dilaton--moduli--vacuum solutions, or $A$
corresponding to the expanding, static-moduli solution.

\subsection{The Case $\Lambda_{\rm M}<0$, $K>0$\label{RRnKp}}

For this case, equation (\ref{rrfriedmanna}) is written in the new variables as
\begin{equation}
0\leq \mu_3^2+\nu_3\leq 1, \qquad \chi_3^2+\zeta_3+\lambda_3=1,
\end{equation}
where the ``$-$'' sign for $\lambda$ and the ``$+$'' sign for $\zeta$
has been chosen in (\ref{TheDefs2}).  The variable $\lambda_3$ is
chosen as the extraneous variable and the system
(\ref{rra}) then reduces to the four-dimensional system:
\beqn
\mainlabel{dmu_3_main}
\label{dmu_3}
\frac{d\mu_3}{d\tau} &=& \left(1-\mu_3^2-\nu_3\right)\left(\sqrt3
	+\mu_3\chi_3\right) +\frac{\sqrt 3}{2}\left(1-\mu_3^2\right)
	\left(1-\chi_3^2-\frac{5}{3}\zeta_3\right), \\
\nonumber
\frac{d\chi_3}{d\tau} &=& -\frac{\sqrt3}{2}\left[ \mu_3\chi_3\left( 1-\chi_3^2
	-\frac{5}{3}\zeta_3\right)\right] -\frac{1}{2}\left(1-\chi_3^2\right)
	\left(1-2\mu_3^2-2\nu_3\right)+\frac{1}{2}\zeta_3, \\ \label{dchi_3} \\
\label{dnu_3}
\frac{d\nu_3}{d\tau} &=&\nu_3\left[ 2\chi_3\left(1-\mu_3^2-\nu_3\right)-\sqrt 3
	\mu_3\left(1-\chi_3^2-\frac{5}{3}\zeta_3\right)\right], \\
\label{dzeta_3}
\frac{d\zeta_3}{d\tau} &=& -\zeta_3 \left[ 2\chi_3\left(\mu_3^2+\nu_3\right) 
	+\frac{1}{\sqrt 3}\mu_3\left(5-3\chi_3^2-5\zeta_3\right)\right].
\eeqn
The invariant sets $\mu_3^2+\nu_3=1$, $\chi_3^2+\zeta_3=1$, $\nu_3=0$
and $\zeta_3=0$ define the boundary of the phase space.  

The
equilibrium sets and their corresponding eigenvalues (denoted by
$\lambda$) are
\beqn
\nonumber
L^\pm : && \chi_3=\pm 1, \mu_3^2+\nu_3=1, \zeta_3=0; \\ \nonumber
	&& (\lambda_1,\lambda_2,\lambda_3,\lambda_4) = \left( 0, \mp\frac{2}
	{\sqrt3}\left[\sqrt3\pm\mu_3\right], \sqrt{3}\left[\mu_3\mp\frac{1}
	{\sqrt3}\right], -2\sqrt 3\left[\mu_3\pm\frac{1}{\sqrt3} 
	\right]\right),\\ \\
\nonumber
R: && \chi_3=-\frac{1}{\sqrt3}, \mu_3=-1, \nu_3=0, \zeta_3=0; \\
   && (\lambda_1,\lambda_2,\lambda_3,\lambda_4) = \frac{1}{\sqrt3}
	\left(1, 2, 6, 10\right), \\
\nonumber
A: && \chi_3=\frac{1}{\sqrt3}, \mu_3=1, \nu_3=0, \zeta_3=0; \\
   && (\lambda_1,\lambda_2,\lambda_3,\lambda_4) = -\frac{1}{\sqrt3}
	\left(1, 2, 6, 10\right).
\eeqn
Here there are two early-time attractors: the point $R$ representing
the $\dot\varphi<0$ static--modulus, static--axion solution
(\ref{at_half}) and the line $L^-$ for $\mu_3^2<\third$.  Likewise,
there are two late-time attractors: the point $A$ representing the
$\dot\varphi>0$ static--modulus, static--axion solution
(\ref{at_half}) and the line $L^+$ for $\mu_3^2<\third$.

\subsubsection{The Invariant Set $\rho=0$ for $\Lambda_{\rm M}<0$, $K>0$}

\begin{figure}[htp]
  \centering
   \epsfxsize=5in
   \includegraphics*[width=5in]{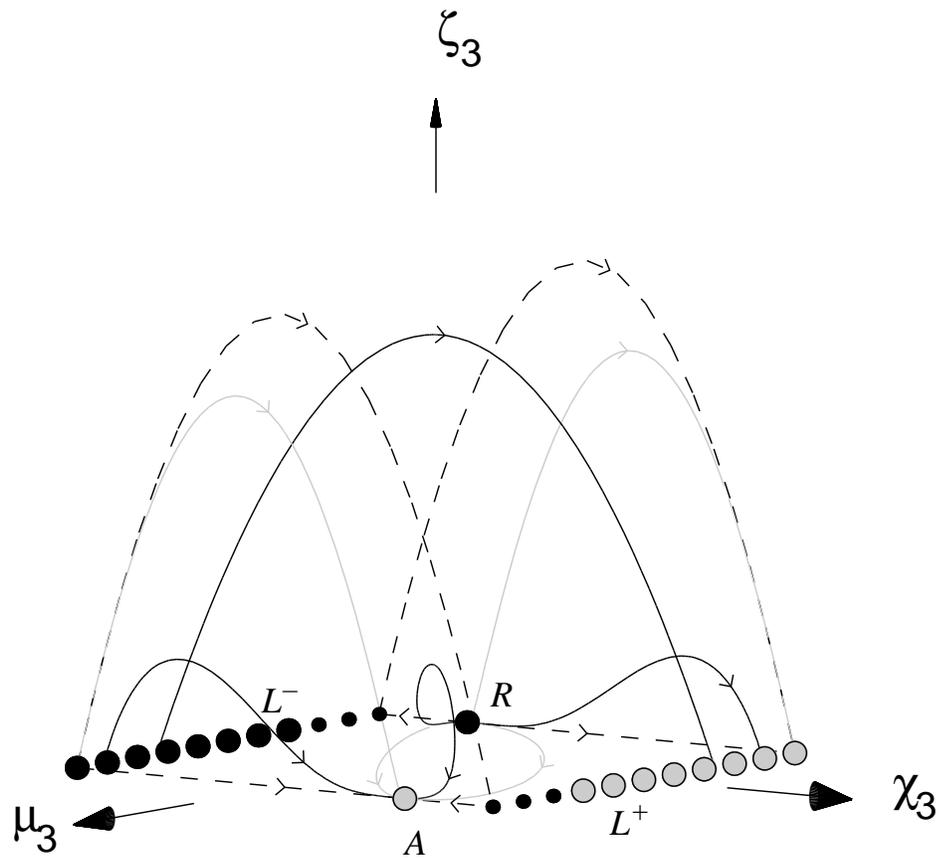}
  \caption[{\em Phase diagram for $\Lambda_{\rm M}<0$, $\rho=0$ and $K>0$}]{{\em Phase diagram of the system
(\ref{dmu_3_main}) in the matter ($\Lambda_{\rm M}<0$)
sector with $\rho=0$ and $K>0$.  Note that $L^+$ and $L^-$ represent
{\em lines} of equilibrium points.  See also caption to figure \ref{fNSpK0} on page \pageref{fNSpK0}.}\label{fRRnKp}}
 \end{figure}

For this invariant set, the four-dimensional system (\ref{dmu_3_main})
reduces to a three-dimen\-sion\-al system involving the coordinates
$\{\mu_3,\chi_3,\zeta_3\}$ ($\nu_3=1-\mu_2^3$).  The equilibrium
points are the same as the full four-dimensional set, but with
eigenvalues ($\lambda_1,\lambda_2,\lambda_3$), and so the line $L^-$
is a source for $\mu_3>-\frac{1}{\sqrt3}$ and the line $L^+$ is a sink
for $\mu_3<\frac{1}{\sqrt3}$.  The function $\chi_3/\sqrt{\nu_3}$ is
{\em monotonically increasing}, and so there are no recurring or
periodic orbits.  Hence, solutions generically asymptote into the past
towards either the $\dot\varphi<0$ dilaton--moduli--vacuum solutions
(\ref{dmv}) for $h_0>-\frac{1}{3}$, or the contracting $\dot\varphi<0$
static--modulus, static--axion solution (\ref{at_half}), represented
by $R$.  Into the future, solutions asymptote towards either the
expanding $\dot\varphi>0$ dilaton--moduli--vacuum solutions
(\ref{dmv}) for $h_0<\third$, or the $\dot\varphi>0$ static--modulus,
static--axion solution (\ref{at_half}), represented by $A$.  Figure
\ref{fRRnKp} depicts this phase space.

\subsubsection{Qualitative Analysis of the Four-Dimensional System}

In the full four-dimensional set, the early-time attractors are the
line $L^-$ for $\mu_3^2<\third$ and the point $R$.  The late-time
attractors are the $L^+$ for $\mu_3^2<\third$ and the point $A$.  All
of these attractors lie in both of the invariant sets $\rho=0$ and
$\tilde K=0$, which is consistent with the analysis of (\ref{dchi}).
The function $\left(\chi_3+\frac{1}{\sqrt3}\mu_3\right)/\sqrt{\nu_3}$
is {\em monotonically increasing}, and so there are no recurring or
periodic orbits.  Hence, solutions generically asymptote into the past
towards either the $\dot\varphi<0$ dilaton--moduli--vacuum solutions
(\ref{dmv}) for $h_0^2<\frac{1}{9}$, or the inflating $\dot\varphi<0$
static--modulus, static--axion solution (\ref{at_half}).  Into the
future, solutions asymptote towards either the $\dot\varphi>0$
dilaton--moduli--vacuum solutions (\ref{dmv}) for $h_0^2<\frac{1}{9}$,
or the $\dot\varphi>0$ static--modulus, static--axion solution
(\ref{at_half}).  The axion field and curvature term are dynamically
significant only at intermediate times.  For early and late times, the
cosmological constant is dynamically significant only when the modulus
field is dynamically trivial (e.g., the points $R$ and $A$) and
vice-versa (e.g., the lines $L^\pm$).

\subsection{The Case $\Lambda_{\rm M}<0$, $K<0$\label{RRnKn}}

For this case, equation (\ref{rrfriedmanna}) is written in the new variables as
\begin{equation}
0\leq \mu_4^2+\nu_4+\zeta_4\leq 1, \qquad \chi_4^2+\lambda_4=1,
\end{equation}
where the ``$-$'' sign for both $\lambda$ and $\zeta$
has been chosen in (\ref{TheDefs2}).  The variable $\lambda_4$ is
chosen as the extraneous variable and the system
(\ref{rra}) then reduces to the four-dimensional system:
\beqn
\mainlabel{dmu_4_main}
\label{dmu_4}
\frac{d\mu_4}{d\tau} &=& \left(1-\mu_4^2-\nu_4\right)\left(\sqrt 3+\mu_4\chi_4
\right)+\frac{\sqrt3}{2}\left(1-\mu_4^2\right)\left(1-\chi_4^2\right)
-\frac{2}{\sqrt3}\zeta_4, \\
\frac{d\chi_4}{d\tau} &=& -\frac{1}{2}\left(1-\chi_4^2\right)\left[ 
	1-2\mu_4^2-2\nu_4+\sqrt3 \mu_4\chi_4\right], \\
\label{dnu_4}
\frac{d\nu_4}{d\tau} &=&\nu_4\left[ 2\chi_4\left(1-\mu_4^2-\nu_4\right)-\sqrt 3
	\mu_4\left(1-\chi_4^2\right)\right], \\
\label{dzeta_4}
\frac{d\zeta_4}{d\tau} &=& -\zeta_4 \left[ 2\chi_4\left(\mu_4^2+\nu_4\right) 
	+\frac{1}{\sqrt 3}\mu_4\left(5-3\chi_4^2\right)\right].
\eeqn
The invariant sets $\mu_4^2+\nu_4+\zeta_4=1$, $\chi_4^2=1$, $\nu_4=0$
and $\zeta_4=0$ define the boundary to the phase space.  The
equilibrium sets and their corresponding eigenvalues (denoted by
$\lambda$) are
\beqn
\nonumber
L^\pm : && \chi_4=\pm 1, \mu_4^2+\nu_4=1, \zeta_4=0; \\ \nonumber
	&& (\lambda_1,\lambda_2,\lambda_3,\lambda_4) = \left( 0, \mp\frac{2}
	{\sqrt3}\left[\sqrt3\pm\mu_4\right], \sqrt{3}\left[\mu_4\mp \frac{1}
	{\sqrt3}\right], -2\sqrt 3\left[\mu_4\pm\frac{1}{\sqrt3} 
	\right]\right),\\ \\
\nonumber
R: && \chi_4=-\frac{1}{\sqrt3}, \mu_4=-1, \nu_4=0, \zeta_4=0; \\
   && (\lambda_1,\lambda_2,\lambda_3,\lambda_4) = \frac{1}{\sqrt3}
	\left(1, 2, 6, 10\right), \\
\nonumber
A: && \chi_4=\frac{1}{\sqrt3}, \mu_4=1, \nu_4=0, \zeta_4=0; \\
   && (\lambda_1,\lambda_2,\lambda_3,\lambda_4) = -\frac{1}{\sqrt3}
	\left(1, 2, 6, 10\right),\\
\nonumber
S^\pm && \chi_4=\pm 1,\mu_4=\frac{\mp 1}{\sqrt3},\nu_4=0,\zeta_4=\frac{2}{3}; \\
      && (\lambda_1,\lambda_2,\lambda_3,\lambda_4) = \frac{2}{3} 
         (\mp 1, \pm 1, \pm 2, 0).
\eeqn
As in the previous case, there are two early-time attractors: the
point $R$ (corresponding to the $\dot\varphi<0$ static--modulus,
static--axion solution (\ref{at_half})) and the line $L^-$ for
$\mu_4^2<\third$ (corresponding to (\ref{dmv}) for
$h_0^2<\frac{1}{9}$).  Likewise, there are two late-time attractors:
the point $A$ (corresponding to the $\dot\varphi>0$ static--modulus,
static--axion solution (\ref{at_half})) and the line $L^+$ for
$\mu_4^2<\third$ (corresponding to (\ref{dmv}) for
$h_0^2<\frac{1}{9}$).

\subsubsection{The Invariant Set $\rho=0$ for $\Lambda_{\rm M}<0$, $K<0$}

\begin{figure}[htp]
  \centering
   \epsfxsize=5in
   \includegraphics*[width=5in]{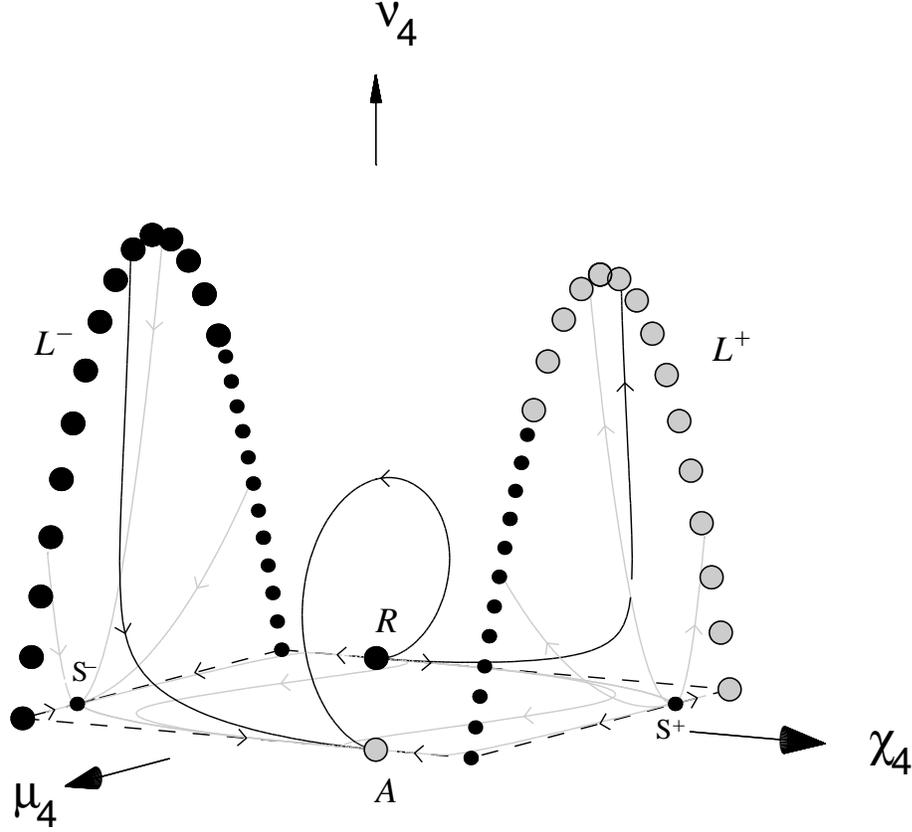}
  \caption[{\em Phase diagram for $\Lambda_{\rm M}<0$, $\rho=0$ and $K<0$}]{{\em Phase diagram of the system
(\ref{dmu_4_main}) in the matter ($\Lambda_{\rm M}<0$)
sector with $\rho=0$ and $K<0$.  Note that $L^+$ and $L^-$ represent
{\em line} of equilibrium points. See also caption to figure \ref{fNSpK0} on page \pageref{fNSpK0}.
}\label{fRRnKn}}
 \end{figure}

For this invariant set, the four-dimensional system (\ref{dmu_4_main})
reduces to a three-dimen\-sion\-al system involving the coordinates
$\{\mu_4,\chi_4,\nu_4\}$ ($\zeta_4=1-\mu_4^3-\nu_4$).  The equilibrium
points are the same as the full four-dimensional set, but with
eigenvalues ($\lambda_1,\lambda_2,\lambda_3$), except now the line
$L^+$ is a sink for $\mu_4<\frac{1}{\sqrt3}$ and the line $L^-$ is a
source for $\mu_4>-\frac{1}{\sqrt3}$.  The function
$\mu_4/\sqrt{\nu_4}$ is {\em monotonically increasing}, and so it is
not possible for periodic orbits to occur.  Hence, solutions
generically asymptote into the past towards either the $\dot\varphi<0$
dilaton--moduli--vacuum solutions (\ref{dmv}) for $h_0>-\frac{1}{3}$,
or the contracting $\dot\varphi<0$ static--modulus, static--axion
solution (\ref{at_half}).  Into the future, solutions asymptote
towards either the $\dot\varphi>0$ dilaton--moduli--vacuum solutions
(\ref{dmv}) for $h_0<\third$, or the expanding $\dot\varphi>0$
static--modulus, static--axion solution (\ref{at_half}).  Figure
\ref{fRRnKn} depicts this phase space.

\subsubsection{Qualitative Analysis of the Four-Dimensional System}

The invariant set $K=0$ is discussed in subsection \ref{RRnKn} (see figure \ref{fRRnK0}).

In the full four-dimensional set, the early-time attractors are the
line $L^-$ for $\mu_4^2<\third$ and the point $R$.  The late-time
attractors are the $L^+$ for $\mu_3^2<\third$ and the point $A$.  All
of these attractors lie in both of the invariant sets $\rho=0$ and
$\tilde K=0$, which is consistent with the analysis of (\ref{dchi}).
The function $\mu_4/\sqrt{\nu_4}$ is {\em monotonically increasing},
and so there are no recurring or periodic orbits.  Hence, solutions
generically asymptote into the past towards either the $\dot\varphi<0$
dilaton--moduli--vacuum solutions (\ref{dmv}) for $h_0^2<\frac{1}{9}$
(corresponding to $L^-$), or the contracting $\dot\varphi<0$
static--modulus, static--axion solution (\ref{at_half}) (represented
by $R$).  Into the future, solutions asymptote towards either the
$\dot\varphi>0$ dilaton--moduli--vacuum solutions (\ref{dmv}) for
$h_0^2<\frac{1}{9}$ (corresponding to $L^+$) or the expanding
$\dot\varphi>0$ static--modulus, static--axion solution
(\ref{at_half}) (represented by $A$).  The axion field and curvature
term are dynamically significant only at intermediate times.  For
early and late times, the cosmological constant is dynamically
significant only when the modulus field is dynamically trivial (e.g.,
the points $R$ and $A$) and vice-versa (e.g., the lines $L^\pm$).

\section{Summary of Analysis in the Jordan Frame\label{summary_II}}

This discussion begins with two tables; Table \ref{term_asymp_II} lists
which terms are the dominant variables for each equilibrium set as
well as the deceleration parameter, $q$, for the corresponding model
and Table \ref{points_asymp_II} lists the attracting behaviour of the equilibrium sets.
\begin{table}[htb]
\begin{center}
\begin{tabular}{|c|cccccccc|c|c|}\hline
Set & \multicolumn{8}{c|}{Dominant Variables}&$q$&H \\
\hline \hline
$L^+$ {\tiny($t<0$)} & $\alpha$ &$\hat\Phi$ & $\dot\beta$ &&&&&
	&$-(1+h_0)h_0^{-1}$ &$h_0(-t)^{-1}$\\
$L^-$ {\tiny($t>0$)} & $\alpha$ &$\hat\Phi$ & $\dot\beta$ &&&&&
	&$(1-h_0)h_0^{-1}$ &$h_0t^{-1}$\\ \hline
$S_1^+$ {\tiny($t<0$)}& $\alpha$ & $\hat\Phi$ && $\dot\sigma$ &&
	& $\Lambda_{\rm M}>0$&&$2$& $\frac{1}{3}t^{-1}$\\
$S_1^-$ {\tiny($t>0$)}& $\alpha$ & $\hat\Phi$ && $\dot\sigma$ &&
	& $\Lambda_{\rm M}>0$&&$2$& $\frac{1}{3}t^{-1}$\\ \hline
$A$ {\tiny($t<0$)} & $\alpha$ & $\hat\Phi$ &&&&& $\Lambda_{\rm M}<0$ &&$-3$
	& $-\frac{1}{2}t^{-1}$\\ 
$R$ {\tiny($t>0$)} & $\alpha$ & $\hat\Phi$ &&&&& $\Lambda_{\rm M}<0$ &&$-3$
	& $-\frac{1}{2}t^{-1}$\\  \hline
$N$ {\tiny($t<0$)} & $\alpha$ & $\hat\Phi$ &&& $k<0$ && $\Lambda_{\rm M}>0$ &
	&$0$&$t^{-1}$ \\ \hline
$S^\pm$     & $\alpha$ &&&& $k<0$ &&&&$0$& $t^{-1}$\\ \hline
\end{tabular}
\end{center}
\caption[{\em Dominant Variables and $q$ for Equilibrium Sets}]{{\em The
dominant variables for each equilibrium set as well as the equilibrium
set's deceleration parameter, $q$, and Hubble parameter $H$.
Inflation occurs when $q<0$ and $H>0$, whereas ``deflation'' occurs
for $q>0$ and $H<0$.  Note that the only ``anisotropic'' solutions are
represented by the lines $L^\pm$, except when
$h_0^2=\third$.}\label{term_asymp_II}}
\end{table}
Note that the only inflationary models are those represented by $L^+$
(for $h_0>0$) and  $A$.  The saddle point $S^+_1$ represents a deflating model
($q>0$, $H<0$).

\begin{table}[ht]
\begin{center}
\begin{tabular}{|ccccc||c|c|c|}\hline
\multicolumn{5}{|c||}{Terms Present} & Early & Intermediate & Late \\
\hline \hline
$\dot\beta$ &     &  & $\Lambda_{\rm M}>0$ &      &$S^+_1$ &$L^+$ &$L^+$ \\
$\dot\beta$ &$k>0$& & $\Lambda_{\rm M}>0$ & &$L^-$&$S^\pm_1$, $L^\pm$ &$L^+$ \\
$\dot\beta$ &$k<0$&  & $\Lambda_{\rm M}>0$ &    &$N$ &$S^+$, $L^+$ & $L^+$\\
$\dot\beta$ & $k$ && $\Lambda_{\rm M}<0$ & &$R$, $L^-$ &$L^\pm$ & $A$, $L^+$ \\
\hline
\end{tabular}
\end{center}
\caption[{\em Summary of Equilibrium Points for Models
considered}]{{\em Summary of the early-time, intermediate, and
late-time attractors for the various models examined.  Note that
$\dot\alpha$, $\hat\Phi$ and $\dot\sigma$ are present in every
model.}\label{points_asymp_II}}
\end{table}

Monotonic functions have been established for each case, which precludes the
existence of recurrent or periodic orbits and thereby allowing the early-time and late-time behaviour of these models to be determined based
upon the equilibrium sets of the system.  In all cases, the
dilaton--moduli--vacuum solutions act as either early-time or
late-time attractors (and, in many of the cases, both).  Because these
solutions lie in both the $\rho=0$ and $\tilde K=0$ invariant sets and
contain no cosmological constant,
and therefore the modulus and dilaton fields are dynamically
important asymptotically.  Furthermore, all early-time and late-time
attracting sets lie in either the $\rho=0$ invariant set, or the
$\tilde K=0$ invariant set, and a majority of these sets lie in both;
thus, there seems to be a generic feature in which the curvature terms and the
axion field are dynamically significant at intermediate times and are
asymptotically negligible at early and at late times. 

When $\Lambda_{\rm M}>0$, the cosmological constant may play a
significant r\^{o}le in the early and late time dynamics.  For
instance, although in the four-dimensional sets there are no repelling
or attracting equilibrium points in which $\Lambda_{\rm M}$ is
dynamically significant, the orbits which are attracted to the $\tilde
K=0$ invariant can spend some time in a heteroclinic sequence which interpolates
between two dilaton--moduli--vacuum solutions (see subsection
\ref{RRpK0}, figure \ref{fRRpK0}).  During this interpolation, the
orbits repeatedly spend time in a region of phase space in which
$\Lambda_{\rm M}$ is dynamically significant (the region $\lambda_1>0$
in figure \ref{fRRpK0}), although most time is spend near the
dilaton--moduli--vacuum saddle points where $\Lambda_{\rm M}$ is
dynamically negligible.  When $\Lambda_{\rm M}<0$, the cosmological
constant can be dynamically significant at both early times and at
late times, since solutions can typically asymptote to static--modulus,
static--axion solutions (\ref{at_half}) (as well as asymptoting to the
dilaton--moduli--vacuum solutions).  For the repelling and attracting
sets of the $\Lambda_{\rm M}<0$ cases, the modulus field is only
dynamically significant when the cosmological constant is {\em not},
and vice versa.

Typically, the curvature term is found to be be dynamically
significant only at intermediate times, and is asymptotically
negligible.  However, the only exceptions to this is the
case $\Lambda_{\rm M}>0$, $\tilde K<0$ in which the repellor $N$
represents the curvature--driven, static--modulus, static--axion
solution (\ref{open_new}).  Finally, cases in which 
$\Lambda_{\rm M}>0$ there exist heteroclinic sequences in the
invariant set $\tilde K=0$; this implies that the qualitative
behaviour associated with heteroclinic sequences described is only
valid for solutions which approach $\tilde K=0$.

The only anisotropic models are the dilaton--moduli-vacuum
solutions (\ref{dmv}).  These solutions are attracting solutions (either into
the past or future) for most models.  However, these are not the only
asymptotic solutions and all other solutions mentioned in this chapter
are isotropic ($\dot\beta=0$).  In particular, the models in which
$\Lambda_{\rm M}<0$ do have isotropic late-time attractors.

The only at\-tracting so\-lu\-tions which are in\-fla\-tion\-ary are
the di\-la\-ton--mo\-du\-li--va\-cuum so\-lu\-tions (the ``$-$''
branch of (\ref{dmv}) for $h_0>0$) and the expanding static--modulus,
static--axion solution (\ref{at_half}) which occur in the pre-big bang
portion of the theory.  Again, the time reverse dynamics of these
models correspond to a post-big bang era in which the solutions are
not inflating. Hence, all late--time attracting solutions for the
post-big bang regime are {\em not} inflationary.

\section{Exact Solutions in the Einstein Frame\label{Einstein_Frame_II}}

Using the transformations (\ref{t2_trans}) and (\ref{a2_trans}) on
page \pageref{t2_trans}, the exact solutions corresponding to the
equilibrium points of this chapter may be transformed to the Einstein
frame.

The solution found in Billyard {\em et al.} \cite{Billyard1999c},
represented by the equilibrium point $S^\pm_1$, in the Einstein frame
becomes
\index{equilibrium sets!$S^\pm_1$}
\index{exact solutions!dilaton--axion}
\begin{eqnarray}
\nonumber
S^\pm_1: && a^* =  a_0^* \left[\half\sqrt{3\Lambda_{\rm M}} t_*
	\right]^{\frac{2}{3}} \qquad \Rightarrow \qquad 
		\up{sf}H=\frac{2}{3}t_*^{-1} 
	\qquad \Rightarrow \qquad \up{sf}q=\half, \\
\nonumber
&& e^{\varepsilon\sqrt2\phi} =\frac{2}{\sqrt{3\Lambda_{\rm M}} t_*}, \\
\nonumber
&& \beta = \beta_0,\\
\nonumber
&& \sigma = \sigma_0\pm \half\sqrt{5\Lambda_{\rm M}} t_*, \\
&&k=0.
\label{E_newsol}
\end{eqnarray}
Note that in the Einstein frame $t_*>0$ for both $S^-_1$ and $S^+_1$,
and hence both equilibrium points represent an expanding, non-inflating
model.  These solutions are saddles in the models for $\Lambda_{\rm
M}>0$, except when $\Lambda=k=0$ in which case $S^+_1$ is a source.

The negative-curvature cosmological model is given by
\index{equilibrium sets!$N$}
\index{exact solutions!negative--curvature driven}
\begin{eqnarray}
\nonumber
N: && a^* =  a^*_0 t_* \qquad \Rightarrow \qquad 
		\up{sf}H=t_*^{-1} 
	\qquad \Rightarrow \qquad \up{sf}q=0, \\
\nonumber
&& e^{\varepsilon\sqrt2\phi} = \left[\sqrt{\Lambda_{\rm M}}t_* \right]^{-1}, \\
\nonumber
&& \beta = \beta_0, \\
\nonumber 
&& \sigma=\sigma_0, \\
&& k = -\frac{3}{4}a^{*2}_0 
\label{E_open_new}
\end{eqnarray}
where $a^*_0=a_0\sqrt{\Lambda_{\rm M}}$, and the time is defined for
$t_*>0$.  This solution arises as an early-time attractor for the case
where $k<0$ and $\Lambda_{\rm M}>0$.

The only two equilibrium points for $\Lambda_{\rm M}<0$ transform to
the Einstein frame to the solutions
\index{equilibrium sets!$R/A$}
\index{exact solutions!dilaton--cosmological constant}
\begin{eqnarray}
\nonumber
R/A: && a^* = a^*_0 t_*^\quart \qquad \Rightarrow \qquad 
	\up{sf}H=\quart t_*^{-1} \qquad \Rightarrow \qquad \up{sf}q=3, \\
\nonumber
&& e^{\varepsilon\sqrt2\phi} = \frac{1}{\sqrt{-2\Lambda_{\rm M}} t_*}, \\
\nonumber
&& \beta=\beta_0, \\
\nonumber
&&\sigma=\sigma_0,\\
&& k=0,
\label{E_at_half}
\end{eqnarray}
where $a^*_0=a_0~2^\frac{1}{8}\left[-\Lambda_{\rm
M}\right]^\frac{7}{8}$.  In the Einstein frame, $t_*>0$ for both
equilibrium points and so both $R$ and $A$ represent an expanding,
non-inflationary model in the Einstein frame.  From the analysis of
the previous sections, whenever $\Lambda_{\rm M}<0$ these solutions
are both late-time and early-time attracting solutions.

Finally, note that the solutions (\ref{dmv}) and (\ref{curv_drive})
were also solutions to this analysis in this chapter.  The forms for
the corresponding solutions in the Einstein frame are give by
equations (\ref{E_dmv}) on page \pageref{E_dmv} and
(\ref{E_curv_drive}) on page \pageref{E_curv_drive}, respectively.

In the Einstein frame, there are no inflationary models, although some
of the pre-big bang ($t<0$) solutions in the Jordan frame do inflate.

\subsection{Mathematical Equivalence to Matter Terms in the Einstein Frame\label{math_II}}

The exact solutions discussed in this chapter will now be transformed
to a theory of general relativity (Einstein frame) containing a matter
field and a scalar field with exponential potential; i.e., $V=V_0
e^{k\phi}$, where either $k^2=8$ or $V_0=0$.  Similar to subsection
\ref{math_I} (page \pageref{math_I}), there will be interaction terms
between the matter and scalar field (see equation
(\ref{E_conserves_I}) on page \pageref{E_conserves_I}).

Again, there are two scenarios from which to choose:
\renewcommand{\labelenumi}{\Alph{enumi})}
\begin{enumerate}
\item $V=0$, ${\cal U}=\half\Lambda_{\rm M}e^{2\sqrt2\varepsilon\phi}$ \\
The interaction term for this case is $\delta=-2\sqrt2\varepsilon\dot\phi\ p$.
\item $V=\half\Lambda_{\rm M} e^{2\sqrt2\varepsilon\phi}$ $(k^2=8)$, 
	${\cal U}=0$ \\
The interaction term for this case is $\delta=-\frac{\sqrt2}{2}
\varepsilon\dot\phi(\mu+3p)$.
\end{enumerate}
For these two scenarios, the matter field is defined by
\beqn
\mu &\equiv & \quart \dot\sigma^2e^{2\sqrt2\varepsilon\phi} + {\cal U} \\
p &\equiv & \quart \dot\sigma^2e^{2\sqrt2\varepsilon\phi} - {\cal U}
\eeqn
and do not in general represent barotropic matter with a linear equation of state [$p=(\gamma-1)\mu$],
although the equations of state are linear  at the equilibrium points (as was the
case in subsection \ref{math_I}).  Tables
\ref{model_A_II} and \ref{model_B_II} list
$\{\mu, p, \gamma, \mu_\phi, p_\phi, \gamma_\phi, \delta\}$ for each
equilibrium set in each of the two scenarios discussed above.
\begin{table}[ht]
\begin{center}
\begin{tabular}{|c|ccccccc|}\hline
\multicolumn{8}{|c|}{\bf Scenario A: \qquad 
	$V=0$, \qquad
	${\cal U}=\half\Lambda_{\rm M}e^{2\sqrt2\varepsilon\phi}$, \qquad
	$\delta=-2\sqrt2\varepsilon\dot\phi\ p$} \\ \hline
\hline
Set & $\mu$ & $p$ & $\gamma$ & $\mu_\phi$ & $p_\phi$ & $\gamma_\phi$ 
		& $\delta$ \\
\hline 
$L^\pm$ & $0$ & $0$ & $-$ & $\frac{(\pm3h_0-1)^2}{9(1\mp h_0)^2}t_*^{-2}$ & 
	$\mu_\phi$ & $2$ & $0$ \\
$S^\pm$ & $0$ & $0$ & $-$ & $0$ & $0$ & $-$ & $0$ \\
$S_1^\pm$ & $\frac{13}{12}t_*^{-2}$& $-\quart t_*^{-2}$ & $\frac{10}{13}$ 
	& $\quart t_*^{-2}$ & $\mu_\phi$ & $2$ 
	& $-\half t_*^{-3}$ \\ 
$N$ & $\half t_*^{-2}$ & $-\mu$ & $0$ & $\quart t_*^{-2}$& $\mu_\phi$ 
	& $2$ & $- t_*^{-3}$ \\ 
$R/A$ & $-\quart t_*^{-2}$ & $-\mu$ & $0$ & $\quart t_*^{-2}$& $\mu_\phi$ 
	& $2$ & $\half t_*^{-3}$\\ 
\hline
\end{tabular}
\end{center}
\caption[{\em Matter Terms for Model A for $\Lambda=Q=0$}]{{\em The matter terms ($\mu$,
$p$, $\gamma$) as well as $\mu_\phi$, $p_\phi$, $\gamma_\phi$ and
$\delta$ for each of the equilibrium sets derived in scenario
A for $\Lambda=Q=0$.}\label{model_A_II}}
\end{table}

From section \ref{summary_II}, the asymptotic behaviour of the
models in the Einstein frame is known and comments on which solutions 
represent asymptotic states in the Einstein frame are equally applicable here.

In scenario A, it is evident that asymptotically the matter field
either vanishes or asymptotes towards a false vacuum (an effective
cosmological constant).  The exception to this is the solution given
by (\ref{newsol}) (represented by $S^+_1$), in which the matter field
asymptotes towards $\gamma=\frac{10}{13}$.  In the case where $k=0$ and
$\Lambda_{\rm M}>0$, this solution is an early-time attracting
solution.  There are no matter scaling solutions at the equilibrium
sets except the trivial case of the saddle Milne model, in which all
sources are zero.

In the heteroclinic sequence found in this chapter for $\Lambda_{\rm
M}>0$, $k=0$ trajectories in the phase space successively approached
two orbits in the invariant set $\dot\beta=\dot\sigma=0$.  The one
trajectory was characterized by $\Lambda_{\rm M}=0$ and the other by
$\Lambda_{\rm M}\neq0$.  In the trajectory for $\Lambda_{\rm M}=0$, we
find that $\mu=p=0$ in scenario A, while the trajectory for
$\Lambda_{\rm M}\neq0$ represents models with $\mu=-p$ ($\gamma=0$).
And so the heteroclinic orbits asymptotically represent solutions in
which the matter field oscillates between a false vacuum ($\gamma=0$)
and a true vacuum ($\mu=p=0$).

\begin{table}[ht]
\begin{center}
\begin{tabular}{|c|ccccccc|}\hline
\multicolumn{8}{|c|}{\bf Scenario B: \qquad 
	$V=\half\Lambda_{\rm M}	e^{2\sqrt2\varepsilon\phi}$,\qquad 
	${\cal U}=0$, \qquad
	$\delta=-\frac{\sqrt2}{2}\varepsilon\dot\phi(\mu+3p)$}\\\hline\hline
Set & $\mu$ & $p$ & $\gamma$ & $\mu_\phi$ & $p_\phi$ & $\gamma_\phi$ 
		& $\delta$ \\
\hline 
$L^\pm$ & $0$ & $0$ & $-$ & $\frac{(\pm3h_0-1)^2}{9(1\mp h_0)^2}t_*^{-2}$ & 
	$\mu_\phi$ & $2$ & $0$ \\
$S^\pm$ & $0$ & $0$ & $-$ & $0$ & $0$ & $-$ & $0$ \\
$S_1^\pm$ & $\frac{5}{12}t_*^{-2}$& $\mu$ & $2$ 
	& $\frac{11}{12}t_*^{-2}$ & $-\frac{5}{12}t_*^{-2}$ & $\frac{6}{11}$ 
	& $\frac{5}{6} t_*^{-3}$ \\ 
$N$ & $0$ & $0$ & $-$ & $\frac{3}{4}t_*^{-2}$& $-\frac{1}{4}t_*^{-2}$ 
	& $\frac{2}{3}$ & $0$ \\ 
$R/A$ & $0$ & $0$ & $-$ & $0$ & $\half t_*^{-2}$ & $-$ & $0 $\\ 
\hline
\end{tabular}
\end{center}
\caption[{\em Matter Terms for Model B for $\Lambda=Q=0$}]{{\em The matter terms ($\mu$,
$p$, $\gamma$) as well as $\mu_\phi$, $p_\phi$, $\gamma_\phi$ and
$\delta$ for each of the equilibrium sets derived in scenario
B for $\Lambda=Q=0$.}\label{model_B_II}}
\end{table}

The situation is similar in scenario B; the matter field typically
asymptotes towards a vacuum solution, except the case of $S^\pm_1$ in
which the matter asymptotes towards a stiff equation of state.  Again
there are no matter scaling solutions except the trivial case of
vacuum solutions.

Here, the heteroclinic orbits asymptote towards solutions in which the
matter field becomes negligible in any part of the cycle.

\chapter{String Models III: The $\Lambda$-$\Lambda_{\rm M}$ Competition ($Q=0$)\label{NSNSRR}}

The purpose of this section is to examine the dynamical evolution of
the field equation (\ref{rr}) when both $\Lambda\neq 0$ and
$\Lambda_{\rm M}\neq 0$ but $Q=0$ in order to investigate which term
dominates at late and early times.  In the last two chapters, it was
evident that either $\dot\beta$ dominated asymptotically (the
corresponding solutions were represented by $L^\pm$) or the
``cosmological'' constant term (either $\Lambda$ or
$\Lambda_{\rm M}$), dominated but there were never instances where both
$\dot\beta$ {\em and} the ``cosmological'' were both the dominant
variables asymptotically.  Hence, a similar result  would be expected to
apply here.  Therefore, this chapter excludes the modulus term in
order to explore which of $\Lambda$ and $\Lambda_{\rm M}$ dominate
asymptotically, and so the metric reduces to that of an FRW model.  In
order to examine a three-dimensional system, a flat FRW model is
assumed.

The chapter is organized as follows.  In section
\ref{Governing_Equations_III}, the field equations (\ref{rr}) and 
(\ref{rrfriedmann}) are reexamined with $Q=0$.  Section
\ref{Analysis_III} proceeds with the analysis of the equations.  The
chapter ends with a summary section and a section which discusses the
corresponding solutions and asymptotic behaviour in the Einstein
frame.  Again, this chapter is primarily confined to the Jordan frame
(except the final section), and so the index ``(st)'' shall be omitted
to save notation, (but must be introduced again in the final section
when both frames are discussed).

\section{Governing Equations\label{Governing_Equations_III}}

For $k=Q=0$, equations (\ref{rr}) can be rewritten as
\beqn
\mainlabel{rr_L}
\label{rr_L1}
\ddot{\alpha} -\dot{\alpha}\dot{\varphi} -\frac{1}{2} 
     \rho +\frac{1}{2}L =0,\\
\label{rr_L2}
2\ddot{\varphi} -\dot{\varphi}^2 -3\dot{\alpha}^2 +\frac{1}{2}\rho +2 \Lambda =0, \\
\label{rr_L3}
\dot L - L\left(\dot\varphi+3\dot\alpha\right) = 0, \\
\label{rr_L5}
\dot{\rho} +6\dot{\alpha} \rho =0, 
\eeqn
where $L\equiv \Lambda_{\rm M} \exp(\varphi+3\alpha)$, and the Friedmann constraint equation (\ref{rrfriedmann}) now becomes
\begin{equation}
\label{rrfriedmann_L}
3\dot{\alpha}^2 -\dot{\varphi}^2 +\frac{1}{2} \rho +2\Lambda + L =0.
\end{equation}

\index{equilibrium sets!$J^\pm$}
\index{exact solutions!de Sitter}
All the exact solutions described in the two preceding chapters are
solutions to (\ref{rr_L}) and (\ref{rrfriedmann_L}), and will be
represented by equilibrium points in the analysis to follow.  In
addition, there also exists the de Sitter solutions for $\Lambda>0$
and $\Lambda_{\rm M}<0$:
\begin{eqnarray}
\nonumber
a & = & a_0 e^{\pm\sqrt{\frac{1}{6}\Lambda}t}, \\
\nonumber
\hat\Phi & = & \hat\Phi_0,  \\
\nonumber 
\Lambda_{\rm M} & = & -\Lambda e^{-\Phi_0}, \\
\nonumber
\beta&=&\beta_0, \\
\nonumber
\sigma&=&\sigma_0,\\
\label{two_lambda}
k&=&0,
\end{eqnarray}
where $\{a_0,\hat\Phi_0,\beta_0,\sigma_0\}$ are constants.  These
solutions are represented in the text by the equilibrium points $J^\pm$,
where the ``$\pm$'' correspond to the $\mp$ solutions.

Through equation (\ref{rrfriedmann_L}), the
variable $\rho$ is eliminated from the field equations, and the following
definitions are made:
\begin{equation}
x\equiv \frac{\sqrt3\dot\alpha}{\xi}, \quad
y\equiv \frac{\pm 2\Lambda}{\xi^2}, \quad
z\equiv \frac{\pm L}{\xi^2}, \quad
u\equiv \frac{\dot\varphi}{\xi}, \quad
\frac{d}{dt}\equiv \xi \frac{d}{dT},\label{TheDefs3}
\end{equation}
where the $\pm$ sign in the definitions for $y$ and $z$ are to ensure
$y>0$ and $z>0$, where necessary.  With these definitions, all
variables are bounded: $0\leq \{x^2,y,z,u^2\}\leq 1$.  The
variable $\xi$ is defined in each of the following four cases by 
(important features to each case are also noted here):
\begin{itemize}
	\item $\Lambda>0$, $L>0$: $\xi=\dot\varphi \geq 0$
	\begin{itemize} \item 
	equation (\ref{rrfriedmann}):
	$ \frac{1}{2}\rho \xi^{-2} = 1- x^2 -y-z \geq 0.$ 
	\item $u^2=1$.  Take $\dot\varphi\geq 0$.
	\end{itemize}
\item $\Lambda<0$, $L>0$: $\xi^2 =\dot\varphi^2-2\Lambda$  
	\begin{itemize} \item 
	equation (\ref{rrfriedmann}):
	$ \frac{1}{2}\rho \xi^{-2} = 1- x^2 -z \geq 0.$ 
	\item $u^2+y=1$.  The variable $y$ is extraneous.
	\end{itemize}
\item $\Lambda>0$, $L<0$: $\xi^2 =\dot\varphi^2-L$  
	\begin{itemize} \item 
	equation (\ref{rrfriedmann}):
	$ \frac{1}{2}\rho \xi^{-2} = 1- x^2 -y \geq 0.$ 
	\item $u^2+z=1$.  The variable $z$ is extraneous.
	\end{itemize}
\item $\Lambda>0$, $L>0$: $\xi^2 =\dot\varphi^2-2\Lambda-L$  
	\begin{itemize} \item 
	equation (\ref{rrfriedmann}):
	$ \frac{1}{2}\rho \xi^{-2} = 1- x^2 \geq 0.$ 
	\item $u^2+y+z=1$.  The variable $z$ is extraneous.
	\end{itemize}
\end{itemize}
For each case, the three-dimensional dynamical system will be constructed,
where each variable will be assigned a subscript to denote the case studied.

\section{Analysis\label{Analysis_III}}
\subsection{The Case $\Lambda>0$, $\Lambda_{\rm M}>0$\label{NSpRRp}}

In this case the definition $\xi = \dot\varphi$ is chosen and the positive
signs for $y_1$ and $v_1$ found in (\ref{TheDefs3}) are explicitly
considered.  The case $\dot\varphi\geq0$ shall be considered only.
The case $\dot\varphi <0$ is equivalent to a time reversal of the
system, and the dynamics are similar.  From the generalized Friedmann
equation one has that
\begin{equation}
0 \leq x_1^2 + y_1 + z_1 \leq 1,\qquad u_1^2=1,
\end{equation}
and therefore the following three-dimensional system of ODEs for
$0\leq \{x_1^2, y_1, z_1\}\leq1$ is considered:
\beqn 
\mainlabel{dx_1_main}
\label{dx_1}
\frac{dx_1}{dT} & = &\sqrt3\left(1-x_1^2-y_1-\frac{3}{2}z_1\right) 
	+x_1\left(1-x_1^2-\frac{1}{2}z_1\right), \\
\label{dy_1}
\frac{dy_1}{dT} &=& -y_1 \left(2x_1^2+z_1\right) < 0,\\
\label{dz_1}
\frac{dz_1}{dT} &=& z_1\left( 1-2x_1^2-z_1 + \sqrt3 x_1\right).
\eeqn
The invariant sets $x_1^2+y_1+z_1=1$, $y_1=0$ and $z_1=0$ define the
boundary of the phase space.  Note that $y_1$ is {\em monotonically
decreasing}.  The equilibrium points and their corresponding
eigenvalues (denoted by $\lambda$) are
\beqn
C^+: & & \left(x_1,y_1,z_1\right)=\left(0,1,0\right); \nonumber \\
   && \left(\lambda_1,\lambda_2, \lambda_3\right) =  \left(0,1,1 \right), \\
S^+_1: & & \left(x_1,y_1,z_1\right)=\left(-\frac{1}{\sqrt{27}},0,
	\frac{16}{27}\right);  \nonumber \\
   && \left(\lambda_1,\lambda_2, \lambda_3\right) =
	\frac{1}{3}\left(1+i\sqrt\frac{77}{27},1-i\sqrt\frac{77}{27},
	-2 \right), \\
L^+_{(\pm)}: & & \left(x_1,y_1,z_1\right)=\left(\pm1,0,0\right);
	 \nonumber \\
   &&  \left(\lambda_1,\lambda_2, \lambda_3\right) =
		\left(\pm\left[\sqrt3\mp1\right],
		\mp2\left[\sqrt3\pm 1\right],-2 \right).
\eeqn
The above symbols all have the superscript ``$+$'' to reflect the fact
that $\dot\varphi>0$ (in subsequent analyses, the $\dot\varphi<0$
analogues will be denoted by a superscript ``$-$'').  For the two points $L^+_{(\pm)}$, the
``$(\pm)$'' subscript, as well as the ``$\pm$'' in the eigenvalues
correspond to $x_1=\pm1$, respectively.

The zero eigenvalue for $C^+$ results from the fact that the
equilibrium point is non-hyperbolic.  However, it is straightforward
to show that $C^+$ is indeed a source.  The eigenvector associated
with the zero eigenvalue lies in the $z_1=0$ plane.  Hence, for
$z_1=0$ (\ref{dx_1}) is negative for
$y_1<(1-x_1^2)(x_1+\sqrt3)/\sqrt3$, and (\ref{dx_1}) is positive for
$y_1>(1-x_1^2)(x_1+\sqrt3)/\sqrt3$.  The line
$y_1=(1-x_1^2)(x_1+\sqrt3)/\sqrt3$ includes the equilibrium point $C^+$.
Hence, the $x_1$-component of the trajectories move {\em away} from
this line.  Furthermore, since $y_1$ is monotonically decreasing,
then trajectories move away from $C^+$ in this invariant set, making $C^+$
a source.  Since the other two eigenvalues are also positive, then
$C^+$ is a source in the full three-dimensional set.

\begin{figure}[htp]
  \centering
   \epsfxsize=5in
   \includegraphics*[width=5in]{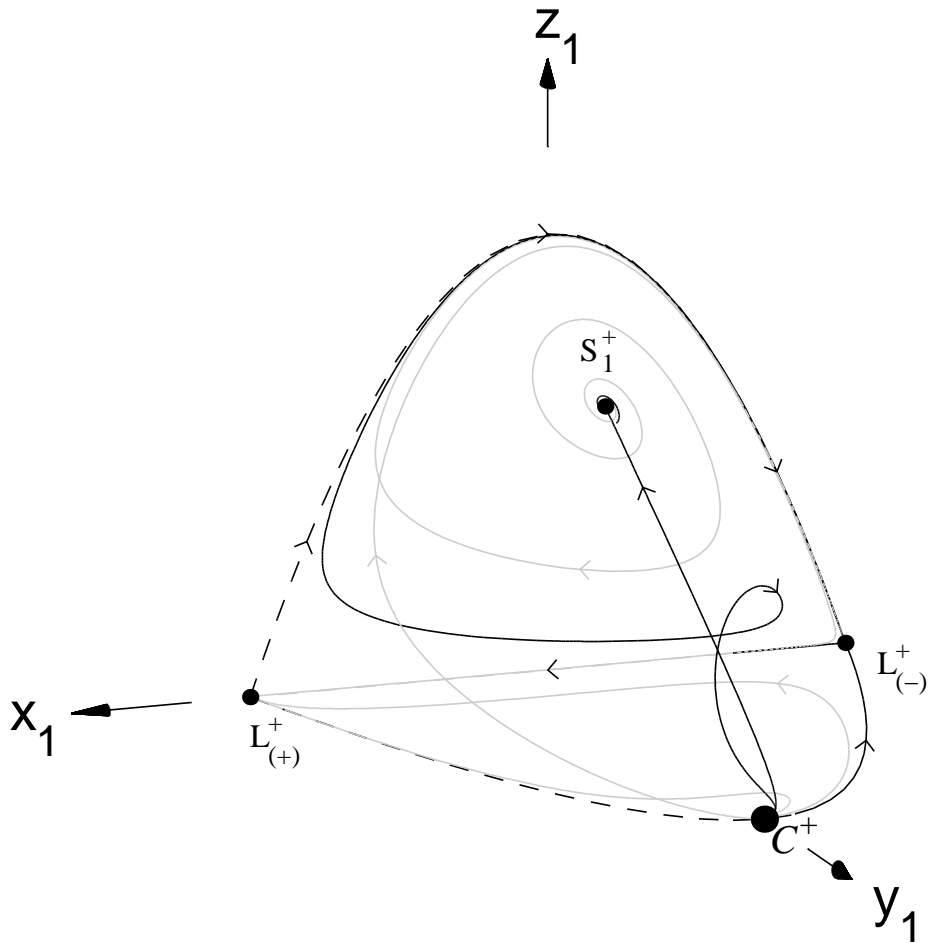}
  \caption[{\em Phase diagram for $\Lambda>0$, $\Lambda_{\rm
M}>0$}]{{\em Phase diagram of the system (\ref{dx_1_main})
($\Lambda>0$, $\Lambda_{\rm M}>0$).  In this phase space,
$\dot\varphi>0$ is assumed.  See also caption to figure \ref{fNSpK0}
on page \pageref{fNSpK0}.} \label{fNSpRRp}}
 \end{figure}

The qualitative behaviour is as follows.  Trajectories generically
asymptote into the past towards $C^+$ which represents the static,
linear dilaton--vacuum solution (\ref{static}).  There is one
trajectory which evolves from $C^+$ towards the solution $S^+_1$.  All
other trajectories spiral about this trajectory towards the $y_1=0$
invariant set, where they end up in a heteroclinic orbit which shadows
the line $x_1^2+z_1=1$, spend considerable time near the saddle point
$L^+_{(-)}$ (which represents the $\dot\varphi>0$ dilaton--vacuum
solution(\ref{dmv}) for $h_0=-\frac{1}{\sqrt3}$), quickly shadows the
line $z_1=y_1=0$ and then spends a large amount of time near the
saddle point $L^+_{(+)}$ (which represents another $\dot\varphi>0$
dilaton--vacuum solution (\ref{dmv}) for $h_0=\frac{1}{\sqrt3}$).
This behaviour of the trajectory is repeated indefinitely, each time
spending more and more time near the saddles $L^+_{(\pm)}$.  The
late-time behaviour is very similar to the early-time behaviour
discussed in subsection \ref{RRpK0} (see figure \ref{fRRpK0b0}),
except here $S^+_1$ is a saddle point in the three-dimensional system.
However the two-dimensional invariant set $z_1$ is identical to the
invariant set $\nu_1=0$ in subsection \ref{RRpK0}.  Figure
\ref{fNSpRRp} depicts typical trajectories in this phase space.

\subsection{The Case $\Lambda<0$, $\Lambda_{\rm M}>0$\label{NSnRRp}}

In this case, the negative sign in (\ref{TheDefs3}) for $y_2$ is
chosen as is the positive sign for $z_2$, and the definition $\xi^2 =
\dot\varphi^2-2\Lambda$ is chosen.  The generalized Friedmann
constraint equation is now written to read
\begin{equation}
0 \leq x_2^2 + z_2 \leq 1,
\end{equation}
and $u_2^2+y_2=1$.  Thus $y_2$ is considered the extraneous variable  and so
three-dimensional system consists of the variables $0\leq \{x_2^2, u_2^2, z_2
\} \leq 1$:
\beqn
\mainlabel{dx_2_main}
\label{dx_2}
\frac{dx_2}{dT} & = & \sqrt 3 \left( 1-x_2^2-\frac{3}{2}z_2  \right) 
		     + x_2u_2 \left( 1-x_2^2-\frac{1}{2}z_2 \right), \\
\label{du_2}
\frac{du_2}{dT} &=& (1-u_2^2) \left(x_2^2+\frac{1}{2}z_2\right) > 0,\\
\label{dz_2}
\frac{dz_2}{dT} &=& z_2 \left[u_2\left( 1-2x_2^2-z_2 \right)+\sqrt3 x_2\right].
\eeqn
The invariant sets $x_2^2+z_2=1$, $z_2=0$, $u_2^2=1$ define the
boundary of the phase space.  Note that $u_2$ is {\em monotonically
increasing}.  The equilibrium sets and their corresponding eigenvalues
(denoted by $\lambda$) are
\beqn
S^\pm_1: & & \left(x_2,u_2,z_2\right)=\left(\frac{\mp1}{\sqrt{27}},\pm 1,\frac{16}{27}\right);
	 \nonumber \\
   && \left(\lambda_1,\lambda_2, \lambda_3\right) =
	\frac{1}{3}\left(\pm1+i\sqrt\frac{77}{27},\pm1-i\sqrt\frac{77}{27},
	\mp2 \right), \\
L^+_{(\pm)}: & & \left(x_2,u_2,z_2\right)=\left(\pm1,1,0\right);
	 \nonumber \\
   &&  \left(\lambda_1,\lambda_2, \lambda_3\right) =
		\left(\pm\left[\sqrt3\mp1\right],
		\mp2\left[\sqrt3\pm 1\right],-2 \right), \\
L^-_{(\pm)}: & & \left(x_2,u_2,z_2\right)=\left(\pm1,-1,0\right);
	 \nonumber \\
   &&  \left(\lambda_1,\lambda_2, \lambda_3\right) =
		\left(\pm\left[\sqrt3\pm1\right],
		\mp2\left[\sqrt3\mp 1\right],2 \right).
\eeqn

Similar to the points $L^+_{(\pm)}$, which represent the two
$\dot\varphi>0$ dilaton--vacuum solutions (\ref{dmv}) for
$h_0=\frac{\pm1}{\sqrt3}$, the two points $L^-_{(\pm)}$ represent the
two $\dot\varphi<0$ dilaton--vacuum solutions (\ref{dmv}) for
$h_0=\frac{\pm 1}{\sqrt3}$.

\begin{figure}[htp]
  \centering
   \epsfxsize=5in
   \includegraphics*[width=5in]{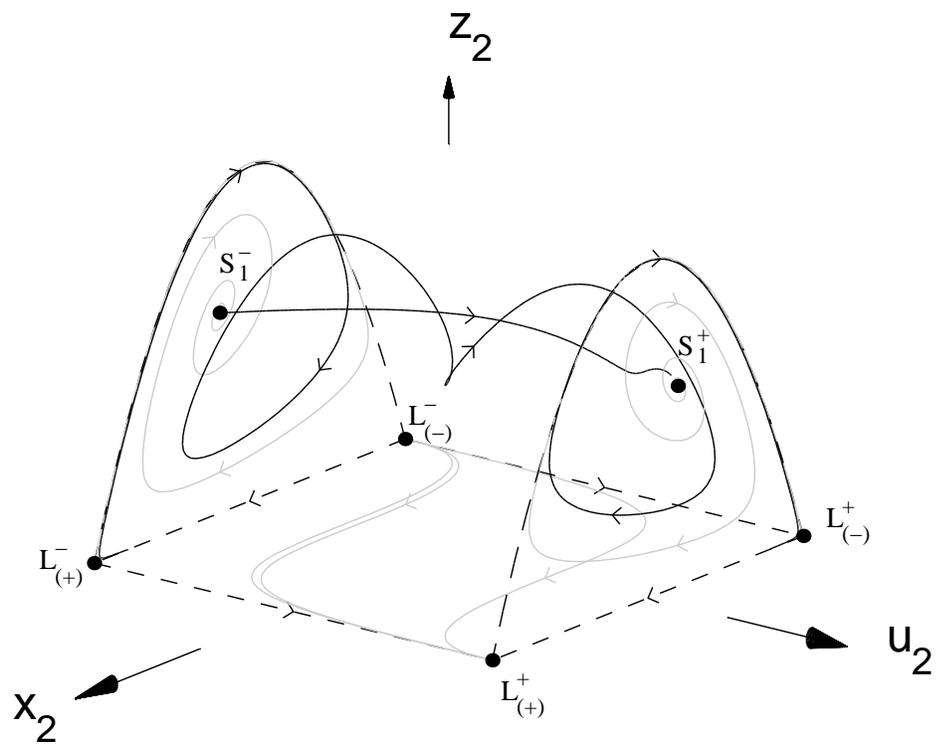}
  \caption[{\em Phase diagram for $\Lambda<0$, $\Lambda_{\rm M}>0$}]{{\em Phase diagram of the system
(\ref{dx_2_main}) ($\Lambda<0$, $\Lambda_{\rm M}>0$). See also caption
to figure \ref{fNSpK0} on page \pageref{fNSpK0}. } \label{fNSnRRp}}
 \end{figure}

In this phase space, there are no repelling/attracting
equilibrium points, but $u_2$ is monotonically increasing.  
The early-time and late-time behaviour is similar to the
late-time behaviour discussed in the previous subsection.  Into the
past, there are heteroclinic orbits which spend considerable time near
the saddle points $L^-_{(-)}$ and $L^-_{(+)}$, in between which they
shadow the lines $u_2=-1$, $x_2^2+z_2=1$ and $u_2=-1$, $z_2=0$.  The
late-time behaviour is again for trajectories to follow heteroclinic
orbits which spend time near the saddles $L^+_{(-)}$ and $L^+_{(+)}$,
in between which they shadow the lines $u_2=+1$, $x_2^2+z_2=1$ and
$u_2=+1$, $z_2=0$.  In the invariant set $u_2=-1$ (which is the
past-attracting invariant set) these orbits spiral towards the
expanding solution $S^-_1$, which represents the $t>0$ solution of
(\ref{newsol}).  In the future-attracting invariant set, $u_2=+1$,
orbits spiral away from the contracting solution $S^+_1$, which
represents the $t<0$ solution of (\ref{newsol}).  Figure \ref{fNSnRRp} depicts
typical trajectories in this phase space.

\subsection{The Case $\Lambda>0$, $\Lambda_{\rm M}<0$\label{NSpRRn}}

For this case, $\xi^2 = \dot\varphi^2-L$ is chosen 
and from (\ref{TheDefs3}) the positive sign for $y_3$ and the negative
sign for $z_3$ are chosen in order for  these variables to be positive definite.  The
generalized Friedmann constraint equation can be rewritten as
\begin{equation}
0 \leq x_3^2 + y_3 \leq 1,\qquad u_3^2+z_3=1,
\end{equation}
and $z_3$ may be eliminated to yield 
the four-dimensional system of ODEs for $0\leq \{x_3^2,y_3, u_3^2\} \leq1$:
\beqn
\mainlabel{dx_3_main}
\label{dx_3}
\frac{dx_3}{dT} & = & x_3u_3\left(1-x_3^2\right)+\sqrt 3 \left[ 
	\frac{1}{2}\left(1-x_3^2\right)\left(3-u_3^2\right)-y_3\right], \\
\label{dy_3}
\frac{dy_3}{dT} &=& -x_3y_3 \left[\sqrt3 \left( 1-u_3^2 
	\right) +2x_3u_3\right] ,\\
\label{du_3}
\frac{du_3}{dT} &=& -\frac{1}{2}\left(1-u_3^2 \right)\left(
	1-2x_3^2+\sqrt3 x_3u_3 \right).
\eeqn
The invariant sets $x_3^2+y_3=1$, $u_3^2=1$, $y_3=0$ define the
boundary of the phase space.  The equilibrium sets and their
corresponding eigenvalues (denoted by $\lambda$) are
\beqn
R: & & \left(x_3,y_3,u_3\right)=\left(1,0,\frac{1}{\sqrt3}\right); \nonumber \\
   & &  \left(\lambda_1,\lambda_2, \lambda_3\right) = -\frac{1}{\sqrt3}
	\left(1,4,10 \right), \\
A: & & \left(x_3,y_3,u_3\right)=\left(-1,0,-\frac{1}{\sqrt3}\right);\nonumber \\
   & &  \left(\lambda_1,\lambda_2, \lambda_3\right) = \frac{1}{\sqrt3}
	\left(1,4,10 \right), \\
C^\pm: & & \left(x_1,y_1,u_1\right)=\left(0,1,\pm1\right); \nonumber \\
   && \left(\lambda_1,\lambda_2, \lambda_3\right) = 
		\left(0,\pm1,\pm1 \right), \\
L^+_{(\pm)}: & & \left(x_3,y_3,u_3\right)=\left(\pm1,0,1\right);
	 \nonumber \\
   &&  \left(\lambda_1,\lambda_2, \lambda_3\right) =
		\left(\pm\left[\sqrt3\mp1\right],
		\mp2\left[\sqrt3\pm 1\right],-2 \right), \\
L^-_{(\pm)}: & & \left(x_3,y_3,u_3\right)=\left(\pm1,0,-1\right);
	 \nonumber \\
   &&  \left(\lambda_1,\lambda_2, \lambda_3\right) =
		\left(\pm\left[\sqrt3\pm1\right],
		\mp2\left[\sqrt3\mp 1\right],2 \right), \\
J^\pm: & & \left(x_3,y_3,u_3\right)=\left(\frac{\pm1}{\sqrt5},\frac{4}{5},
	\mp\sqrt\frac{3}{5}\right); \nonumber \\
   && \left(\lambda_1,\lambda_2, \lambda_3\right) = \frac{1}{\sqrt 5} \left(
	\frac{\sqrt{19}\mp\sqrt3}{2\sqrt5},\frac{-\sqrt{19}\mp\sqrt3}{2\sqrt5},
	\mp2\sqrt\frac{3}{5}\right).
\eeqn
Again, the points $C^\pm$ are non-hyperbolic, but it can be shown that
$C^+$ is a source whilst $C^-$ is a sink.  The zero eigenvalue
corresponds to a eigenvector lying within the invariant set $u_3=+1$.
It is easily seen from (\ref{dy_3}) that $y_3$ is monotonically
decreasing within this invariant set, moving away from $C^+$.
Similarly for this invariant set, equation (\ref{dx_3}) is negative
for $y_3>(1-x_3^2)(x_3+\sqrt3)/\sqrt3$, and (\ref{dx_3}) is positive for
$y_1<(1-x_3^2)(x_3+\sqrt3)/\sqrt3$.  The line
$y_1=(1-x_3^2)(x_3+\sqrt3)/\sqrt3$ includes the equilibrium point $C^+$.
Hence, the $x_1$-component of the trajectories move {\em away} from
this line.  Since $y_3$ is monotonically decreasing, then
trajectories move away from $C^+$ in this invariant set, making $C^+$
a source.  For points very close to $x_3=0$ equation (\ref{du_3}) is
negative in the full three-dimensional set.  Hence, $C^+$ is a source
for the three-dimensional system.  Similarly, $C^-$ is a sink.  These
two points represent the linear dilaton--vacuum solutions
(\ref{static}).  The point $R$, which acts as a source, represents the ``$+$''
solution of equation (\ref{at_half}), and the point $A$, which acts as a
sink, is the ``$-$'' solution of equation (\ref{at_half}).  The four points
$L^\pm_{(\pm)}$ are saddle points, as are the two equilibrium points
$J^\pm$, representing the de Sitter-like solutions (\ref{two_lambda}).

\begin{figure}[htp]
  \centering
   \epsfxsize=5in
   \includegraphics*[width=5in]{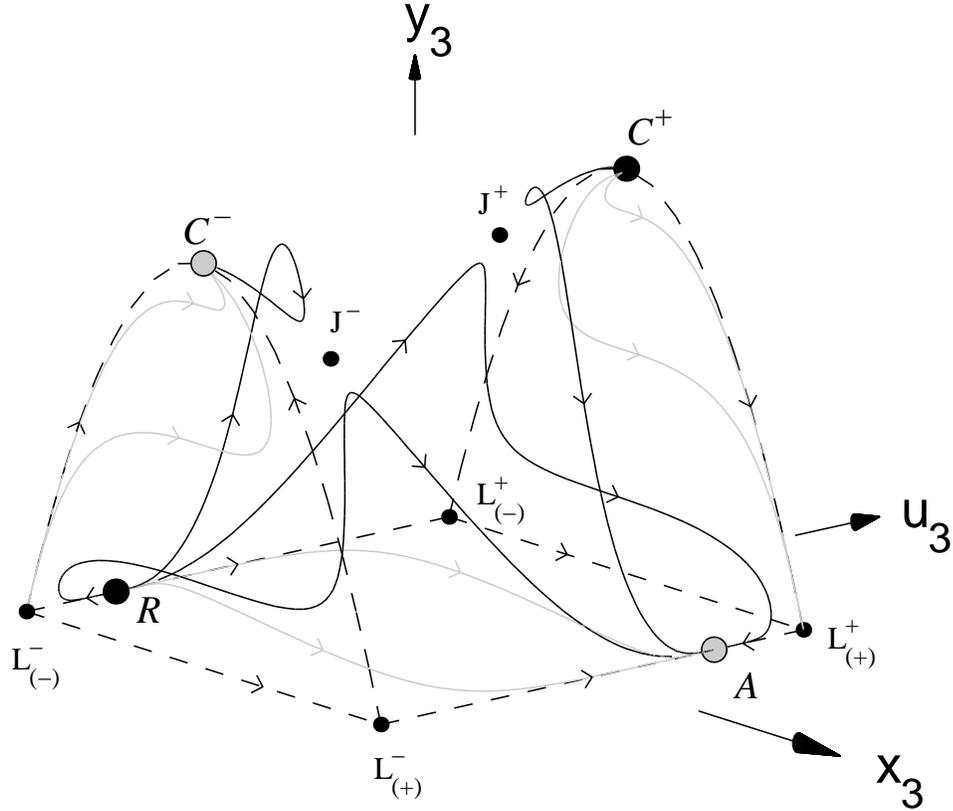}
  \caption[{\em Phase diagram for $\Lambda>0$, $\Lambda_{\rm M}<0$}]{{\em Phase diagram of the system
(\ref{dx_3_main}) ($\Lambda>0$, $\Lambda_{\rm M}<0$).  See also caption
to figure \ref{fNSpK0} on page \pageref{fNSpK0}. } \label{fNSpRRn}}
 \end{figure}

Hence, solutions asymptote into the past towards either $C^+$ or $R$
and asymptote into the future towards either $C^-$ or $A$.  Figure
\ref{fNSpRRn} depicts typical trajectories in this phase space.

\subsection{The Case $\Lambda<0$, $\Lambda_{\rm M}<0$\label{NSnRRn}}

\begin{figure}[htp]
  \centering
   \epsfxsize=5in
   \includegraphics*[width=5in]{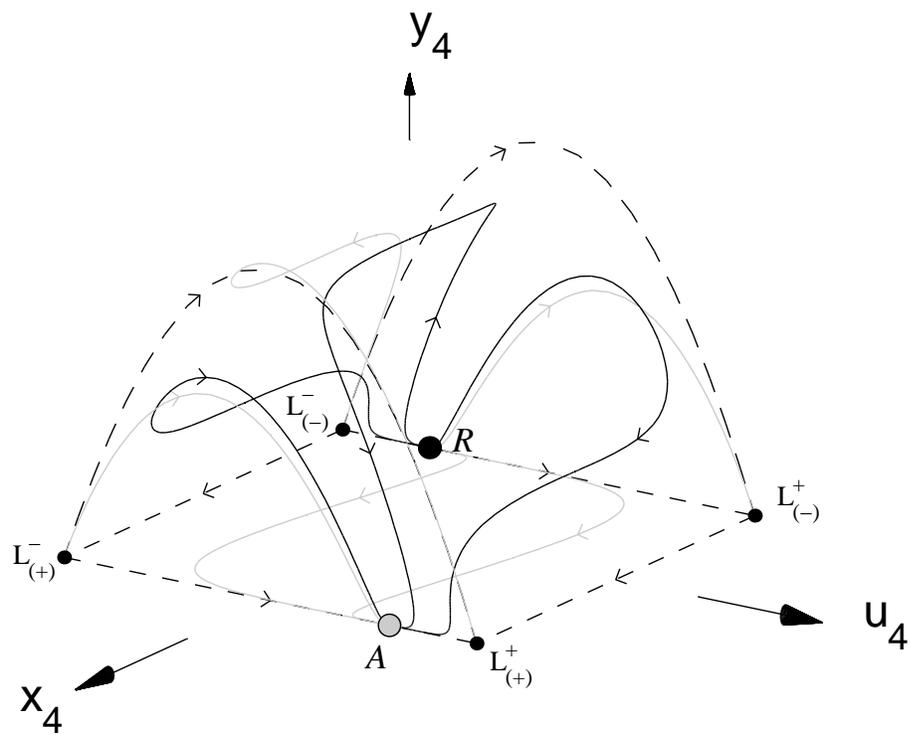}
  \caption[{\em Phase diagram for $\Lambda<0$, $\Lambda_{\rm M}<0$}]{{\em Phase diagram of the system
(\ref{dx_4_main}) ($\Lambda<0$, $\Lambda_{\rm M}<0$).  See also caption
to figure \ref{fNSpK0} on page \pageref{fNSpK0}. } \label{fNSnRRn}}
 \end{figure}

The negative signs for both $y_4$ and $z_4$ from (\ref{TheDefs3}) are
chosen in this case, as well as the definition $\xi^2 =
\dot\varphi^2-2\Lambda-L$.  The generalized Friedmann constraint
equation is now written to read
\begin{equation}
0 \leq x_4^2 \leq 1, \qquad u_4^2+z_4+y_4=1,
\end{equation}
Again, $z_4$ is con\-si\-dered to be the ex\-tra\-neous vari\-able, re\-sul\-ting in
the three-di\-men\-sion\-al system consisting of the variables $0\leq \{x_4^2, 
y_4, u_4^2\} \leq 1$:
\beqn
\mainlabel{dx_4_main}
\label{dx_4}
\frac{dx_4}{dT} & = & \frac{1}{2}\left(1-x_4^2\right)
	\left[2x_4u_4 + \sqrt 3\left(3-y_4-u_4^2\right) \right]>0, \\
\label{dy_4}
\frac{dy_4}{dT} &=& -x_4y_4 \left[2x_4u_4 +\sqrt3 \left(1-y_4-u_4^2\right)
	\right],\\
\label{du_4}
\frac{du_4}{dT} &=& -\frac{1}{2} \left(1-u_4^2\right) \left( 1-2x_4^2 +\sqrt3
	x_4u_4\right) +y_4x_4^2 .
\eeqn
The invariant sets $x_4^2=1$, $u_4^2+y_4=1$, $y_4=0$ define the
boundary of the phase space.  The variable $x_4$ is {\em monotonically
increasing}.  The equilibrium sets and their corresponding eigenvalues
(denoted by $\lambda$) are
\beqn
R: & & \left(x_4,y_4,u_4\right)=\left(1,0,\frac{1}{\sqrt3}\right); \nonumber \\
   & &  \left(\lambda_1,\lambda_2, \lambda_3\right) = -\frac{1}{\sqrt3}
	\left(1,4,10 \right), \\
A: & & \left(x_4,y_4,u_4\right)=\left(-1,0,-\frac{1}{\sqrt3}\right);\nonumber \\
   & &  \left(\lambda_1,\lambda_2, \lambda_3\right) = \frac{1}{\sqrt3}
	\left(1,4,10 \right), \\
L^+_{(\pm)}: & & \left(x_4,y_4,u_4\right)=\left(\pm1,0,1\right);
	 \nonumber \\
   &&  \left(\lambda_1,\lambda_2, \lambda_3\right) =
		\left(\pm\left[\sqrt3\mp1\right],
		\mp2\left[\sqrt3\pm 1\right],\mp2 \right), \\
L^-_{(\pm)}: & & \left(x_4,y_4,u_4\right)=\left(\pm1,0,-1\right);
	 \nonumber \\
   &&  \left(\lambda_1,\lambda_2, \lambda_3\right) =
		\left(\pm\left[\sqrt3\pm1\right],
		\mp2\left[\sqrt3\mp 1\right],\pm2 \right).
\eeqn
The global source for this system is the point $R$, which corresponds
to the ``$+$'' solution of (\ref{at_half}), and the global sink is
the point $A$, which corresponds to the ``$-$'' solution of
(\ref{at_half}).  Figure \ref{fNSnRRn} depicts typical trajectories in this
phase space.

\section{Summary of Analysis in the Jordan Frame\label{summary_III}}

This discussion begins with two tables; Table \ref{term_asymp_III} lists
which terms are the dominant variables for each equilibrium set as
well as the deceleration parameter, $q$, for the corresponding model
and Table \ref{points_asymp_III} lists the attracting behaviour of the equilibrium sets occur as
early-, intermediate- and late-time attractors.
\begin{table}[htb]
\begin{center}
\begin{tabular}{|c|ccccccc|c|c|}\hline
Set & \multicolumn{7}{c|}{Dominant Variables}&$q$&H \\
\hline \hline
$L^+_{(\pm)}$ {\tiny($t<0$)} & $\alpha$ &$\hat\Phi$ &&&&&
	&$1\mp \sqrt3$ &$\pm\frac{1}{\sqrt3}(-t)^{-1}$\\
$L^-_{(\pm)}$ {\tiny($t>0$)} & $\alpha$ &$\hat\Phi$ &&&&&
	&$\pm\sqrt3-1$ &$\pm\frac{1}{\sqrt3}t^{-1}$\\ \hline
$S_1^+$ {\tiny($t<0$)}& $\alpha$ & $\hat\Phi$ & $\dot\sigma$ &&
	& $\Lambda_{\rm M}>0$&&$2$& $\frac{1}{3}t^{-1}$\\
$S_1^-$ {\tiny($t>0$)}& $\alpha$ & $\hat\Phi$ & $\dot\sigma$ &&
	& $\Lambda_{\rm M}>0$&&$2$& $\frac{1}{3}t^{-1}$\\ \hline
$A$ {\tiny($t<0$)} & $\alpha$ & $\hat\Phi$ &&&& $\Lambda_{\rm M}<0$ &&$-3$
	& $-\frac{1}{2}t^{-1}$\\ 
$R$ {\tiny($t>0$)} & $\alpha$ & $\hat\Phi$ &&&& $\Lambda_{\rm M}<0$ &&$-3$
	& $-\frac{1}{2}t^{-1}$\\  \hline
$J^-$  & $\alpha$ &&&& $\Lambda>0$ & $\Lambda_{\rm M}<0$ &
	&$-1$ & $-\sqrt{\frac{1}{6}\Lambda}$\\
$J^+$  & $\alpha$ &&&& $\Lambda>0$ & $\Lambda_{\rm M}<0$ &
	&$-1$ & $\sqrt{\frac{1}{6}\Lambda}$\\ \hline
$C^\pm$     & & $\hat\Phi$ &&& $\Lambda>0$ &&&$0$&$0$\\
$L_1$       & &$\hat\Phi$ & $\dot\sigma$ & $k\geq 0$ & $\Lambda>0$&&&$0$&$0$\\
\hline
$N$ {\tiny($t<0$)} & $\alpha$ & $\hat\Phi$ && $k<0$ && $\Lambda_{\rm M}>0$ &
	&$0$&$t^{-1}$ \\ \hline
\end{tabular}
\end{center}
\caption[{\em Dominant Variables and $q$ for Equilibrium Sets}]{{\em The
dominant variables for each equilibrium set as well as the equilibrium
set's deceleration parameter, $q$, and Hubble parameter $H$.
Inflation occurs when $q<0$ and $H>0$, whereas ``deflation'' occurs
for $q>0$ and $H<0$.  Note that for $L^+_{(\pm)}$ and $L^-_{(\pm)}$, 
$h_0=\frac{\pm1}{\sqrt3}$}\label{term_asymp_III}}
\end{table}
Note that the only inflationary models are those represented by
$L^+_{(-)}$, $A$ and $J^+$.  The saddle point $S^+_1$ represents a
deflating model ($q>0$, $H<0$).

\begin{table}[ht]
\begin{center}
\begin{tabular}{|cc||c|c|c|}\hline
\multicolumn{2}{|c||}{Case}&Early & Intermediate & Late \\
\hline \hline
$\Lambda>0$ & $\Lambda_{\rm M}>0$ &$C^+$ &$S_1^+$, $L^+_{(\pm)}$  &$-$ \\
$\Lambda<0$ & $\Lambda_{\rm M}>0$ &$-$ &$S^\pm_1$, $L^\pm_{(\pm)}$ & $-$\\
$\Lambda>0$ & $\Lambda_{\rm M}<0$ &$C^+$, $R$ &$L^\pm_{(\pm)}$, $J^\pm$ &$C^-$, $A$ \\
$\Lambda<0$ & $\Lambda_{\rm M}<0$ &$R$  &$L^\pm_{(\pm)}$ &$A$ \\
\hline
\end{tabular}
\end{center}
\caption[{\em Summary of Equilibrium Points for Models
considered}]{{\em Summary of the early-time, intermediate, and
late-time attractors for the various models examined.  Note that
$\dot\alpha$, $\hat\Phi$ and $\dot\sigma$ are present in every
model.}\label{points_asymp_III}}
\end{table}

Monotonic functions have been established in all cases which precludes
the existence of recurrent or periodic orbits, thereby allowing the early-time and late-time behaviour conclusions to be made based
upon the equilibrium sets of the system.

For $\Lambda>0$ and $\Lambda_{\rm M}>0$, solutions generically
asymptote in the future towards the $y_1=0$ invariant set which
includes the heteroclinic orbit seen previously in the $\Lambda=0$,
$\Lambda_{\rm M}>0$ model, although the solution represented by
$S^+_1$ is no longer a source to the system, but rather a saddle. Since
there are no sinks, orbits evolve to the future toward $y_1=0$ (with
$y_1$ monotonically decreasing) and shadow
the heteroclinic orbit mimicking the the behaviour of figure
\ref{fNSpK0}, whereby solutions alternate between the dilaton--vacuum
solutions corresponding to the saddle points $L^+_{(-)}$ and $L^+_{(+)}$,
where the central charge deficit and $\Lambda_{\rm M}$ are dynamically
negligible.  Solutions generically asymptote in the past towards the
static, linear dilaton--vacuum solution corresponding to the global
source $C^+$, at which $\Lambda_{\rm M}$ is dynamically negligible.

In the case where $\Lambda<0$ and $\Lambda_{\rm M}>0$, there are no
sinks nor sources in the full 3-dimensional phase-space! The variable
$u_2$ (and hence $\dot\varphi$) is monotonically increasing, evolving
from $u_2=-1$ to $u_2=+1$.  Solutions generically asymptote in both
the past and future towards invariant sets which include an
heteroclinic orbit; i.e., they have a similar asymptotic behaviour at
both early and late times to the late-time behaviour of the previous
cases of $\Lambda_{\rm M}>0$, alternating between the dilaton--vacuum
solutions corresponding to the saddle points $L^-_{(\pm)}$ in the past
and the dilaton--vacuum solutions corresponding to the saddle points
($L^-_{(-)}$ represents an inflationary model) $L^+_{(\pm)}$ to the
future. Note that $\Lambda_{\rm M}$ is dynamically significant at both
early and late times.

In the case where $\Lambda>0$ and $\Lambda_{\rm M}<0$, solutions
asymptote into the past towards either the $\dot\varphi>0$ linear
dilaton--vacuum solution (\ref{static}) or the contracting,
static-moduli solution (\ref{at_half}).  Similarly, solutions
asymptote into the future either the $\dot\varphi<0$ linear
dilaton--vacuum solution (\ref{static}) or the expanding,
static-moduli solution (\ref{at_half}).  Hence both $\Lambda$ and
$\Lambda_{\rm M}$ can be dynamically significant at both early and
late times.

When $\Lambda<0$ and $\Lambda_{\rm M}<0$ solutions only asymptote
towards the the contracting/expanding, static-moduli solution to the
past/future.  Because $x_4$ monotonically increases, it would seem
that a bouncing cosmology from a contracting phase to an expanding
phases is typical within this model.  The central charge deficit is
only dynamically significant at intermediate times, whereas
$\Lambda_{\rm M}$ is dynamically significant at both early and late
times.

In general, there are only a few cases where $\Lambda$ and
$\Lambda_{\rm M}$ are both dynamically significant at early times or
at late times.  In particular, $\Lambda$ is only dynamically
significant at early/late times when $\Lambda>0$, whereas
$\Lambda_{\rm M}$ is typically dynamically significant for both early
and late times.  Because most phase diagrams exhibited some sort of
transition from $\dot\alpha<0$ to $\dot\alpha>0$ (or vice versa) it
has been found that bounce cosmologies are fairly typical within these class
of models.

\section{Exact Solutions in the Einstein Frame\label{Einstein_Frame_III}}

The de Sitter solutions transform to 
\index{equilibrium sets!$J^\pm$}
\index{exact solutions!de Sitter}
\begin{eqnarray}
\nonumber
J^\mp: && a^* = a^*_0 e^{\pm\sqrt{\frac{1}{6}\Lambda}t_*} 
	\qquad \Rightarrow \qquad 
	\up{sf}H=\pm \sqrt{\sixth \Lambda} \qquad \Rightarrow 
	\qquad \up{sf}q=-1, \\
\nonumber
&& \phi = \phi_0,  \\
\nonumber 
&& \Lambda_{\rm M} = -\Lambda e^{-\varepsilon\sqrt2\phi_0}, \\
\nonumber
&& \beta=\beta_0, \\
\nonumber
&& \sigma=\sigma_0,\\
&& k=0,
\label{E_two_lambda}
\end{eqnarray}
Note that $J^+$ represents the only inflationary model (of the exact
solutions) in the Einstein frame.  In the analysis for $\Lambda>0$,
$\Lambda_{\rm M}<0$ it was shown that both de Sitter solutions arise
as saddles and are dynamically important for  intermediate times.

All of the other exact solutions have been specified in the Einstein
frame in sections \ref{Einstein_Frame_I} (page
\pageref{Einstein_Frame_I}) and \ref{Einstein_Frame_II} (page
\pageref{Einstein_Frame_II}).

\subsection{Mathematical Equivalence to Matter Terms in the Einstein Frame\label{math_III}}

The exact solutions discussed in this chapter will now be transformed
to a theory of general relativity (Einstein frame) containing a matter
field and a scalar field with exponential potential; i.e., $V=V_0
e^{k\phi}$, where either $k^2=2$ or $k^2=8$.  Similar to subsection
\ref{math_I} (page \pageref{math_I}), there will be interaction terms
between the matter and scalar field (see equation
(\ref{E_conserves_I}) on page \pageref{E_conserves_I}).

Just like in the previous two chapters, there are two scenarios from which to 
choose:
\renewcommand{\labelenumi}{\Alph{enumi})}
\begin{enumerate}
\item $V=\Lambda e^{\sqrt2\varepsilon\phi}$  $(k^2=2)$, 
	${\cal U}=\half\Lambda_{\rm M}e^{2\sqrt2\varepsilon\phi}$ \\
The interaction term for this case is $\delta=-2\sqrt2\varepsilon\dot\phi\ p$.
\item $V=\half\Lambda_{\rm M} e^{2\sqrt2\varepsilon\phi}$ $(k^2=8)$
	${\cal U}=\Lambda e^{\sqrt2\varepsilon\phi}$ \\
The interaction term for this case is $\delta=-\frac{\sqrt2}{2}
\varepsilon\dot\phi(\mu+3p)$.
\end{enumerate}
For these two scenarios, the matter field is defined by
\beqn
\mu &\equiv & \quart \dot\sigma^2e^{2\sqrt2\varepsilon\phi} + {\cal U} \\
p &\equiv & \quart \dot\sigma^2e^{2\sqrt2\varepsilon\phi} - {\cal U}
\eeqn
and do not in general represent barotropic matter with linear equations of state [$p=(\gamma-1)\mu$],
although the equations of state are linear at the equilibrium points. Tables
\ref{model_A_III} and \ref{model_B_III} list $\{\mu, p, \gamma, \mu_\phi, p_\phi, \gamma_\phi, \delta\}$ for
each equilibrium set in each of the two scenarios discussed above.
\begin{table}[hb]
\begin{center}
\begin{tabular}{|c|ccccccc|}\hline
\multicolumn{8}{|c|}{\bf Scenario A: \qquad 
	$V=\Lambda e^{\sqrt2\varepsilon\phi}$, \qquad
	${\cal U}=\half\Lambda_{\rm M}e^{2\sqrt2\varepsilon\phi}$, \qquad
	$\delta=-2\sqrt2\varepsilon\dot\phi\ p$} \\ \hline
\hline
Set & $\mu$ & $p$ & $\gamma$ & $\mu_\phi$ & $p_\phi$ & $\gamma_\phi$ 
		& $\delta$ \\
\hline 
$L^\pm_{(\pm)}$ & $0$ & $0$ & $-$ & $\third t_*^{-2}$ & 
	$\mu_\phi$ & $2$ & $0$ \\
$C^\pm$ & $0$ & $0$ & $-$ & $3t_*^{-2}$ & $-t_*^{-2}$ & $\frac{2}{3}$ & $0$ \\
$S_1^\pm$ & $\frac{13}{12}t_*^{-2}$& $-\quart t_*^{-2}$ & $\frac{10}{13}$ 
	& $\quart t_*^{-2}$ & $\mu_\phi$ & $2$ 
	& $-\half t_*^{-3}$ \\ 
$R/A$ & $-\quart t_*^{-2}$ & $-\mu$ & $0$ & $\quart t_*^{-2}$& $\mu_\phi$ 
	& $2$ & $\half t_*^{-3}$\\ 
$J^\pm$ & $-\half\Lambda$ &$-\mu$ & $0$ & $\Lambda$ & $-\mu_\phi$ 
	& $0$ & $0$\\ \hline
\end{tabular}
\end{center}
\caption[{\em Matter Terms for Model A}]{{\em The matter terms ($\mu$,
$p$, $\gamma$) as well as $\mu_\phi$, $p_\phi$, $\gamma_\phi$ and
$\delta$ for each of the equilibrium sets derived in scenario
A.}\label{model_A_III}}
\end{table}

From section \ref{summary_III}, the asymptotic behaviour of the
models in the Einstein frame is known and comments on which solutions 
represent asymptotic states in the Einstein frame are equally applicable here.

In scenario A, solutions asymptote to models in which the
matter field is either vacuum ($\mu=p=0$) or are in false vacuum
state ($\gamma=0$).  As was the case in subsection
\ref{math_II} (\pageref{math_II}), the heteroclinic sequences
asymptote to solutions in which the matter field oscillates between a
false vacuum ($\gamma=0$) and a true vacuum ($\mu=p=0$).

\begin{table}[ht]
\begin{center}
\begin{tabular}{|c|ccccccc|}\hline
\multicolumn{8}{|c|}{\bf Scenario B: \qquad 
	$V=\half\Lambda_{\rm M}	e^{2\sqrt2\varepsilon\phi}$,\qquad 
	${\cal U}=\Lambda e^{\sqrt2\varepsilon\phi}$, \qquad
	$\delta=-\frac{\sqrt2}{2}\varepsilon\dot\phi(\mu+3p)$}\\\hline\hline
Set & $\mu$ & $p$ & $\gamma$ & $\mu_\phi$ & $p_\phi$ & $\gamma_\phi$ 
		& $\delta$ \\
\hline 
$L^\pm_{(\pm)}$ & $0$ & $0$ & $-$ & $\third t_*^{-2}$ & 
	$\mu_\phi$ & $2$ & $0$ \\
$C^\pm$ & $2t_*^{-2}$ & $-\mu$ & $0$ & $t_*^{-2}$ & $\mu_\phi$ 
	& $2$ & $-4t^{-3}$ \\
$S_1^\pm$ & $\frac{5}{12}t_*^{-2}$& $p$ & $2$ 
	& $\frac{11}{12}t_*^{-2}$ & $-\frac{5}{12}t_*^{-2}$ & $\frac{6}{11}$ 
	& $\frac{5}{6} t_*^{-3}$ \\
$R/A$ & $0$ & $0$ & $-$ & $0$ & $\half t_*^{-2}$ & $-$ & $0 $\\ 
$J^\pm$ & $\Lambda$ &$-\mu$ & $0$ & $-\half\Lambda$ & $-\mu_\phi$ 
	& $0$ & $0$\\ \hline
\end{tabular}
\end{center}
\caption[{\em Matter Terms for Model B}]{{\em The matter terms ($\mu$,
$p$, $\gamma$) as well as $\mu_\phi$, $p_\phi$, $\gamma_\phi$ and
$\delta$ for each of the equilibrium sets derived in scenario
B.}\label{model_B_III}}
\end{table}

Similar to scenario A, in scenario B solutions asymptote
to models in which the matter field is either absent ($\mu=p=0$) or
is a false vacuum  ($\gamma=0$), although the
equation of state is reversed to that in scenario A for the solutions
at the asymptotic equilibrium points.  Here, the heteroclinic orbits
asymptotically represent solutions in which the matter field becomes
negligible in any part of the cycle.

\chapter{String Models IV: Ramond--Ramond Term ($\Lambda_{\rm
M}=0$)\label{Qsection}}

In this chapter, the field equations (\ref{rr}) are considered when
$Q\neq0$.  Of primary interest is how these equations evolve for flat
FRW models with neither a central charge deficit nor a $\Lambda_{\rm
M}$ term.  However, it is straightforward to perform a perturbation
analysis to the curvature term and the $\Lambda$ term, allowing a
comment to be made about the stability of these models to curvature or
$\Lambda$ perturbations.  It should be stressed that within the string
context, the central charge deficit does not naturally arise in the
type IIA supergravity theory where the Ramond--Ramond term arises, but
it will be included in section \ref{Q_5D} as a perturbation parameter,
and as a source for the exponential potential in the Einstein frame.
If $k<0$ and $\Lambda>0$, the variables used lead to a compact
five-dimensional phase space and so the global qualitative analysis
shall be restricted to this case.  Note that all the local results
(e.g., stability of equilibrium points) obtained will also be true for
the case $k>0$ (as well as $\Lambda>0$).

The chapter is organized as follows.  In section
\ref{Governing_Equations_IV}, the field equations (\ref{rr}) and 
(\ref{rrfriedmann}) are reexamined with $\Lambda_{\rm M}=0$.  Section
\ref{Analysis_IV} proceeds with the analysis of the equations.  The
chapter ends with a summary section and a section which discusses the
corresponding solutions and asymptotic behaviour in the Einstein
frame.  Again, this chapter is primarily confined to the Jordan frame
(except the final section), and so the index ``(st)'' shall be omitted
to save notation (but must be introduced again in the final section
when both frames are discussed).

\section{Governing Equations \label{Governing_Equations_IV}}

For $Q\neq0$, $\Lambda=\Lambda_{\rm M}=0$, there exists the solution
not found in the previous chapters:
\index{equilibrium sets!$T$}
\index{exact solutions!curved dilaton--moduli--vacuum}
\begin{eqnarray}
a & = & a_0 \left(\pm \varsigma t\right), \\
\nonumber
\hat{\Phi} & = & \hat{\Phi}_0 -\frac{4}{5}\ln\left(\pm \varsigma t\right),  \\
\nonumber 
\beta&=&\beta_0 + \frac{1}{5}\ln\left(\pm \varsigma t\right), \\
\nonumber
\sigma&=&\sigma_0,\\
\label{CS-soln}
k&=&-\frac{3}{2}Q^2e^{\hat{\Phi}_0+2\alpha_0-6\beta_0},
\end{eqnarray}
where $\varsigma = \frac{5}{2\sqrt 7}Q
\mbox{exp}\left\{\half(\hat{\Phi}_0 -6\beta_0)\right\}$ and
$\{a_0=e^{\alpha_0},\hat{\Phi}_0,\beta_0,\sigma_0\}$ are constants.
The ``$-$'' solutions correspond to $t<0$ whilst the ``$+$''
correspond to the $t>0$ solutions.  These solutions are represented in
the text by the equilibrium points $T$.  Since the non-negligible
terms are the same as equation (\ref{dmv}) with the addition of a curvature
term, this solution will be referred to as the ``curved
dilaton--moduli--vacuum'' solutions, although it does {\em not} reduce
to (\ref{dmv}) when $k=0$.

By introducing a new time variable via 
\begin{equation}
  \frac{d}{d\eta} = {\rm e}^{-\frac{1}{2}(-6\beta + \varphi +
  3\alpha)}\frac{d}{dt},
\end{equation}
the field equations (\ref{rr}) may be written
\beqn
\mainlabel{rrQ}
  \alpha'' &=& \frac{1}{2}\alpha'\varphi' +3\alpha'\beta' -
  \frac{9}{2} \left(\alpha'\right)^2 + \left( \varphi' \right)^2 -
  6\left(\beta'\right)^2 +4k{\rm e}^{-(5\alpha +
    \varphi-6\beta)} \nonumber \\
	&&\qquad \qquad - 2\Lambda e^{-(3\alpha + \varphi -6\beta)} +
  \frac{3}{4}Q^2, \\  
  \varphi'' &=& 3\left(\alpha'\right)^2 + 6\left(\beta'\right)^2 +
  \frac{1}{4}Q^2 - \frac{1}{2}\left( \varphi' \right)^2 +
  3\beta'\varphi' - \frac{3}{2}\alpha'\varphi', \\
  \beta'' &=& \frac{1}{2}\beta'\varphi' +
  3\left(\beta'\right)^2 - \frac{3}{2}\alpha'\beta' +
  \quart Q^2,
\eeqn
and the Friedman constraint may be written
\be
\label{QFriedmann}
\frac{1}{2}\rho e^{-3\alpha -\varphi +6\beta}= \left( \varphi' \right)^2 
	- 3\left(\alpha'\right)^2 -  6\left(\beta'\right)^2 
	+ 6ke^{-(5\alpha +\varphi-6\beta)} - 2\Lambda e^{-(3\alpha 
	+ \varphi -6\beta)}- \half Q^2.  
\ee

Furthermore, these equations may be further reduced by defining the variables
\begin{equation}
\label{Q_def}
  x = \frac{\sqrt{3}\alpha'}{\varphi'} \ , \ y =
  \frac{\sqrt{6}\beta'}{\varphi'}\ , \ z = \frac{\half Q^2}{(\varphi')^2}\ , \
  u = -\frac{6ke^{-(5\alpha + \varphi-6\beta)}}{(\varphi')^2}\ , \
  v =\frac{2\Lambda e^{-(3\alpha + \varphi -6\beta)}}{(\varphi')^2} \ , 
\end{equation}
and a new reduced time variable, $\tau$,
\begin{equation}
  \frac{d}{d\tau} = \left(\varphi'\right)^{-1}\frac{d}{d\eta},
\end{equation}
(where it is assumed that $\varphi'>0$) to yield the equations
\beqn
\mainlabel{dxdtau_main}
\label{dxdtau}
  \frac{dx}{d\tau} &=& \left(1-x^2-y^2-z\right)(x+\sqrt{3}) +
  \frac{1}{2}z\left(x-\sqrt{3}\right)- \frac{2}{\sqrt{3}}u - \sqrt{3}v, \\ 
\label{dydtau}
  \frac{dy}{d\tau} &=& \left(1-x^2-y^2-z\right)y + \frac{1}{2}z\left(y +
    \sqrt{6}\right), \\
  \frac{dz}{d\tau} &=& z\left[z-1-\sqrt{6}y + \sqrt{3}x +
    2\left(1-x^2-y^2-z\right)\right], \\
  \frac{du}{d\tau} &=& -u\left(\frac{2}{\sqrt{3}}x +
    z+2x^2+2y^2\right), \\
	\label{dvdtau}
  \frac{dv}{d\tau} &=& -v(2x^2+2y^2 + z).
\eeqn
The Friedmann equation (\ref{QFriedmann}) is now written 
\begin{equation}
  \frac{1}{2}\rho e^{-3\alpha -\varphi +6\beta} = 1-x^2-y^2-z-u-v,
\end{equation}
so that it is apparent for $\Lambda>0$, $k<0$ that the variables $0
\leq \left\{ x^2, y^2, z , u, v\right\} \leq 1$ are indeed bounded
(this is apparent from equation (\ref{QFriedmann}); the right side is
positive definite and therefore $(\varphi')^2$ must be larger than the
other terms if $\Lambda>0$, $k<0$).  From equation (\ref{dvdtau}) it is
apparent that $v$ is a {\em monotonically decreasing function}.

\section{Analysis\label{Analysis_IV}}

The equilibrium sets to the system (\ref{dxdtau_main}) and their
corresponding eigenvalues are
\beqn
\mainlabel{main_eigens}
L^+: && x^2 + y^2=1, z=u=v=0; \nonumber \\ &&
	\left(\lambda_1,\lambda_2,\lambda_3,\lambda_4,\lambda_5\right)=\nonumber
	\\ && 
	\left(0,-2[1+\sqrt3 x],[\sqrt3x-\sqrt6y-1],-\frac{2}{\sqrt3}
	\left[x+\sqrt3\right],-2 \right), \nonumber \\ && \label{Qdmv} \\
S^+: && x=-\frac{1}{\sqrt3}, y=z=0, u=\frac{2}{3}, v=0; \nonumber \\ &&
	\left(\lambda_1,\lambda_2,\lambda_3,\lambda_4,\lambda_5\right)=
	\left(\frac{4}{3},-\frac{2}{3},\frac{2}{3},\frac{2}{3},-\frac{2}{3}
	\right), \\ \nonumber \\
C^+: && x=y=z=u=0, v=1; \nonumber \\ &&
	\left(\lambda_1,\lambda_2,\lambda_3,\lambda_4,\lambda_5\right)=
	\left(1,0,1,1,0\right),\\ \nonumber \\
T: && x=-\frac{5\sqrt3}{19}, y=-\frac{\sqrt6}{19}, z=\frac{28}{361},
	u=\frac{252}{361}, v=0; \nonumber \\ &&
	\left(\lambda_{1 \atop 2},\lambda_3,\lambda_4,\lambda_5\right)=
	\left(\frac{1}{19}\left[7\pm i\sqrt{119}\right],
	\frac{14}{19}, \frac{20}{19}, -\frac{10}{19}\right).
\eeqn
The equilibrium set $L^+$ represent the $\dot\varphi>0$
dilaton--moduli--vacuum so\-lu\-tions (\ref{dmv}) and are sinks in
this five dimensional set for $x>-\frac{1}{\sqrt3}$ and $\sqrt2
y>x-\frac{1}{\sqrt3}$.  The one zero eigenvalue results from the fact
that $L^+$ is a {\em line} of equilibrium points.  The equilibrium
point $S^+$ represents the Milne solution (\ref{curv_drive}) and is a
saddle.  The equilibrium point $C^+$ is the linear dilaton--vacuum
solution and will be shown to be a source in the five-dimensional
system in section
\ref{Q_5D}.  Finally, the equilibrium point $T$ represents a new solution, 
referred to as the ``curved dilaton--moduli--vacuum'' solution
(\ref{CS-soln}) due to its similarity to the dilaton--moduli--vacuum
solutions, and is a source in the four--dimensional system $\Lambda=0$
but a saddle in the full five--dimensional system.

\subsection{Two-Dimensional Invariant Set, $z=u=v=0$ ({$Q=k=\Lambda=0$})\label{Q_2Da}}

\begin{figure}[htp]
  \centering
   \includegraphics*[width=3in]{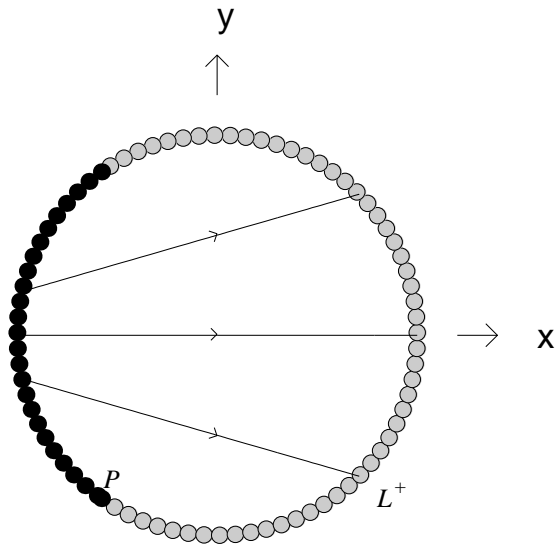}
  \caption[{\em Phase portrait for $Q=k=\Lambda=0$}]{{\em Phase portrait of the system
(\ref{dxdtau_main}) for $Q=k=\Lambda=0$.  Note that $L^+$ represents a line of equilibrium points.  See also caption to figure \ref{fNSpK0} on page \pageref{fNSpK0}. } \label{fQ_2Da}}
 \end{figure}

The equilibrium points $C^+$, $S^+$ and $T$ do not exist in this
invariant set, and therefore the line $L^+$ is the only equilibrium
set, with eigenvalues $\lambda_1$ and $\lambda_2$ given in equation
(\ref{Qdmv}).  The trajectories in this invariant can be completely
solved:
\begin{equation}
  y = \frac{y_0(x+\sqrt{3})}{x_0+\sqrt{3}}\ ,
\end{equation}
where $(x_0, y_0)$ is the initial point of the orbit. The function $x$
is a {\em monotonically increasing} function in this invariant set.  Figure
\ref{fQ_2Da} depicts this two-dimensional phase plane.  The point $P$
in this portrait corresponds to the point on the line in which both
eigenvalues are zero.  In the three-dimensional ($\{x,y,z\}$) analysis
below, it will be shown to be a source and so it is a source here,
albeit non-hyperbolic.

\subsection{Two-Dimensional Invariant Set, $z=1-x^2-y^2$, $u=v=0$ ({$\rho=k=\Lambda=0$})\label{Q_2Db}}

\begin{figure}[htp]
  \centering
   \includegraphics*[width=3in]{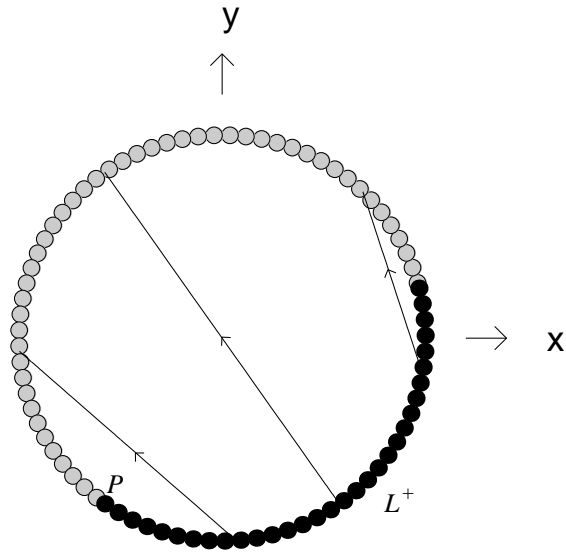}
  \caption[{\em Phase portrait for $\rho=k=\Lambda=0$}]{{\em Phase
portrait of the system (\ref{dxdtau_main}) for $\rho=k=\Lambda=0$.
Note that $L^+$ represents a line of equilibrium points.  See also
caption to figure \ref{fNSpK0} on page \pageref{fNSpK0}. }
\label{fQ_2Db}}
 \end{figure}

Again, the line $L^+$ is the only equilibrium set, but this time with
eigenvalues $\lambda_1$ and $\lambda_3$ given in equation
(\ref{Qdmv}).  The point $P$ on the line $L^+$ is again
non-hyperbolic, but is a source since it is a source in the
three--dimensional system $k=\Lambda=0$.  The trajectories in this
invariant can also be completely solved:
\begin{equation}
  y = -\sqrt{6} +
  \frac{(y_0+\sqrt{6})(x-\sqrt{3})}{x_0-\sqrt{3}}\ ,
\end{equation}
where $(x_0, y_0)$ is again the initial point of the orbit. In this
invariant set, $x$ is {\em monotonically decreasing} while $y$ is {\em
monotonically increasing}.  Figure \ref{fQ_2Db} depicts this two-dimensional 
phase plane.  

\subsection{Three-Dimensional Invariant Set, $u=v=0$ ($k=\Lambda=0$)\label{Q_3D}}

The equilibrium points $C^+$, $S^+$ and $T$ do not exist in this
invariant set, and therefore the line $L^+$ is the only equilibrium
set, with eigenvalues $\lambda_1$, $\lambda_2$ and $\lambda_3$ given in
equation (\ref{Qdmv}).  From these eigenvalues, it has been determined that
this line is a sink for $x>-\frac{1}{\sqrt3}$ and $\sqrt2 y>x-\frac{1}{\sqrt3}$.  The
lines $x=-\frac{1}{\sqrt3}$ and $\sqrt2 y=x-\frac{1}{\sqrt3}$ intersect on $L^+$ at
the point $P: (x,y)=\left(-\frac{1}{\sqrt3},-\sqrt\frac{2}{3}\right)$,
at which all three eigenvalues are zero.  It shall now be demonstrated
that this point is indeed a source to the three-dimensional system.

\begin{figure}[htp]
  \centering
   \includegraphics*[width=5in]{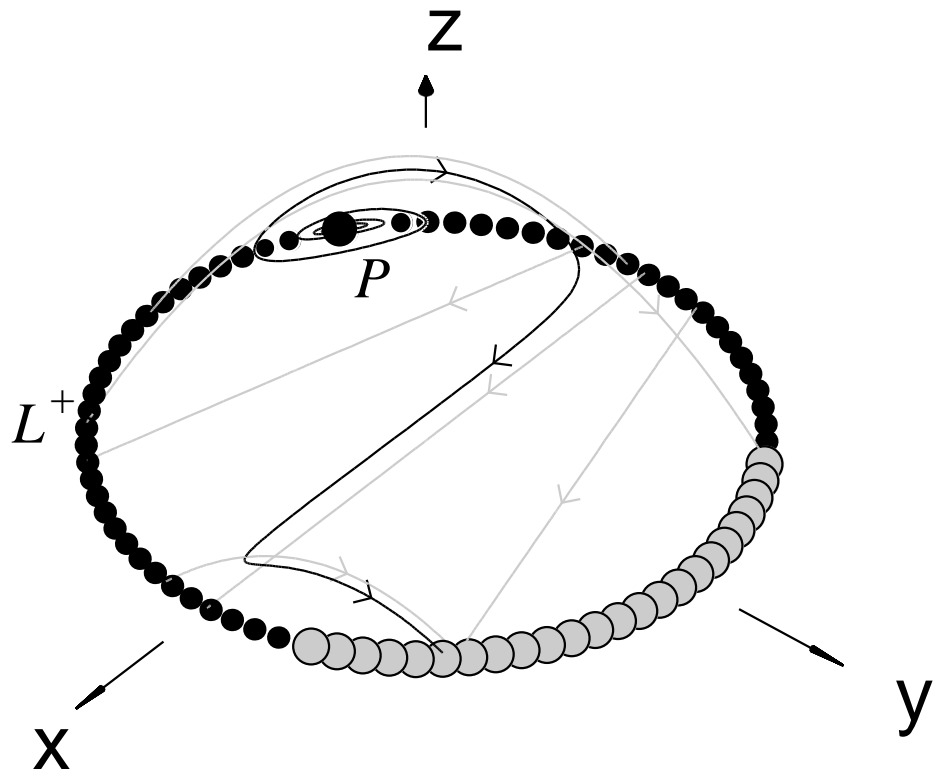}
  \caption[{\em Phase diagram for $Q\neq0$}]{{\em Phase diagram of the
system (\ref{dxdtau_main}) ($Q\neq0$).  Note that $L^+$ represents a
line of equilibrium points.  Note that the trajectories in figures
\ref{fQ_2Da} and \ref{fQ_2Db} are depicted in this figure along $z=0$
and $z=1-x^2-y^2$, respectively.  See also caption to figure
\ref{fNSpK0} on page \pageref{fNSpK0}. } \label{Qfig}} \end{figure}

\begin{figure}[htp]
  \centering
   \includegraphics*[width=5in]{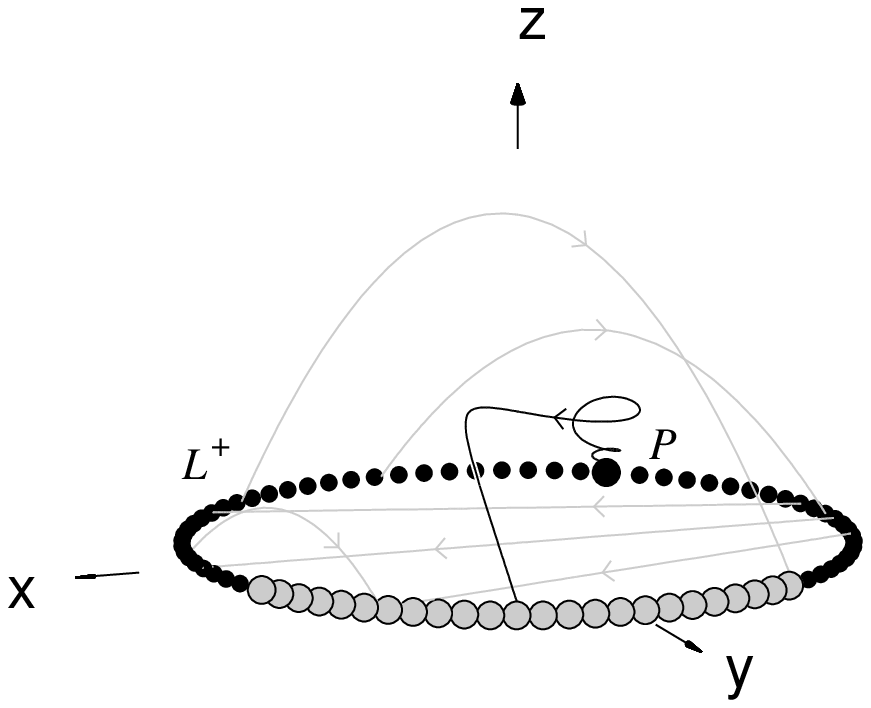}
  \caption[{\em Phase diagram for $Q\neq0$: alternative view}]{{\em An alternative trajectory within the phase space of the system (\ref{dxdtau_main}) ($Q\neq0$).  Note that $L^+$ represents a line of equilibrium points.  See also caption
to figure \ref{fNSpK0} on page \pageref{fNSpK0}. } \label{Qfig2}}
 \end{figure}

From equations (\ref{dxdtau}) and (\ref{dydtau}) for $u=v=0$, is can be shown that
\be
\label{pseudomono}
\frac{d}{d\tau}\left(x+\sqrt2y+\sqrt3\right) = \left(x+\sqrt2 y+\sqrt3\right)
	\left(1-x^2-y^2-\half z\right),
\ee
and hence $x+\sqrt2y+\sqrt3$ is a {\em
monotonically increasing} function.  Note that the right hand side of
(\ref{pseudomono}) is zero only at $P$ and positive everywhere else in
the phase space.  Also, the line tangent to the unit circle at $P$ is
$x+\sqrt2 y=-\sqrt3$ and so this monotonic function contains this
tangent line as part of its class.  Hence, the monotonic function
asymptotically approaches the tangent line to the unit circle at point
$P$ into the past and hence $P$ is a source for the three-dimensional
system.  Numerical calculations have verified this point to be a
source and the three-dimensional system is depicted in figures
\ref{Qfig} and \ref{Qfig2} (the first figure depicts a trajectory which stays near the boundary, whilst the second depicts a more internal trajectory).

\subsection{Four- and Five-Dimensional System and Perturbations\label{Q_5D}}

The fact that $v$ is monotonically decreasing suggests that $C^+$ is a
source in the five-dimensional system.  Similar to the arguments of
subsection \ref{NSpKn} (page \pageref{NSpKn}), solutions asymptote
into the past towards larger values of $v$, the maximum value of which
(i.e. $v=1$) coincides with the equilibrium point $C^+$ and so it is a
source for the system (\ref{dxdtau_main}).  In the invariant set
$v=0$, the corresponding eigenvalues to the equilibrium point are
$\lambda_1$, $\lambda_2$, $\lambda_3$ and $\lambda_4$ found in
equations (\ref{main_eigens}) and hence the point $T$ represents a
source in the four-dimensional set $v=0$.  Unfortunately, due to the
lack of any monotonic function found in this invariant set, the
possibility of recurring orbits cannot be excluded and so it is not
possible to say that the attracting set on $L^+$ is the sole
attracting set of the four-dimensional system.

Although the compactness of the phase space depends on the fact that
$k<0$ and $\Lambda>0$, arbitrary signs of $k$ and $\Lambda$ can be
assumed and then (\ref{main_eigens}) is used in order to determine the
local stability of the three-dimensional set $u=v=0$.  Specifically,
the eigenvalues $\lambda_4$ and $\lambda_5$ are those related to the
eigenvectors extending into the $u$ and $v$ directions, and for $L^+$
these eigenvalues are negative.  Hence, the line $L^+$ is locally
attracting trajectories from the $u\neq0$ and $v\neq0$ portion of the
phase space and so the point $P$ is a only a saddle point in the
extended space (for either sign of $k$ and $\Lambda$), whereas the
sink portion of $L^+$ (i.e. points on $L^+$ for $x>-\frac{1}{\sqrt3}$ and
$\sqrt2 y>x-\frac{1}{\sqrt3}$) is a sink for the extended space (for either
sign of $k$ and $\Lambda$).  Therefore, the dilaton--moduli--vacuum
solutions for $3\dot\alpha>-\dot\varphi$ and
$6\dot\beta>3\dot\alpha-\dot\varphi$ will be attracting solutions for
models containing either a curvature term or a central charge deficit.

\section{Summary of Analysis in the Jordan Frame\label{summary_IV}}

This discussion begins with two tables; Table \ref{term_asymp} lists
which terms are the dominant variables for each equilibrium set as
well as the deceleration parameter, $q$, for the corresponding model
and Table \ref{points_asymp} lists the attracting behaviour of the  equilibrium sets.
\begin{table}[htb]
\begin{center}
\begin{tabular}{|c|cccccccc|c|c|}\hline
Set & \multicolumn{8}{c|}{Dominant Variables}&$q$&H \\
\hline \hline
$L^+$ {\tiny($t<0$)} & $\alpha$ &$\hat\Phi$ & $\dot\beta$ &&&&&
	&$-(1+h_0)h_0^{-1}$ &$h_0(-t)^{-1}$\\ \hline
$C^+$     & & $\hat\Phi$ &&&& $\Lambda>0$ &&&$0$&$0$\\
$S^+$     & $\alpha$ &&&& $k<0$ &&&&$0$& $t^{-1}$\\ \hline
$T$   & $\alpha$ &$\hat\Phi$&$\dot\beta$&& $k<0$ &&&$Q^2$&$0$ 
	&$t^{-1}$\\
\hline
\end{tabular}
\end{center}
\caption[{\em Dominant Variables and $q$ for Equilibrium Sets}]{{\em The
dominant variables for each equilibrium set as well as the equilibrium
set's deceleration parameter, $q$, and Hubble parameter $H$.
Inflation occurs when $q<0$ and $H>0$, whereas ``deflation'' occurs
for $q>0$ and $H<0$.  Note that the only ``anisotropic'' solutions are
represented by the lines $L^\pm$, except when
$h_0^2=\third$.}\label{term_asymp}}
\end{table}

\begin{table}[ht]
\begin{center}
\begin{tabular}{|cccc||c|c|c|}\hline
\multicolumn{4}{|c||}{Terms Present} & Early & Intermediate & Late \\
\hline \hline
$\dot\beta$ &       &             & $Q^2$ & $L^+(P)$ & $L^+$             & $L^+$\\
$\dot\beta$ & $k<0$ &             & $Q^2$ & $T$   & $S^+$, $L^+$      & $L^+$\\
$\dot\beta$ & $k<0$ & $\Lambda>0$ & $Q^2$ & $C^+$ & $T$, $S^+$, $L^+$ & $L^+$\\
\hline
\end{tabular}
\end{center}
\caption[{\em Summary of Equilibrium Points for Models
considered}]{{\em Summary of the early-time, intermediate, and
late-time attractors for the various models examined.  Note that
$\dot\alpha$, $\hat\Phi$ and $\dot\sigma$ are present in every
model.  In the first row, $L^+(P)$ refers to the equilibrium point $P$ which lies in the line $L^+$.}\label{points_asymp}}
\end{table}

For the three-dimensional system ($k=\Lambda=0$), a monotonic
functions had been established which precludes the existence of
recurrent or periodic orbits, thereby allowing the
early-time and late-time conclusions of these models to be based upon the
equilibrium sets of the system.  For the four dimensional case
($\Lambda=0$), a monotonic function could not be determined.  Hence,
although the equilibrium points in this phase space are still
appropriate attracting sets, there may exist periodic orbits in
the higher dimensions which may act as asymptotic attracting sets.
Therefore, it cannot be asserted, for instance, that the four dimensional
phase space asymptotes towards the three dimensional phase space at
late times.  A monotonically decreasing function exists for the
five-dimensional phase space and hence this phase space {\em does}
asymptote to the four dimensional phase space ($\Lambda=0$) at late
times.

In all cases, the $\dot\varphi>0$ dilaton--moduli--vacuum solutions
(\ref{dmv}) act as a late-time attractor, and as a source (for one
particular point on $L^+$ where $h_0=-\frac{1}{3}$).  The source
represents a ten--dimensional isotropic solution
($\dot\alpha=\dot\beta$).  In the presence of a curvature term or a
central charge deficit, the late-time behaviour {\em may} be the same
(again when $k<0$ there was no monotonic function found to ensure that
this is true), but evolve from some model in which $k$ or $\Lambda$ is
dominant [for example, if $k<0$ and $\Lambda>0$ then possibly either
the $\dot\varphi>0$ linear dilaton--vacuum solution (\ref{static}) or
the curved dilaton--moduli--vacuum solution (\ref{CS-soln})].

As discussed above, the time-reversed dynamics of the above class of
models is deduced by interchanging the sources and sinks and
reinterpreting expanding solutions in terms of contracting ones, and
vice--versa. Thus, the late-time attractor for the time--reversed
system is the expanding, isotropic, ten--dimensional cosmology located
at point $P$ and all late--time attracting solutions in the post-big bang regime are
{\em not} inflationary.

\section{Exact Solutions in the Einstein Frame\label{Einstein_Frame}}

For $Q\neq0$, $\Lambda=\Lambda_{\rm M}=0$, the curved
dilaton--moduli--vacuum solution transforms to the Einstein frame to
the solution
\index{equilibrium sets!$T$}
\index{exact solutions!curved dilaton--moduli--vacuum}
\begin{eqnarray}
T: && a^* = a^*_0 \left|t_*\right| \qquad \Rightarrow \qquad 
	\up{sf}H=t_*^{-1} \qquad \Rightarrow \qquad \up{sf}q=0, \\
\nonumber
&& e^{\varepsilon\sqrt2\phi} = e^{\varepsilon\sqrt2\phi}\left[ 
	{\cal A}^4\left|\varsigma t_*\right|^\frac{4}{7}\right]^{-1},  \\
\nonumber 
&& e^\beta=e^{\beta_0} \left[{\cal A}\left|\varsigma t_*\right|^\frac{1}{7}
	\right],\\
\nonumber
&& \sigma = \sigma_0,\\
&& k=-\frac{6}{7}a^{*2}_0,
\label{E_CS-soln}
\end{eqnarray}
where ${\cal A} =
\left[\frac{7}{5}\exp\{\varepsilon\phi_0/\sqrt2\}\right]^\frac{1}{7}$,
$a_0^*=\frac{7}{5}\varsigma a_0$ and $\varsigma = \frac{5}{2\sqrt 7}Q
\mbox{exp}\left\{\half(\hat{\Phi}_0 -6\beta_0)\right\}$.
The analysis in the Jordan frame
shows that this solution arises only as an early-time attractor in the
presence of negative curvature, otherwise it is a saddle in a
higher-dimensional analysis (including a central charge deficit) or
does not exist at all when no curvature is present.

In addition to this solution, this chapter also has as equilibrium
sets which represent solutions found in earlier chapters; the
solutions in the Einstein frame for these equilibrium points have also
been given previously: the linear--dilaton--vacuum solution (equation
(\ref{E_dmv}) on page \pageref{E_dmv}), the Milne solutions (equation
(\ref{E_curv_drive}) on page \pageref{E_curv_drive}), and the linear
dilaton--vacuum solution (equation (\ref{E_static}) on page
\pageref{E_static}).

There are no inflationary models in the Einstein frame.

\subsection{Mathematical Equivalence to Matter Terms}

When considering the mathematical equivalence to matter fields in the
Einstein frame, the case $Q\neq0$ differs from the previous cases.  In
particular, the presence of the $Q$ term does {\em not} allow for an
equivalence to occur between certain Bianchi models with or without a
modulus term and curved FRW models with a modulus field.  Therefore,
the $\beta$ term is explicitly the modulus field and not a combination
of the modulus field with shear terms (as described in equation
(\ref{Betadefine}) on page \pageref{Betadefine}).  However, in this
case, matter fields can be constructed in the Einstein frame from both
the modulus field and the axion field, leading to the conservation
equations
\beqn
\mainlabel{E_conserves}
\dot\phi\left(\ddot\phi+3\up{sf}H\dot\phi +\frac{dV}{d\phi}\right) 
	 &=& -\delta = 
	\sqrt2\varepsilon\dot\phi\left[\left(\mu_2-\mu_1\right)+
	\left(p_2+p_1\right)\right],\label{E_phi_conserves}\\
\dot\mu_1 + 3\up{sf}H\left(\mu_1+p_1\right) &=& 0\label{E_phi_conserves_a}\\
\dot\mu_2 + 3\up{sf}H\left(\mu_2+p_2\right) &=& \delta = 
	-\sqrt2\varepsilon\dot\phi\left[\left(\mu_2-\mu_1\right)+
	\left(p_2+p_1\right)\right],\label{E_phi_conserves_b}
\eeqn
where
\beqn
\mu_1 & \equiv &  3\dot\beta_m^2 +{\cal U}, \\
p_1 & \equiv & 3\dot\beta_m^2 -{\cal U},\\
\mu_2&=&p_2=\quart \dot{\sigma}^2 e^{2\sqrt2\varepsilon\phi} \qquad (\gamma_2=2),
\eeqn
and 
\be
{\cal U} = \quart Q^2 e^{-6\beta_m +2\sqrt2\varepsilon\phi}.
\ee
Again, the definition 
\beqn
\mu_\phi & = & \half\dot\phi^2+V,\\
  p_\phi & = & \half\dot\phi^2-V,
\eeqn
where
\be
V = \Lambda e^{\sqrt2\varepsilon\phi} \qquad (k^2=2),
\ee
allows a comparison between
$\gamma_\phi\equiv(\mu_\phi+p_\phi)/\mu_\phi$ and the $\gamma$ of each
matter field.

Note that the first matter field ($\mu_1$, $p_1$) will not represent
 barotropic matter with a linear equation of state [$p=(\gamma-1)\mu$] in general, although they are at the equilibrium sets. Table
\ref{model_A_IV} lists $\{\mu_1, p_1, \gamma_1, \mu_\phi, p_\phi, \gamma_\phi, \mu_2, \delta\}$ for each equilibrium set.

\begin{table}[htb]
\begin{center}
\begin{tabular}{|c|cccccccc|}\hline
\multicolumn{9}{|c|}{\bf $V=\Lambda e^{\sqrt2\varepsilon\phi}$, \qquad
	${\cal U}=\quart Q^2 e^{-6\beta_m +2\sqrt2\varepsilon\phi}$, \qquad
	$\delta=-\sqrt2\varepsilon\dot\phi\left[\left(\mu_2-\mu_1\right)+
	\left(p_2+p_1\right)\right]$} \\ \hline
\hline
Set & $\mu_1$ & $p_1$ & $\gamma_1$ & $\mu_\phi$ & $p_\phi$ & $\gamma_\phi$ 
	& $\mu_2=p_2$ & $\delta$ \\
\hline 
$L^+$ & $\frac{2(1-3h_0^2)}{9(1+h_0)^2}t_*^{-2}$ & $\mu_1$ & $2$ & $\frac{(3h_0+1)^2}{9(1+ h_0)^2}t_*^{-2}$ & 
	$\mu_\phi$ & $2$ & 0  & $0$ \\
$L^+(P)$ & $t_*^{-2}$ & $\mu_1$ & $2$ & $0$ & 
	$0$ & $-$ & 0  & $0$ \\
$S^+$ & $0$ & $0$ & $-$ & $0$ & $0$ & $-$ & $0$ & $0$ \\
$C^+$ & $0$ & $0$ & $-$ & $3t_*^{-2}$ & $-t_*^{-2}$ & $\frac{2}{3}$ & $0$ & $0$ \\
$T$   & $\frac{10}{49}t_*^{-2}$ & $-\frac{4}{49}t_*^{-2}$ & $\frac{3}{5}$ & $\frac{4}{49}t_*^{-2}$ & $\mu_\phi$ & $2$ & $0$ & $-\frac{8}{49}t_*^{-3}$ \\
\hline
\end{tabular}
\end{center}
\caption[{\em Matter Terms for $Q\neq0$}]{{\em The matter terms ($\mu_1$,
$p_1$, $\gamma_1$, $\mu_2$) as well as $\mu_\phi$, $p_\phi$, $\gamma_\phi$ and
$\delta$ for each of the equilibrium sets.  Note that $p_2=\mu_2$ and hence $\gamma_2=2$ for all sets.}\label{model_A_IV}}
\end{table}

As is evident from table \ref{model_A_IV}, the second fluid ($\mu_2$)
approaches a vacuum solution asymptotically (either to the future or
the past) and so arises only at intermediate times in which case it is
always a stiff fluid.  A late-time attractor is the set $L^+$ (this is
the only late-time attractor when $k=\Lambda=0$).  Hence, to the
future the first fluid asymptotes towards a stiff equation of state,
as does the scalar field contribution.  This is the only example in
which a matter scaling solution arises from the exact solutions of the
string theory.  For $P$, which represents an early-time attracting
solution in the three-dimensional set, the only energy contribution is
from the first fluid.  For the past attractor represented by $C^+$
(five--dimensional phase space), all matter fields asymptote to a
vacuum solution, whereas for the past attractor represented by $T$
(four-dimensional phase space) the first fluid asymptotes towards a
model of negative pressure ($\gamma_1=\frac{3}{5}$).  In the
three-dimensional set, the point $P$ (belonging to the set $L^+$) is
an early-time attractor, and the corresponding matter fields are all
vacuum.

\chapter{{\em Apodeixis}}

\section{Summary}

The goal of this thesis was to ascertain the asymptotic properties of
cosmological models containing a scalar field, both in the Einstein
frame (general relativity minimally coupled to a scalar field) and the
Jordan frame (Brans-Dicke theory, scalar-tensor theory and string
theory).  A formal mathematical equivalence between the two frames was
first discussed, explicitly deriving exact solutions in the
Brans-Dicke theory from asymptotically stable solutions obtained from
a model containing a scalar field with an exponential potential in the
Einstein frame.  This equivalence was utilized in subsequent chapters
to comment on the asymptotic properties of models in both frames.  Furthermore,
it naturally provides a form for an interaction term studied in the
Einstein frame for scalar field models with an exponential potential.  In general, the
isotropization of the models in the two frames are identical; should a
model isotropize in one frame, it will do so in the other.  However,
inflation is frame dependent and through these mathematical
transformations the asymptotic inflationary behaviour in the other
frame can be calculated.

In the Einstein frame, scalar fields with an exponential potential
(and matter terms) were considered in several contexts.  In all cases,
the field equations could be written in terms of first order,
non-linear, autonomous, ordinary, differential equations and the
variables chosen led to a compact phase space; hence, a complete
qualitative analysis could be performed.  First, the flat FRW matter
scaling solutions were subjected to curvature and shear perturbations
to examine their stability.  Subsequently, a more detailed analysis
was performed which examined the asymptotic properties of the
spatially-homogeneous Bianchi class B models containing a scalar field
with an exponential potential and matter.  Monotonic functions were
found in many instances, in which cases the global asymptotic results
could be fully determined.  When monotonic functions could not be
found, plausible conjectures were presented about the asymptotic
properties.  Finally, within the context of flat FRW models,
interaction terms were introduced in order to determine if they had a
significant effect on the late-time dynamics, particularly with
regards to inflation.

The remainder of the thesis was devoted to certain string cosmologies,
within the context of both flat FRW models and a class of 
spatially-homogeneous Bianchi models.  Although these models are of
great physical interest in their own right, it was shown that they are
also related to general relativity containing a scalar field with an
exponential potential and matter with a non-linear equation of state,
thereby complimenting analyses in previous chapters.  In these chapters
the qualitative effects of the physical fields arising from
string theory, as well as the qualitative effects of curvature and
shear, were studied.  In each case, a choice of variables leading to
a compact phase space was established, thereby allowing a complete
analysis.  In all of the Bianchi models studied monotonic functions
were found allowing global results to be obtained.  For the flat FRW
models, monotonic functions were found in most cases, and plausible
conjectures were made for those cases in which a monotonic function
could not be found.

\section{Conclusions}

For the Bianchi class B models containing a scalar field with an
exponential potential and matter (and without an interaction term),
{\bf solutions generically asymptote into the past towards anisotropic
models in which matter is negligible and the scalar field is massless}
($V=0$); these solutions are a generalization of Jacob's anisotropic
Bianchi I solutions to include scalar fields (represented by the line
${\cal K}$ on page \pageref{general_end}).  Towards the future, {\bf
all models for $k^2<2$ asymptote towards an inflationary, isotropic
scalar field model in which matter is negligible}.  Since the Bianchi
types VI$_h$ and VII$_h$ are both a one parameter family of models
they represent an open class of models to which the other Bianchi
types are of zero measure; hence generic behaviour to these two types
represent generic behaviour for all Bianchi models.  For $k^2>2$,
Bianchi
type VI$_h$ models asymptote to either anisotropic solutions in which
matter is negligible (one asymptotic solution is represented by the
point $P_{\cal S}^\pm(VI_h)$ on page \pageref{BVIh_end2} and the other
by the line ${\cal L}_l(VI_h)$ on page \pageref{BVIh_end1}) or to the
anisotropic matter scaling solution (represented by the point ${\cal
A}_{\cal S}(VI_h)$ on page
\pageref{BVIh_end3}).  However, this class of models does not admit an isotropic subgroup and isotropization was not expected.  For $k^2>2$,  Bianchi type
VII$_h$ models generically asymptote towards a curved, isotropic model
in which the matter terms are negligible (either represented by point
$P_{\cal S}(V)$ on page \pageref{BVIIh_end2} or by the point
P$^{\pm}_{\cal S}(VII_h)$ on page \pageref{BVIIh_end}).  Therefore,
{\bf there exists an open set in the Bianchi class B models which {\em
do isotropize} to the future and which do not inflate asymptotically}.
Furthermore, the flat FRW matter scaling solutions are {\em not}
stable attracting solutions, which was proven both in the perturbation
analysis (chapter
\ref{scaling}) and in the Bianchi class B analysis (chapter
\ref{BianchiB}).  In these models, curvature can play a dynamic
r\^{o}le at late times (note that the none of Bianchi class B models
studied have positive spatially curvature).

It was shown that the presence of an interaction term can considerably
change the dynamics the models.  In particular, it was shown that when
forms for the interaction term discussed in this thesis are used then
the flat FRW matter scaling solutions ($\gamma_\phi=\gamma$) cannot be
represented by an equilibrium point and thus cannot be an asymptotic
solution in the models considered here.  However, the examples in this
thesis have shown that analogous solutions arise, in which both the
scalar field and matter
terms are non-negligible, but with $\gamma_\phi\neq\gamma$.
Furthermore, it has been shown to be  possible that {\bf the
power-law inflationary model need {\em not} be a late-time attracting
solution for $k^2<2$}; hence, it is possible to
have an inflationary model without driving the matter content to zero
at late times.  The examples given have shown that the stable solutions
(which are inflationary) for the same parameter value are 
spiral nodes representing an {\bf oscillating scalar field}. 

{\bf In the string models examined, there are {\em no} late-time
attracting solutions which inflate} in the post-big bang regime.
Furthermore, the curvature terms are generically dominant only at
intermediate times, although there are exceptions in which a
positively-curved model is an attractor (\ref{static_general}) to the
past and future.  Therefore, {\bf there is no flatness problem for the
negatively-curved string models}.  The ``many--bounce'' cosmologies
represented by heteroclinic sequences in the phase space exist only
when the curvature becomes negligible.  The axion field is dynamically
significant only at intermediate times.  When a negative central
charge deficit is present, it is dynamically negligible both at early
and late times.  When it is positive, it can be significant at early
and late times only in the case of positive curvature (i.e., the
solutions represented by (\ref{static_general})).  The constant
$\Lambda_{\rm M}$ can play a major r\^{o}le asymptotically; when it is
positive there exists a heteroclinic sequence in the zero--curvature
invariant set in which the dynamical significance of $\Lambda_{\rm M}$
is ``pseudo-cyclic'' in nature.  When both $\Lambda$ and $\Lambda_{\rm
M}$ are present, there are no asymptotic solutions in which both are
dynamically significant, and in general $\Lambda_{\rm M}$ dominates
asymptotically.  In the three-dimensional FRW models including a
three--form gauge potential ($Q\neq0$), it was shown that this
three-form potential is dynamically significant only at intermediate
times.

When the string models with $Q=0$ are mathematically transformed to
theories containing a scalar field with an exponential potential
(either $k^2=2$ when $\Lambda\neq0$ or $k^2=8$ when $\Lambda_{\rm
M}\neq 0$) and a fluid with a non-linear equation of state, {\bf there
are no equilibrium points which represent inflationary solutions}.  In
general, the matter terms will either approach a vacuum solution or a
false vacuum solution.  In the cases where $\Lambda$ and $\Lambda_{\rm
M}$ are separately considered, there are cases in which the matter can
approach linear equations of state asymptotically as either stiff
fluids ($\gamma=2$) or fluids with $\gamma <1$.  When $Q\neq 0$, the
string theory is mathematically equivalent to a theory containing a
scalar field with a $k^2=2$ exponential potential and two fluids, one
stiff ($\gamma=2$) and one with a non-linear equation of state.
Asymptotically, the latter fluid becomes negligible and the asymptotic
solution represents a matter scaling solution with
$\gamma_\phi=\gamma=2$.

Finally, within the string models considered, the variables were 
normalized by $\dot\varphi>0$ (explicitly, subsections \ref{NSpK0},
\ref{NSpKn}, \ref{RRpK0}, \ref{RRpKn} and \ref{NSpRRp}, as well as chapter
\ref{Qsection}); in such cases the dynamics for
$\dot\varphi<0$ are related to the dynamics studied in this thesis by a time reversal
and a reflection of the Hubble parameter ($\dot\alpha\rightarrow
-\dot\alpha$).  Such dynamics lead to the same conclusions that these
models do not inflate at late times and an open set of models will
asymptote towards flat spacetimes.  However, in the $\dot\varphi<0$ case
isotropization is more typical.\footnote{For instance, in such cases
solutions will asymptote into the past towards the ``$-$''
dilaton--moduli--vacuum solutions \eref{dmv}, and asymptote to the
future towards isotropic solutions: for $\Lambda>0$ and $\tilde K \leq
0$ towards the ``$-$'' linear--dilaton vacuum solutions \eref{static},
for $\Lambda_{\rm M} >0$ and $K= 0$ towards the ``$-$'' solution given
by \eref{newsol}, for $\Lambda_{\rm M} >0$ and $K< 0$ towards the
curvature dominated solution \eref{open_new}, and for
$\Lambda>0$ and $\Lambda_{\rm M}>0$ solutions asymptote from the
bouncing cosmologies represented by heteroclinic sequence towards the
``$-$'' linear--dilaton vacuum solutions \eref{static}.}  Hence, the
results quoted above apply to a more general class of models.

Note that the string models asymptote to the future towards
non-inflationary models.  This behaviour is also evident for the GR
scalar field Bianchi class B models for $k^2>2$. However, the string
models usually asymptote towards anisotropic models (although there
are cases in which an open set of solutions will isotropize), whereas it
has been demonstrated that there is an open class of GR scalar field
Bianchi class B models which isotropize.  A predominant difference
between the GR scalar field Bianchi class B models of chapter
\ref{BianchiB} and the string models of chapters \ref{string} - 
\ref{Qsection} (transformed into the Einstein frame) is
that the fluid in one case has a linear equation of state and the
other does not.  In fact, in the latter theory, the fluid terms
typically asymptote to equations of state which were excluded from the
stability analysis of chapter \ref{BianchiB} (e.g., $\gamma=0$ or
$\gamma=2$).

\section{Future Work}

The results obtained from the analysis involving the interaction terms
proved {\em very} interesting, and it is imperative to perform a more
detailed analysis.  For example, it would be important to determine if
physically motivated interaction terms could be found which could lead
to a non-inflationary late-time solution in which both the scalar
field and the matter terms are both non-negligible.  Another extension
would be to investigate how these interaction terms affect
isotropization using techniques in this thesis.  It would also be
interesting to study GR scalar field models with other potentials
(i.e., not exponential).

A much more important long-term goal would be to apply techniques
similar to those used in this thesis to determine the qualitative
properties of inhomogeneous models.  For instance, the r\^{o}le of
scalar fields in such models needs to be explored, perhaps by
exploiting the known transformations between the Einstein frame and
the Jordan frame (\ref{SFtoSTtrans}); e.g., in Billyard {\em et al.}
\cite{Billyard1999g} the asymptotic behaviour of
inhomogeneous G$_2$ models within the Brans-Dicke theory were derived
from known vacuum G$_2$ models containing a scalar field with an exponential
potential,
\cite{Ibanez1998a}, and it was shown that these models generically
homogenize into the future.

Finally, the techniques used in this thesis have permitted a
comprehensive analysis of the asymptotic properties of various string
cosmological models and can be extended to more general string
cosmologies and cosmologies in other fundamental theories of gravity.

\sappendix

\chapter{Brief Survey of Techniques in Dynamical Systems\label{OD}}

 The asymptotic states of various solutions of the EFEs (\ref{EFE}) are of 
special importance in the study of cosmology, as these represent possible 
states of the universe at {\it important} times - i.e. at early and late 
times.  Dynamical systems theory is especially suited to determining the possible 
asymptotic states, especially when the governing equations are a finite system of 
autonomous ODEs.  This section will review some of the results of dynamical 
systems theory which will be used throughout the thesis in the analysis of 
the solutions of the EFEs (\ref{EFE}).  

The following are definitions of terms in dynamical systems theory which will be used 
throughout the thesis:

\noindent
{\bf Definition 1}
A {\it singular point} of a system of autonomous, ordinary differential equations 
\begin{equation} \dot{x} = f(x)\label{DE}\end{equation}
 is a point $\bar{x} \in {\mbox{\boldmath R}}^n$ such that $f(\bar{x})=0$.

\noindent
{\bf Definition 2}
Let $\bar{x}$ be a singular point of the DE (\ref{DE}).  The point $\bar{x}$ is 
called a {\it hyperbolic} singular point if $Re(\lambda_i) \neq 0$ for all 
eigenvalues, $\lambda_i$, of the Jacobian of the vector field $f(x)$ evaluated 
at $\bar{x}$.  Otherwise the point is called {\it non-hyperbolic}.

\noindent
{\bf Definition 3}
Let $x(t)=\phi_a(t)$ be a solution of the DE (\ref{DE}) with initial condition 
$x(0)=a$.  The flow $\{g^t\}$ is defined in terms of the solution function 
$\phi_a(t)$ of the DE by
\begin{center}
$g^ta=\phi_a(t).$
\end{center}

\noindent
{\bf Definition 4} The orbit through $a$, denoted by $\gamma(a)$ is defined by
\begin{center}
$\gamma(a)=\{x\in \mR^n | x=g^ta, $ for all $t \in \mR\}$.
\end{center}

\noindent
{\bf Definition 5} Given a DE (\ref{DE}) in \mR$^n$, a set 
$S \subseteq \mR^n$ is called an invariant set for the DE 
if for any point $a \in S$ the orbit through $a$ lies entirely 
in $S$, that is $\gamma(a) \subseteq S$.

\noindent
{\bf Definition 6}  Given a DE (\ref{DE}) in \mR$^n$, with flow 
$\{g^t\}$, a subset $S \subseteq \mR^n$ is said to be a trapping 
set of the DE if it satisfies:
\begin{enumerate}
\item $S$ is a closed and bounded set,
\item $a \in S$ implies that $g^ta \in S$ for all $t \geq 0$.
\end{enumerate}

Qualitative analysis of a system begins with the location of singular
points.  Once the singular points of a system of ODEs are obtained, it
is of interest to consider the dynamics in a local neighbourhood of
each of the points.  Assuming that the vector field $f(x)$ is of class
$C^1$ the process of determining the local behaviour is based on the
linear approximation of the vector field in the local neighbourhood of
the singular point $\bar{x}$.  In this neighbourhood
\be
{f(x) \approx Df(\bar{x})(x-\bar{x})}\label{linapp}
\ee
where $Df(\bar{x})$ is the Jacobian of the vector field at the singular 
point $\bar{x}$.  The system (\ref{linapp}) is referred to as the 
{\it linearization of the DE at the singular point}.
Each of the singular points can then be classified according to the 
eigenvalues of the Jacobian of the linearized vector field at the point.

The classification then follows from the fact that if the singular point 
is hyperbolic in nature the flows of the non-linear system and it's linear 
approximation are {\it topologically equivalent} in a neighbourhood of the 
singular point.  This result is given in the form of the following theorem:

{\bf Theorem 1:  Hartman-Grobman Theorem}
Consider a DE: $\dot{x}=f(x)$, where the vector field $f$ is of class $C^1$.  
If $\bar{x}$ is a hyperbolic singular point of the DE then there exists a 
neighbourhood of $\bar{x}$ on which the flow is topologically equivalent to 
the flow of the linearization of the DE at $\bar{x}$.

Given a linear system of ODEs: 
\be{\dot{x}=Ax,}\label{linsys}\ee
where A is a matrix with constant coefficients, it is a straightforward matter to 
show that if the eigenvalues of the matrix A are all positive the solutions 
in the neighbourhood of $\bar{x}=0$ all diverge from that point.  This point 
is then referred to as a source.  Similarly, if the eigenvalues all have negative 
real parts all solutions converge to the singular point $\bar{x}=0$, and the 
point is referred to as a sink.  Therefore, it follows from topological 
equivalence that if all eigenvalues of the Jacobian of the vector field 
for a non-linear system of ODEs have positive real parts the point is 
classified as a source (and all orbits diverge from the singular point), 
and if the eigenvalues all have negative real parts the point is classified 
as a sink.

In most cases the eigenvalues of the linearized system (\ref{linapp}) will 
have eigenvalues with both positive, negative and/or zero real parts.  In 
these cases it is important to identify which orbits are attracted to the 
singular point, and which are repelled away as the independent variable 
(usually $t$) tends to infinity.

For a linear system of ODEs, (\ref{linsys}), the phase space \mR$^n$ is spanned 
by the eigenvectors of $A$.  These eigenvectors divide the phase space into 
three distinct subspaces; namely:
\begin{center}
\begin{tabular}{ll}
The {\it stable subspace}     & $E^s=$ span$(s_1, s_2, ... s_{ns})$ \\
The {\it unstable subspace}   & $E^u=$ span$(u_1, u_2, ... u_{nu})$ 
\end{tabular}
\end{center}
and 
\begin{center}
\begin{tabular}{ll}
The {\it centre subspace} & $E^c=$ span$(c_1, c_2, ... c_{nc})$
\end{tabular}
\end{center}
where $s_i$ are the eigenvectors who's associated eigenvalues have negative 
real part, $u_i$ those who's eigenvalues have positive real part, and $c_i$ 
those who's eigenvalues have zero eigenvalues.  Flows (or orbits) in the stable 
subspace asymptote in the future to the singular point, and those in the 
unstable subspace asymptote in the past to the singular point.

In the non-linear case, the topological equivalence of flows allows 
for a similar classification of the singular points.  The equivalence 
only applies in directions where the eigenvalue has non-zero real parts.  
In these directions, since the flows are topologically equivalent, there 
is a flow {\it tangent} to the eigenvectors.  The phase space is again 
divided into stable and unstable subspaces (as well as centre subspaces).  
The {\it stable manifold} $W^s$ of a singular point is a differential 
manifold which is tangent to the stable subspace of the linearized 
system ($E^s$).  Similarly, the {\it unstable manifold} is a differential 
manifold which is tangent to the unstable subspace ($E^u$) at the singular 
point.  The centre manifold, $W^c$, is a differential manifold which is 
tangent to the centre subspace $E^c$.  It is important to note, however, 
that unlike the case of a linear system, this centre manifold, $W^c$ will 
contain all those dynamics not classified by linearization (i.e., the 
non-hyperbolic directions).  In particular, this manifold may contain regions 
which are stable, unstable or neutral.  The classification of the dynamics in 
this manifold can only be determined by utilizing more sophisticated methods, 
such as centre manifold theorems or the theory of normal forms (see \cite{Wiggins1990a}).

Unlike a linear system of ODEs, a non-linear system allows for
equilibrium structures which are more complicated than that of the
singular points fixed lines or periodic orbits.  These structures
include, though are not limited to, such things as heteroclinic and/
or homo-clinic orbits, non-linear invariant sub-manifolds, etc (for
definitions see \cite{Wiggins1990a}).  The set of non-isolated
singular points will figure into the analysis of solutions in this
thesis, and therefore it's stability will be examined more rigorously.

{\bf Definition 7}:  A set of non-isolated singular points is said to be normally 
hyperbolic if the only eigenvalues with zero real parts are those whose 
corresponding eigenvectors are tangent to the set.

Since by definition any point on a set of non-isolated singular points will 
have at least one eigenvalue which is zero, all points in the set are 
{\it non-hyperbolic}.  A set which is normally hyperbolic can, however, 
be completely classified as per it's stability by considering the signs 
of the eigenvalues in the remaining directions  (i.e. for a curve, in the 
remaining $n-1$ directions) \cite{Aulbach1984a}.

The local dynamics of a singular point may depend on one or more 
arbitrary parameters.  
When small continuous
changes in the parameter result in dramatic
changes in the dynamics, the singular point is said to undergo a {\em bifurcation}\label{bifurcation}.
The values of the parameter(s) which result in a bifurcation at the singular
point can often be located by examining the linearized system.  Singular point
bifurcations will only occur if one (or more) of the eigenvalues of the 
linearized systems are a function of the parameter.  The bifurcations
are located at the parameter values for which the real part of an eigenvalue
is zero. 

There are several different types of
singular point bifurcations, which are classified
according to the particular nature of the change in the dynamics.  Some of
the more common bifurcations are:
\begin{itemize}
\item {\bf Saddle-node bifurcation}: A saddle-node bifurcation is characterized
by the non-existence of a singular point on one side of the bifurcation
value and the existence of two singular points on the other side of
the bifurcation value.  At the bifurcation value, a singular point in two
(or higher) dimensions has a saddle-node structure.
\item{\bf Transcritical bifurcation}: A transcritical bifurcation is 
characterized by the \lq\lq exchange'' of stability.  By passing through
the bifurcation value the stability of two singular points interchange.  Once
again, in two-dimensional phase space, the
singular point has a saddle-node structure.
\item {\bf Poincare-Andronov-Hopf (PAH) bifurcation}:  In the preceding examples,
the bifurcation occurs when a single eigenvalue is identically zero.  In
contrast, a PAH bifurcation occurs when there is a pair of eigenvalues
whose {\bf real} part becomes zero.  In this case, the singular point
on either side of the bifurcation value is a spiral (either attracting
or repelling). 
\end{itemize}
A complete classification of singular point bifurcations can be found
in \cite{Wiggins1990a}.

The future and past asymptotic states of a non-linear system 
may be represented by any singular or periodic structure.
In the case of a plane system (i.e. in two-dimension phase 
space), the possible asymptotic states can be given explicitly.  This result 
is due to the limited degrees of freedom in the space, and the fact that the 
flows (or orbits) in any dimensional space cannot cross.  The result is given 
in the form of the following theorem:

{\bf Theorem 2: Poincare-Bendixon Theorem}:  Consider the system of ODEs 
$\dot{x}=f(x)$ on \mR$^2$, with $f \in C^2$, and suppose that there 
are at most a finite number of singular points (i.e. no non-isolated 
singular points).  Then any compact asymptotic set is one of the following:
\begin{enumerate}
\item a singular point
\item a periodic orbit
\item the union of singular points and heteroclinic or homo-clinic orbits.
\end{enumerate}

This theorem has a very important consequence in that if the existence of 
a closed (i.e. periodic, heteroclinic or homo-clinic) orbit can be ruled 
out it follows that all asymptotic behaviour is located at a singular point.

The existence of a closed orbit can be ruled out by many methods, the most 
common is to use a consequence of Green's Theorem, as follows:

{\bf Theorem 3: Dulac's Criterion}:  If $D \subseteq R^2$ is a simply 
connected open set and 
$\nabla(Bf) = \frac{\partial}{\partial x_1}
	(Bf_1)+\frac{\partial}{\partial x_2}(Bf_2) > 0, $ or $(<0)$ 
for all $x \in D$ where $B$ is a $C^1$ function, then the DE $\dot{x}=f(x)$ 
where $f \in C^1$ has no periodic (or closed) orbit which is contained in $D$.

A fundamental criteria of the Poincare-Bendixon theorem is that the phase 
space is two-dimensional.  When the phase space is of a higher dimension 
the requirement that orbits cannot cross does not result in such a decisive 
conclusion.  The behaviour in such higher-dimensional spaces is known to be 
highly complicated, with the possibility of including such phenomena as 
recurrence and strange attractors (see, for example, \cite{Guckenheimer1983a}).
  For that reason, the analysis of non-linear systems in 
spaces of three or more dimensions cannot in general progress much further 
than the local analysis of the singular points (or non-isolated singular sets).  
The one tool which does allow for some progress in the analysis of higher 
dimensional systems is the possible existence of monotonic functions.  Since in this
thesis there will be the need to analyse three-dimensional phase spaces the
tools for higher dimensional spaces will now be outlined.

{\bf Theorem 4: LaSalle Invariance Principle}:  Consider a DE 
$\dot{x}=f(x)$ on \mR$^n$.  Let $S$ be a closed, bounded and 
positively invariant set of the flow, and let $Z$ be a $C^1$ monotonic 
function.  Then for all $x_0\in S$, 
\begin{center}
$w(x_0) \subset \{x \in S | \dot{Z}=0\}$
\end{center}
where $w(x_0)$ is the forward asymptotic states for the orbit with 
initial value $x_0$; i.e. a $w$-limit set \cite{Tavakol1997a}.

\noindent
This principle has been generalized to the following result:

{\bf Theorem 5: Monotonicity Principle} (see \cite{LeBlanc1995a}).  
Let $\phi_t$ be a flow on \mR$^n$ with $S$ an invariant set.  
Let $Z : S \rightarrow \mR$ be a $C^1$ function whose range 
is the interval $(a,b)$, where $a \in \mR\cup \{-\infty\}$, 
$b \in \mR \cup \{\infty\}$ and $a<b$.  If Z is decreasing 
on orbits in $S$, then for all $X \in S$,
\begin{center}
$\omega(x) \subseteq \{s \in \bar{S}\setminus S | lim_{y\rightarrow s} Z(y) \neq b\},$ \\
$\alpha(x) \subseteq \{s \in \bar{S}\setminus S | lim_{y\rightarrow s} Z(y) \neq a\},$
\end{center}
where $\omega(x)$ and $\alpha(x)$ are the forward and backward limit set of 
$x$, respectively(i.e., the $w$ and $\alpha$ limit sets.)

\chapter{Restoring non-Geometerized Units \label{restore}}

This thesis explicitly uses geometerized units in which $c=8\pi
G=\hbar=1$.  Below is a list of how these constants are reintroduced when
non-geometerized units are used.  First, the constants in the coordinates and their velocities are restored:
\beqn
t & \longrightarrow & ct, \\
\frac{\di}{\di t} & \longrightarrow & c^{-1} \frac{\di}{\di t}, \\
x^{j} & \longrightarrow & x^{j}, ~~~(j=1,2,3), \\
\frac{\di}{\di x^j } & \longrightarrow & \frac{\di}{\di x^j}, \\
u^{\alpha} & \longrightarrow & u^{\alpha} /c.
\eeqn
Next, constituents of the energy-momentum tensor,
\begin{equation}
T_{\alpha \beta} \longrightarrow \frac{8\pi G}{c^4} T_{\alpha \beta}.
\end{equation}
Reintroducing the constants into the matter fields yield
\beqn
\mu & \longrightarrow & \frac{8\pi G}{c^4} \mu c^2, \\
p & \longrightarrow & \frac{8\pi G}{c^4} p, \\
q_{\alpha} & \longrightarrow & \frac{8\pi G}{c^3}q_{\alpha},  \\
\pi_{\alpha \beta} & \longrightarrow & \frac{8\pi G}{c^4} \pi_{\alpha \beta},
\eeqn
Reintroducing the constants for the electromagnetic fields yields
\begin{equation}
F_{\alpha \beta} \longrightarrow  \sqrt{\varepsilon_0} F_{\alpha \beta}.
\end{equation}
Reintroducing the constants into the scalar fields yield
\beqn
\phi & \longrightarrow & \frac{\sqrt{8\pi G}}{c^2} \phi, \\
\grad_\alpha \phi & \longrightarrow & \frac{\sqrt{8\pi G}}{c^2} \grad_\alpha \phi,\\
V & \longrightarrow & \frac{8\pi G}{c^4} V.
\eeqn
Reintroducing the constants into the  various forms of the potential $V$ yield
\beqn
V=V_0 e^{k\phi} & \longrightarrow & V=V_0 e^{\frac{\sqrt{8\pi G}}
		{c^2}k\phi}, \\
V=\half m\phi^2 & \longrightarrow & V=\half m\phi^2, \\
V=\quart \lambda\phi^4 & \longrightarrow & V=\quart \lambda\phi^4,
\eeqn
where $k$ is a unitless constant, $m$ is a constant with units {\em
length}$^{-2}$, and $\lambda$ is a constant with units {\em
mass}$^{-1}${\em length}$^{-3}${\em time}$^2$.

Finally, in considering interaction terms between the scalar field and
the matter terms, namely,
\beqn
&& \dot\phi \left(\ddot\phi +3 H \dot\phi +c^2
	\frac{dV}{d\phi}\right) =-\delta \\
&& c^2\left[\dot\mu c^2+3H\left(\mu c^2+p\right)\right] = +\delta,
\eeqn
where $\dot\phi\equiv d\phi/dt$, etc., there were two mathematical forms
for $\delta$:
\beqn
\delta & = & a c^2 \sqrt{8\pi G} \dot\phi \mu, \\
\delta & = & a c^4 \mu H,
\eeqn
where $a$ is a unitless constant.

\section[GR Scalar Field Theory $\leftrightarrow$ Scalar-Tensor
Theory]{Transformations between GR Scalar Field \\ Theory and
Scalar-Tensor Theory \label{GRtoST}}

There is a freedom in the definitions for the transformation between
GR with a scalar field and scalar-tensor theories with respect to the
placement of $G$, and described here are the two commonly used forms,
explicitly using non-geometerized units.  In both forms, the scalar
field $\Phi$ is unitless.  In the first form, the transformation
between the two theories can be written
\beqn
\mainlabel{GRtoST1}
\up{sf}g_{\alpha \beta} & = & G \Phi \up{st}g_{\alpha \beta}, \\
\frac{\sqrt{8\pi G}}{c^2} d\phi &=& \pm 
			\sqrt{\omega+\frac{3}{2}}\frac{d\Phi}{\Phi},\\
\up{sf}T_{\alpha \beta} & = & 
			\frac{\up{st}T_{\alpha \beta}}{G\Phi}, \\
\up{sf}{\cal L}_{\rm M} &=& 
			\frac{\up{st}{\cal L}_{\rm M}}{(G\Phi)^2},\\
V & = & \frac{c^4 U}{8\pi G^2\Phi^2},
\eeqn
where $\omega=\omega(\Phi)$, and the superscripts (sf) and (st) refer
to GR scalar field theory and ST theory, respectively.
These transformations lead to the mathematical equivalence of the
following equations.
\beqn
\up{sf}S & = & \int d^4x \sqrt{-\up{sf}g} \left\{ 
	\frac{\up{sf}Rc^4}{16\pi G} -\half \grad^\alpha\phi 
	\grad_\alpha\phi -V + \up{sf}{\cal L}_{\rm M}\right\}, \\
\up{st}S & = & \int d^4x \sqrt{-\up{st}g} \left\{ \frac{c^4}
	{16 \pi}\left[ 
	\Phi \up{st}R -\frac{\omega}{\Phi} \grad^\alpha\Phi 
	\grad_\alpha\Phi -2U\right] + \up{st}{\cal L}_{\rm M}\right\}.
\eeqn
\beqn
\up{sf}G_{\alpha \beta} &=& \frac{8\pi G}{c^4}\left[
	\up{sf}T_{\alpha \beta} + \grad_\alpha\phi 
	\grad_\beta\phi +\up{sf}g_{\alpha \beta}V\right],  \\
\up{st}G_{\alpha \beta} &=& \frac{8\pi}{\Phi c^4} 
	\up{st}T_{\alpha \beta} + \frac{\omega}{\Phi^2} \left[ 
	\grad_\alpha\Phi \grad_\beta\Phi- \half \up{st}g_{\alpha \beta}
	 \grad_\gamma\Phi \grad^\gamma\Phi \right] 
	+\frac{\grad_\alpha\grad_\beta\Phi}{\Phi} \nonumber \\ 
	&&-\up{st}g_{\alpha \beta}
	\left(\frac{U}{\Phi}+\frac{\Box\Phi}{\Phi} \right).
\eeqn
\beqn
\grad^\alpha \up{sf}T_{\alpha \beta} &+ & \grad_\beta\phi\left(\Box 
	\phi - \frac{dV}{d\phi} \right)=0,\\
\frac{8\pi}{\Phi c^4} \grad^\alpha \up{st}T_{\alpha \beta} & + & 
	\half \frac{\grad_\beta\Phi}{\Phi} \left[ 2\omega\frac{\Box\Phi}{\Phi}
	+\frac{\grad_\gamma\Phi \grad^\gamma\Phi}{\Phi}\left(
	\frac{d\omega}{d\Phi} -\frac{\omega}{\Phi}\right) -2 \frac{dU}{d\Phi}
	+\up{st}R\right]. \nonumber \\
\eeqn
The scalar-tensor equations do not change when one chooses $8\pi G=1$,
although there will be terms of $8\pi$ throughout the equations.  This
is the form which was originally used by Brans and Dicke
\cite{Brans1961a}, for $\omega=\omega_0$ (a constant).

Alternatively, one may choose the transformations
\beqn
\mainlabel{GRtoST2}
\up{sf}g_{\alpha \beta} & = & 8\pi G \Phi \up{st}g_{\alpha \beta}, \\
\frac{\sqrt{8\pi G}}{c^2} d\phi &=& \pm 
			\sqrt{\omega+\frac{3}{2}}\frac{d\Phi}{\Phi},\\
\up{sf}T_{\alpha \beta} & = & 
			\frac{\up{st}T_{\alpha \beta}}{8\pi G\Phi}, \\
\up{sf}{\cal L}_{\rm M} &=& 
			\frac{\up{st}{\cal L}_{\rm M}}{(8\pi G\Phi)^2},\\
V & = & \frac{c^4 U}{(8\pi G\Phi)^2}
\eeqn
(these are the same as (\ref{GRtoST1}) when $\Phi\rightarrow 8\pi \Phi$ and $U
\rightarrow 8\pi U$).
These transformations lead to the mathematical equivalence of the
following equations.
\beqn
\up{sf}S & = & \int d^4x \sqrt{-\up{sf}g} \left\{ 
	\frac{\up{sf}Rc^4}{16\pi G} -\half \grad^\alpha\phi 
	\grad_\alpha\phi -V + \up{sf}{\cal L}_{\rm M}\right\}, \\
\up{st}S & = & \int d^4x \sqrt{-\up{st}g} \left\{ \half c^4 \left[ 
	\Phi \up{st}R -\frac{\omega}{\Phi} \grad^\alpha\Phi 
	\grad_\alpha\Phi -2U\right] + \up{st}{\cal L}_{\rm M}\right\}.
\eeqn
\beqn
\up{sf}G_{\alpha \beta} &=& \frac{8\pi G}{c^4}\left[
	\up{sf}T_{\alpha \beta} + \grad_\alpha\phi 
	\grad_\beta\phi +\up{sf}g_{\alpha \beta}V\right],  \\
\up{st}G_{\alpha \beta} &=& \frac{\up{st}T_{\alpha \beta}}{\Phi c^4} 
	+ \frac{\omega}{\Phi^2} \left[ 
	\grad_\alpha\Phi \grad_\beta\Phi- \half \up{st}g_{\alpha \beta}
	 \grad_\gamma\Phi \grad^\gamma\Phi \right] 
	+\frac{\grad_\alpha\grad_\beta\Phi}{\Phi} \nonumber \\ 
	&&-\up{st}g_{\alpha \beta}
	\left(\frac{U}{\Phi}+\frac{\Box\Phi}{\Phi} \right).
\eeqn
\beqn
\grad^\alpha \up{sf}T_{\alpha \beta} &+ & \grad_\beta\phi\left(\Box 
	\phi - \frac{dV}{d\phi} \right)=0,\\
\frac{\grad^\alpha \up{sf}T_{\alpha \beta}}{\Phi c^4}  & + & 
	\half \frac{\grad_\beta\Phi}{\Phi} \left[ 2\omega\frac{\Box\Phi}{\Phi}
	+\frac{\grad_\gamma\Phi \grad^\gamma\Phi}{\Phi}\left(
	\frac{d\omega}{d\Phi} -\frac{\omega}{\Phi}\right) -2 \frac{dU}{d\Phi}
	+\up{st}R\right].
\eeqn
Here, the scalar-tensor equations do not change when one chooses $8\pi
G=1$, and there are no factors of $8\pi$ throughout.

\chapter{Kaluza-Klein Reduction to Four Dimensions\label{KKreduced}}

In Billyard and Coley \cite{Billyard1997a}, the mathematical
relationship between vacuum ST theories and GR scalar field theories with
higher--dimensional vacuum Kaluza--Klein theories was discussed,
mainly to elucidate the fact that previously solutions in one theory
had been ``discovered'' after the corresponding solutions in another
theory already existed.  In particular, five--dimensional vacuum theories are
mathematically equivalent to $\omega=0$ Brans--Dicke theories.  In
\cite{Holman1991a}, such a correspondence between a (4+N)-dimensional
vacuum theory and $\omega=\omega(N)$ Brans-Dicke theory was derived,
where it was assumed that the extra dimensions were described by an
N-dimensional maximally symmetric space of constant positive
curvature.  Below, we generalize this by assuming the N extra
dimensions are composed of $m$ maximally symmetric submanifolds of
constant (arbitrary sign) curvature.  Unlike the analysis in
\cite{Holman1991a}, a dependence on the extra coordinates will be
included.  The analysis below shows that such models are
mathematically equivalent to GR scalar field theories with an exponential potential.

We begin this section by considering a ($D=4+N$)-dimensional manifold
as a product of $m$ submanifolds,
$\mM_{4+N}=\mM_{N_0}\times\mM_{N_1}\times \cdots \times \mM_{N_m}$,
$m-1$ of which are spaces of constant curvature with a conformal scale
factor depending on the coordinates of the other submanifold,
$\sigma_{(i)}=\sigma_{(i)}(x^{a_{j\neq i}})$ (in this way, the only
dependence on the coordinates of the submanifold {\em in} the
submanifold itself is through the metric of constant curvature).  Each
submanifold is of dimension $N_i$ (so that $\sum N_i=4+N$), where
$N_0=4$.  This is not the most general higher-dimensional manifold one
can consider, but it is the one of the most general one can assume in
order to reduce the action to an effective four-dimensional action.
One may include cross-terms
between the submanifolds and this would induce terms in the action
which look like a Maxwellian source (for a five-dimensional example of
this, see \cite{Billyard1997a}).

With the above assumptions, the line element may be written as
\be
ds^2=\up{D}\tilde{g}_{AB}dx^Adx^B = \sum^m_{i=0}
e^{2\sigma_{(i)}}\gamma_{a_ib_i}dx^{a_i}dx^{b_i}, \label{whatmetric}
\ee
(Einstein's summation notation is {\em not} used on sub-indices here) where for
$i>0$, $\gamma_{a_ib_i}$ is the metric of a manifold of constant
curvature:
\be
\up{i}R_{a_ib_i}=\pm (N_i-1)K_i\gamma_{a_ib_i}. \label{constant_curvature}
\ee 
The ``$\pm$'' arises from the definition of the Ricci tensor:
\be
R_{bc} = \pm \left\{\Gamma^a_{bc,a} -\Gamma^a_{ba,c}
+\Gamma^a_{da}\Gamma^d_{bc} -\Gamma^a_{db}\Gamma^d_{ac}\right\}.
\ee
The notation here is as follows.  Indices $\{A,B,\ldots\}$ range 0 to $(3+N)$, $\{a_0,b_0,\ldots\}$ range 0 to 3, and $\{a_i,b_i,\ldots\}$ range $(4+N_1+\cdots +N_{i-1})$ to $(3+N_1+\cdots+ N_i)$.  

Note the following assumptions are made, namely, 
\be
\up{D}\tilde{g}_{a_ib_j} = e^{2\sigma_{(i)}}\gamma_{a_ib_i}\delta_{ij}, \quad 
\up{D}\tilde{g}^{a_ib_j} = e^{-2\sigma_{(i)}}\gamma^{a_ib_i}\delta_{ij},
\ee
and therefore
\be
\up{D}\tilde{g}_{a_ib_i}\up{D}\tilde{g}^{a_jc_j} = \delta^{a_i}_{c_j}\delta_{ij} = \delta^{a_i}_{c_i},
\ee
then the relevant geometrical quantities may be written
\beqn
\mainlabel{first_breakdown}
\up{D}\Gamma^{a_i}_{b_j c_k} & = & \up{i}\Gamma^{a_i}_{b_j c_k}
   \delta_{ijk} + \grad_{c_k}\sigma_{(j)} \delta^{a_i}_{b_j}\delta_{ij}
+ \grad_{b_j}\sigma_{(k)} \delta^{a_i}_{c_k}\delta_{ik} \nonumber \\
&& \qquad \qquad \qquad-e^{\left[2\sigma_{(j)}-2\sigma_{(i)}\right]}
	\gamma_{b_jc_k}\grad^{a_i}
	\sigma_{(j)} \delta_{jk}, \\  \nonumber \\ \nonumber 
\pm\up{D}R_{a_ib_j} & = & \delta_{ij}\left\{\pm \up{i}R_{a_ib_j} 
	-\sum^m_{l=0}N_l\grad_{a_i}\grad_{b_j}\sigma_{(l)}
	-e^{2\sigma_{(i)}}\gamma_{a_ib_j} \tilde{\cal Z}_j\right\} \\
	&& \nonumber
	-2 \grad_{a_i}\sigma_{(j)}\grad_{b_j}\sigma_{(a_i)} \\
	&& \nonumber
	+\sum^m_{l=0} N_l \left[
	 \grad_{a_i}\sigma_{(j)}\grad_{b_j}\sigma_{(l)} 
	+\grad_{a_i}\sigma_{(l)}\grad_{b_j}\sigma_{(i)} 
	-\grad_{a_i}\sigma_{(l)}\grad_{b_j}\sigma_{(l)} \right], 
	\\  && \label{Rab} \\ \nonumber
\pm\up{D}R &=&\sum^m_{i=0}\left\{\frac{\pm\up{i}R}{e^{2\sigma_{(i)}}}
	-e^{-2\sigma_{(i)}}\sum^m_{l=0}N_l\sBox{i}\sigma_{(l)} - 
	N_i\tilde{\cal Z}_i \right.
	\nonumber \\ &&  \qquad \qquad
        \left.-\sum^m_{l=0} N_l e^{-2\sigma_{(i)}}\grad_{a_i}\sigma_{(l)}
	\grad^{a_i}\sigma_{(l)}\right\}, \label{R}
\eeqn
where $\grad_{a_i}$ and $\sBox{i}\equiv \grad_{a_i}\grad^{a_i}$ are
the covariant derivative and d'Alembertian operator, respectively,
defined on the $i^{\mbox{th}}$ submanifold, and the ``extended''
Kronecker-Delta function $\delta_{ijk}=1$ if all three sub-indices are
equal (zero otherwise) and $\tilde{\cal Z}_i$ is defined by
\be
\tilde{\cal Z}_i = \sum^m_{l=0}e^{-2\sigma_{(l)}} \left[
\sBox{l}\sigma_{(j)} +\grad^{d_l}\sigma_{(j)} \sum^m_{n=0}
N_n \grad_{d_l}\sigma_{(n)}\right]
\ee

The fol\-lowing de\-com\-po\-si\-tion is now per\-formed.  Let
$\{a_0,b_0,\cdots\}\equiv\{\alpha,\beta,\cdots\}$, $\gamma_{\alpha\beta}
\equiv g_{\alpha\beta}$, and 
$\sigma_{(i)} \equiv y_i+\ln\Upsilon_i$ where $\Upsilon_0=1$,
$\Upsilon_i=\Upsilon_i(x^\alpha)$ and $y_i=y_i(x^{j\neq i\neq\alpha})$.  Using
(\ref{constant_curvature}) for $i>0$, equations (\ref{Rab}) and
(\ref{R}) can be reduced to the following forms:
\beqn
\mainlabel{second_breakdown}
\pm\up{D}R_{\alpha\beta} & = & \pm R_{\alpha\beta} -\sum^m_{l=1} 
	N_l\frac{\grad_\alpha\grad_\beta\Upsilon_l}{\Upsilon_l} - e^{2y_{_0}}
	g_{\alpha\beta} \bar{\cal Z}_0, \\
\pm\up{D}R_{\alpha b_j} & = &2\frac{\grad_\alpha\Upsilon_j}{\Upsilon_j}
	\grad_{b_j}y_{_0} +\sum^m_{l=1} N_l \left[
	 \frac{\grad_\alpha\Upsilon_j}{\Upsilon_j}\grad_{b_j}y_{_l}+ \frac{\grad_\alpha
	\Upsilon_l}{\Upsilon_l}\grad_{b_j}y_{_0} - \frac{\grad_\alpha\Upsilon_l}{\Upsilon_l}
	\grad_{b_j}y_{_l}\right], \nonumber \\ \\
\pm\up{D}R_{a_i b_i} & = & \gamma_{a_ib_i}\left\{
	\left(N_i-1\right)K_i -\frac{e^{2y_{_i}}\Upsilon_i^2}{e^{2y_{_0}}}\left[
	\frac{\Box\Upsilon_i}{\Upsilon_i}\!-\!\frac{\grad^\alpha\Upsilon_i}{\Upsilon_i} \left(
	\frac{\grad_\alpha\Upsilon_i}{\Upsilon_i}\!-\!\sum^m_{l=1}N_l\frac{\grad_\alpha
	\Upsilon_l}{\Upsilon_l}\right)\right]\right.\nonumber \\  && \qquad \qquad 
	\left. -e^{2y_{_i}}\Upsilon_i^2 \bar{\cal Z}_i\right\} 
	- \sum^m_{l=0}N_l\left(
	\grad_{a_i}\grad_{b_i}y_{_l}+\grad_{a_i}y_{_l}\grad_{b_i}y_{_l}\right),
	\\
\pm\up{D}R_{a_i b_j} & = & -2\grad_{a_i}y_{_j}\grad_{b_j}y_{_i} +
	\sum^m_{l=0}N_l\left[\grad_{a_i}y_{_j}\grad_{b_j}y_{_l} +
	\grad_{a_i}y_{_l}\grad_{b_j}y_{_i} - 
	\grad_{a_i}y_{_l}\grad_{b_j}y_{_l}\right],\nonumber \\ \\
\pm\up{D}R & = & e^{-2y_{_0}}\left\{ \pm R -\sum^m_{l=1} N_l\left[
	\frac{2\Box\Upsilon_l}{\Upsilon_l} -\frac{\grad^\alpha\Upsilon_l}
	{e^{2y_{_0}}\Upsilon^l} \left(\frac{\grad_\alpha\Upsilon_l}{\Upsilon_l} -
	\sum^m_{n=1}N_n\frac{\grad_\alpha\Upsilon_n}{\Upsilon_n}\right) \right]\right\}
	\nonumber \\ && -\bar{\cal Z}^* 
	+\sum^m_{l=1}\frac{N_l\left(N_l-1\right)K_l}{e^{2y_{_l}}\Upsilon_l^2},
\eeqn
where
\beqn
\bar{\cal Z}_i & = & \tilde{\cal Z}_i - e^{-2y_{_0}}\left[\frac{\Box\Upsilon_i}
	{\Upsilon_i}\!-\!\frac{\grad^\alpha\Upsilon_i}{\Upsilon_i} 
	\left(\frac{\grad_\alpha\Upsilon_i}	{\Upsilon_i}\!-\!\sum^m_{l=1}N_l
	\frac{\grad_\alpha \Upsilon_l}{\Upsilon_l}\right)\right] \nonumber\\
 & = & \sum^m_{l=1}\Upsilon^{-2}_l e^{-2y_{_l}}\left[\sBox{l}y_{_i} 
	+\grad^{d_l}y_{_i}\sum^m_{n=0} N_n\grad_{d_l}y_{_n}\right], \\
\bar{\cal Z}^* & = & \sum^m_{n=0} N_n \left [ \bar{\cal Z}_n +\sum^m_{l=1} 
	\Upsilon^{-2}_l e^{-2y_{_l}}\left( \sBox{l}y_{_n} + \grad^{d_l}y_{_n}
	\grad_{d_l}y_{_n}\right)\right] \nonumber\\
   & = & \sum^m_{l=1} \Upsilon^{-2}_l e^{-2y_{_l}} \sum ^m_{n=0}
	N_n\left[2\sBox{l}y_{_n} 
	+\grad^{d_l}y_{_n}\left(\grad_{d_l}y_{_n}+\sum^m_{q=0}N_q\grad_{d_l}y_{_q}\right)\right].
\eeqn

At this point, one needs further assumptions on the metric's form in
order to establish a mathematical equivalence between the above theory
with scalar-tensor theory or vacuum general relativity with a
cosmological constant.  Therefore, assume that the scale
factors depending on the external space, $\Upsilon_i$, are equal;
i.e. the entire internal space has one conformal factor which depends
on the external coordinates,
$\Upsilon_i=\Upsilon\equiv\left[\up{D}G\Phi\right]^{1/N}$ (where
$\up{D}G$ is Newton's constant in D-dimensions). By using the definitions
\beqn
\omega=-\frac{(N-1)}{N} & \Leftrightarrow & N = \frac{1}{1+\omega},\\
{\cal Z}_i &= & \Upsilon^2\bar{\cal Z}_i, \\
{\cal Z}^* &= & \Upsilon^2\bar{\cal Z}^*, 
\eeqn
then equations (\ref{second_breakdown}) reduce to the following:
\beqn
\mainlabel{third_breakdown}
\pm \up{D}R_{\alpha\beta} & = &  \pm R_{\alpha\beta} - \frac{\grad_\alpha
	\grad_\beta\Phi}{\Phi} -\omega \frac{\grad_\alpha\Phi
	\grad_\beta\Phi}{\Phi^2} - \frac{e^{2y_{_0}}g_{\alpha\beta}{\cal Z}_0}
	{\left[\up{D}G\Phi\right]^{2(1+\omega)}}, \label{Field_4D}\\
\pm \up{D}R_{\alpha b_j} & = &  -\omega \frac{\grad_\alpha
	\Phi}{\Phi}\grad_{b_j}y_{_0}, \label{JordanOrEinstein}\\
\pm \up{D}R_{a_i b_i} & = & -\left[\up{D}G\Phi\right]^{2(1+\omega)}
	e^{2y_{_i}}\gamma_{a_i b_i} \left\{\frac{e^{-2y_{_0}}}{N} 
	\frac{\Box\Phi}{\Phi} + \frac{{\cal Z}_i- e^{-2y_{_i}}\left(N_i-1\right)K_i}
	{\left[\up{D}G\Phi\right]^{2(1+\omega)}} \right\} \nonumber \\
   &&	- \sum^m_{l=0}N_l\left[\grad_{a_i}\grad_{b_i}y_{_l} +\grad_{a_i}
	y_{_l}\grad_{b_i}y_{_l}\right],\\
\pm \up{D}R_{a_i b_j} & = & \sum^m_{l=0}\left[ \grad_{a_i}y_{_j}
	\grad_{b_j}y_{_l}\! +\!\grad_{a_i}y_{_l}\grad_{b_j}y_{_i} 
	\!-\!\grad_{a_i}y_{_l}\grad_{b_j}y_{_l}\right] 
	- 2\grad_{a_i}y_{_j}	\grad_{b_j}y_{_i},\label{NeededConstraint}\\
\pm \up{D}R & = & e^{-2y_{_0}}\left\{\pm R 
	- 2\frac{\Box\Phi}{\Phi}-\omega \frac{\grad_\gamma\Phi
	\grad^\gamma\Phi}{\Phi^2}\right\} -\frac{2U_0} 
	{\left[\up{D}G\Phi\right]^{2(1+\omega)}},\label{constraints}
\eeqn
where
\be
U_0 \equiv\half\left\{{\cal Z}^*-\sum^m_{l=1}e^{-2y_{_l}}N_l(N_l-1)K_l\right\}.
\ee
The determinant of the metric now has the form
\be
\sqrt{\left|\up{D}g\right|} = \frac{e^{4y_{_0}}e^{N_1y_{_1}}\cdots 
	e^{N_my_{_m}}}	{Q_1^{N_1}\cdots Q_m^{N_m}} \Phi \sqrt{\left|g\right|}
	\equiv {\cal A} \Phi \sqrt{\left|g\right|},
\ee
where $Q_i\equiv 1+\frac{1}{4}K_i\sum^{N_i}_{a_i=1}(x^{a_i})^2$,
and the D-dimensional action
\benonumber
\int d^{(4+N)}x \sqrt{\left|\up{D}g\right|}\frac{\up{D}Rc^4}{16\pi \up{D}G},
\eenonumber 
reduces to 
\be
\int {\cal A}d^Nx \int d^4x \sqrt{\left|g\right|}\frac{c^4\Phi}{16\pi}\left\{ 
	R \mp \left[2\frac{\Box\Phi}{\Phi}+\omega 
	\frac{\grad_\gamma\Phi\grad^\gamma\Phi}{\Phi^2} 
	+\frac{2U_0} {\left[\up{D}G\Phi\right]^{2(1+\omega)}}\right]\right\}.
	\label{action_reduced}
\ee

Now it will show that a D-dimensional vacuum theory
\be
\up{D}R_{AB}=0 \label{theVacuum}
\ee 
corresponds to either vacuum general relativity with a cosmological
constant or a Brans-Dicke theory (scalar tensor) with a potential.
Equation (\ref{constraints}) will not be explicitly given in what
follows, although it is important to note that a reduction from a
higher-dimensional theory of gravity will have such constraint
equations if there is dependence on the extra coordinates.  

First, note
that contraction of equations (\ref{NeededConstraint}), using equation
(\ref{theVacuum}) yields the relation
\be
\frac{4{\cal Z}_0}{\left[\up{D}G\Phi\right]^{2(1+\omega)}}
	= \frac{2U_0}{\left[\up{D}G\Phi\right]^{2(1+\omega)}}
	- e^{-2y_{_0}}\frac{\Box\Phi}{\Phi}, \label{J_Friedmann}
\ee
which is the Friedmann constraint in the Brans-Dicke theory (Jordan
frame).  Furthermore, equations (\ref{JordanOrEinstein}) determines
whether the reduced theory is a theory of general relativity
($\grad_\alpha\Phi=0$), in which case 
$\Phi=\up{D}G^{-1}$ can be set without loss of generality, or a Brans-Dicke
theory in which $y_0=\mbox{constant}\equiv0$ (without loss of
generality).  What these equations physically mean is that in a
higher-dimensional vacuum theory with ansatz associated with line
element (\ref{whatmetric}), it is not possible to simultaneously have
the size of the external space dictated by the internal dimensions
{\em and} the size of the internal space dictated by the external
dimensions.

If $\Phi=\mbox{constant}=\up{D}G^{-1}$, then (\ref{J_Friedmann})
reduces to
\be 
2U_0=4{\cal Z}_0,
\ee
and the remaining equations of (\ref{third_breakdown}) finally reduce to
\beqn
R_{\alpha\beta} &=&\pm g_{\alpha\beta} e^{2y_{_0}}{\cal Z}_0, \\
R & = & \pm 4 e^{2y_{_0}}{\cal Z}_0,
\eeqn
which are the field equations for a vacuum solution in general
relativity with a cosmological constant, $\Lambda\equiv
e^{2y_{_0}}{\cal Z}_0$.  The action (\ref{action_reduced}) reduces to 
\be
\int {\cal A}d^Nx \int d^4x \sqrt{\left|g\right|}\frac{c^4}{16\pi\up{D}
	G}\left\{ R \mp 4e^{2y_{_0}}{\cal Z}_0\right\}.
\ee

If $y_{_0}=0$, then ${\cal Z}_0=0$ and (\ref{J_Friedmann}) reduces to
\be
\frac{U}{\Phi} \equiv \frac{U_0}{\left[\up{D}G\Phi\right]^{2(1+\omega)}} = 
	-\frac{1}{2} e^{-2y_{_0}}\frac{\Box \Phi}{\Phi},
\ee
and the remaining equations of (\ref{third_breakdown}) reduce to
\beqn
R_{\alpha \beta} &=& \pm \left\{ \frac{\grad_\alpha
	\grad_\beta\Phi}{\Phi} +\omega \frac{\grad_\alpha\Phi
	\grad_\beta\Phi}{\Phi^2}\right\}, \\
R & = & \pm \left\{ 2\frac{\Box\Phi}{\Phi} +\omega \frac{\grad_\gamma\Phi
	\grad^\gamma\Phi}{\Phi^2} + 2 \frac{U}{\up{D}G\Phi}\right\},
\eeqn which are the field equations of the Brans-Dicke theory 
containing a
scalar potential $U\propto \Phi^{-(1+2\omega)}$.  They can be derived
from the action (\ref{action_reduced}) which now reduces to the form
\be
\int {\cal A}d^Nx \int d^4x \sqrt{\left|g\right|}\frac{c^4\Phi}{16\pi}\left\{ 
	R \mp \left[2\frac{\Box\Phi}{\Phi}+\omega \frac{\grad_\gamma\Phi
	\grad^\gamma\Phi}{\Phi^2} +\frac{2U}{\up{D}G\Phi}\right]\right\}.
\ee

Of course, when one transforms to a the Einstein frame, one sees that such reductions lead to an exponential potential.  Using the transformation equations (), one obtains the field equations
\beqn
\bar R_{\alpha\beta}& = & \pm \frac{8\pi G}{c^4} \left\{\frac{1}{2}
	\grad_\alpha\varphi \grad_\beta\varphi +V\right\}, \\
\bar \Box\varphi & = & \pm \frac{dV}{d\varphi},
\eeqn
and the action
\be
\int {\cal A}d^Nx \int d^4x \sqrt{\left|g\right|}\left\{ \frac{c^4\bar R}
	{16\pi G} \mp \left[\frac{1}{2}\grad_\gamma\varphi
	\grad^\gamma\varphi +V\right]\right\},
\ee
where
\be
V = V_0 \exp\left\{ \frac{\sqrt{8\pi G}}{c^2}k\varphi\right\}
\ee
and $k\equiv \mp 2\sqrt{\omega+3/2}$ (note that $2\leq k^2 \leq 8$
since $-1\leq \omega \leq 0$ for $N\in[1,\infty]$).

\begin{theindex}

  \item action, \bb{8}
    \subitem scalar field, 8
    \subitem scalar-tensor, 9
    \subitem string, 11

  \indexspace

  \item critical density parameter ($\Omega$), 2, \bb{6}

  \indexspace

  \item deceleration parameter ($q$), 1, \bb{7}

  \indexspace

  \item Einstein's field equations (EFE), \bb{4}
  \item equilibrium sets
    \subitem $C^\pm$, 66, 87
    \subitem $J^\pm$, 116, 126
    \subitem $L^\pm$, $L^{\pm}_{(\pm)}$, 65, 87
    \subitem $L_1$, 67, 88
    \subitem $N$, 92, 112
    \subitem $P(I)$, 41
    \subitem $R/A$, 92, 112
    \subitem $S^\pm$, 66, 87
    \subitem $S^\pm_1$, 91, 111
    \subitem $T$, 129, 137
    \subitem ${\cal A}_S(II)$, 43
    \subitem ${\cal A}_S(VI_h)$, 43
    \subitem ${\cal F}_S(I)$, ${\cal F}_S$, 30, 43
    \subitem ${\cal K}_{\cal M}$, ${\cal K}$, ${\cal K}^\pm$, 41, 42, 
		53
    \subitem ${\cal L}^\pm_l$, 42
    \subitem ${\cal N}$, 55
    \subitem ${\cal N}_1$, 59
    \subitem ${\cal N}_2$, 56
    \subitem P$^\pm$(II), 41
    \subitem P$^\pm_{\cal S}(VI_h)$, 40
    \subitem P$^{\pm}_{\cal S}(II)$, 40
    \subitem P$^{\pm}_{\cal S}(VII_h)$, 40
    \subitem P$_{\cal S}(I)$, 40, 53
    \subitem P(VI$_h$)), 41
  \item exact solutions
    \subitem 10D isotropic, frozen axion, 66
    \subitem curved dilaton--moduli--vacuum, 129, 137
    \subitem de Sitter, 116, 126
    \subitem dilaton--axion, 91, 111
    \subitem dilaton--cosmological constant, 92, 112
    \subitem dilaton--moduli--axion, 65
    \subitem dilaton--moduli--vacuum, 65, 87
      \subsubitem dilaton--vacuum, 65
    \subitem linear dilaton--vacuum, 66, 87
      \subsubitem generalized linear dilaton--vacuum, 67, 88
    \subitem Milne, 66, 87
    \subitem negative--curvature driven, 92, 112
    \subitem rolling radii, 66

  \indexspace

  \item flatness problem, 2
  \item Friedmann--Robertson--Walker (FRW), \bb{6}

  \indexspace

  \item horizon problem, 2

  \indexspace

  \item inflation paradigm, 3--4

  \indexspace

  \item spatial homogeneity, 5
    \subitem Bianchi models, \bb{5}
    \subitem cosmological issue, 2
  \item spatial isotropy, \bb{5}
    \subitem isotropy issue, 2

\end{theindex}

\end{document}